\documentclass[iop,revtex4]{emulateapj}
\usepackage{txfonts}
\usepackage{graphicx}
\usepackage{lscape}
\usepackage{longtable}

\usepackage{natbib}
\usepackage{url}
\usepackage{fancyref}

\usepackage[hidelinks,hyperfootnotes=true,naturalnames=true,pdfstartview=FitH,pdfpagemode=UseNone,citecolor=citeRGB,colorlinks=false,pdfborder={0 0 0}]{hyperref}

\newcommand{\kms}{${\rm km \; s^{-1}}$}

\shorttitle{A Spectroscopic Survey of the Fields of 28 Strong Gravitational Lenses}
\shortauthors{Momcheva et al.}

\begin{document} 

\submitted{Accepted for publication in ApJS.}
\title{A Spectroscopic Survey of the Fields of 28 Strong Gravitational Lenses:\\ The Redshift Catalog \footnotemark[1]}

\footnotetext[1]{This paper includes data gathered with the 6.5 meter Magellan Telescopes located at Las Campanas Observatory, Chile and the 6.5 meter MMT located in Arizona.}
\author{Ivelina G. Momcheva\altaffilmark{2,3,4}, Kurtis A. Williams\altaffilmark{5}, Richard J. Cool\altaffilmark{4}, Charles R. Keeton\altaffilmark{6}, Ann I. Zabludoff\altaffilmark{4}}
\altaffiltext{2}{Yale University, 260 Whitney Ave., New Haven, CT 06511}
\altaffiltext{3}{Observatories, Carnegie Institution of Washington, Pasadena, CA 91101}
\altaffiltext{4}{Steward Observatory, University of Arizona, Tucson, AZ 85721}
\altaffiltext{5}{Physics and Astronomy Department, Texas A\&M University, Commerce, Texas, 75429}
\altaffiltext{6}{Department of Physics and Astronomy, Rutgers University, Piscataway, NJ 08854}
\email{ivelina.momcheva@yale.edu}

\begin{abstract}
We present the spectroscopic redshift catalog from a wide-field survey of the fields of 28 galaxy-mass strong gravitational lenses. We discuss the acquisition and reduction of the survey data, collected over 40 nights of 6.5m MMT and Magellan time, employing four different multi-object spectrographs. We determine that no biases are introduced by combining datasets obtained with different instrument/spectrograph combinations. Special care is taken to determine  redshift uncertainties using repeat observations. The redshift catalog consists of 9768 new and unique galaxy redshifts. 82.4\% of the catalog redshifts are between $z=0.1$ and $z=0.7$, and the catalog median redshift is $z_{med}=0.36$. The data from this survey will be used to study the lens environments and line-of-sight structures to gain a better understanding of the effects of large scale structure on lens statistics and lens-derived parameters.  
\end{abstract}
\keywords{catalogs---galaxies: distances and redshifts---cosmology: distance scale---cosmology: gravitational lensing---galaxies:individual:(ER 0047-2808, CTQ 0414 , HE0435-1223,  CLASS B0712+472, MG J0751+2716, FBQS J0951+2635, BRI0952-0115, CTS J03.13, HE1104-1805, PG 1115+080, MG J1131+0456, RXS J1131-1231, CLASS B1152+199, HST J12531-2914, LBQS 1333+0133, CTQ 0327, HST J14113+5211, [HB89] 1413+117, CLASS B1422+231, SBS 1520+530, MG J1549+3047, CLASS B1600+434, CLASS B1608+656, MG J1654+1346, PMN J2004-1349, WFI J2033-4723, CLASS B2114+022, HE2149-2745)}

\section{Introduction}

\begin{deluxetable*}{llllllccccc}
\tablecolumns{11}
\tabletypesize{\footnotesize}
\tablecaption{Gravitational Lens Galaxies \label{lenssample}}
\tablehead{
\colhead{Lens} & \colhead{RA} & \colhead{Dec} & \colhead{$z_l$} & \colhead{F814W} & \colhead{$z_s $} & \colhead{$\Delta t$\tablenotemark{a,b}} &
\colhead{Images\tablenotemark{c}} &\colhead{$N_{\rm grp}$\tablenotemark{a}} & \colhead{$\sigma_r$} & \colhead{$kT_{X}$\tablenotemark{a}} \\
& \colhead{(J2000)} & \colhead{(J2000)} &\colhead{} & \colhead{[mag]} &\colhead{} &
\colhead{[days]} &\colhead{} & \colhead{}  & \colhead{[\kms]} &\colhead{[KeV]}}
\startdata
Q 0047$-$2808  	& 00:49:41.89 & 	$-$27:52:25.7  	& 0.485  &  20.05 & 3.595  		& \nodata 	& 4ER &  \nodata 		& \nodata  & \nodata  \\
Q J0158$-$4325  	& 01:58:41.44 &	$-$43:25:04.2  	& 0.317  & 18.97  & 1.290  		& \nodata 			& 2      & \nodata		& \nodata  &  \nodata \\
HE 0435$-$1223  	& 04:38:14.90  & 	$-$12:17:14.4  	& 0.455  & 18.05\tablenotemark{e} 	& 1.689 	& 7.8$\pm$0.8  	& 4      &   	\nodata	& \nodata  &  \nodata \\
B0712+472       		& 07:16:03.58 &	$+$47:08:50.0   & 0.406  & 19.56  & 1.34    	& \nodata 			& 4      & \nodata 		& \nodata  &  \nodata \\
MG 0751+2716 	   	& 07:51:41.46 & 	$+$27:16:31.4  & 0.349   & 21.26  & 3.20    	& \nodata			& R      &  13  & $320^{+170}_{-110}$ & \nodata \\
FBQ 0951+2635   	& 09:51:22.57 &	$+$26:35:14.1   & 0.260  & 19.67  & 1.24    & 16$\pm$2			& 2      & \nodata 		& \nodata  &  \nodata \\
BRI 0952$-$0115 	& 09:55:00.01 & 	$-$01:30:05.0  	& 0.632  & 21.21  & 4.50    &  \nodata 			& 2      & 5 & $170^{+150}_{-100}$ & \nodata 		\\
Q 1017$-$207    	& 10:17:24.13 &	$-$20:47:00.4  	&(0.78)\tablenotemark{d}   & 21.82  & 2.545   & \nodata			& 2      & \nodata 		& \nodata  &  \nodata \\
HE 1104$-$1805  	& 11:06:33.45 &	$-$18:21:24.2  	& 0.729  & 20.01  & 2.319  & 161$\pm$7				& 2      & \nodata 		& \nodata  &  \nodata \\
PG 1115+080     	& 11:18:17.00 & 	$+$07:45:57.7   & 0.310  & 18.92  & 1.722  & 25.0$^{+3.3}_{-3.38}$	& 4      & 13  & $440^{+90}_{-80}$ & 0.8$\pm$0.2 \\
RX J1131$-$1231 	& 11:31:51.60 &	$-$12:31:57.0   	& 0.295  & 17.88  & 0.658  & 87$\pm$8  		& 4      &  	\nodata	& \nodata  & \nodata  \\
MG 1131+0456    	& 11:31:56.48 &	$+$04:55:49.8   & 0.844  & 21.21  & (2.0) \tablenotemark{d}   & \nodata 				& 2R    & \nodata 		& \nodata  & \nodata  \\
B1152+200       		& 11:55:18.30 &	$+$19:39:42.2   & 0.439  & 19.26  & 1.019  &  \nodata 				& 2      & \nodata 		& \nodata  & \nodata  \\
HST12531$-$2914 	& 12:53:06.70 &	$-$29:14:30.0  	&(0.69)\tablenotemark{d}   & 21.83  &  \nodata         &  \nodata 				& 4      & \nodata 		& \nodata  & \nodata  \\
LBQ 1333+0113   	& 13:35:34.79 &	$+$01:18:05.5  	& 0.440  & 20.05  & 1.570  &  \nodata 				& 2      & \nodata 		& \nodata  & \nodata  \\
Q 1355$-$2257   	& 13:55:43.38 &	 $-$22:57:22.9   & 0.702  & 19.04  & 1.370  &  \nodata 				& 2      & \nodata 		& \nodata  & \nodata  \\
HST14113+5211   	& 14:11:19.60 &	$+$52:11:29.0   & 0.465  & 19.99  & 2.811  &  \nodata 				& 4      & \nodata 		& \nodata  & \nodata  \\
H1413+117       	& 14:15:46.40 &	$+$11:29:41.4   & (0.9)\tablenotemark{d}    & 18.61  & 2.550  & 23$\pm$4 				& 4      & \nodata 		& \nodata  & \nodata  \\
B1422+231       		& 14:24:38.09 & 	$+$22:56:00.6   & 0.338  & 19.66  & 3.620  & 8.2$\pm$2.0 	& 4      & $470^{+100}_{-90}$ & 1.0$^{+\inf}_{-0.3}$ & 16  \\
SBS1520+530     	& 15:21:44.83 &	$+$52:54:48.6   & 0.710  & 20.16  & 1.855  & 125.8$\pm$2.1 	& 2      & \nodata 		& \nodata  & \nodata  \\
MG 1549+3047    	& 15:49:12.37 &	$+$30:47:16.6   & 0.111  & 16.70  & 1.170  &  \nodata 	& R      & \nodata 		& \nodata  & \nodata  \\
B1600+434       		& 16:01:40.45 &	$+$43:16:47.8   & 0.414  & 20.78  & 1.589  & 51.0$\pm$4.0   	& 2      & \nodata 		& \nodata  & \nodata  \\
B1608+656       		& 16:09:13.96 &	$+$65:32:29.0   & 0.630  & 19.02  & 1.394  & $77.0^{+2}_{-1}$				& 4      & \nodata 		& \nodata  & \nodata  \\
MG 1654+1346    	& 16:54:41.83 & 	$+$13:46:22.0   & 0.254  & 17.90    & 1.740  &  \nodata 			& R      & 7  & $200^{+120}_{-80}$ & \nodata	 	 \\
PMN J2004$-$1349	& 20:04:07.07 & 	$-$13:49:30.7  	&  \nodata         &  \nodata         &  \nodata       &  \nodata 				& 2      & \nodata 		& \nodata  & \nodata  \\
WFI 2033$-$4723 	& 20:33:42.08 &	$-$47:23:43.0    & 0.661  & 19.71\tablenotemark{e}  & 1.660 & $62.6^{+4.1}_{- 2.3}$ & 4  & \nodata 		& \nodata  & \nodata  \\\
B2114+022       		& 21:16:50.75 & 	$+$02:25:46.9   & 0.316  & 18.63   &  \nodata        &  \nodata				& 2+2  & 5  & $110^{+170}_{-80}$  & \nodata 		 \\
                		&             	      &                		& 0.588  &            &          & 				&        &  		&    &   \\
HE 2149$-$2745 	& 21:52:07.44 & 	$-$27:31:50.2  	& 0.495  & 19.56   & 2.033 & 103.0$\pm$12.0	& 2       & \nodata 		& \nodata  & \nodata  \\
\enddata
\tablecomments{Data from the CASTLES website, unless otherwise noted (\url{http://cfa-www.harvard.edu/castles/}). The columns are as follows: full lens ID, right ascension (J2000), declination (J2000), spectroscopic redshift of the lens galaxy, $I$-band magnitude of the lens galaxy,  spectroscopic redshift of the source, time delay, image configuration, number of spectroscopically identified group members associated with the lens galaxy including the lens, velocity dispersion of the group, temperature of diffuse X-ray emission of the group. }
\tablenotetext{a}{References in Appendix.}
\tablenotetext{b}{If multiple time delays are measured, the longest one is listed.}
\tablenotetext{c}{R: Einstein ring; E: extended; 2+2: two pairs of images of two different sources.}
\tablenotetext{d}{Photometric redshifts.}
\tablenotetext{e}{Magnitudes are SDSS $i'$.}
\end{deluxetable*}

Strong gravitational lenses hold the promise to constrain cosmological parameters such as the Hubble constant $H_0$ \citep[e.g.,][]{refsdal, kochanek04} and the dark energy density $\Omega_{\Lambda}$ \citep{chae2, mitchell}, to determine the properties and evolution of dark matter halos \citep[e.g., ][]{koopmans09, barnabe}, and to uncover substructure in those halos \citep[e.g.,][]{mao, metcalf, dalal}. Inspite of ever-increasing sample sizes and improved observational data, systematic issues continue to plague current lens analyses. One such remaining source of systematics is the influence of the large scale environments in which lenses reside \citep[e.g.,][]{kz2004}. This point is borne out in the detailed analysis of the lens B1608+656 by \citet{suyu10}; the authors find that the largest sources of uncertainty on the derived cosmological parameters are the mass distributions at the lens and along the line of sight \citep[also see ][]{ken}.

Both theoretical and observational work suggest that galaxy-scale gravitational lenses lie in complex environments. Statistical arguments imply that at least $25\%$ of lens galaxies lie in groups or clusters \citep{kcz2000}. Spectroscopic observations have confirmed several groups \citep[MG0751+2716, PG1115+080, B1422+231 and B1600+434,][]{kundic97, kundic97b, tonry1999, fassnacht06, my, auger} and clusters \citep[RXJ0911+0551, Q0957+561, HST14113+5211, and MG2016+112,][]{knieb00, young, fischer, soucail} around lens galaxies. The analysis of \citet{my} showed that $\sim50\%$ of lenses in their sample are in groups and that a wide range of environments ---small groups to massive clusters --- host lenses. Furthermore, indirect, yet unconfirmed, evidence for the existence of complex environments comes from the large tidal shears required to explain the image configurations in many four-image lenses \citep[e.g., ][also see the Appendix]{keeton97, lehar1997, morgan, morgan2006, witt}. 

Gravitational lensing is produced by the integral of the mass between the source and the observer and, therefore, foreground and background structures may also influence the lensing potential. Observations show that line-of-sight structures are common \citep[B0712+472, MG1131+0456, B1608+656, MG0751+2716,][]{fassnacht2002, tonry1131, fassnacht06, my}, and although the requirement that the structure is located at a small impact parameter, is fairly close to the lens in redshift, and/or has significant mass would suggest that only some of these structures should have a major impact on the lens, a detailed census of line-of-sight structures is needed to understand their frequency and significance.

We have carried out a spectroscopic redshift survey to map the environments and line-of-sight structures for 28 lenses. Here we present the data acquisition and resulting galaxy redshifts from this survey; subsequent papers will discuss in detail the lens environments, lines of sight, and the implications for lensing studies. \citet{ken} have already explored the effects of the local environments and line-of-sight structures on the lensing potential. 

Apart from lensing, this survey will produce a large sample of spectroscopically confirmed groups at intermediate redshifts that can be used for studies of galaxy evolution in group environments. Group catalogs based on this survey  are presented in \citet{my_thesis} and Wilson et al. (in prep.).  

In this paper we describe the acquisition and reduction of data for the spectroscopic survey, which spanned over 40 nights of 6.5-meter telescope time and incorporates redshift data from four different spectrographs.  In Section 2 we outline the sample of lens fields surveyed. In Section 3 we present the target selection, observations, and data reduction. In Section 4 we present the master galaxy redshift catalog, and discuss the errors and the completeness of the redshift catalogs for each field. Throughout this paper we adopt the WMAP year seven cosmological parameters \citep{jarosik} based on the combined constraints from WMAP, $H_{0}$, and baryonic acoustic oscillations: $H_0$=70.4 \kms, $\Omega_m = 0.272$, and $\Omega_l = 0.728$.

\section{The Sample}

The goal of this survey is to study the local environments of and lines of sight to strong gravitational lens galaxies, and to quantify the effects of any structures on the lensing potential.  We selected our sample of galaxy-mass lenses from the CASTLES\footnote{\url{http://www.cfa.harvard.edu/castles/}}, a survey that obtained \textit{Hubble Space Telescope} (\textit{HST}) images in the H, I and V bands of galaxy-scale lenses and binary quasars with the NICMOS and WFPC2 cameras (GO: 7495, 7887; PI: Falco). The goals of CASTLES are to obtain photometric redshifts for lens galaxies, to measure their $M/L$ ratios and compare the distribution of dark matter and stellar light, to probe the ISM of lens galaxies, and to identify simple lens systems for measuring  the Hubble constant. Including archival observations and \textit{HST} follow-up by other groups, the current CASTLES sample consists of 100 lenses and 18 binary quasars. 

In order to identify targets for spectroscopy, we obtained wide-field two-band imaging for 69 CASTLES fields between 2002 and 2005.   The general observing and reduction strategy are given in Williams et al. (2006), which also presented initial photometric results. A subset of 28 lens fields was selected from the 69 imaged fields for follow-up spectroscopy; an upcoming paper will discuss the photometry of these 28 fields.  The primary criteria for spectroscopic follow-up were: that the lens galaxy had a known redshift $z\leq 0.83$, approximately where the 4000\AA~ break leaves the R band, thereby removing our ability to estimate red sequence redshifts photometrically; and that the field had already been imaged and the photometry reduced prior to the spectroscopic run, as our imaging and spectroscopic runs were interleaved.  In early observing runs, this latter criterion introduces bias toward lens fields we considered ``interesting": those with evidence of complex environments (i.e., large shear), or with poor lens models, as well as a bias against lens environments that had been well characterized by previous spectroscopic studies.  

As our project progressed, additional lens fields were added to create larger samples of 2- and 4-image lenses, lenses with and without observed time delays, and lenses with previous redshift surveys near the lens but little or no spectroscopy $\gtrsim 3\arcmin$\ from the lens. We did retain the redshift criterion $z_{\rm lens}\leq 0.9$ for every field except PMN 2004 (and, as discussed below, this field did not produce a useful sample of redshifts).  We therefore conclude that while our lens sample is not unbiased, it covers the range of fundamental lens parameters.

Table \ref{lenssample} presents the 28 lens systems. References and detailed information on each system are given in the Appendix. The redshifts of the lens galaxies range from $0.1$ to $0.9$. Time delays have been measured for twelve of the lenses (HE0435, FBQ0951, HE1104, PG1115, RXJ1131, H1413, B1422, SBS1520, B1600, B1608, WFI2033 and HE2149, see the Appendix for references). 

Of the 28 lenses, 12 are two-image systems, ten are four-image systems, and three have Einstein rings. Three systems have more complicated image morphologies: four extended images with an Einstein ring (Q0047), two images and a ring (MG1131), and a system with possibly two pairs of images of two different sources (B2114). Most lens galaxies have early type morphologies, with the following exceptions: B1600 and PMN2004 are lensed by spiral galaxies; the B1152 lens galaxy has a late-type spectrum; the MG1549 and WFI2033 lenses have S0 morphology (MG1549 is barred, an SB0); the B1608 lens is a pair of interacting early-type galaxies; and the H1413 lens morphology is unknown. 

Previous studies have shown that seven of the lens galaxies in this sample are in groups (MG0751, PG1115, B1422, B1600, B1608, MG1654 and B2114), and one is in a cluster (HST14113). Photometric observations and lens models that require large shears have led to suggestions that another ten lens galaxies may be in groups (Q0158, HE0435, FBQ0951, HE1104, MG1131, RXJ1131, HST12531, H1413, SBS1520 and WFI2033). Line-of-sight structures have been found in the fields of B0712, MG0751, MG1131, and HE2149. Photometric observations have suggested such structures in FBQ0951 and RXJ1131.

\section{The Data}
\subsection{Spectroscopic Target Selection}

We obtained deep I and either V or R images of each field between May 2002 and January 2006 using the $36\arcmin\times36\arcmin$\ MOSAIC imagers on the 4-meter Cerro-Tololo Inter-American Observatory (CTIO) Blanco telescope for the southern fields and the 4-m Kitt Peak National Observatory (KPNO) Mayall telescope for the northern fields. These images were reduced using standard IRAF\footnote{IRAF is distributed by the National Optical Astronomy
Observatories, which are operated by the Association of Universities or Research in Astronomy, Inc., under cooperative agreement with the National Science Foundation.} and SExtractor \citep{sex} routines. For a more detailed description of the photometric reduction and analysis, see \citet{kurtis, kurtis2}. 

We determined fiber and slitlet placement based on a prioritization scheme that considered observational constraints, projected distances of galaxies from the lens galaxy, and galaxy colors.  For runs at Magellan, our observational constraints considered the instrumental setup, including the desired wavelength range, the orientation of the slit mask, and the total time a target was observable.  Slit mask orientations were typically either near the expected parallactic angle or at a fixed east-west slit orientation if an atmospheric dispersion corrector was in operation (after May 2004 for Magellan/IMACS).  

As the significance of a mass concentration for a lens model depends on the impact parameter, we prioritized slit mask centering and slitlet placement around the lens system.   In early observing runs, slit masks tended to be centered on the lens system, while in later observing runs, more effort was made to tile a larger region around the lens, although always with the lens galaxy in the field so the region within a few arcminutes of the lens would be more completely covered than at larger radii.  Within a field, galaxies within 3\arcmin\ of the lens always had highest priority, followed by galaxies situated 3\arcmin - 5\arcmin\ from the lens, and then by more distant galaxies.  The limiting magnitude for spectroscopic follow-up was $I=21.5$ in all observing runs prior to Spring 2006 and $I=20.5$ for the subsequent runs.

\begin{figure}
\figurenum{1}
\includegraphics[angle=270,width=0.49\textwidth]{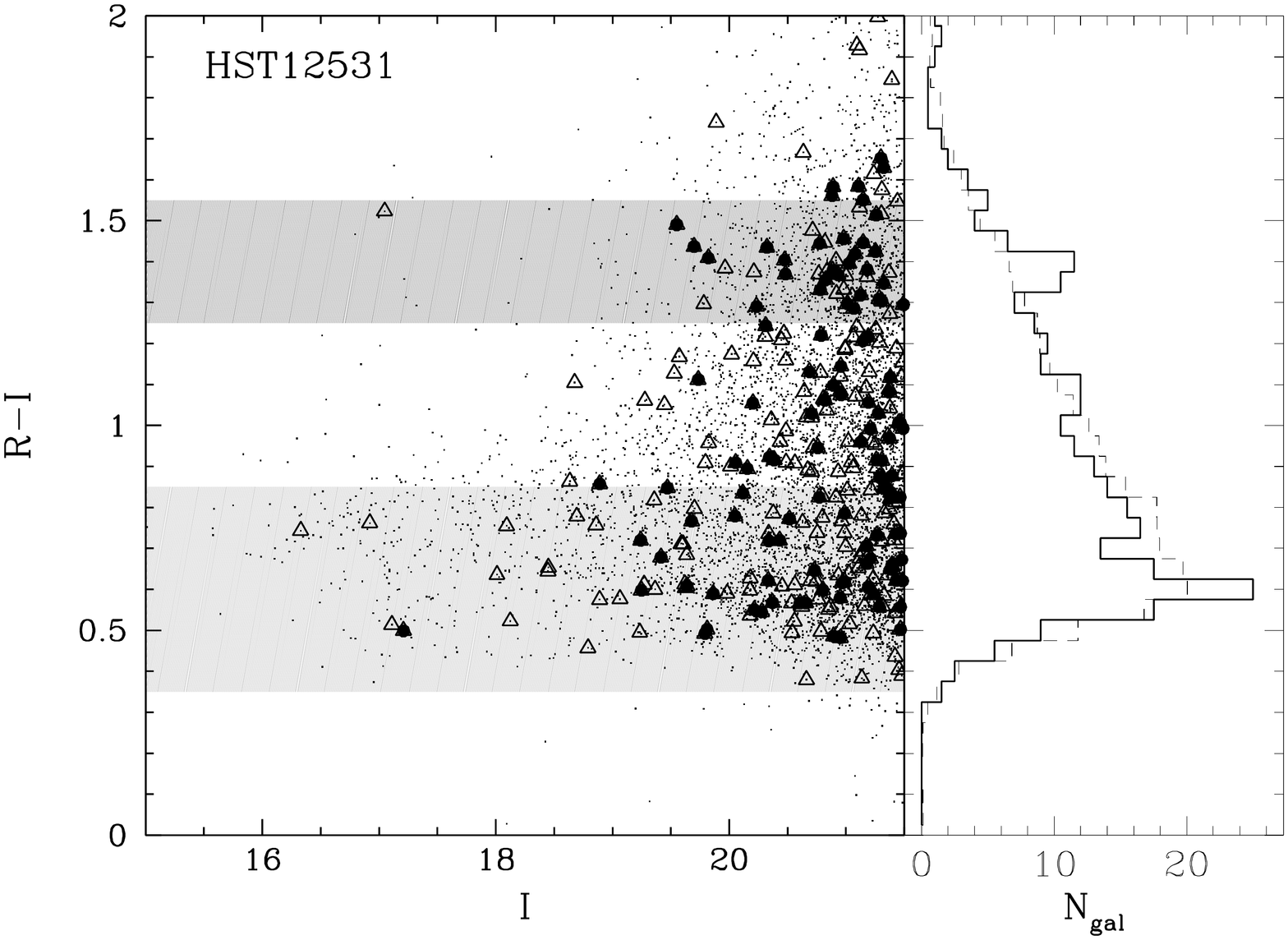}\vspace{-5mm}
\includegraphics[angle=270,width=0.49\textwidth]{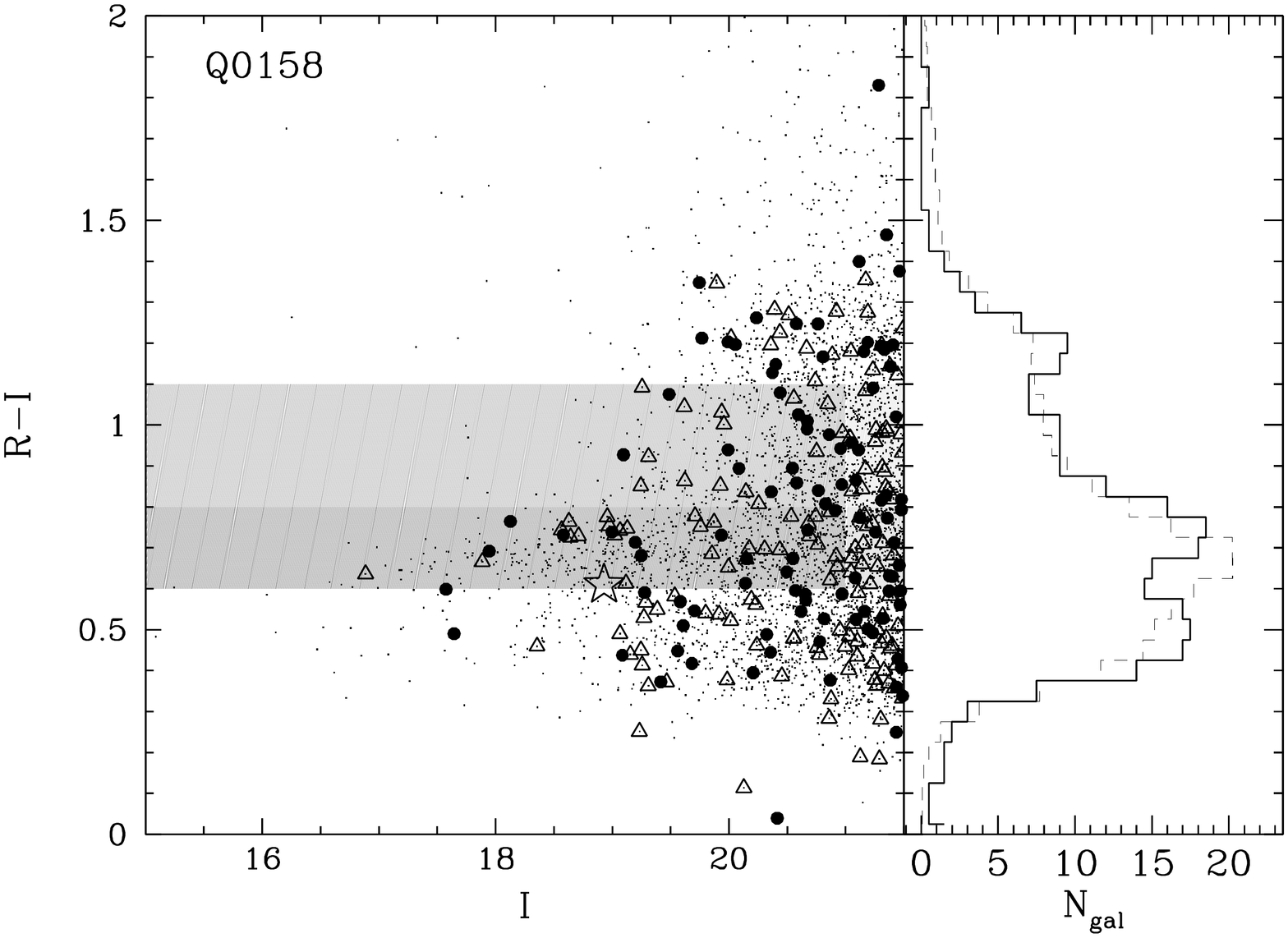}
\caption{\footnotesize Color-magnitude diagrams (left) and histogram of galaxy colors (right) for the lens fields HST12531 (top) and Q0158 (bottom).  Small points indicate the position of all galaxies in the photometric catalog, filled triangles are galaxies located between 3\arcmin\ and 5\arcmin\ of the lens galaxy, filled circles are galaxies located within 3\arcmin\ of the lens galaxy. The Q0158 lens is indicated with a star; the HST12531 lens is too faint to appear on the plot ($21.52$). Shaded regions indicate prioritized objects, with darker shading implying higher priority. Slitlet placement for galaxies within 5\arcmin\ of the lens galaxy received the highest priority (darkest shaded regions).  The histograms show the color distribution of galaxies within 5\arcmin\ of the lens (solid line) and for the entire field (dashed line, renormalized to the solid histogram).   From these we see that slitlet priorities cover broad sections of color space, not just regions around apparent red sequences.\label{fig.prioritization} }
\end{figure}

Prioritization also depended on galaxy colors. Evolved single-burs models with formation redshift $z=5$ were used to determine the colors of galaxies as a function of redshift \citep{kurtis}. In cases where a red sequence was obvious with a color close to that expected for a single-burst population at the lens redshift, galaxies with colors near the red sequence were prioritized. These color selection ranges were typically broad, including red sequence and blue cloud galaxies at the lens redshift. Second priority was given to regions in color space near any other suspected red sequences; this weaker priority would also be assigned to colors near an expected red sequence at the lens redshift if no red sequence was detected around the lens. Lowest priorities were given to galaxies of other colors. For spectroscopic runs which occurred prior to the finalization of the red-sequence finding algorithm discussed in \citet{kurtis},  galaxies along poorly-populated red sequences were not prioritized.
Two examples of this prioritization scheme used in single observing runs are shown in Figure \ref{fig.prioritization}.  The figure shows color-magnitude diagrams (left panels) and color histograms (right panels) for the galaxies in the imaged regions around HST12531 (top) and Q0158 (bottom). 

In the field of HST 12531, two potential sequences are visible in the histogram; one near $R-I\approx 0.6$ and a second near $R-I\approx 1.4$. The lens galaxy is in the latter peak but is too faint to appear in the plot ($I=21.52$, $R-I=1.35\pm0.04$). Therefore, galaxies with $1.25 \leq R-I \leq 1.55$ received highest priority for slitlet assignment, and those with $0.35\leq R-I\leq 0.85$ were assigned second priority; this latter region was broadened to include several bright galaxies between 3\arcmin\ and 5\arcmin\ from the lens.  As can be seen in Figure 11m, these prioritized regions correspond to a rich structure at $z\approx 0.05$ (bluer region) and a pair of structures near $z\approx 0.66$ (redder region).  

The field around Q0158 presents a more problematic case.  For this field, the first run of spectroscopic observations began before the photometric reduction pipeline was finalized.  The panels in Figure \ref{fig.prioritization} show final photometry, including evidence of color concentrations near $R-I\approx 0.5$ and $R-I\approx 1.2$.  However, different regions of color space were prioritized in the slit masks (shaded regions). This was due to quality control issues in the initial photometric reduction, which also led to a low redshift success rate (Table 5%\ref{tab:data}
). Later runs covered a broader color space. As can be seen in Figure 11a, the resulting galaxy redshift distribution is clearly incomplete, but galaxies spanning a wide range of redshifts were still observed successfully.

From these examples, we can see that slit and fiber placement priorities are highly variable in terms of color.  The breadth of the prioritized colors purposefully covers large fractions of color space, in order to include star-forming galaxies in groups, although we gain an added benefit of reducing the bias against finding groups with weak/no red sequences at a wide range of redshifts.  The issue is further complicated by each instrument's slit mask or fiber configuration software, which weighs the assigned priorities differently and the observing sequence for each field. We examine the final observed completeness as a function of color, magnitude and projected distance from the lens in \S\ 4.2.

\subsection{Observations}

\begin{deluxetable}{llcc}
\tablecolumns{4}
\tablewidth{0pc}
\tabletypesize{\footnotesize}
\tablecaption{Observing Runs \label{data1}}
\tablehead{
\colhead{Date}  & \colhead{Instrument}  & \colhead{\# of Nights}    & \colhead{\# of Masks or Configs}}
\startdata
March 2003      & LDSS-2                 & 4                         & 20                     \\   
August 2003     & LDSS-2                 & 4                         & 18                     \\ 
March 2004      & IMACS                 & 4                         & 21                  \\    
November 2004   & IMACS                 & 4                         & 8                 \\    
March 2005      & IMACS                 & 4                        & 5                     \\    
April 2005      & IMACS                 & 4 (partial)               & 16                 \\   
September 2005  & LDSS-3                 & 4                         & 7                  \\   
February 2006   & LDSS-3                 & 2                         & 10                 \\   
August 2006     & LDSS-3                 & 2                         & 9                   \\   
September 2006  & LDSS-3                 & 2                         & 9                   \\ 
\hline
Spring 2004     & Hectospec             & 4                         & 11                     \\
Fall 2004       & Hectospec             & 4                         & 5                        \\ 
Spring 2005     & Hectospec             & 4                         & 6                      \\ 
Summer 2005     & Hectospec             & 2                         & 4                    \\
Spring 2006     & Hectospec             & 2                         & 4                      \\
\enddata
\end{deluxetable}

We carried out the follow-up spectroscopy using the Magellan 6.5-meter telescopes for the southern targets and the MMT 6.5-meter telescope for the northern targets. B2114 was observed with both Magellan and the MMT. We used multi-object spectrographs that have a field of view $\geq 5\arcmin$\ to permit uniform coverage out to at least 0.5 Mpc at the lens redshift, i.e., a group virial radius, in every pointing. On Magellan, we employed the LDSS-2, LDSS-3, and IMACS multi-slit spectrographs. On the MMT, we used the Hectospec multi-fiber spectrograph. The observations were carried out between March 2003 and September 2006. Table \ref{data1} summarizes the observing runs. Table 5
 shows a breakdown of the observations for every field, including the number of masks or fiber configurations, $N_{m}$, used for the field, and the total number of slitlets, $N_{s}$, used or target object fibers, $N_{f}$. The relevant details of the observing setup for each of the four spectrographs are described below.

LDSS-2 (Low Dispersion Survey Spectrograph - 2) is a multi-object spectrograph in use on the Magellan II Clay telescope between 2001 and 2004 \citep{ldss2, ldss2-2}. We used the medium blue, 300 lines/mm grism blazed at 5000 \AA, which provides 5.3 \AA/pixel dispersion, and the medium red, 300 l/mm grism blazed at 8000 \AA,  ~which provides 5.1 \AA/pix dispersion. The field of view of the slit masks was $5\arcmin\times5\arcmin$. The slit widths were 0.9\arcsec, giving a resolution of $\sim$15\AA. Observations were carried out during two observing runs in 2003 March and 2003 August. A total of 38 masks containing 858 slitlets were observed. These data were first presented in \citet{my} but are re-analyzed here.

IMACS (Inamori-Magellan Areal Camera and Spectrograph) is a wide-field camera and spectrograph on the Magellan I Baade telescope \citep{imacs}. Data were obtained during four observing runs: 2004 March, 2004 November, 2005 March and 2005 April. We were among the first users of IMACS after its commissioning and saw gradual improvements in the instrument performance throughout 2004 and 2005. We used the f/2 Short Camera, which has a 27$\arcmin$\ diameter field of view, combined with the 200 lines/mm grating, which provides 2.037 \AA/pixel dispersion and 10 \AA ~resolution. During the 2005 March observing run, the 300 lines/mm grating was used instead, giving a 1.34 \AA/pixel dispersion and a 5 \AA ~resolution. We used slitlet widths of 0.9$\arcsec$. The wavelength coverage is 4000 to 9000 \AA ~for the data from the 2004 March and 2004 November runs and 5000 to 8000 \AA ~for the 2005 March and 2005 April observing runs, when the WB4800 filter was used to remove first order contamination. A total of 50 masks and 10,713 slitlets were observed.

LDSS-3 (Low Dispersion Survey Spectrograph - 3) is an upgraded version of LDSS-2 with larger field of view (8.3$\arcmin$\ diameter) and higher throughput installed on Magellan II Clay in 2005. We used the VPH-Blue grating, which has 1019 lines/mm and provides dispersion of 0.682 \AA/pixel at 5200 \AA, which gives a spectral resolution of 2.7\AA. We used 0.9$\arcsec$ slitlets. We chose this grism to achieve sensitivity in the 4500 to 6500 \AA\ interval where we expected to find the 4000 \AA ~break for targets between redshifts of 0.125 and 0.625. However, a shortcoming of this setup was that the spectra only span 2550 \AA. Data were taken during four observing runs: 2005 September, 2006 February, 2006 August, and 2006 September. A total of 35 masks containing 597 slitlets were observed.

Hectospec is a multi-object optical spectrograph fed by 300 fibers on the MMT \citep{hecto}. Data were obtained in queue scheduling mode during five observing trimesters: 2004 Spring and Fall, 2005 Spring and Summer, and 2006 Spring. We used the 270 lines/mm grating, blazed at 5000 \AA~ resulting in 1.21 \AA/pixel dispersion and spectral coverage of $4000$ to $9770$ \AA. Each fiber subtends 1.5\arcsec, which gives a spectral resolution of 6 \AA. 29 different fiber configurations were observed and a total of 7036 objects were targeted. 

\subsection{Data Reduction}

All data were reduced using standard methods and employed instrument-specific, publicly-available reduction software whenever possible. Here we provide a brief description of the main reduction steps.

The data obtained with LDSS-2 were previously presented in \citet{my}. In this paper, we use the previous IRAF reduction and extraction but redo the flux calibration and redshift determination.

The data obtained with LDSS-3 and IMACS were all reduced using the COSMOS data reduction package \citep{cosmos}. COSMOS relies on a detailed optical model of the spectrograph that allows for an accurate prediction of the positions of the spectral features on the detector. The reduction proceeds in the following steps: (1) alignment of the slit mask relative to the focal plane based on several bright and isolated comparison arc lines; (2) perfection of the alignment using one or more comparison arc images (we chose to fit the offsets along the slit and along the wavelength direction with first and second order polynomials, respectively); (3) image reduction, which includes bias subtraction, flat-fielding, and two-dimensional sky-subtraction \citep{kelson}; (4) extraction of the two-dimensional spectra; and (5) co-addition of the separate images with cosmic ray removal; (6) extraction of 1-dimensional spectra; (7) heliocentric velocity correction; (8) redshift determination.

We extracted the 1-dimensional spectra using a procedure written by A. Marble (personal communication), which implements an optimal extraction method similar to that described by \citet{optimal}. We fit a polynomial of order one (for IMACS) or two (LDSS-3) to the flux-weighted center of the spatial profile along the wavelength axis. The spatial profile is fit by a spline curve and the profile values are weighted by their propagated variance. 

There are two caveats in applying an optimal extraction algorithm to spectra of galaxies. First, optimal extraction is primarily intended for point sources. However, our routine does not assume that the spatial profile is gaussian, and the spline fits to the spatial profiles of the galaxies (which subtend 1\arcsec\ to 2$\arcsec$) are excellent. Second, the optimal extraction algorithm assumes that the spectral signature is constant for all illuminated rows. This is generally true for the continuum light, but may not be true for emission lines, which can have different spatial profiles and, occasionally, resolved rotation curves. The weighting used by the optimal extraction may lead to bulge dominated spectra. For that reason, we also extract a traced but non-weighted spectrum of each object. The optimally extracted spectra, which have slightly higher signal-to-noise, are used for determining redshifts. The non-optimally extracted spectra, paired with the redshifts determined from the optimally-extracted counterparts, will be used for all other applications. 

The Hectospec data are reduced using HSRED\footnote{\url{http://mmto.org/~rcool/hsred/}}, an IDL reduction package based on the SDSS spectroscopic pipeline \citep{sdss, hsred, cool2008}. For every observing night, a bias, a dome-flat, a sky-flat, and a combined arc frame are produced by combining all relevant images taken over the course of the night. The uniformly illuminated dome flats are used to take out the high-frequency flat-field variations and fringing, while the twilight sky flats (when available) provide a correction for the low-frequency fiber-to-fiber variations. The comparison arc lamp spectra are extracted and the centroids of the lines are measured and fit with a fifth order Legendre polynomial. After the bias subtraction, the object and sky fibers in each science exposure are traced with tweaking from the flat-field trace and optimally extracted. The extracted spectra are flat-fielded, and wavelength calibrated with slight tweaking to match the positions of selected skylines. Heliocentric velocity corrections are applied to the wavelength solution. An over-sampled supersky vector is constructed using the sky fibers. For each object fiber, the sky is re-sampled at every pixel and subtracted from the object spectrum. Finally, all exposures done with the same fiber configuration are co-added.

\subsection{Flux Calibrations}

The instrument response must be removed from the object spectra before applying the cross-correlation routine to determine the redshifts. For the purpose of the work presented here, absolute fluxing is not required.

Spectra of spectrophotometric standards were not taken during every night of the LDSS-2 2003 March and September runs. Only one spectrum of the spectrophotometric standard star CD-32 9927 was taken during the two runs (in March 2003). All LDSS-2 data were fluxed using this spectrum. During the IMACS and LDSS-3 observations, we took spectra of at least one spectrophotometric standard star during each observing night. We use the standards taken during the night of the observations to flux the spectra, with the exception of the February 2006 LDSS-3 run. 
%(standard spectra were corrupted and taken with the wrong longslit)
 For this run, we use the sensitivity curves from the August 2006 run instead, when the same instrument setup was used. The standard star spectra are reduced and extracted in the same manner as the spectra of the science objects, corrected for atmospheric extinction (using the standard IRAF CTIO extinction curve), and used to produce a sensitivity function for the night using noao.onedspec.calibrate and noao.onedspec.sensfunc in IRAF. Multiple standard star observations on a given night are averaged to produce a single average response function. The sensitivity function is applied to the spectra to convert them to units  of $f_{\lambda}$. The fluxing is not absolute because we do not correct for slit losses in either the standard or the science spectra.

We did not observe F stars along with our scientific targets \citep{papovich, ages}, nor did the Hectospec operation provide for taking standard star spectra during the night in any other way. We therefore choose to ``pseudoflux" our spectra by applying an average flux vector derived from the AGES observations of F-stars over several observing runs \citep{ages}. No extinction correction is applied. This method is adequate for our purposes as it removes the instrumental signature in the spectra. Again, the flux calibration is relative.

\subsection{Redshift Determination}

For LDSS-3, IMACS, and Hectospec, we determine redshifts using a routine based on the SDSS redshift-finding algorithm \citep{cool2008}. The redshifts for the spectra taken with LDSS-2 were already measured as part of \citet{my}. However, for consistency, here we re-determine all LDSS-2 redshifts in the same manner as for LDSS-3, IMACS and Hectospec. The \citet{cool2008} routine uses $\chi^2$ minimization to compare each object spectrum to a library of galaxy and QSO model templates that are linear combinations of eigenspectra, as well as to a library of stellar spectra. Each template is shifted through a range of redshifts and a $\chi^2$ is calculated. The minimum $\chi^2$ yields the object redshift, a spectral classification from the best fit template, and a flag for the quality of the fit. 

Each redshift was visually inspected to eliminate obvious failures. All IMACS and LDSS-3 sky-subtracted, two-dimensional spectra were also visually inspected alongside the redshift inspection to insure that the spectral features are real. Spectra that did not yield redshifts were flagged and discarded. Spectra for which the redshift was not convincing ($\sim7.5\%$ of all spectra) were flagged and further inspected by a second person. Finally, for some spectra, the routine failed to yield a correct redshift measurement, typically due to misidentification of poorly subtracted sky-lines with emission lines or to misidentification of breaks. In these cases, we examined not only the lowest $\chi^2$, but the lowest ten $\chi^2$ templates. If none of these yielded a correct redshift measurement, the spectrum is also flagged and we do not attempt to obtain correct redshifts for the object. In total, we have obtained 410 redshifts with LDSS-3, 4451 redshifts with IMACS, and  4743 redshifts with Hectospec. Table 5 details the number of redshifts $N_{z}$ in the individual fields.

\begin{deluxetable}{lrccllllll}
\tablecolumns{7}
\tablewidth{0pc}
\tabletypesize{\footnotesize}
\tablecaption{LDSS-2 Redshifts Not Included in the Current Redshift Catalog \label{ldss_zs}}
\tablehead{\colhead{Field} & \colhead{ID} & \colhead{RA} & \colhead{Dec} & \colhead{d} & \colhead{$z$\tablenotemark{b}} & \colhead{$z_{err}$} \\
\\
\colhead{} & \colhead{} & \colhead{[J2000]} & \colhead{[J2000]} & \colhead{[\arcmin]} & \colhead{}  & \colhead{} 
}
\startdata
MG0751 	& 9809   & 07.85839 & +27.24532 &     3.10 &          0.56066  &    	8.0E-05 \\
  		& 8385   & 07.86227 & +27.27772 &     0.61 &        0.37507  &    	6.6E-04 \\
  	 	& 9606   & 07.85897 & +27.21637 &     4.10 &        0.35677  &    	3.2E-04 \\
PG1115    & 11240 & 11.30232   & +07.72276 &     3.37 &       0.31095  &           3.5E-04 \\
B1422 	& 14140 & 14.40848   &+22.92461 &     1.82 &        0.06195    &         2.1E-04  \\
  		& 11479 &  14.41359  &+22.92659 &     2.53 &         0.35796   &          3.5E-04 \\
MG1654 	& 16977 &  16.91175	  &+13.83011 &     3.45 &      0.44661   &          4.7E-04 \\
  		& 19266 &   16.91498 &+13.78022 &     2.98 &        0.12443    &         5.5E-04 \\
PMN2004 & 00001\tablenotemark{a} &   20.06752 & --13.81756 &    1.60 &       0.50824     &     	2.5E-04 \\
		& 00002\tablenotemark{a} &   20.07425 & --13.81511 &    4.38 &        0.23134       &    	4.0E-04 \\
		& 21081 &   20.07356 & --13.80754 &    3.89 &       0.45221     &     	7.6E-05 \\
		& 00003\tablenotemark{a} &   20.06358 & --13.77533 &    5.81 &        0.44997    &      	3.8E-04 \\
B2114 	&  8228  &  21.27930  &+02.40109 &    2.17 &       0.30624    &  	3.8E-04  \\
 		& 10686 &  21.28443  &+02.45915 &    3.73 &       0.45850    &  	5.0E-04 \\
HE2149 	&  8733  &  21.87054  &  --27.52074 &   1.55 &        0.20430     &     	2.6E-04  \\
		& 7534   &  21.86752  &  --27.51397 &   1.38 &        0.46437     &     	7.7E-04 \\
		& 6742   &  21.86554  &  --27.49273 &   3.41 &       0.25014      &      2.5E-04 \\
 		& 7976   &  21.86867  &  --27.52636 &   0.25 &      0.27707      &      7.9E-04 \\
\enddata
\tablenotetext{a}{Objects which are not in our final photometric catalog.}
\tablenotetext{b}{These redshifts were previously published in \citet{my}. We recommend that readers do not use them.}
\end{deluxetable}

The old and new LDSS-2 redshift measurements are consistent with each other but for a small {\bf systematic} offset ($\sim100$ \kms). The new LDSS-2 redshift measurements are more consistent with those from other instruments for the same galaxies.
%Comparison of redshift measurements for the same objects made with LDSS-2 and other instruments show that the new measurements are more consistent with the rest of the data set. 
Thus we use the new redshift measurements throughout this paper. A total of 340 redshifts are added from LDSS-2. For 18 of the LDSS-2 objects, we could not recover the redshift from \citet{my}. The coordinates, photometric properties and original LDSS-2 redshifts for these objects are listed in Table \ref{ldss_zs}. Most were not recovered because the low signal-to-noise and poor quality of the spectra prevented any of the template fits from yielding a believable redshift. Four of these objects are potential group members based on our follow-up group catalogs. However, three of these groups have at least 10 members so the effect of removing the redshifts from the catalog is not significant.  The objects in Table~\ref{ldss_zs} are not included in the current redshift catalog, and we recommend that readers do not use these redshifts.
%However, only four of these objects would have been included as members in groups with at least five members, and, further, only one of these groups has less than 10 members, thus the
%Exclusion of these objects from the redshift catalog does not significantly change the group catalog presented in $\S$ 7.

Our success rate at obtaining redshifts, i.e., the ratio of redshifts obtained to objects targeted, varies with observing run. The overall success rate by instrument is 44\% for LDSS-2, 68\% for LDSS-3, 42\% for IMACS and 67\% for Hectospec. The IMACS success rate is heavily influenced by the first two runs, prior to improvements in the instrument: 34\% for 2004 March and 35\% for 2004 November (the latter was further worsened by poor weather). The IMACS success rate went up to 61\% during our last run in 2005 April. The LDSS-3 success rate shows less run-to-run variation; the best (75\%) was the 2006 August run, which had excellent conditions and 0.5\arcsec~ seeing, and the worst (59\%) was the September 2006 run, which was marked by variable conditions and sub-par seeing ($\ge 1\arcsec$). Hectospec delivered a gradually improving success rate as the instrument and our observing strategies improved, starting with 58\% in 2004 Fall and going up to 79\% for 2005 Summer. The overall success rate for our observations is 52\%; we positioned 19212 slits and fibers and obtained 10044 redshifts, including 276 repeated measurements of the same object ($\S$ \ref{errors}). In the end, we have determined redshifts for a total of 9768 unique objects.

\begin{figure*}[ht]
\figurenum{2}
\label{intra}
\epsscale{1}
\plotone{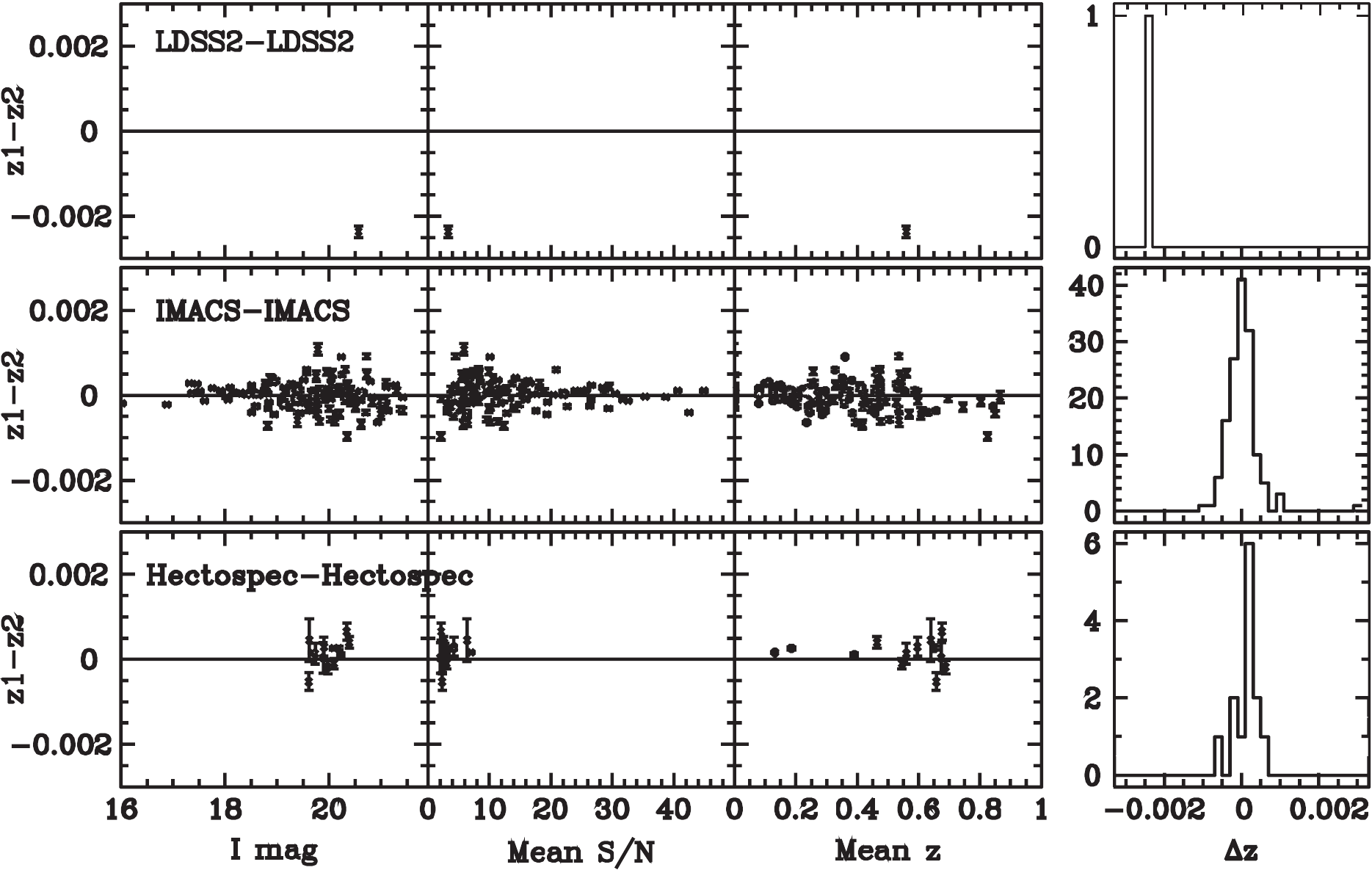}
\caption{\footnotesize Comparison of repeated redshift measurements taken with the same instrument. We show $\Delta z = z_1 - z_2$ as a function of $I$-band magnitude (first column), mean S/N of the spectrum (second column), and mean redshift (third column). We also show the overall $\Delta z$ distribution (rightmost panel). No repeat redshifts were measured with LDSS-3. The IMACS-IMACS and Hectospec-Hectospec distributions above are generally symmetric about zero. We conclude that no systematic errors are being introduced by merging data-sets from different observing runs with the same instrument. Table \ref{comp} lists median $\Delta z$ and $\sigma_{\Delta z}$ for these distributions. }
\end{figure*}

\subsection{Additional Redshifts}

In addition to the redshifts from our observations, we also collect redshifts from the literature. The main goal of adding these objects to the spectroscopic catalogs is to increase the membership of small/undersampled structures. We query the NASA/IPAC Extragalactic Database (NED) for galaxies with spectroscopically measured redshifts within a 20\arcmin\ radius of each field center and match the NED redshift to both our spectroscopic and photometric catalogs. The objects matched to our spectroscopic catalog are used for estimating the external redshift errors (see $\S$ 4). Objects that do not have counterparts in our spectroscopic catalog are added to the spectroscopic catalogs. We do not add objects not present in our photometric catalogs, as these are typically high-redshift and/or low-luminosity galaxies. NED objects without published redshift errors ($\sim10$ objects) are not included either. In total, we add redshifts for 870 objects in 27 fields with a median redshift $z=0.43$. The cross-matching with NED was done in 2013 November and includes any redshifts from SDSS and 2dfGRS in these fields.  

\begin{figure*}[ht]
\figurenum{3}
\label{inter}
\epsscale{1}
\plotone{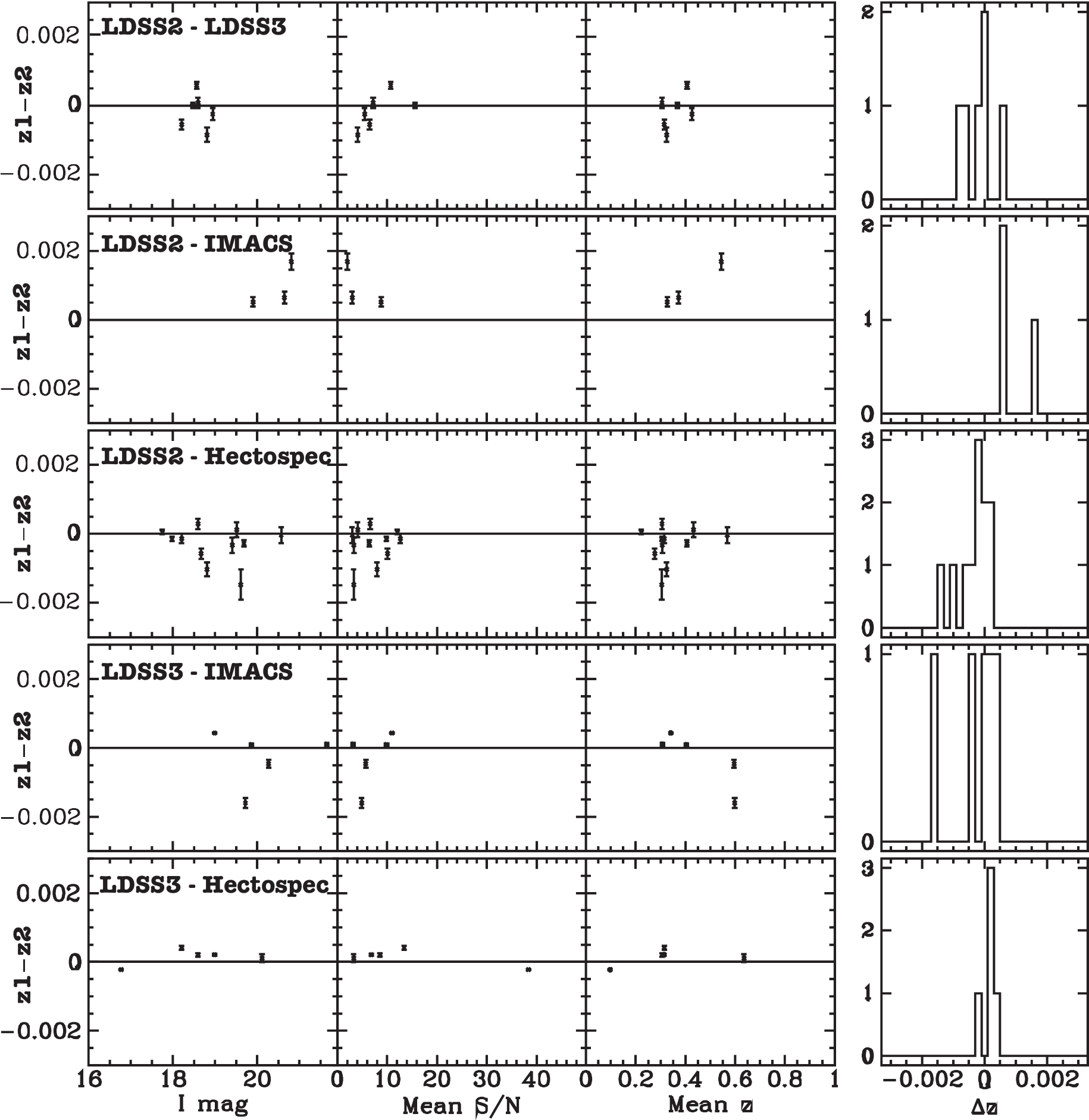}
\caption{\footnotesize Comparison of repeated redshift measurements taken with different instruments. We show $\Delta z = z_1 - z_2$ as a function of $I$-band magnitude (first column), mean S/N of the spectrum (second column), mean redshift (third column), and the overall $\Delta z$ distribution (rightmost column). Table \ref{comp} lists the median $\Delta z$ and the  $\sigma_{\Delta z}$ for these distributions. In all cases, the distributions are consistent with zero within $1\sigma$. We conclude that there are no systematic offsets among the data sets taken with different instruments.}
\end{figure*}

The number of redshifts added to each field is given in Table 5. Notably, we add 240 redshifts to in the field of H12531, which contains a $z=1.237$ supercluster \citep{demarco}. Of these, 113 galaxies are at $z>1$. We also add 133 objects in the field of PG1115, which contains the RXJ1117.4+0743 cluster at $z=0.485$. In the field of HST14113, which includes the 3C 295 cluster at $z=0.46$ \citep{dressler, thimm}, we identify 108 objects with NED redshifts.  However 72 of these objects have redshift errors $\Delta z >0.001$, potentially introducing significant uncertainties in the derived properties of structures in the field. Thus only 36 of these redshifts are added to the catalog. 

The NED redshifts we add are flagged and will be treated differently throughout our analysis;  they are not included in the completeness estimates or used in the assessment of the spectroscopic properties of individual galaxies.

\citet{auger08} measure redshifts for 28 objects between $z=0.71$ and 0.83 in the field of SBS1520. These redshifts are not included in NED, but, due to their importance in characterizing the environment of the lens, we add 26 of the 28 to our catalog. Of the remaining two, one is already in our catalog (ID=10411) and the other is the lens galaxy. The lens galaxy redshift is uncertain and is not added to the catalog (see Appendix). Errors of $\Delta z=0.0004$ are assigned to these redshifts, as suggested by the authors. These redshifts are treated in the  same manner as the NED redshifts.

\begin{figure*}
\figurenum{4}
\label{ned}
\epsscale{1.0}
\plotone{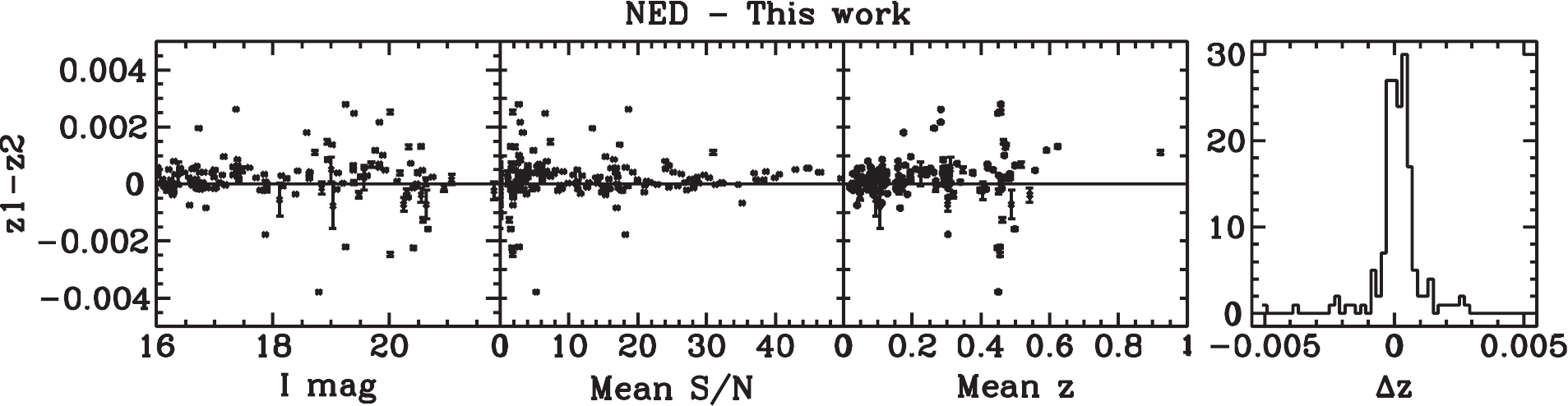}
\caption{\footnotesize Comparison between our redshift measurements and those from NED, as a function of $I$-band magnitude, S/N, and redshift, as well as the overall $\Delta z$ distribution. Table \ref{comp} lists the median $\Delta z$ and $\sigma_{\Delta z}$ of the distribution. Our redshifts are consistent with those in the literature, and there are no systematic offsets. Therefore  the addition of redshifts from NED to our redshift catalogs does not introduce systematic errors.}
\end{figure*}

\subsection{Systematic Redshift Errors\label{errors}}

The data presented in this paper were taken during 15 observing runs over a period of four years. 14 of the fields were observed during more than one run and 13 of the fields were observed with more than one instrument. Despite our efforts to minimize any possible systematics by reducing all data in a uniform manner, some systematics may still be present. In this section we consider three possible sources of systematic errors: (1) systematic offset of the zero point of the redshift determination; (2) systematic offsets between redshifts measured with the same instrument during different observing nights/runs, i.e., the intra-instrument errors; and (3) systematic offsets between redshifts measured with different instruments, i.e., inter-instrument errors. 

Due to errors in the wavelength calibration and/or instrument and telescope flexure that cannot be removed during the reduction, the wavelength calibrations of the spectra may be offset from rest. In addition, the zero point may change between masks, fiber configurations, nights and runs. We estimate the redshift zero points by measuring the velocities of the night sky lines in each spectrum. In \citet{my}, we found that no correction is necessary for the LDSS-2 redshifts based on zeropoints determined from skylines and serendipitous stars. For the rest of the data, we extract the sky spectra in the same manner as the science target spectra. For Hectospec, the distribution of the sky spectra velocities is consistent with zero. For IMACS and LDSS-3, the zero points vary from mask to mask, and we apply a correction equal to the median zero point offset on a per mask basis.

We compare repeated redshift measurements from spectra taken during different observing runs to assess measurement errors. Some objects were observed repeatedly during different observing runs because the first observation did not yield a satisfactory S/N. However for 254 of these targets, we were able to obtain more than one redshift measurement. They were compared after making the zeropoint corrections above. 

The intra-instrument errors are the variations among observations with one instrument on different nights or different runs. Figure \ref{intra} and Table \ref{comp} present the comparison or repeat redshift measurements obtained with the same instrument. The mean differences are consistent with zero. Thus we conclude that no systematic errors will be introduced as a result of combining data from different observing runs.

The {\it rms} dispersion of these repeated redshift measurements allow us to compare our results to those of larger surveys carried with the same instruments and the same instrumental setups. The Arizona CDFS Environmental Survey \citep[ACES, ][]{cooper2012}, carried out with IMACS, finds a dispersion of $\sigma_{z}c\sim75$\kms~ based on 2438 pairs of repeat redshift measurements. Our value is $\sigma_{z}c\sim92$\kms~ ($\sigma_{z}=3.06e-04$)  based on 136 repeat measurements.  The Smithsonian Hectospec Lensing Survey \citep[SHELS, ][]{geller2014} find scatter of $\sigma(\Delta z/(1+z))c = 48$\kms~ for 1651 pairs of absorption-line objects and 24 \kms~ for 238 pairs of emission line objects. The scatter of our sample of 13 objects (both emission- and absorption-line), normalized by $(1+z)$, is 54 \kms. In both cases our scatter is $\sim25\%$ larger. This difference is likely accounted for by differences in depth, S/N, balance between absorption and emission line objects and our smaller sample size.

The inter-instrument errors are the potential systematic variations from instrument to instrument. Figure \ref{inter} and Table \ref{comp} present the number of objects in common for each instrument combination as well as the median and standard deviation of the redshift measurement differences. In all cases, the median difference between repeat measurements is consistent with zero within the standard deviation.

Both the inter- and the intra-instrument dispersions in redshift measurements are an order of magnitude higher than the errors produced by our formal fitting procedure. We use our findings here to define better errors in Section 4.

To perform an external cross-check on our redshifts, we compare them to redshift measurements found in NED. Figure \ref{ned} and Table \ref{comp} show the distribution of redshift measurement differences between our spectroscopic catalog and NED as a function of magnitude, S/N, and redshift. The distribution is broad, as expected from the heterogeneity of the sample, but the mean is small ($cz=52\pm290$ \kms). We conclude that, in comparison to redshifts in the literature, our redshift catalog has no significant systematic errors. One of our redshifts is clearly wrong; object 9136 in the PG1115 field is misidentified as a star in our spectroscopy, while it is a $z=0.14$ galaxy in NED. Visual inspection of the image confirms that the object is extended and the coordinate match is correct. The failure is due to a wrong redshift generated by the automated fitting and not caught in the follow-up visual inspection. Based on this literature comparison, we conclude that our catastrophic failure rate is  $\sim0.54\%$ (1 failure out of 186 matches). Assuming this rate is correct, we may expect $\sim50$ such failures in the entire survey. No catastrophic outliers are found internally because conflicts between repeated redshift measurements for objects with multiple observations were resolved by the visual inspection of the spectra.

\begin{figure*}[ht]
\figurenum{5a}
\begin{tabular}{@{}c@{\hspace{-5.0mm}}@{\hspace{-5.0mm}}c@{\hspace{-5.0mm}}@{\hspace{-5.0mm}}c@{\hspace{-5.0mm}}@{\hspace{-5.0mm}}c@{\hspace{0mm}}}
\includegraphics[width=0.28\textwidth]{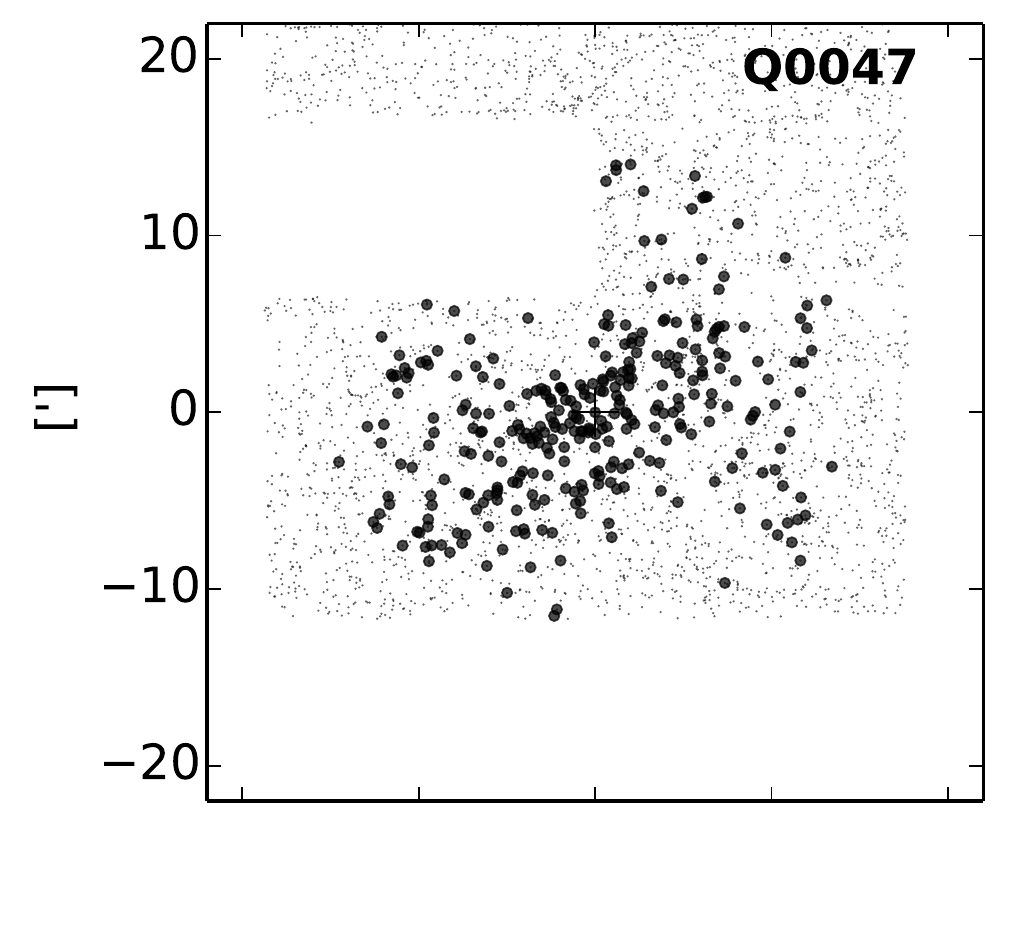} &
\includegraphics[width=0.28\textwidth]{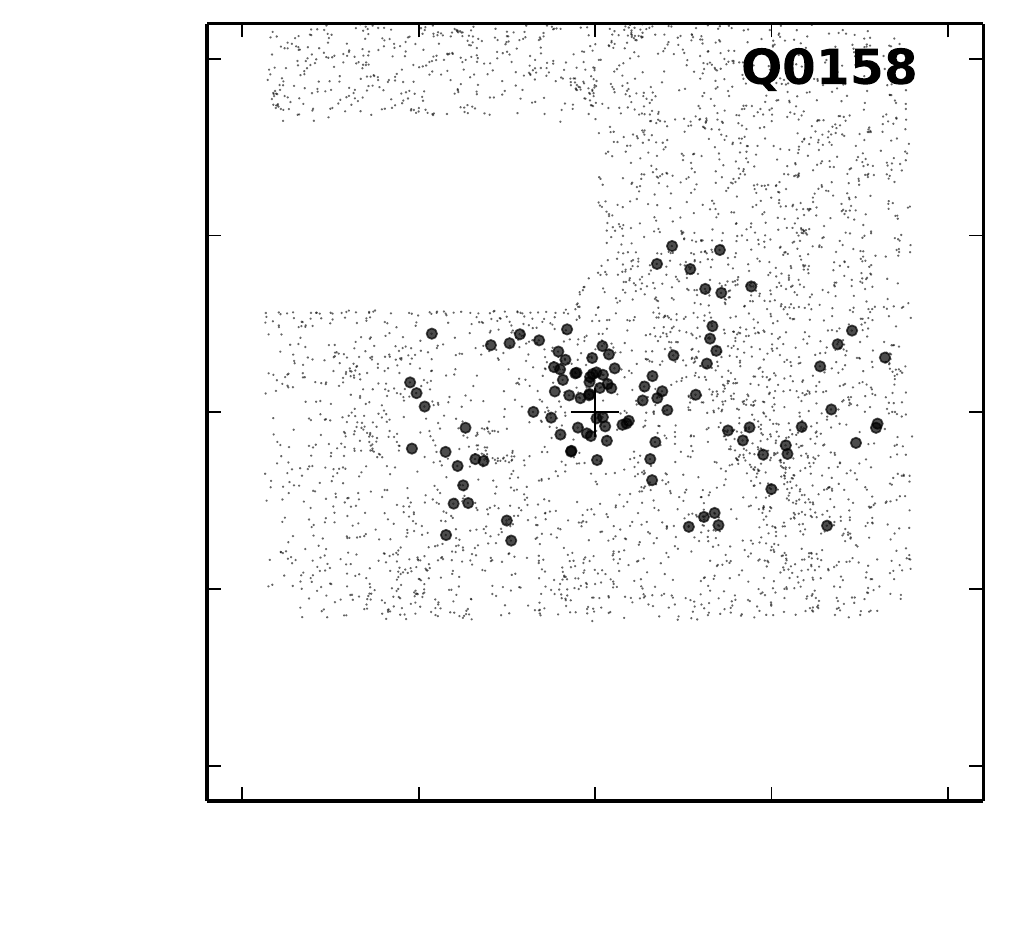} &
\includegraphics[width=0.28\textwidth]{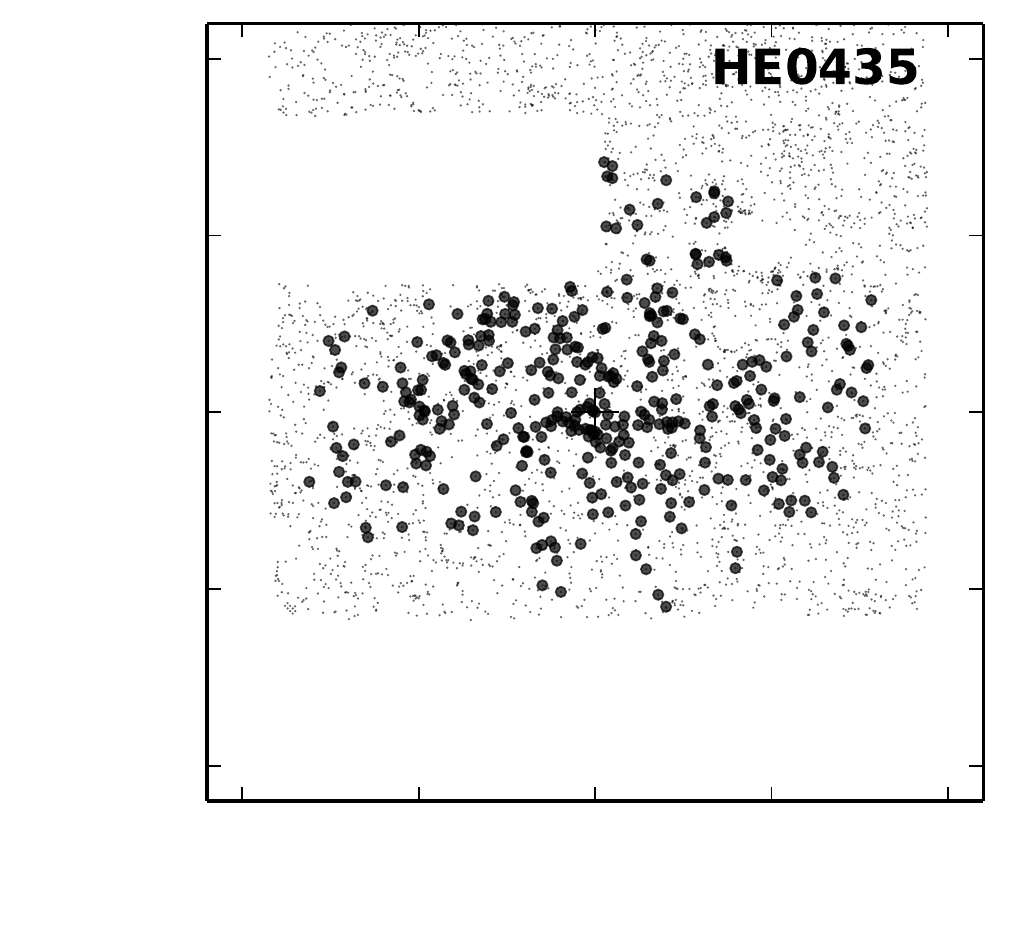} &
\includegraphics[width=0.28\textwidth]{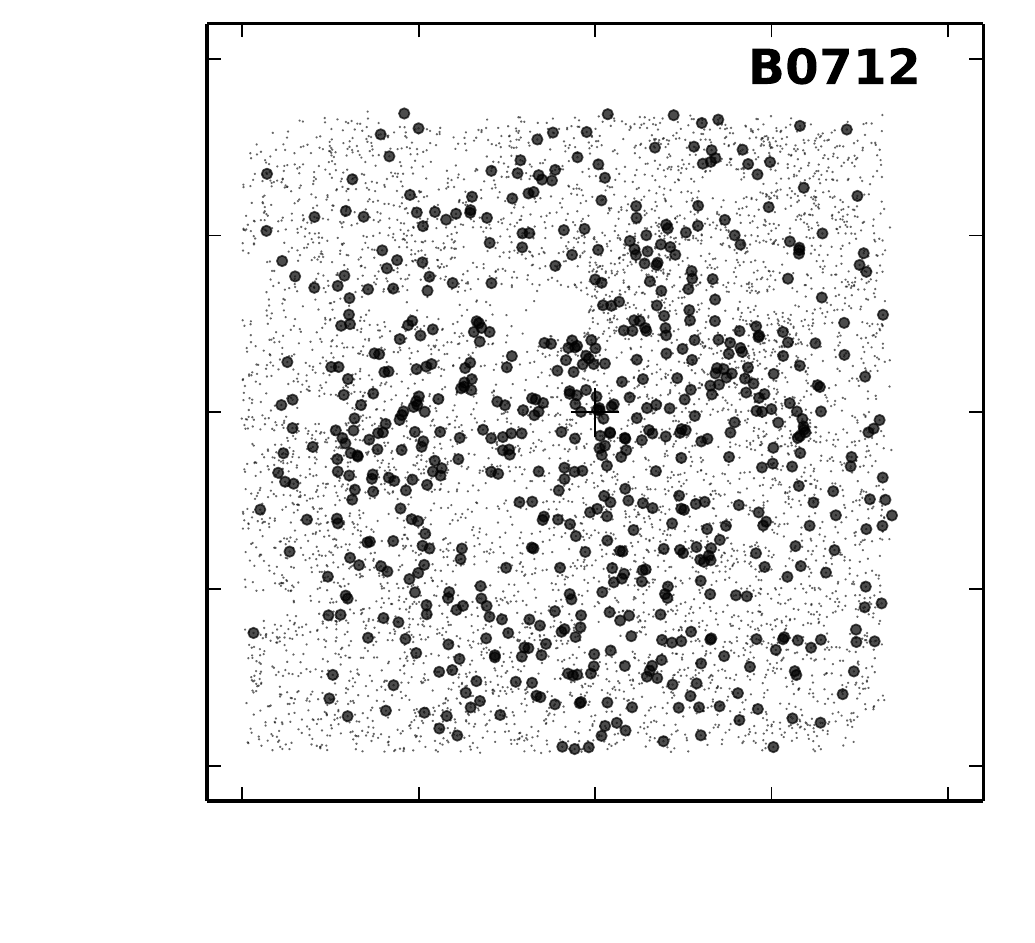} \\ [-20pt]
\includegraphics[width=0.28\textwidth]{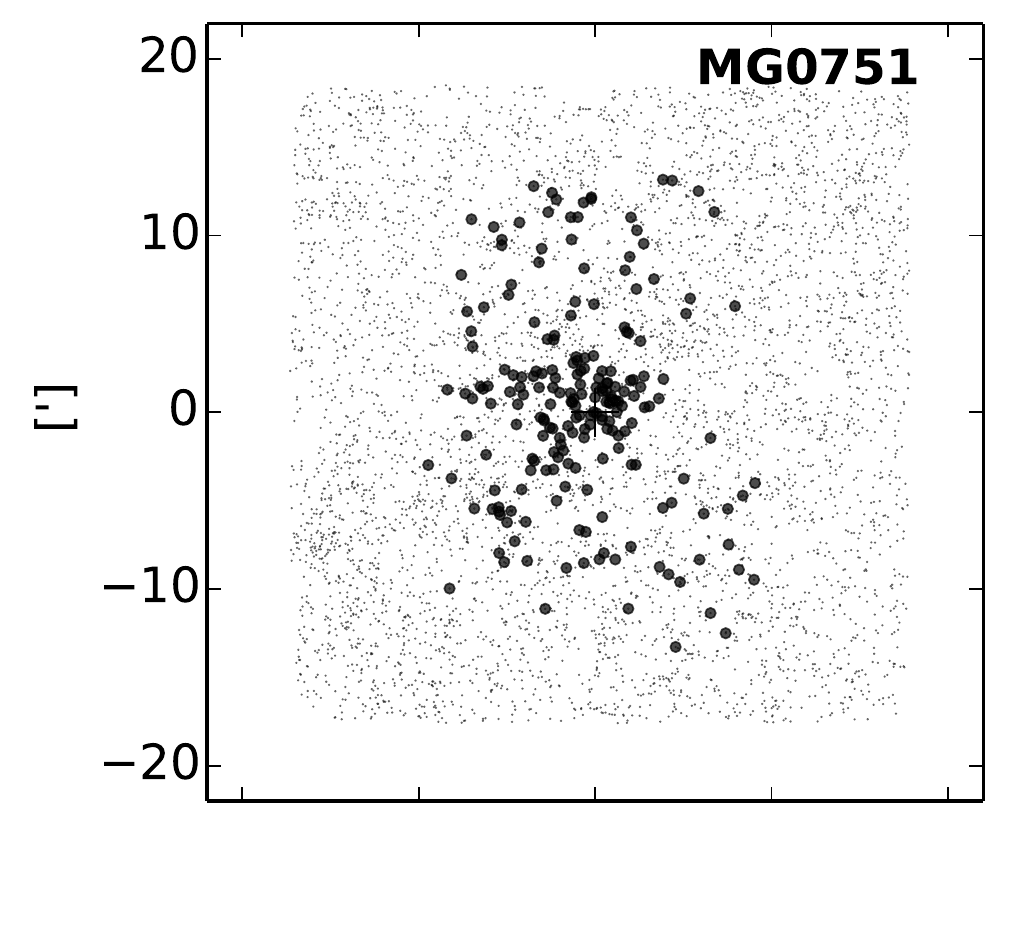} &
\includegraphics[width=0.28\textwidth]{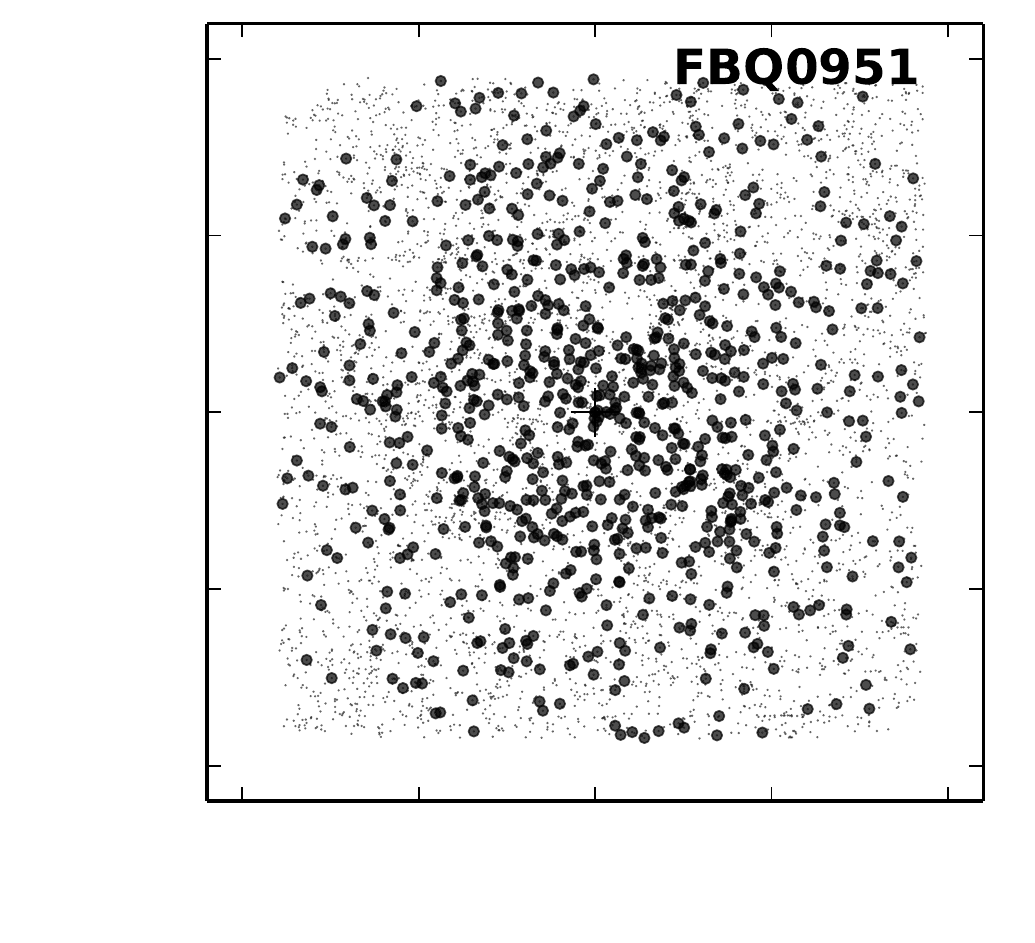} &
\includegraphics[width=0.28\textwidth]{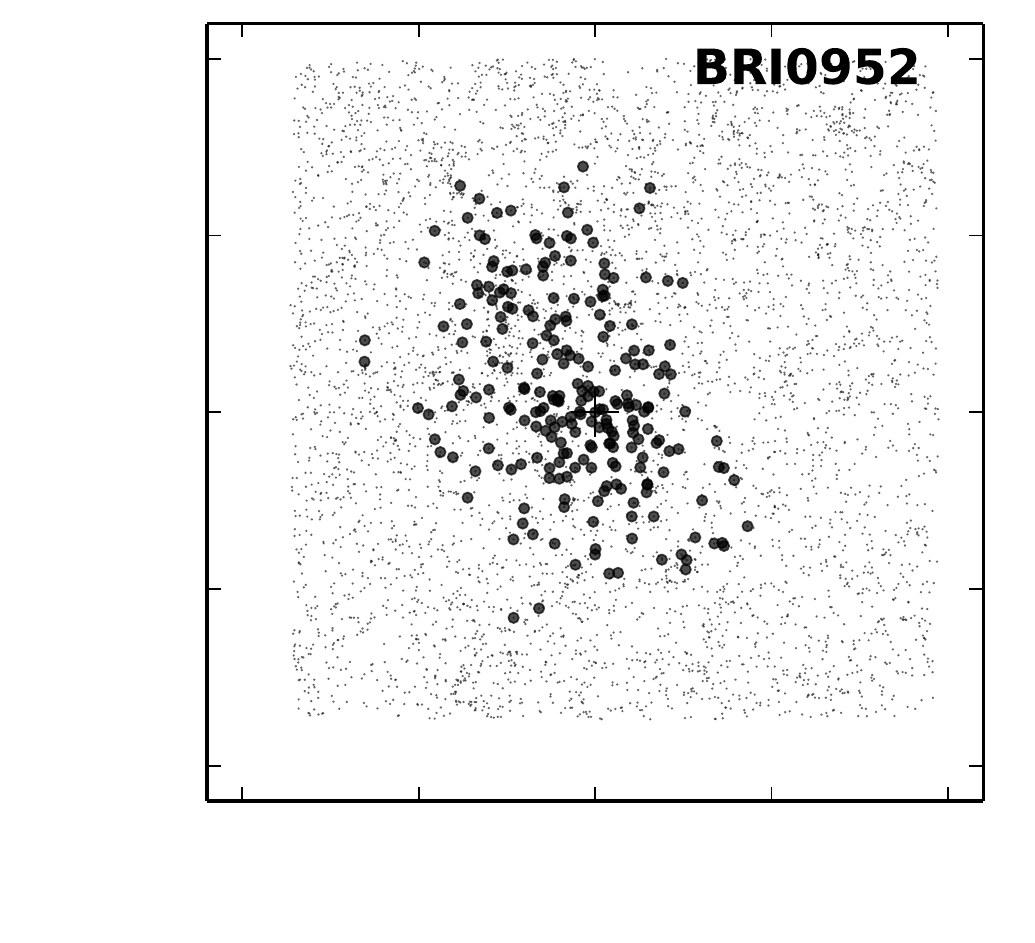} &
\includegraphics[width=0.28\textwidth]{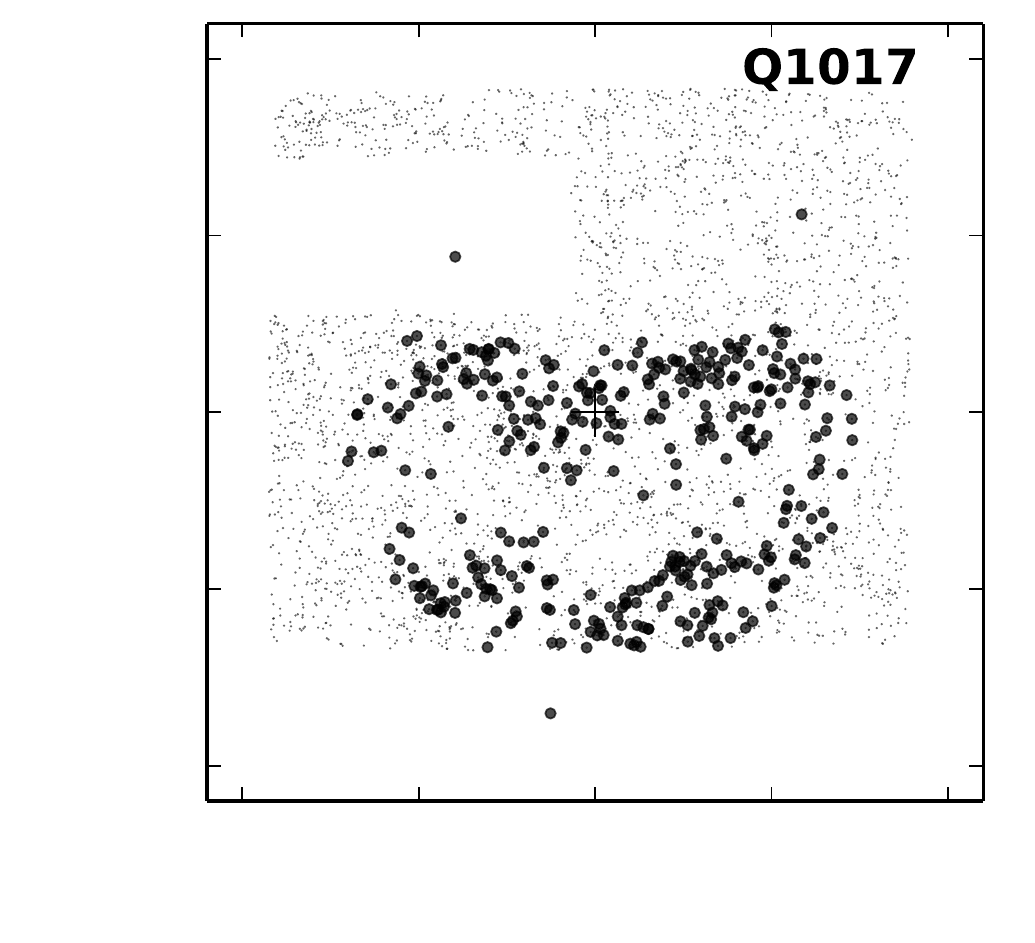} \\ [-20pt]
\includegraphics[width=0.28\textwidth]{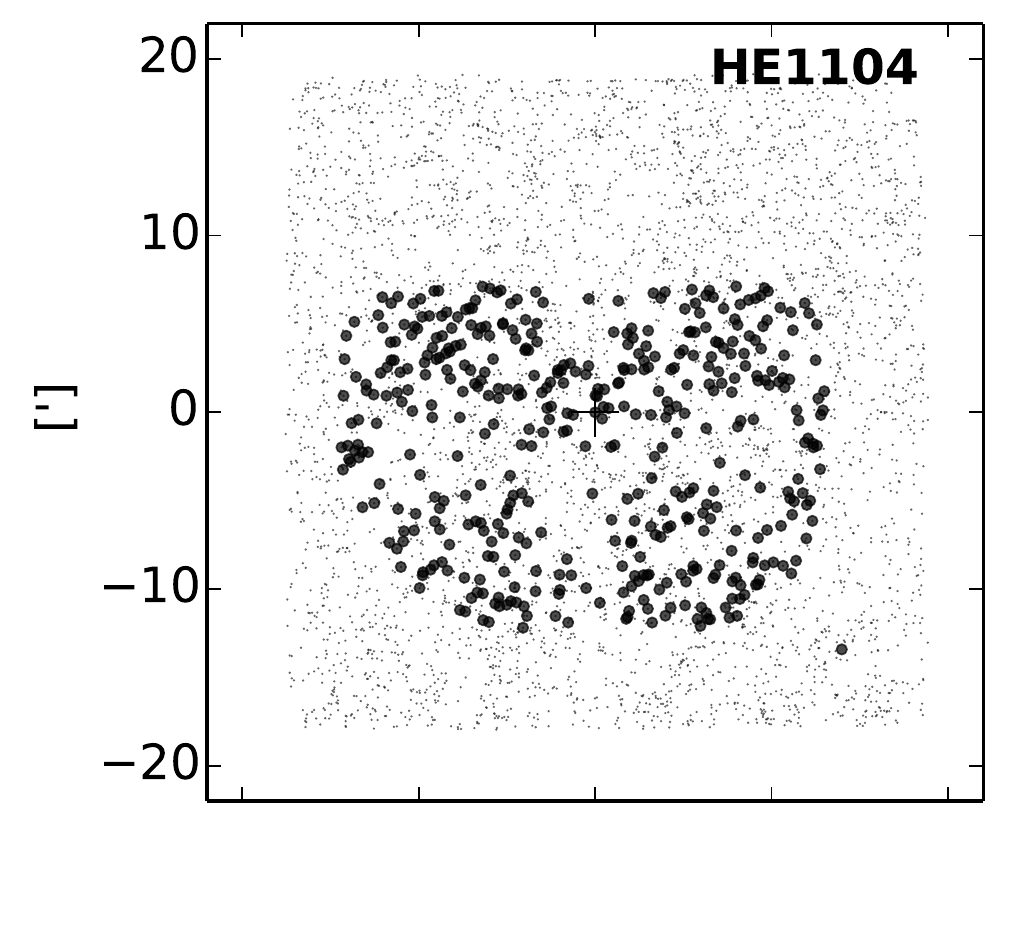} &
\includegraphics[width=0.28\textwidth]{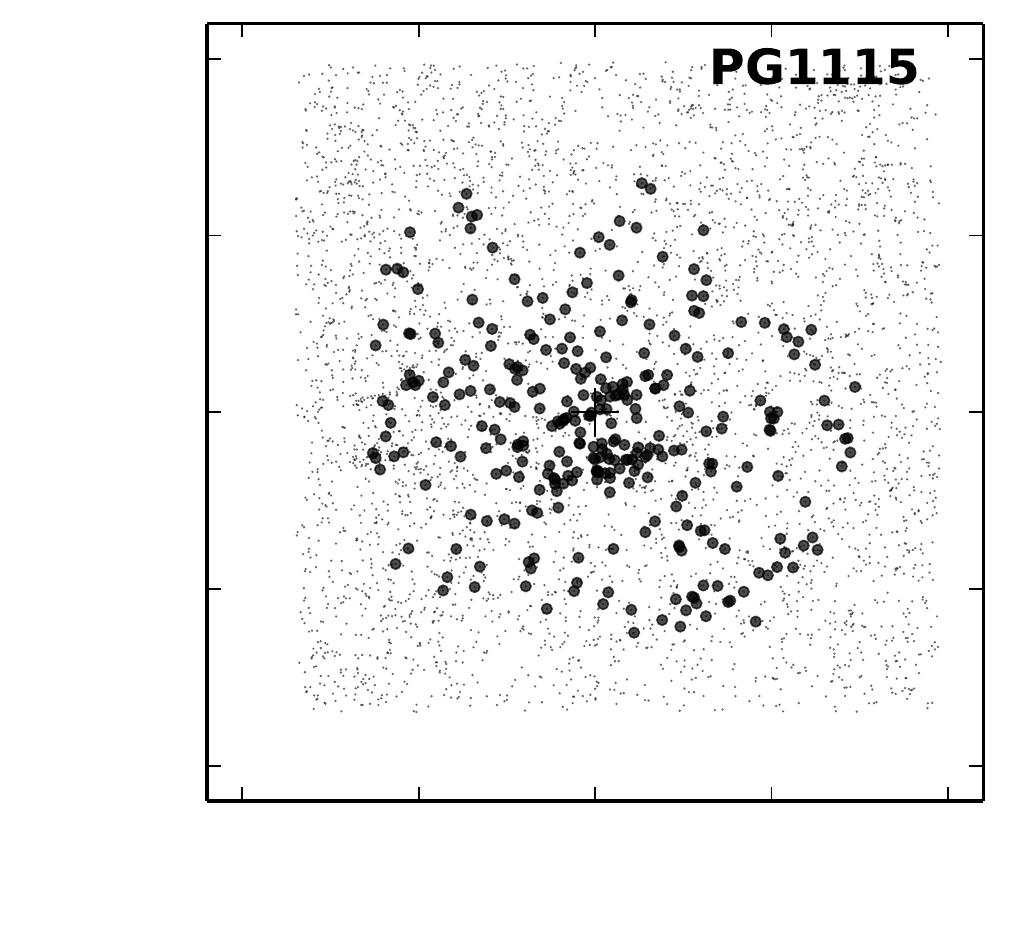} &
\includegraphics[width=0.28\textwidth]{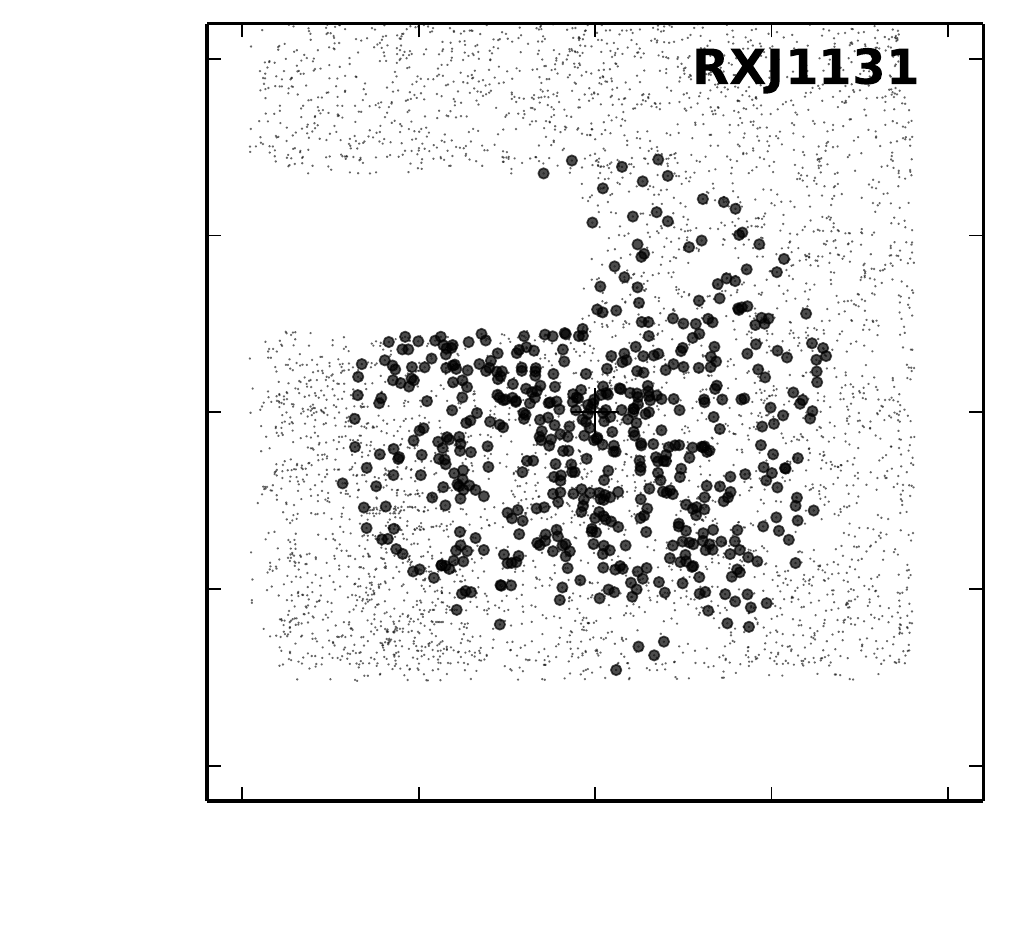} &
\includegraphics[width=0.28\textwidth]{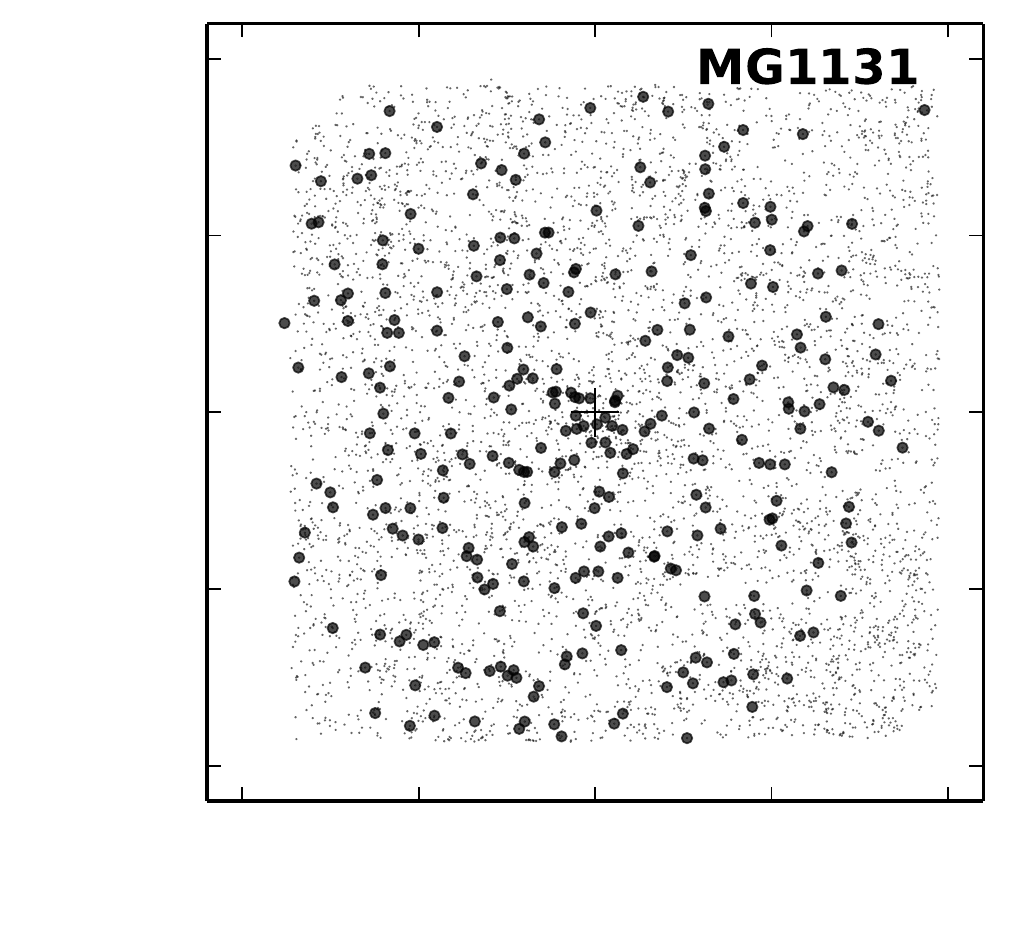} \\ [-20pt]
\includegraphics[width=0.28\textwidth]{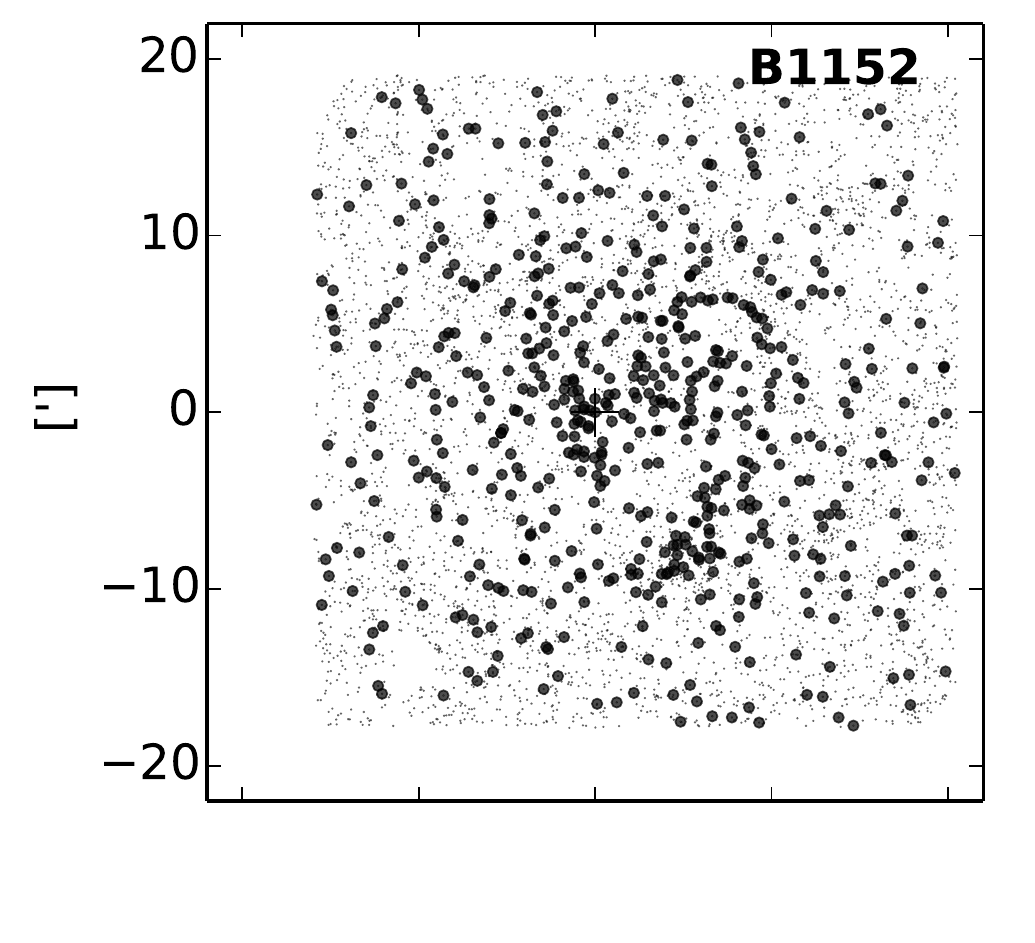} &
\includegraphics[width=0.28\textwidth]{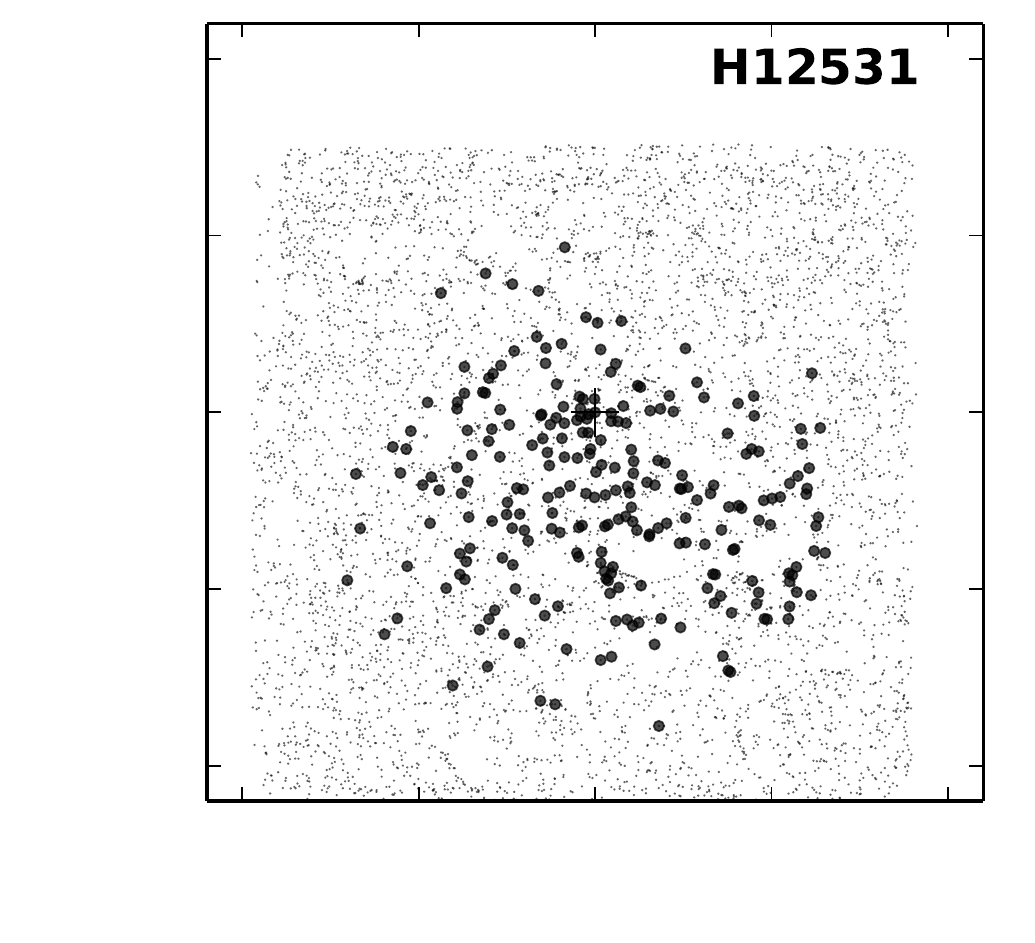} &
\includegraphics[width=0.28\textwidth]{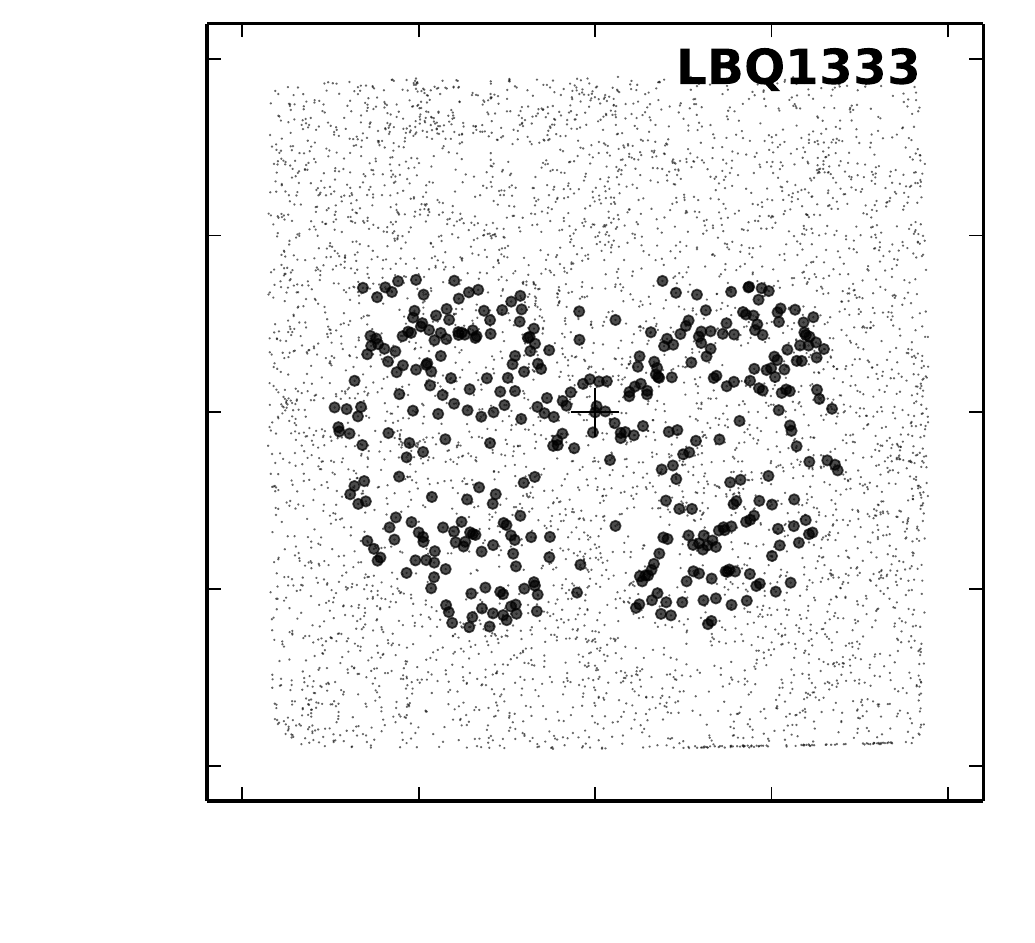} &
\includegraphics[width=0.28\textwidth]{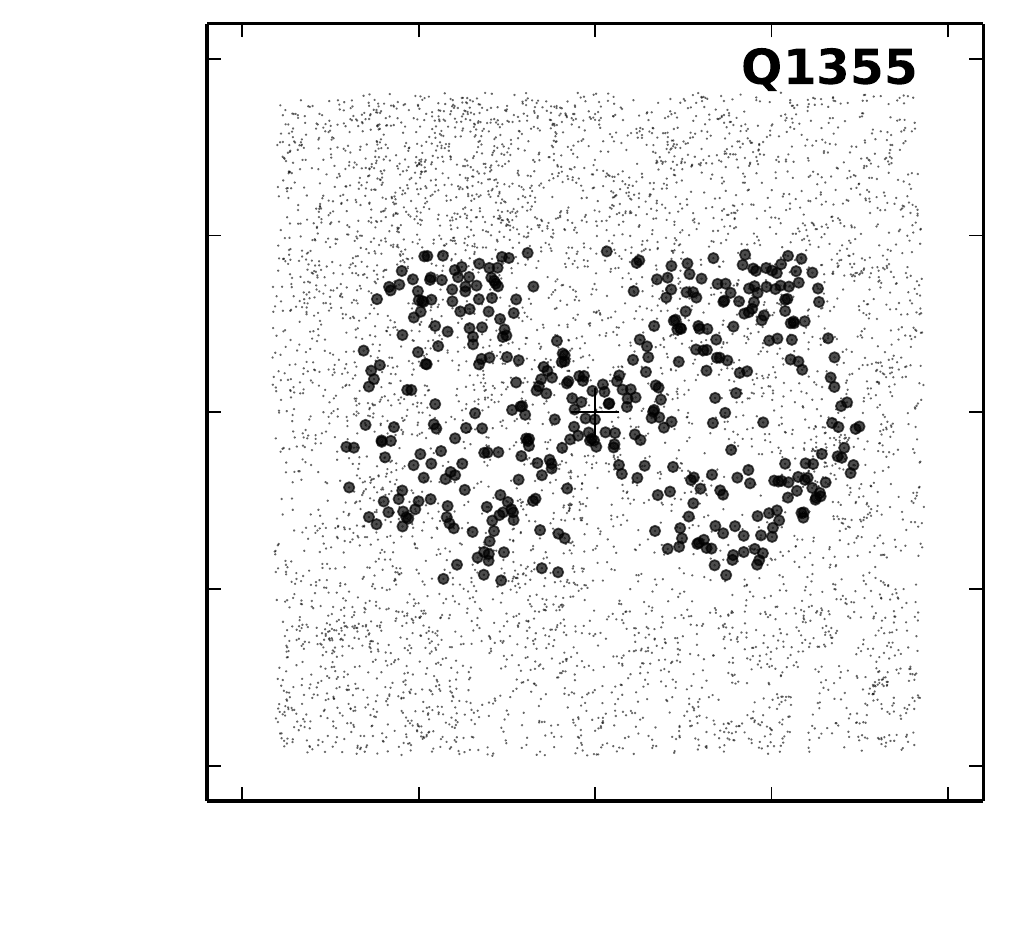} \\ [-20pt]
\includegraphics[width=0.28\textwidth]{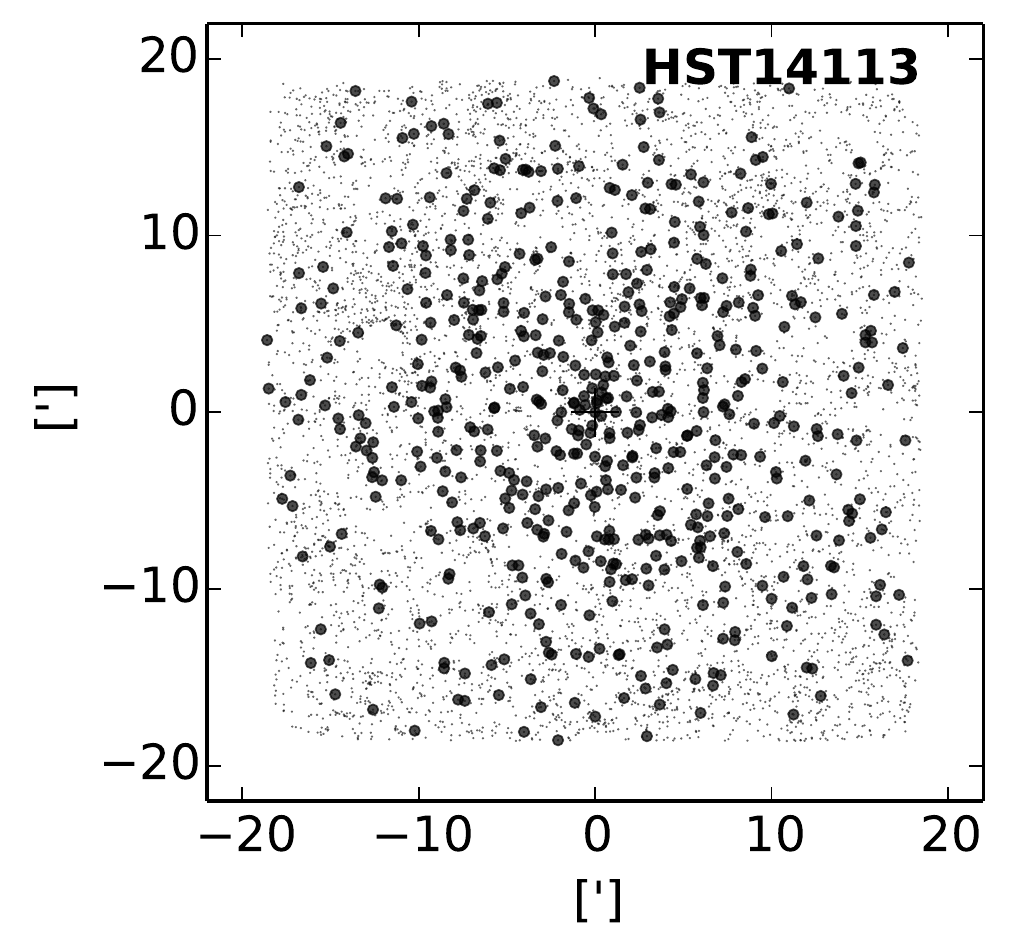} &
\includegraphics[width=0.28\textwidth]{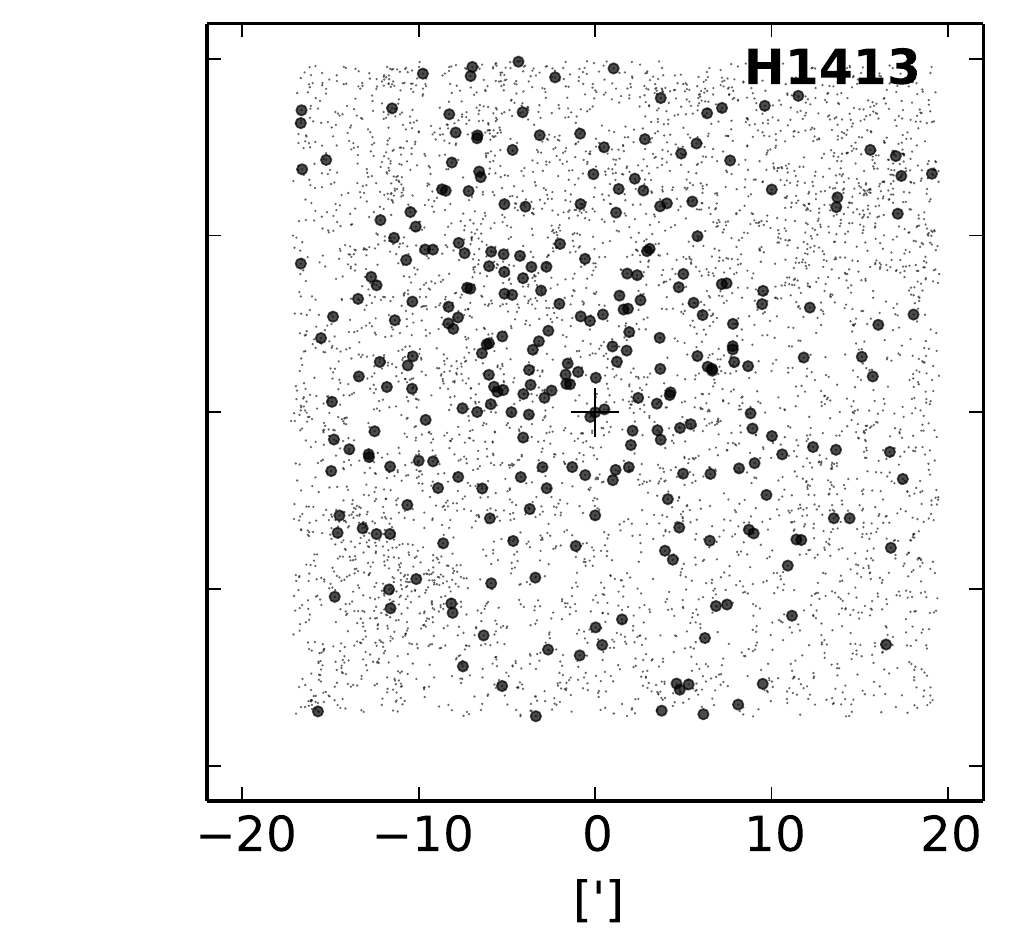} &
\includegraphics[width=0.28\textwidth]{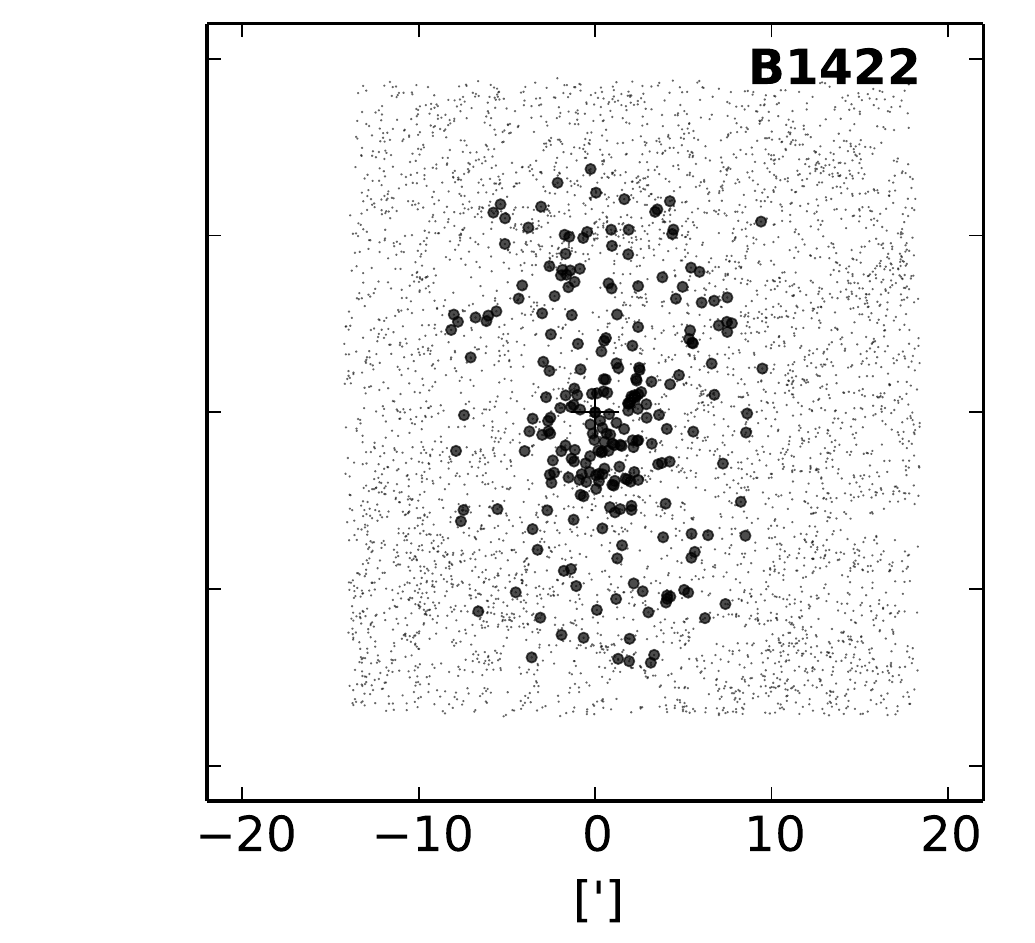} &
\includegraphics[width=0.28\textwidth]{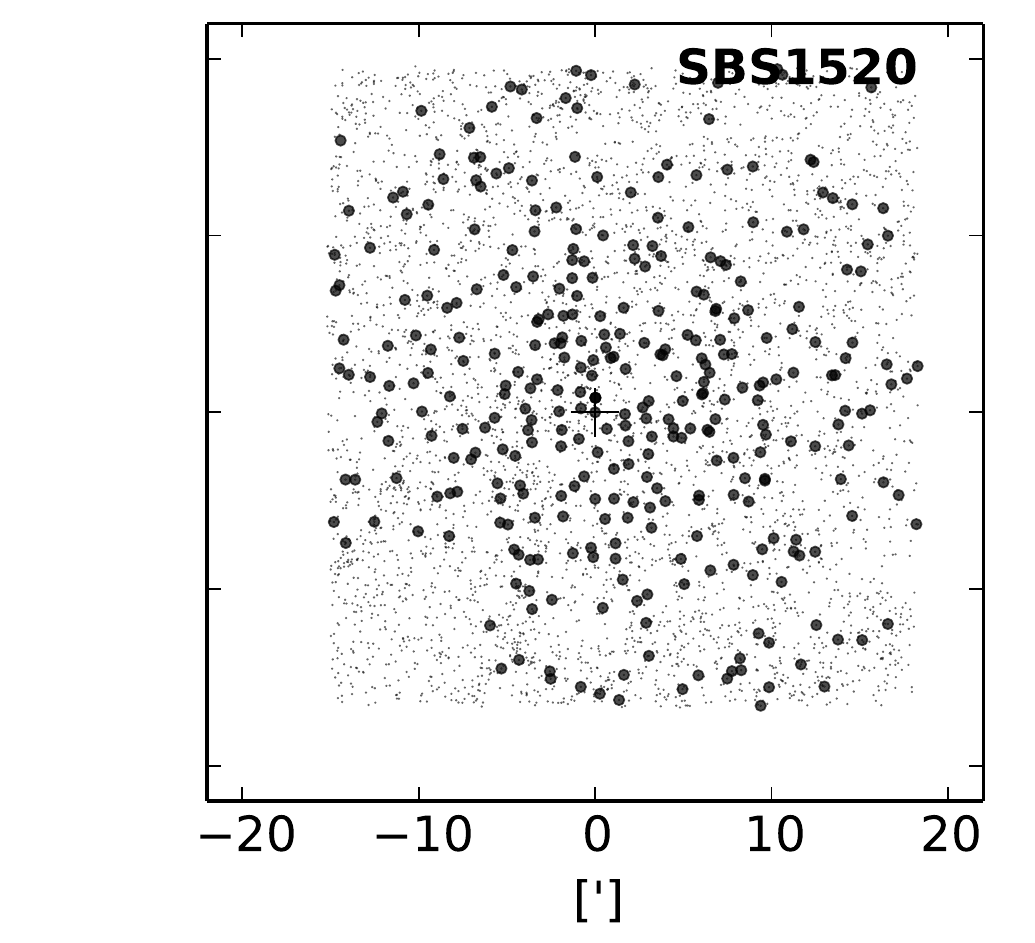} \\ [-20pt]
\end{tabular}
\vspace{7mm}	
\caption{\footnotesize Projected distributions of the objects in the final spectroscopic catalog (black points) relative to the objects in the full photometric catalog down to $I=21.5$ (gray points). The positions of objects are in arcminutes relative to the lens galaxy (black cross). The areas without photometric coverage in the fields of Q0047, Q0158, HE0435, Q1017, RXJ1131, and B1608 are due to dead CCDs in the MOSAIC imager. The images in the B2114 field were dithered with too small of a step which causes the blank stripes between the detectors. Objects added from literature are not included in this plot. For the spatial distribution of those objects, see Figures \ref{fig:radec_extra_a} and \ref{fig:radec_extra_b} at the end of the paper. \label{fig:radec_a}}
\end{figure*}

\begin{figure*}[ht]
\figurenum{5b}
\begin{tabular}{@{}c@{\hspace{-5.0mm}}@{\hspace{-5.0mm}}c@{\hspace{-5.0mm}}@{\hspace{-5.0mm}}c@{\hspace{-5.0mm}}@{\hspace{-5.0mm}}c@{\hspace{0mm}}}
\includegraphics[width=0.28\textwidth]{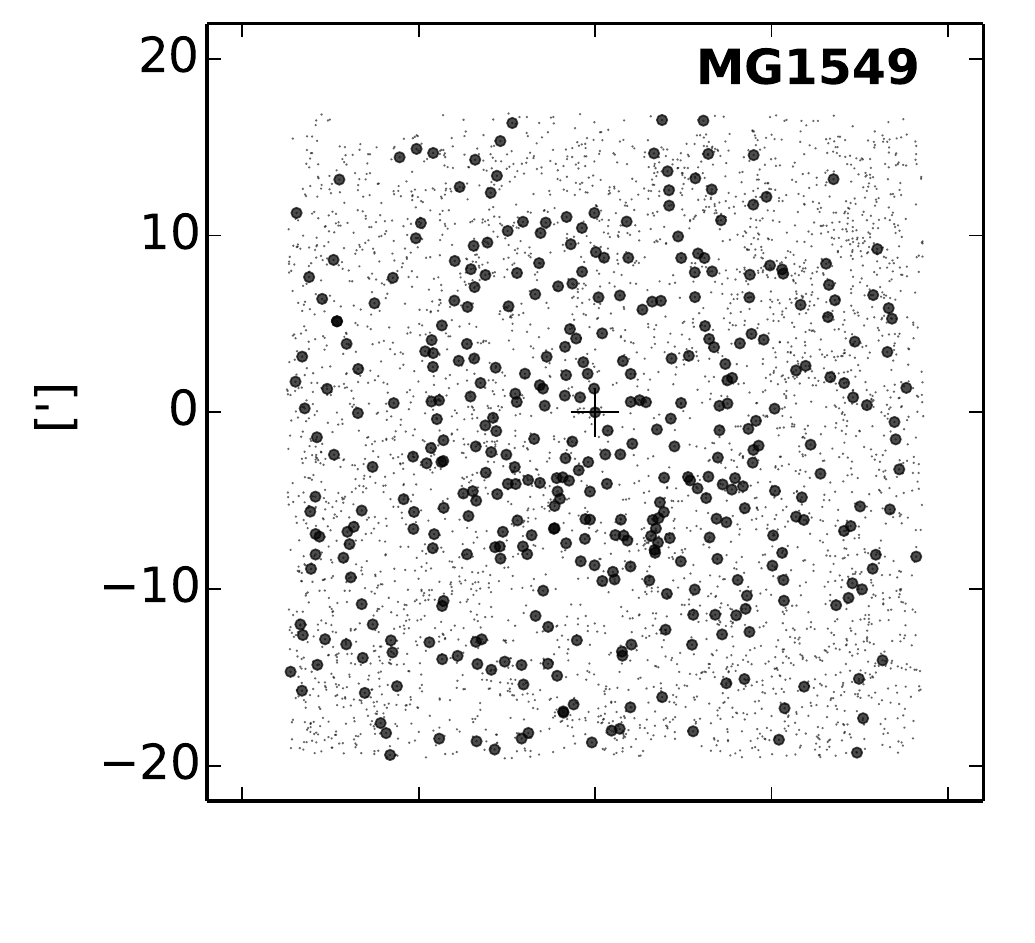} &
\includegraphics[width=0.28\textwidth]{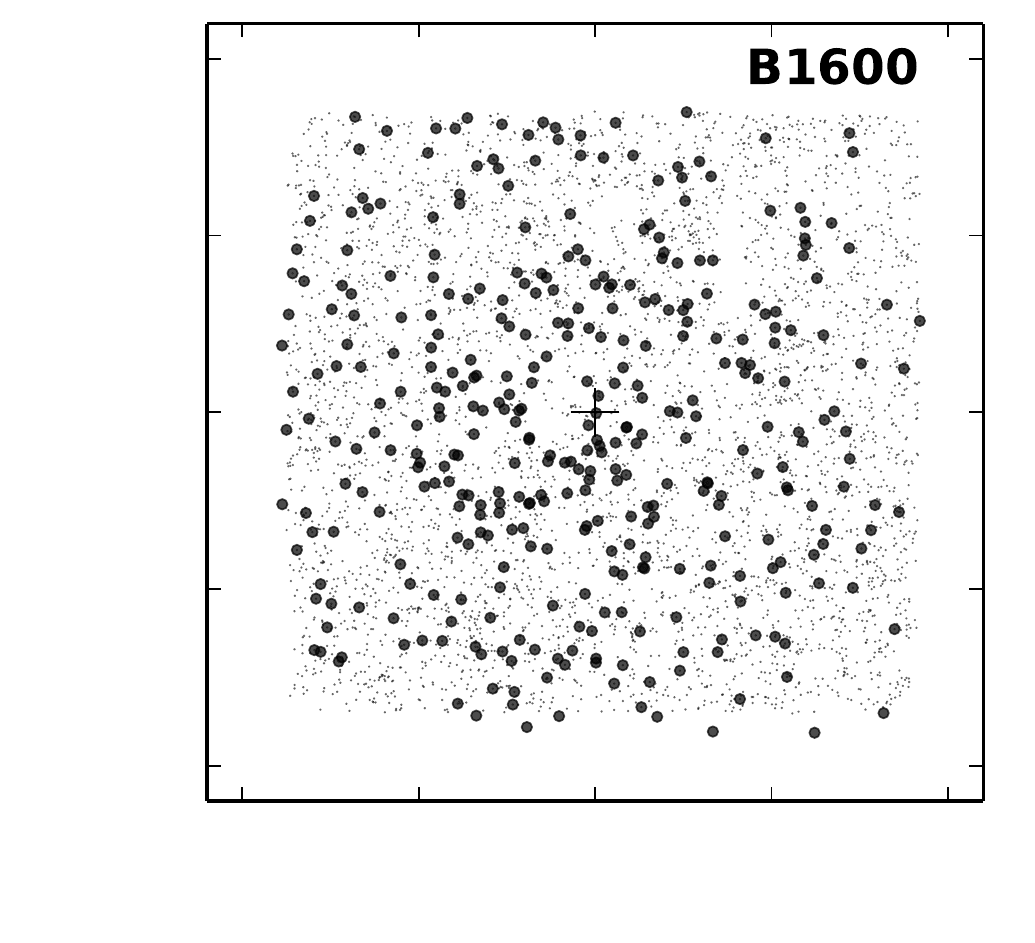} &
\includegraphics[width=0.28\textwidth]{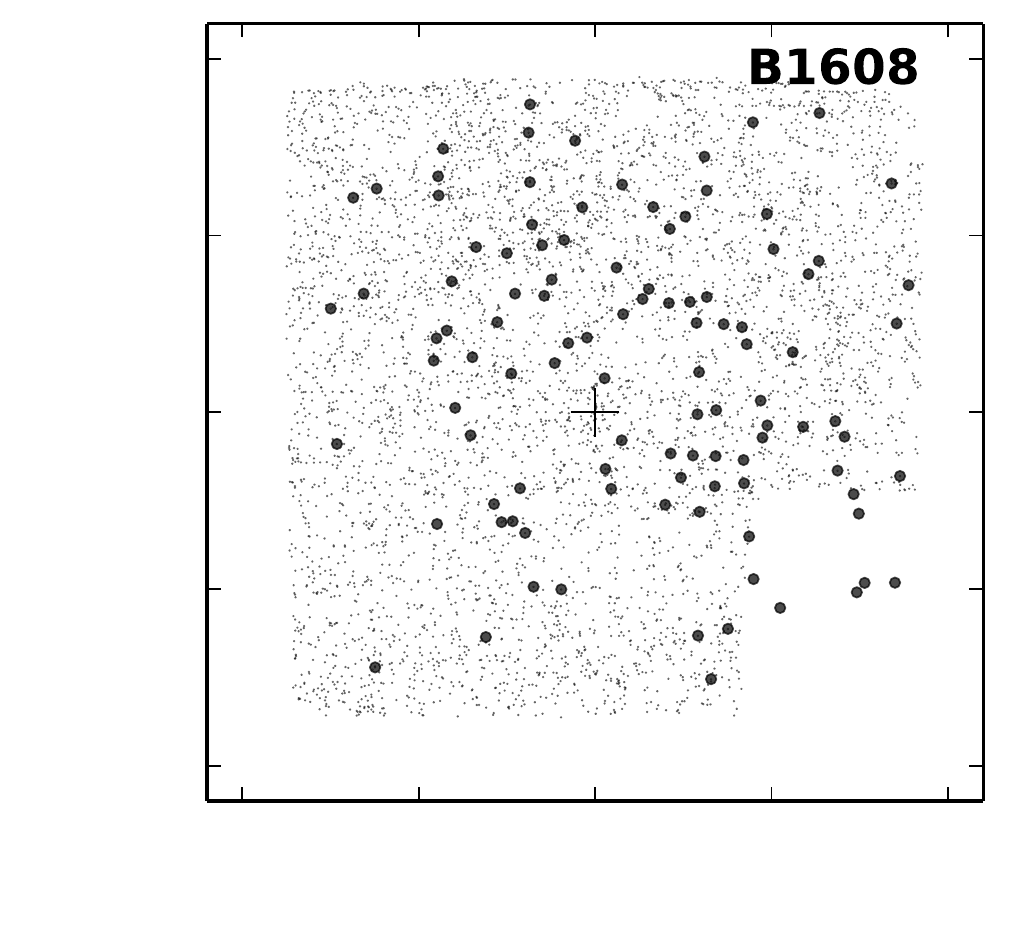} &
\includegraphics[width=0.28\textwidth]{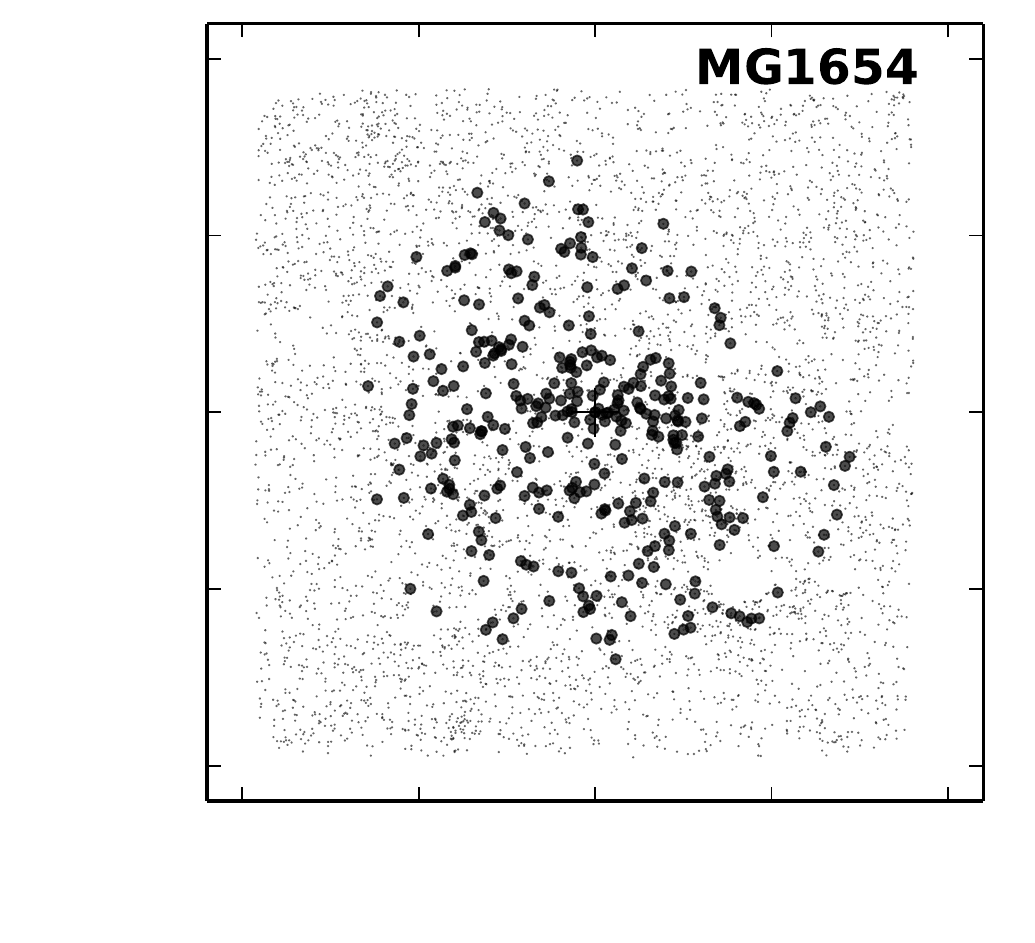} \\ [-20pt]
\includegraphics[width=0.28\textwidth]{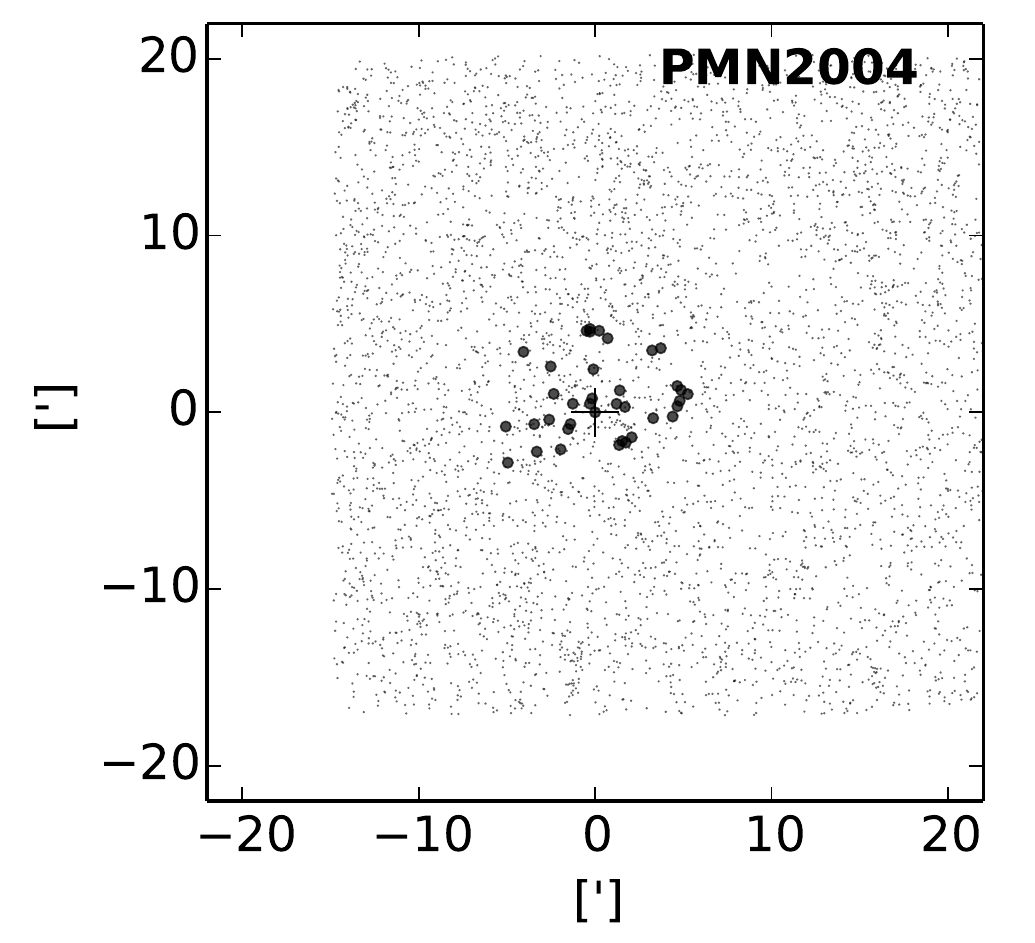} &
\includegraphics[width=0.28\textwidth]{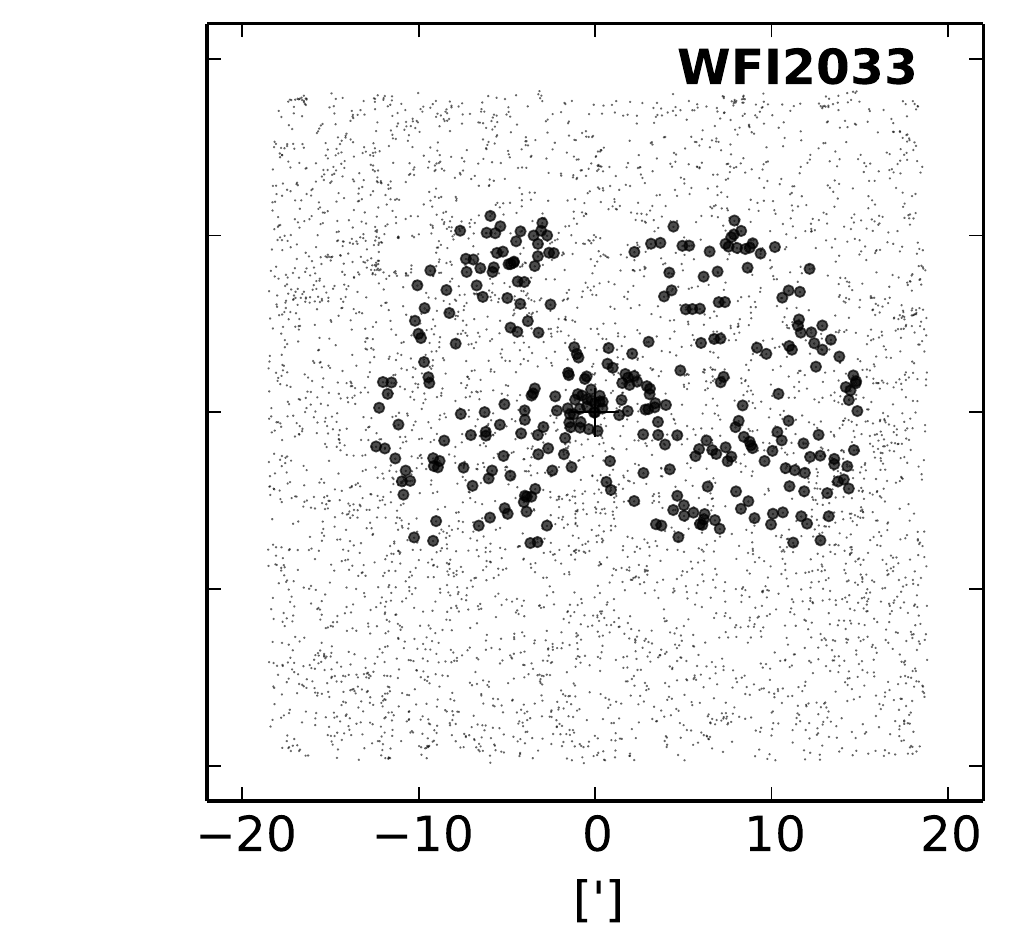} &
\includegraphics[width=0.28\textwidth]{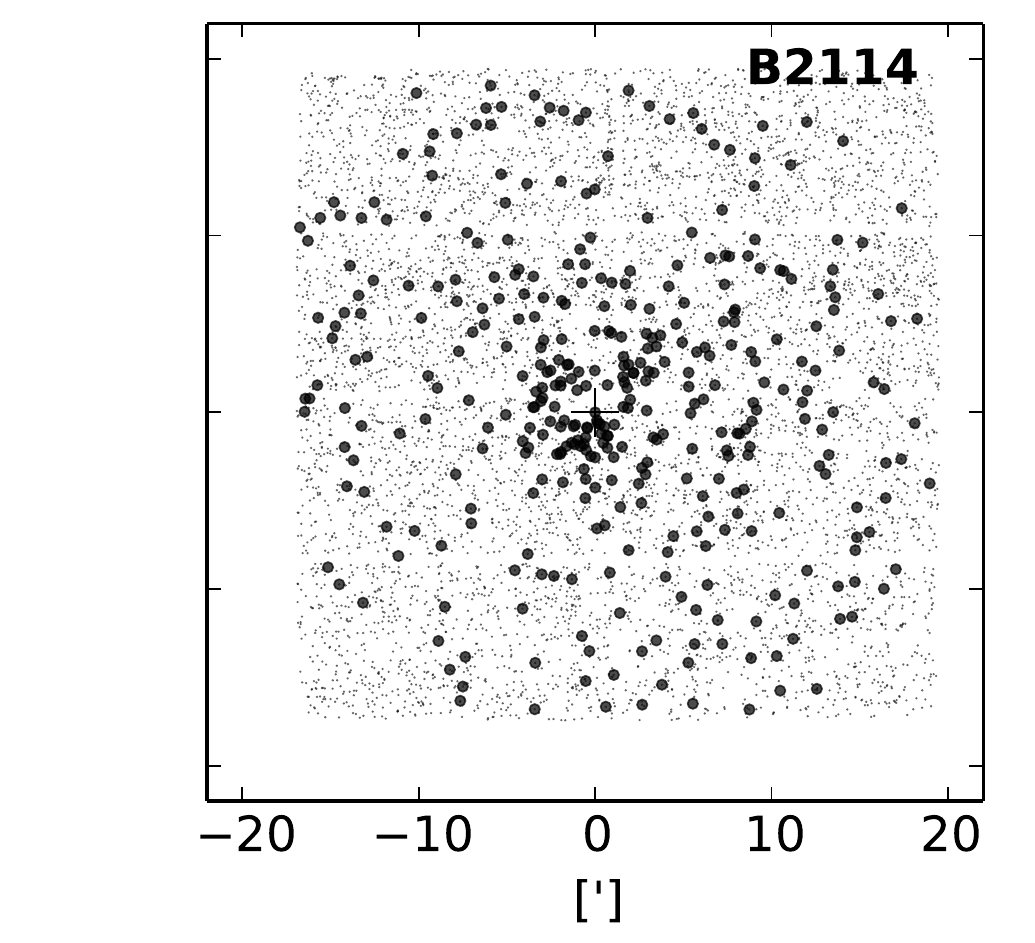} &
\includegraphics[width=0.28\textwidth]{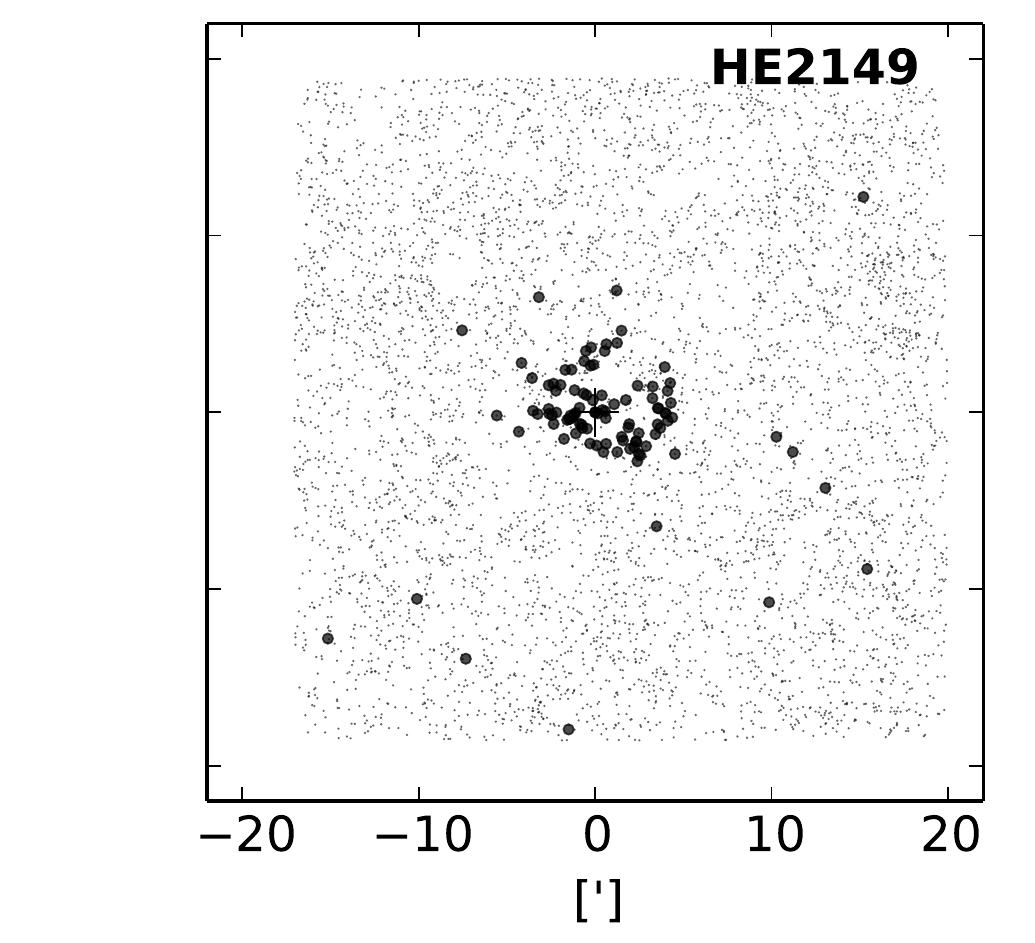} \\ [-20pt]
\end{tabular}
\vspace{7mm}
\caption{\footnotesize Continued from Figure \ref{fig:radec_a}. \label{fig:radec_b}}
\end{figure*}

\begin{figure}
\figurenum{6}
\epsscale{1.}
\plotone{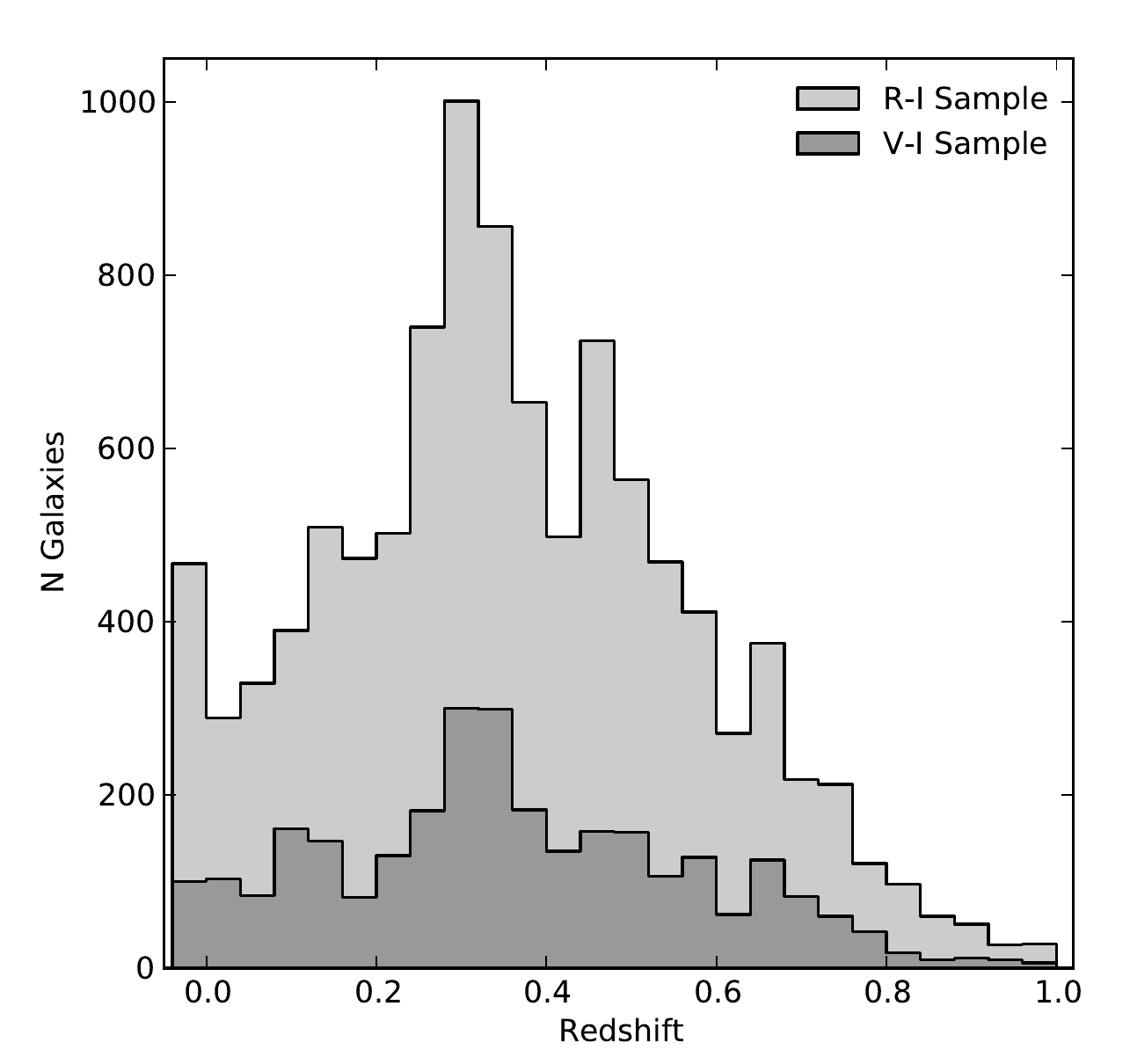}
\caption{\footnotesize Redshift distribution of the galaxies in the final spectroscopic redshift catalogs. The relative contributions from the $R-I$ and $V-I$ samples are shown in light and dark gray, respectively. The redshift distributions are qualitatively similar, with medians $z=0.370$ ($R-I$) and $z=0.366$ ($V-I$).\label{fig.redshifts}}
\end{figure}

\begin{figure*}
\figurenum{7}
\epsscale{1.2}
\label{field-first}
\plotone{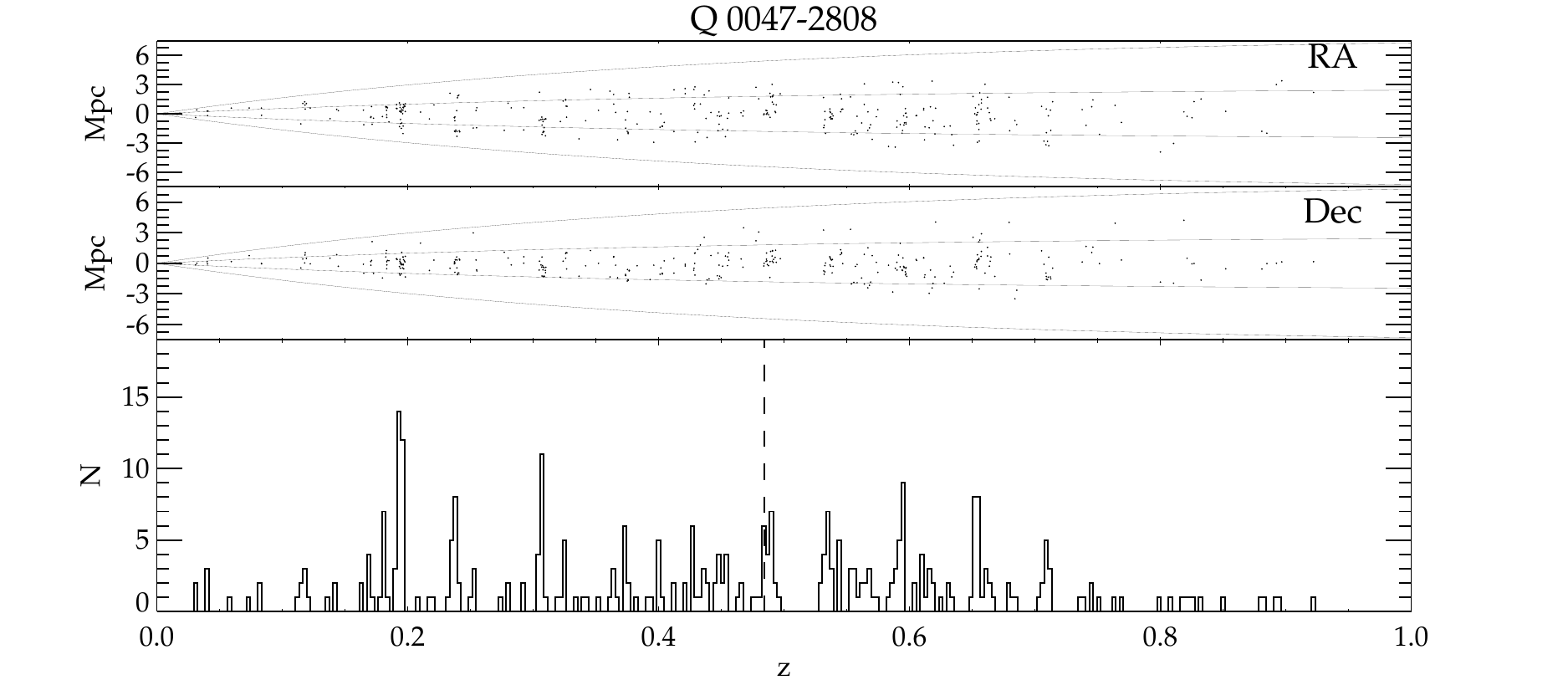}
\caption{\footnotesize Galaxy redshift distribution in the field of Q0047. The top two panels show the redshift distributions of galaxies projected in RA (top) and Dec (middle) in units of proper Mpc from the center of the field. This projection exaggerates the opening angle of the beam ($30\arcmin$), which causes structures to be stretched perpendicular to the redshift direction but makes them easier to see. The horizontal lines represent linear separations of $5\arcmin$\ (inner two lines) and $15\arcmin$\ (outer lines) from the center of the field. The bottom panel shows the distribution in the form of a histogram with a binsize of $\delta z = 0.001$. The vertical dashed line indicates the spectroscopic redshift of the lens galaxy. The histogram includes all galaxies with spectroscopic redshifts. The analogous figures for the remaining 27 fields are shown in Figure 11.}
\end{figure*}

\section{Redshift Catalog}

Table \ref{catalog} presents the spectroscopic catalog in the field of Q0047 as an example. For each entry, we give the catalog number, J2000.0 coordinates in degrees calibrated to USNO-B2.0, projected distance to the lens in arc-minutes, redshift $z$, and redshift error $\delta z$. The positional errors are $\sim0\farcs2$ based on the scatter in the positions of stars in USNO-B2.0 and our catalog. We do not include photometry (magnitudes and colors), as these will be published later.  We provide a spectroscopic flag with the following values: Flag=1 for objects with redshifts that are not in our final photometric catalog; Flag=2 for data obtained with LDSS-2; Flag=3 for data obtained with LDSS-3; Flag=4 for data obtained with IMACS; Flag=5 for data obtained with Hectospec; Flag=6 for NED objects.. The spectroscopic redshifts of the lens galaxy and source are listed first in each field (where available). Following them, galaxies are ordered within each field by ascending order of RA. Serendipitously observed stars are included at the end of the redshift catalog for each field for completeness, but radial velocities for them are not included. Finally, objects from NED are listed.

Two of the fields in our sample, B1608 and PMN2004,  have only limited spectroscopic observations (a single mask for B1608) and therefore very low completeness. The spectroscopic catalog includes 106 new redshifts in the field of B1608. Redshifts in the field of PMN2004 have already been published in \citet{my}, but we update them here. Due to the sparse coverage, these fields will not be considered in the completeness discussion in $\S$ 4.2. We warn users of this catalog that there are insufficient data to obtain meaningful constraints on the environments for these lenses. 

The final spectroscopic catalog includes 9768 unique redshift measurements. The projected positions on the sky of of these objects relative to the objects in the photometric catalog ($I\leq21.5$) are presented in Figures \ref{fig:radec_a} and \ref{fig:radec_b}. As designed, the objects in the spectroscopic catalogs are concentrated around the lens galaxy but extend over the full field for most lenses. The completeness as a function of radius is examined in \S\ 4.2. Spatial distributions for objects from NED are shown in Figures \ref{fig:radec_extra_a} and \ref{fig:radec_extra_b} in the Appendix.

The redshift distribution of the galaxies in the catalog is shown in Figure \ref{fig.redshifts}. The median redshift for this sample is $z_{med} = 0.360$ (stars excluded) and 82.4\% (8045) of the objects are between $z=0.1$ and $z=0.7$. There are 622 objects (6.5\% of the sample) at $z>0.7$ and only 30 objects (0.3\%) at $z>1.0$. Serendipitously observed stars represent 6.0\% of the sample or 646 objects.  Including the objects from NED, the size of the final catalog is 10638 unique stars and galaxies. NED contributes a large fraction of  the high redshift galaxies, adding 192 objects at $z>1$. The median redshift for the full catalog (stars excluded) is $z=0.361$, and 79.4\% of all galaxies are between $z=0.1$ and $z=0.7$. 

To illustrate the data, in Figure \ref{field-first} we present the redshift distribution of galaxies in the field of Q0047. The top two panels show the projection of this distribution in Right Ascension and Declination as a function of redshift. The opening angle of the beams ($0.5$ degrees) is exaggerated in this projection, which causes structures to be stretched perpendicular to the redshift direction but makes them easier to see. The diverging horizontal lines indicate angular distances of $5\arcmin$\ and $15\arcmin$\ from the center of the field. The bottom panel shows the redshift histogram. The vertical dashed line shows the spectroscopic redshift of the lens galaxy. Analogous figures for the remaining 27 fields are shown in Figure 11 in the Appendix.

\subsection{Redshift Uncertainties}

\begin{deluxetable}{lllll}
\tablecolumns{5}
\tablewidth{0pc}
\tabletypesize{\footnotesize}
\tablecaption{Redshift Comparisons For Objects With Multiple Observations\label{comp}}
\tablehead{
\colhead{Instruments} & \colhead{$N_{match}$} & \colhead{$N_{field}$} & \colhead{$| \Delta z_{med} |$} & \colhead{$\sigma_{\Delta z}$}
}
\startdata
\cutinhead{Intra-instrument comparison}
LDSS-2-LDSS-2 &   1   &   1   &  \nodata &   \nodata \\
LDSS-3-LDSS-3 &   0   &   0   & \nodata   &   \nodata  \\
IMACS-IMACS	&	136	&	3	&	9.450e-6	&	3.063e-4 \\ 
Hecto-Hecto	&	13	&	2	&	1.644e-4	&	2.905e-4 \\ 
\cutinhead{Inter-instrument comparison}
LDSS-2-LDSS-3	&	6	&	3	&	1.173e-4	&	4.622e-4 \\ 
LDSS-2-IMACS	&	3	&	2	&	6.440e-4	&	5.244e-4 \\ 
LDSS-2-Hecto	&	11	&	1	&	1.500e-4	&	5.008e-4 \\ 
LDSS-3-IMACS	&	65	&	7	&	1.710e-5	&	6.887e-4 \\ 
LDSS-3-Hecto	&	7	&	1	&	1.050e-4	&	2.267e-4 \\ 
IMACS-Hecto &   0   &   0   &   \nodata   &   \nodata   \\
\cutinhead{External comparison}
NED-This work	&	167	&	18	&	1.739e-4	&	9.674e-4 \\ 
\enddata
\end{deluxetable}

The redshift errors in our catalog deserve separate attention. The errors output by the cross-correlation routine, which rely only on the goodness of fit, underestimate the true errors in the redshift measurements, because the scatter in $\Delta z = z_1 - z_2$ for objects with repeated redshift measurements is much greater than the typical output redshift errors. We use the scatter in $\Delta z$ to estimate the true redshift errors. Figures \ref{intra} and \ref{inter} show that high S/N spectra yield more accurate redshift measurements. Objects with strong emission lines also exhibit lower $\Delta z$ scatter. Because of its large size, we use the IMACS-IMACS sample in Figure \ref{intra} to estimate the magnitude of the errors as a function of emission line strength. We fit the fluxed spectra with stellar population models, subtract the continuum fit, and measure the equivalent widths ($EW$) of five prominent emission lines: OII [3727], H[$\beta$], OIII 4959, OIII 5007 and H[$\alpha$] \citep{tremonti}. We split the galaxy sample into three sub-samples: (1) galaxies with at least two lines, each with $EW>10$ \AA~ and S/N in both emission lines greater than 3, (2) galaxies with only one emission line with $EW>10$ \AA~ and $S/N>3$, and (3) galaxies without any of these lines under the imposed requirements. We further split each sub-sample into three bins by mean continuum S/N and determine the standard deviation in $\Delta z$ in each of the resulting nine bins. We divide the standard deviations by $\sqrt{2}$ to account for the fact that $\Delta z$ is based on two redshifts. We use these results, listed in Table 7, to apply error-bars to our redshift catalog. The overall scatter in Table \ref{comp} for all instrument combinations is similar. We apply the errors from Table 7 to all spectra obtained with IMACS, Hectospec, and LDSS-3. The cross-correlation errors are not added in quadrature because they are an order of magnitude smaller.

The fluxing of the LDSS-2 spectra did not have sufficient accuracy for them to be fit with stellar population models. Only one standard star was observed during the course of the two observing runs and, while this was sufficient to remove the overall instrument response, large deviations remain in the continuum precluded us from measuring line fluxes. We assign redshift errors based on the spectrum S/N from the ``full sample'' values in Table 7. The majority of LDSS-2 spectra have S/N$<5$ and their corresponding errors are $\sigma_{z} = 3.6\times10^{-4}$,  consistent with the RVSAO errors in \citet{my}.

Finally, for the redshifts added from NED, we use their published errors. In summary, the redshift errors are generally $c\Delta z \lesssim100$ \kms  or $\Delta z \lesssim 3.33\times10^{-4}$.

\subsection{Completeness}

\begin{figure*}
\figurenum{8}
\label{gals_figure}
\epsscale{1.15}
\plotone{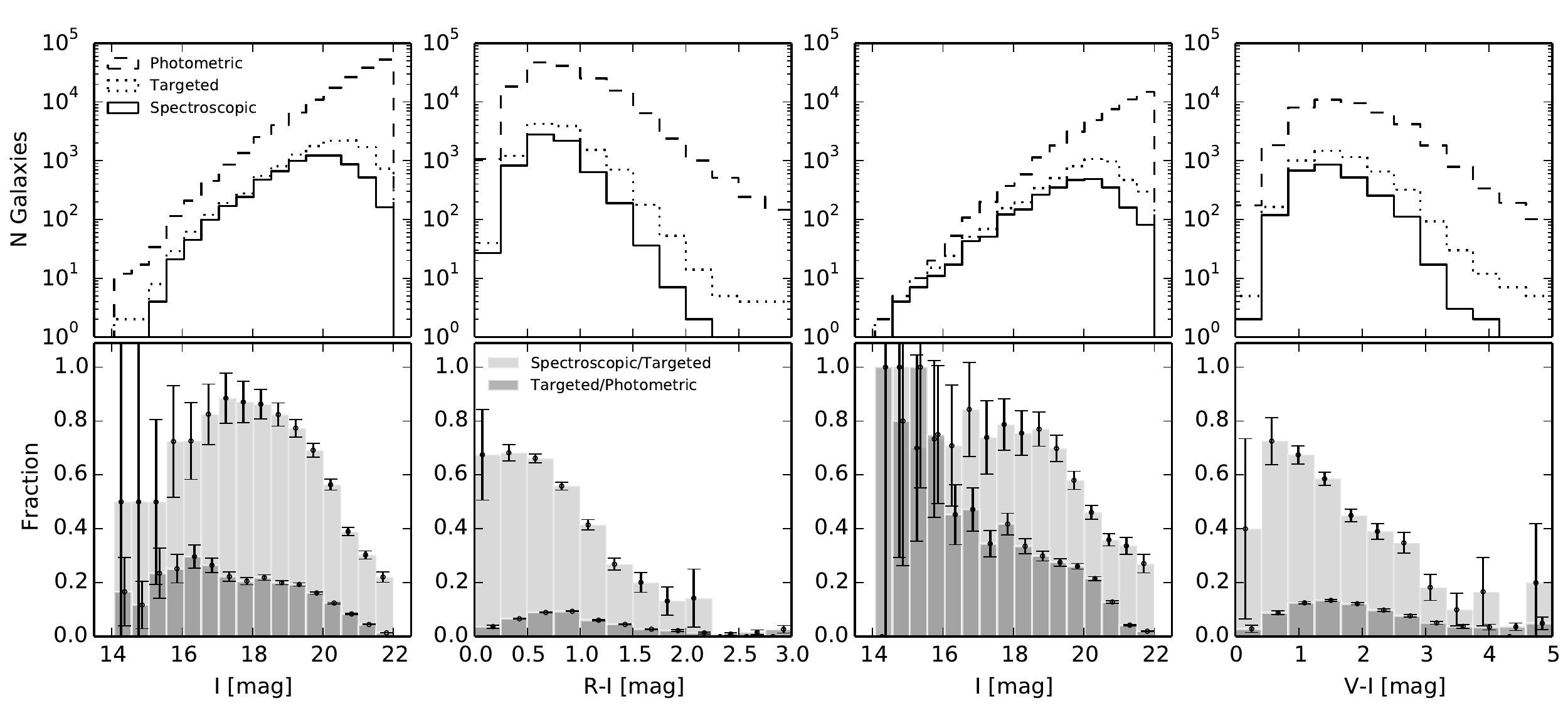}
\caption{\footnotesize {\it Top row:} Distribution of the galaxies in the photometric (dashed line), the target (dotted line), and the final spectroscopic redshift (solid line) catalogs as a function of magnitude (first and third column), and color (second and fourth column). The left two columns shows the fields with $R-I$ color and the right two columns shows the fields with $V-I$ color. The magnitude distributions of the spectroscopic catalogs mirror the distributions of the photometric catalogs well down to $I=20.0$. The color distributions of the galaxies in the three catalogs (bottom row) are also qualitatively similar, although discrepancies appear at the blue and the red end. {\it Bottom row:} Targeted completeness (dark grey) and spectroscopic success rate (light gray) histograms as a function of magnitude and color. The distributions in the bottom row are constructed as ratios of the distributions in the top row. Error bars reflect the Poisson errors in each bin. The error bars are slightly offset from the centers of the bins for clarity.
}
\end{figure*}

\begin{figure*}[ht]
\figurenum{9a}
\begin{tabular}{@{}c@{\hspace{-4.5mm}}@{\hspace{-4.5mm}}c@{\hspace{0mm}}}
  \includegraphics[width=0.52\textwidth]{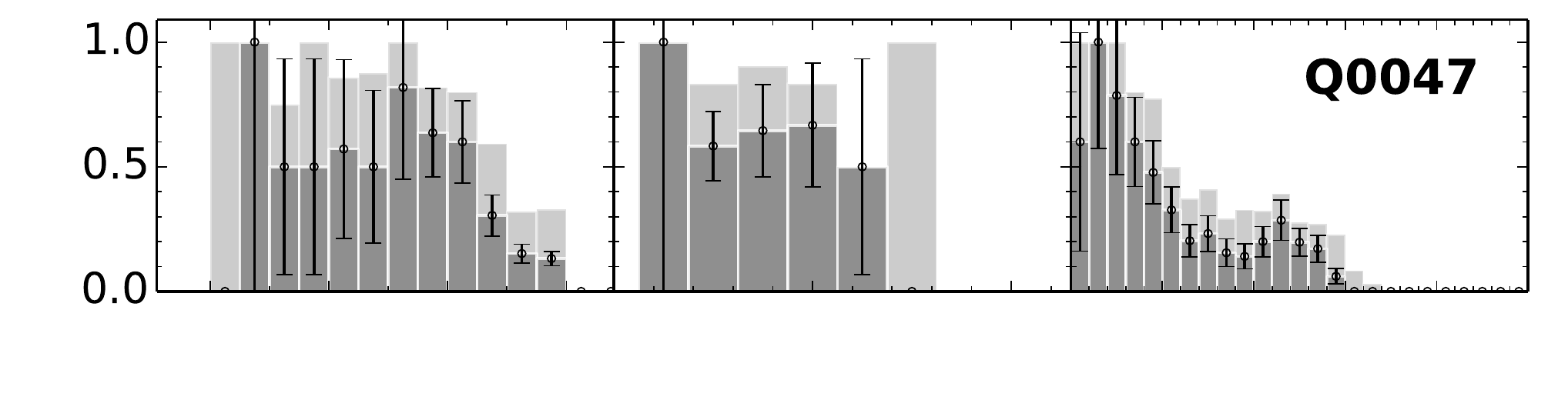} &    \includegraphics[width=0.52\textwidth]{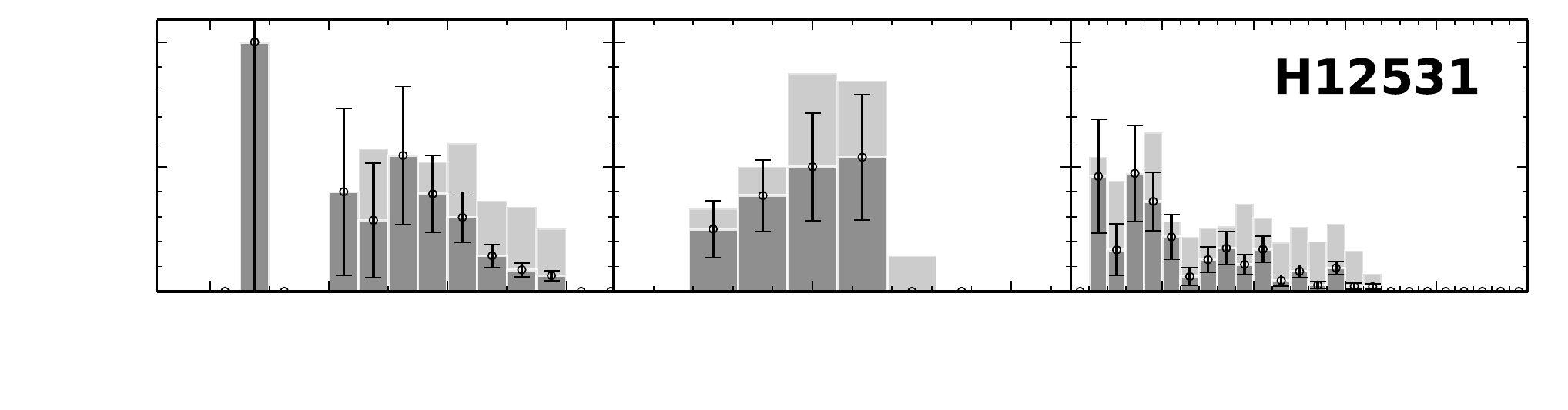} \\[-22pt]
  \includegraphics[width=0.52\textwidth]{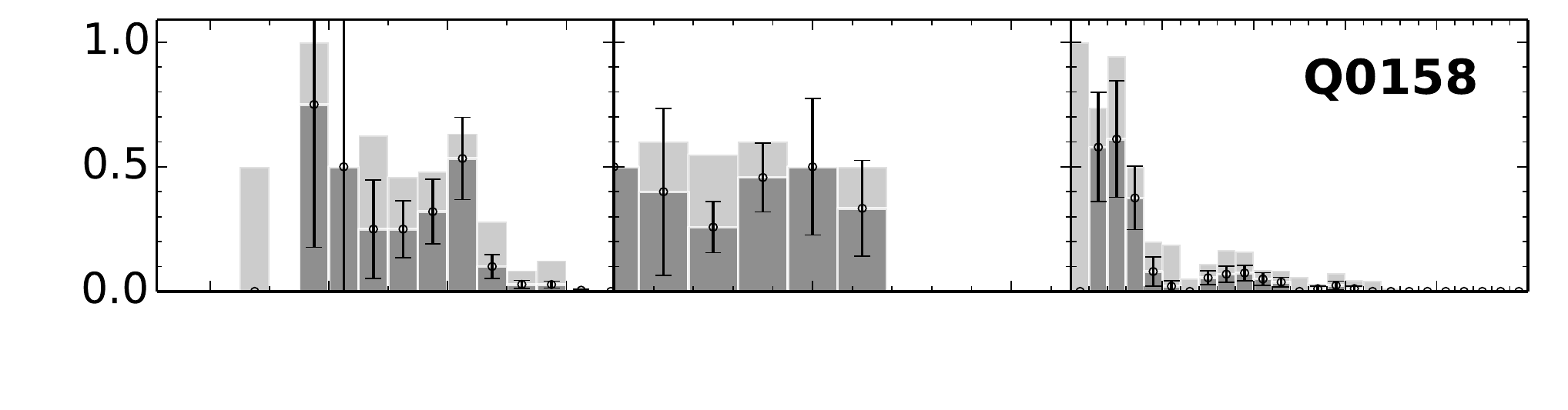} &     \includegraphics[width=0.52\textwidth]{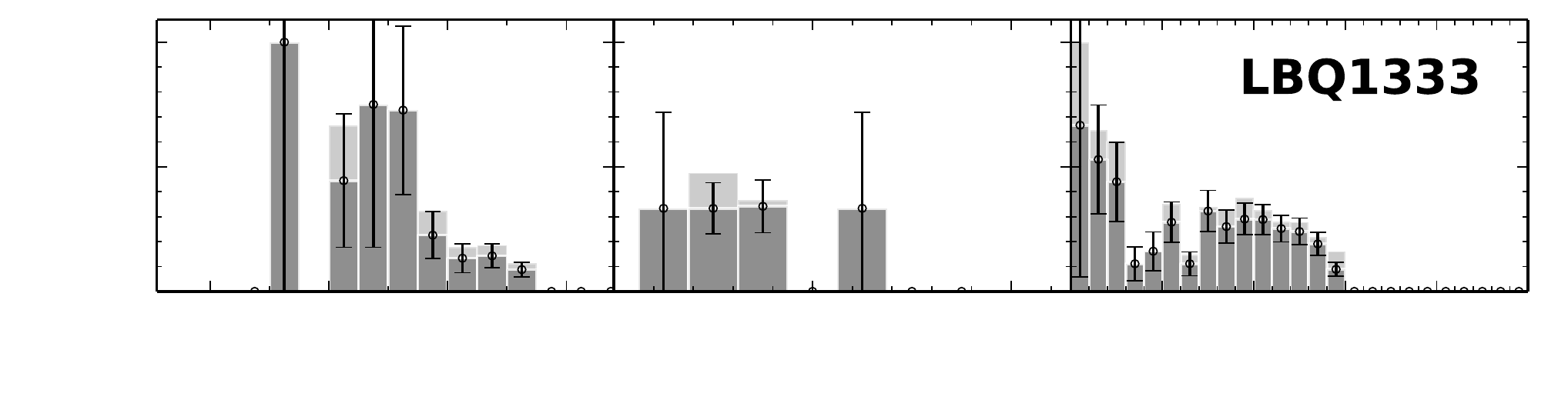} \\[-22pt]
  \includegraphics[width=0.52\textwidth]{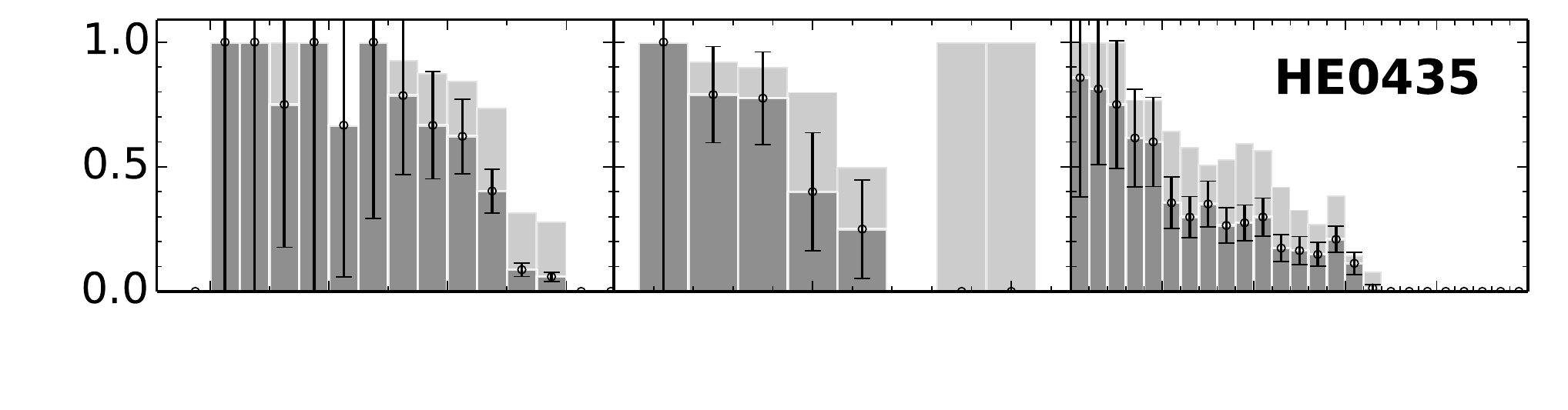} &      \includegraphics[width=0.52\textwidth]{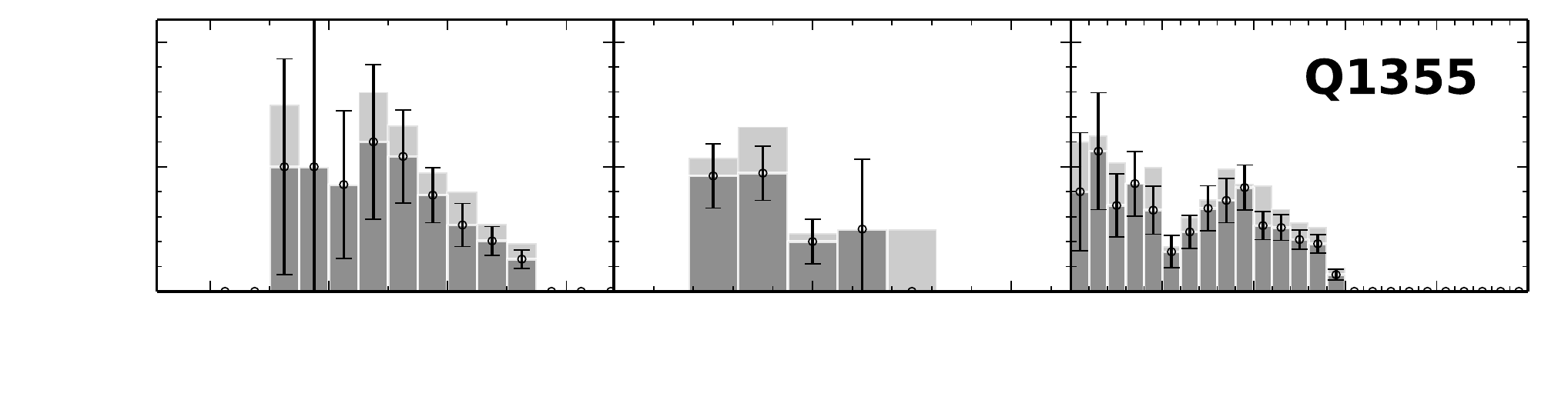} \\[-22pt]
  \includegraphics[width=0.52\textwidth]{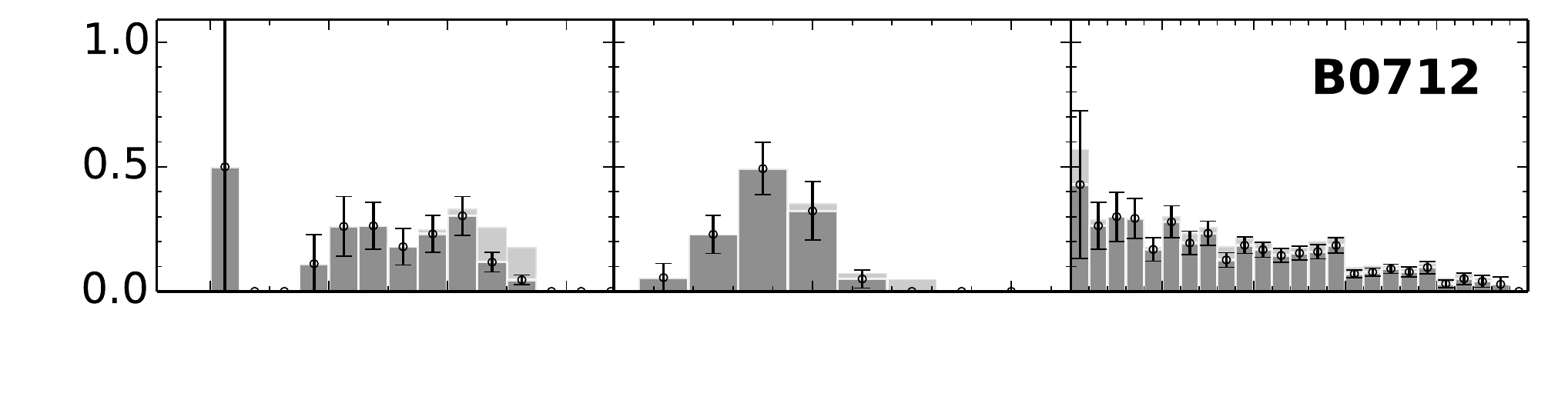} &   \includegraphics[width=0.52\textwidth]{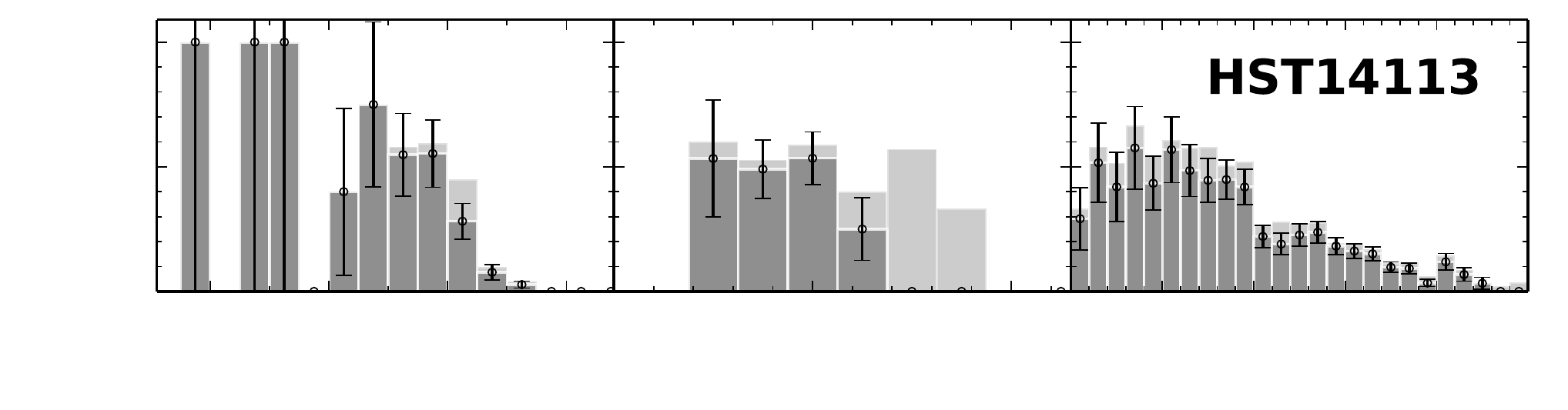} \\[-22pt]   
  \includegraphics[width=0.52\textwidth]{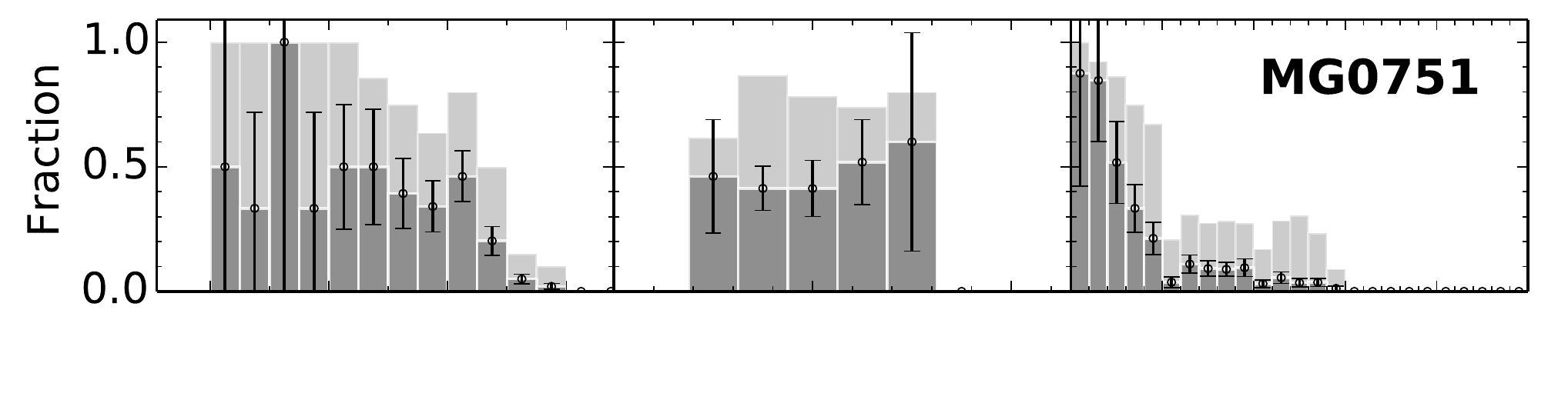} &   \includegraphics[width=0.52\textwidth]{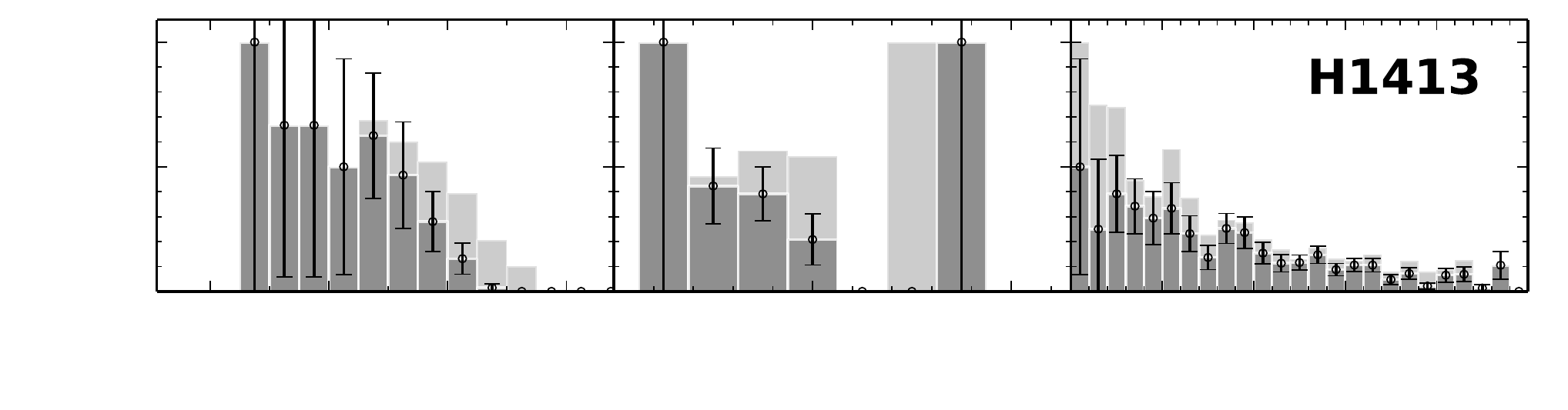} \\[-22pt]  
  \includegraphics[width=0.52\textwidth]{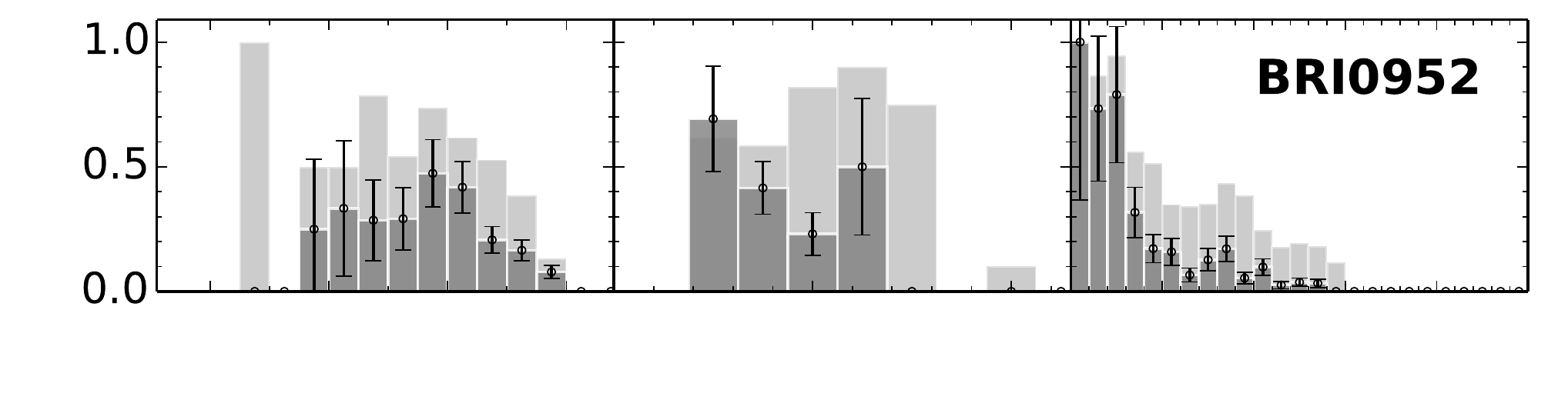} &   \includegraphics[width=0.52\textwidth]{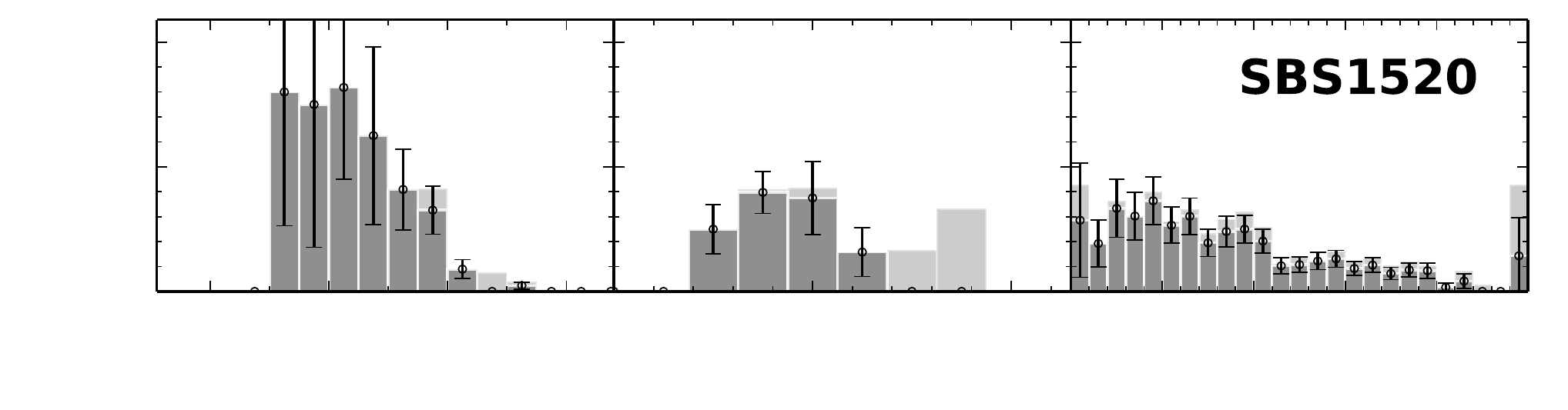} \\[-22pt]
 \includegraphics[width=0.52\textwidth]{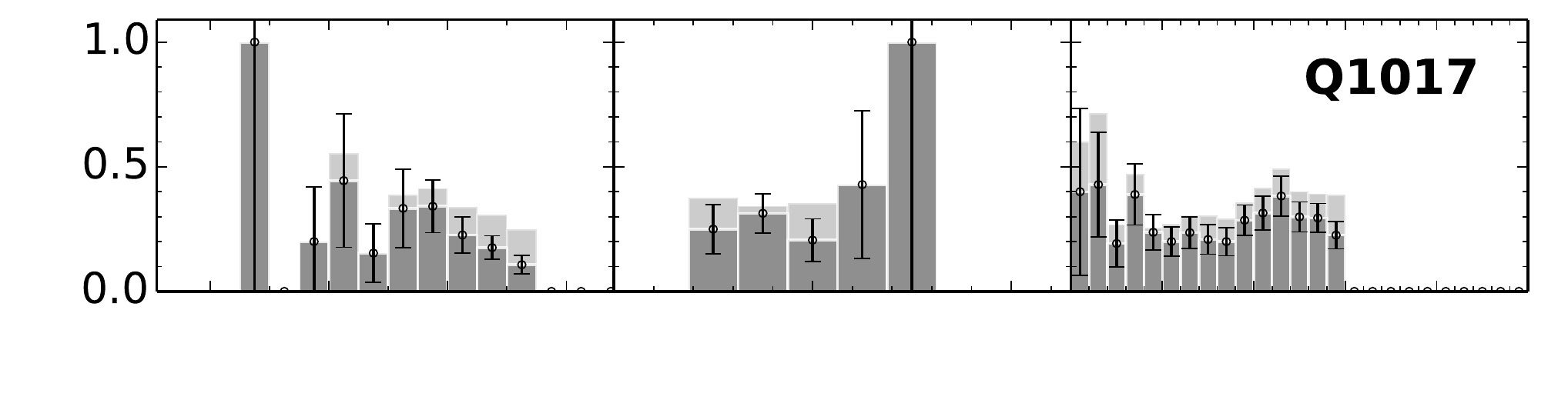} &   \includegraphics[width=0.52\textwidth]{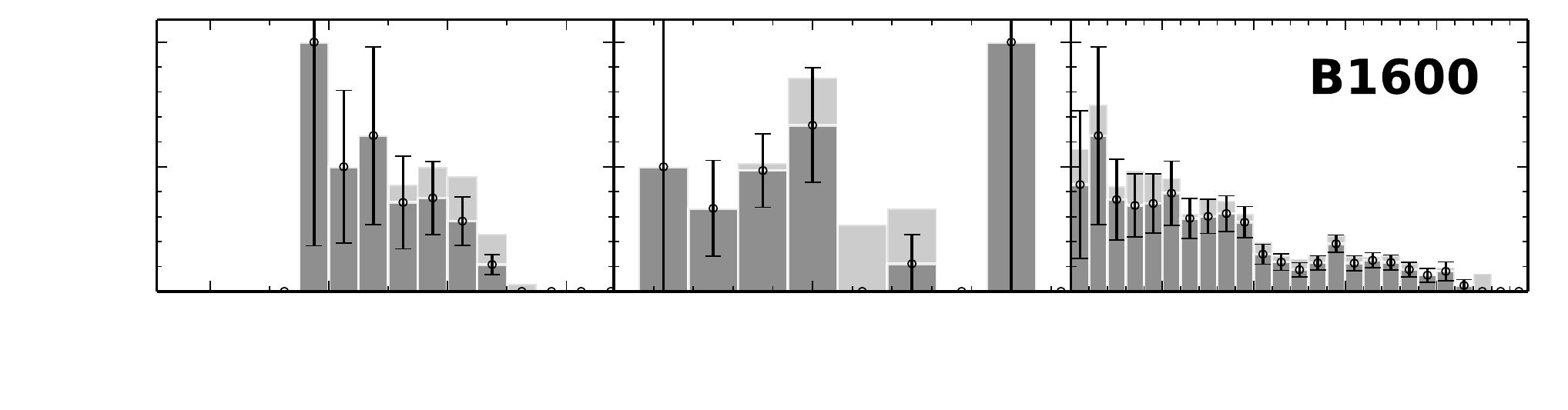}  \\[-22pt]
 \includegraphics[width=0.52\textwidth]{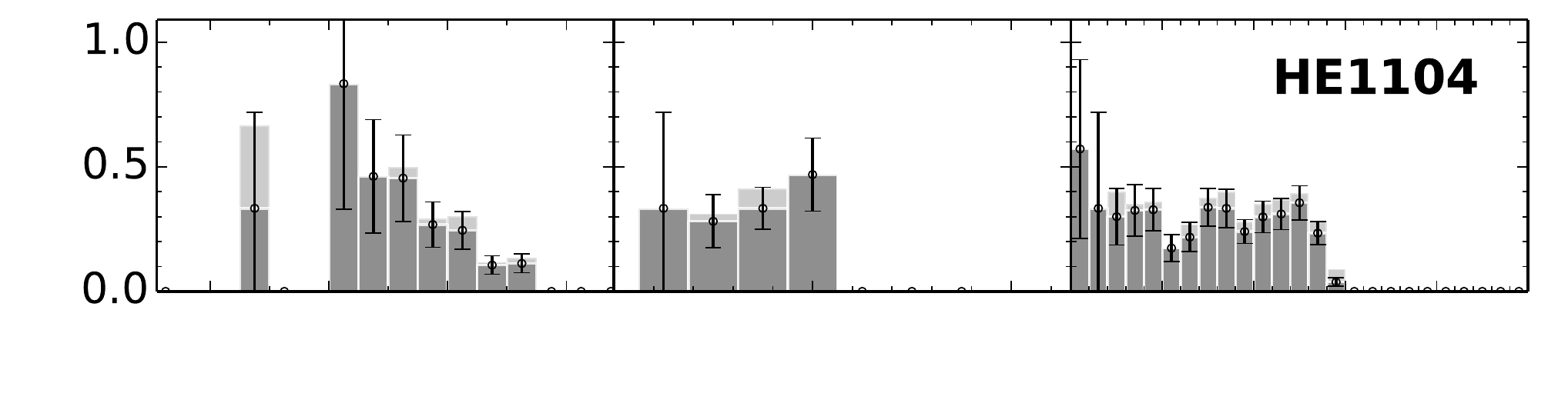} & \includegraphics[width=0.52\textwidth]{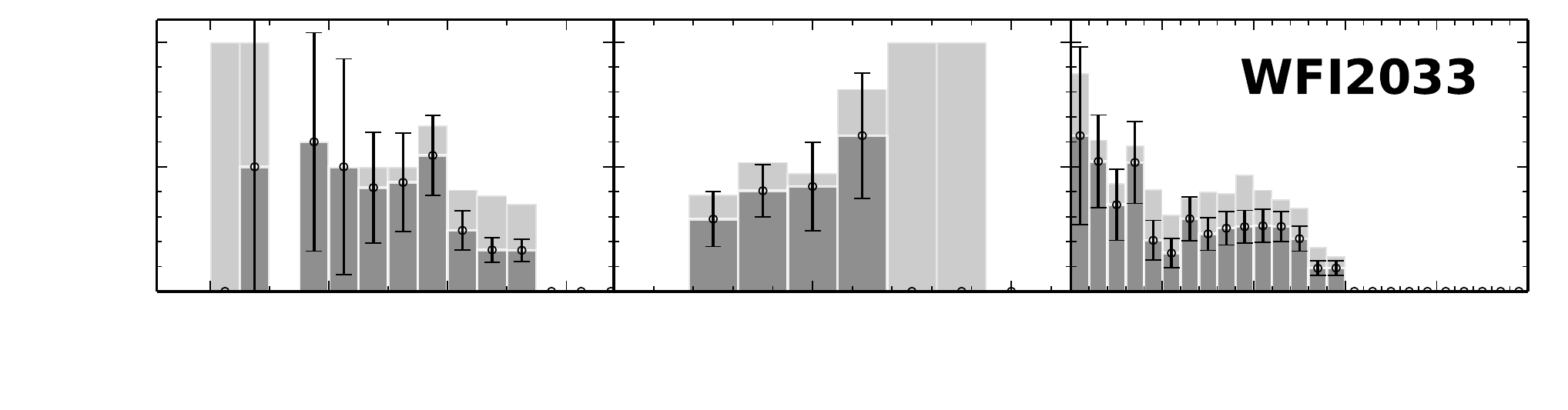} \\[-22pt]
 \includegraphics[width=0.52\textwidth]{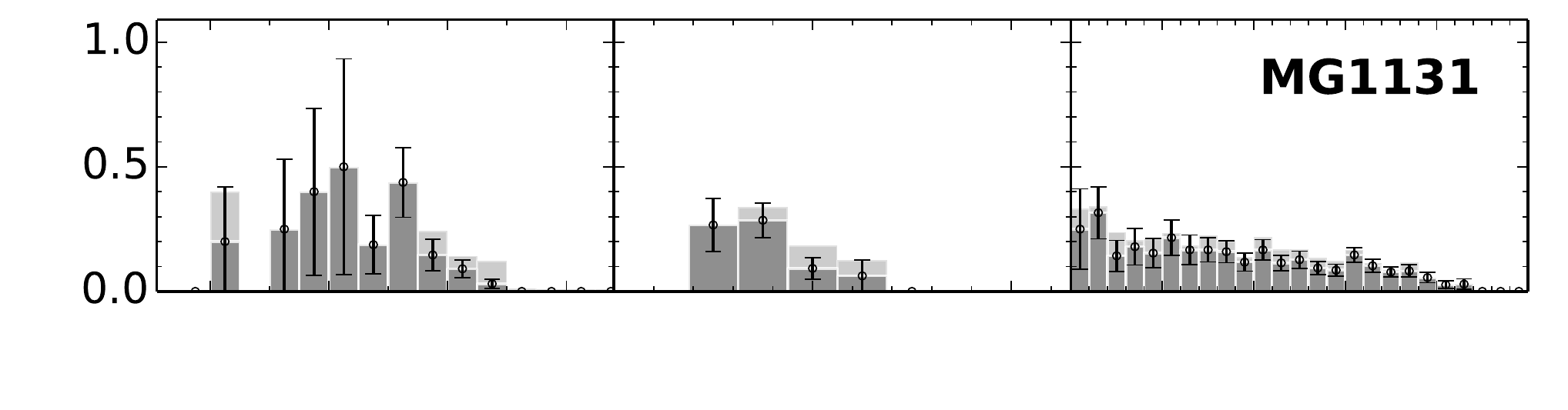} & \includegraphics[width=0.52\textwidth]{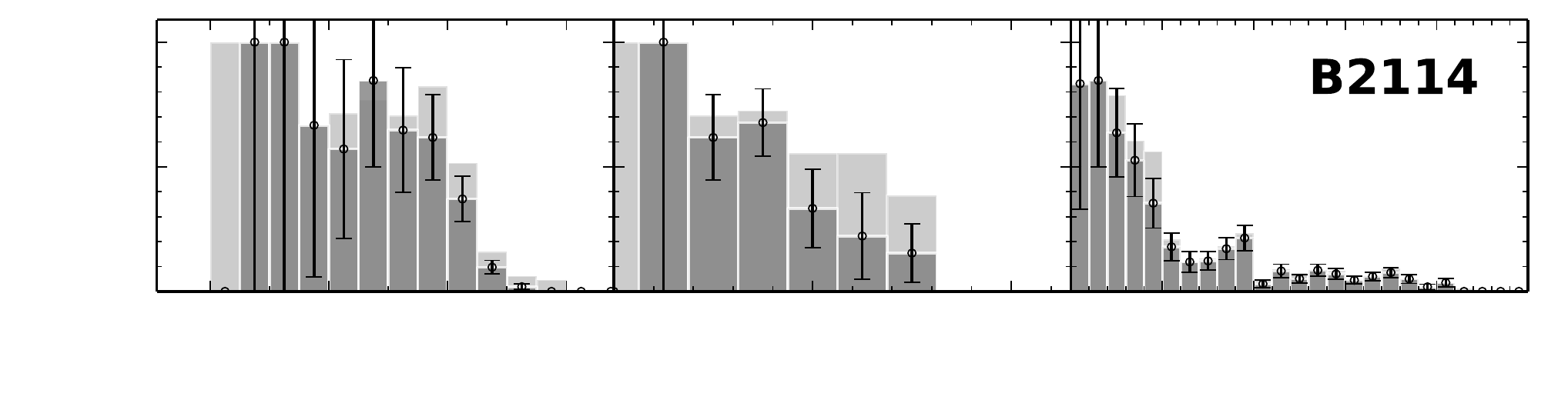} \\[-22pt]
  \includegraphics[width=0.52\textwidth]{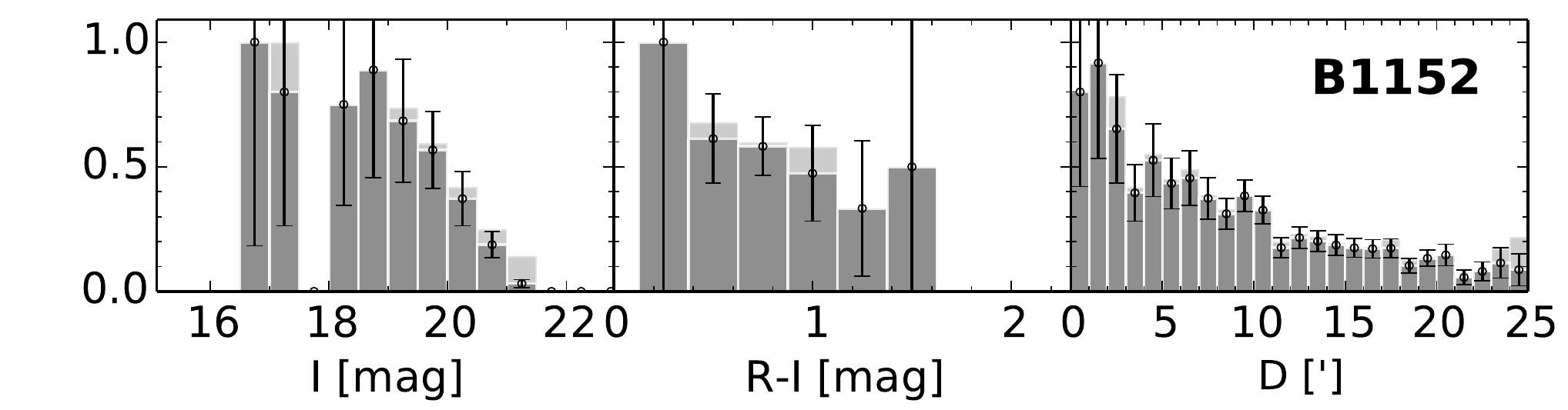} & \includegraphics[width=0.52\textwidth]{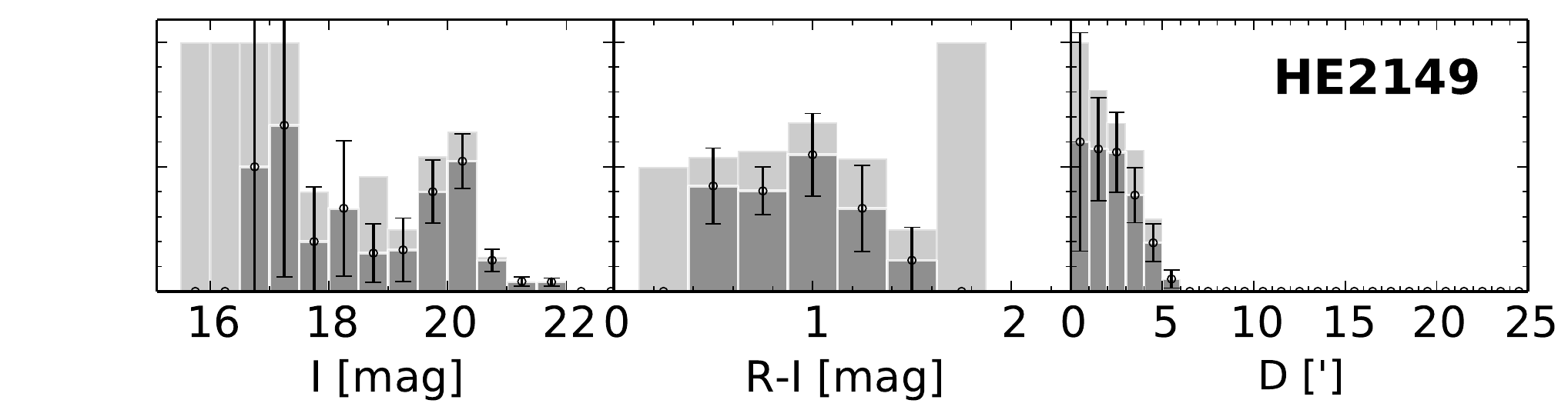} \\[-22pt]
\end{tabular}
\vspace{7mm}
\caption{\footnotesize Fraction of objects from the photometric catalog that we have targeted (light shaded histograms) and obtained redshifts for (dark shaded histograms) as a function of magnitude (within 5\arcmin, left), $R-I$ color (within 5\arcmin\ and brighter than $I=20.5$, center) and distance from the lens (brighter that $I=20.5$, right).The errors are based on Poisson statistics. Overall, the targeted and final spectroscopic sample mirror each other well. There are no systematic differences, implying that the spectroscopic sample is representative of the targeted population. The completeness relative to the photometric catalog is high, typically $>40\%$ within the central 5\arcmin\ of the lens and down to $I=20.5$. The completeness as a function of magnitude drops below $30\%$ at $I>20.5$ for all lenses, indicating the limitations of the spectroscopy. \label{complete1}}
\end{figure*}

\begin{figure*}[ht]
\figurenum{9b}
\begin{tabular}{@{}c@{\hspace{-4.5mm}}@{\hspace{-4.5mm}}c@{\hspace{0mm}}}
  \includegraphics[width=0.52\textwidth]{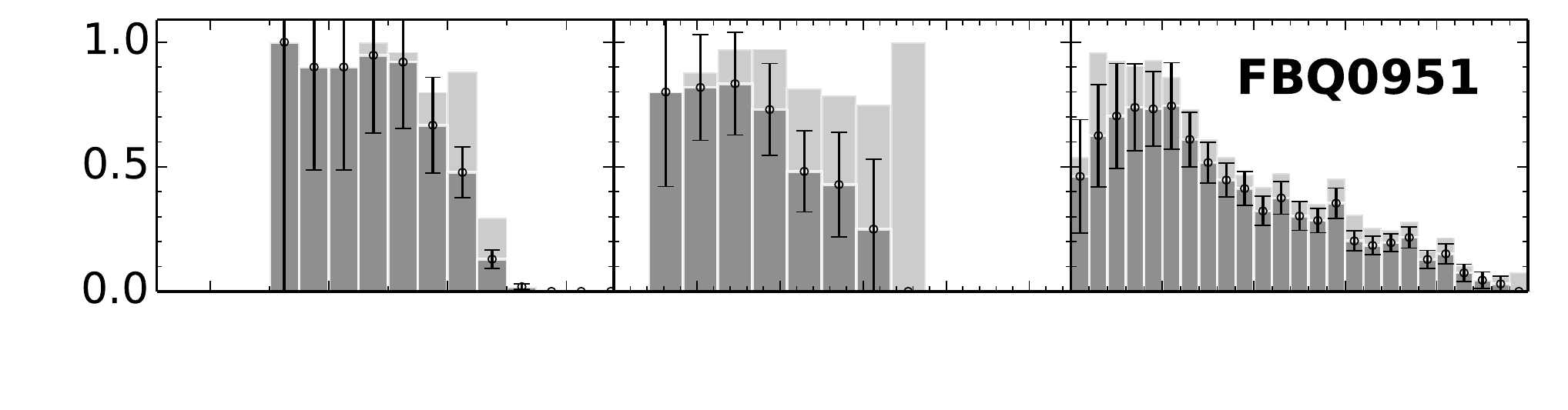} &  \includegraphics[width=0.52\textwidth]{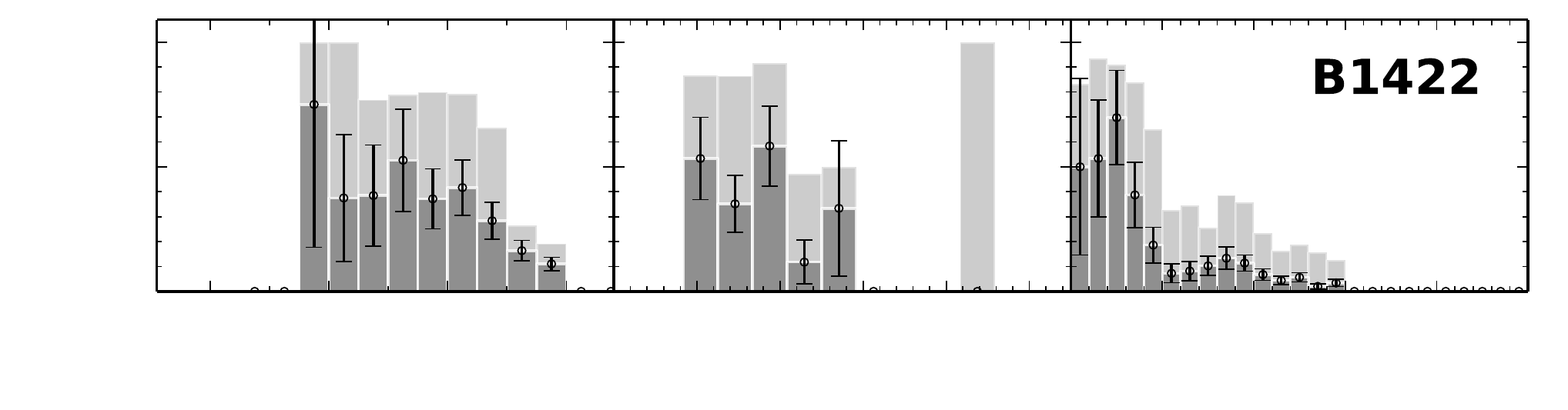} \\[-22pt]
 \includegraphics[width=0.52\textwidth]{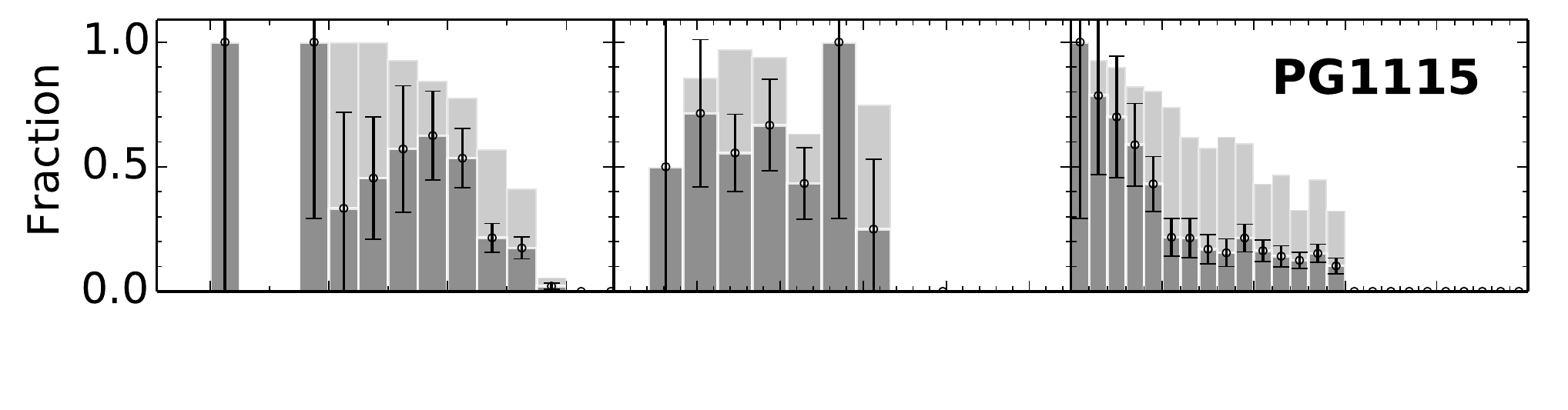} &   \includegraphics[width=0.52\textwidth]{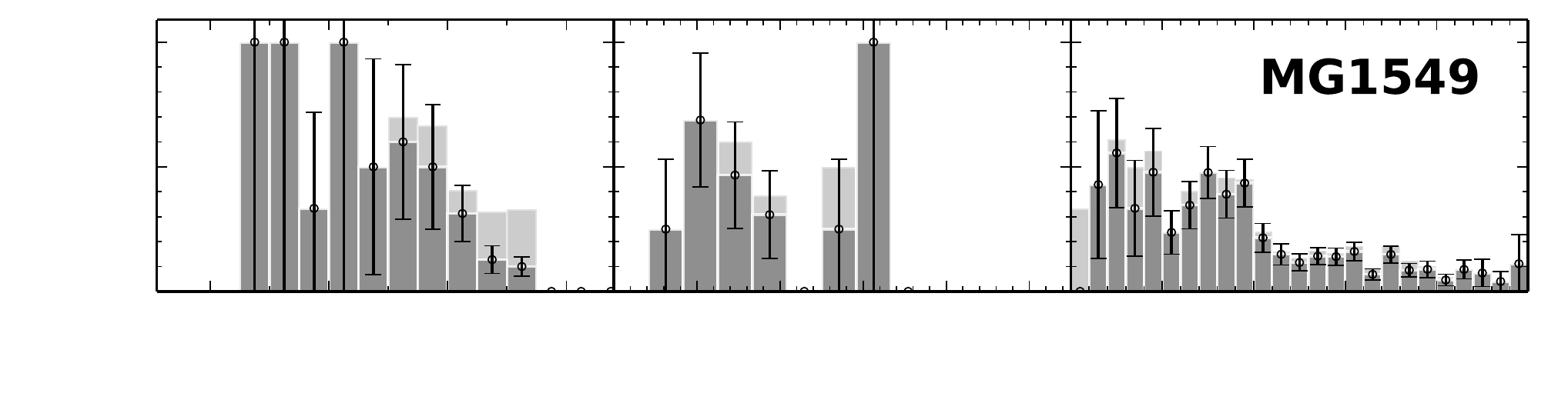} \\[-22pt]
 \includegraphics[width=0.52\textwidth]{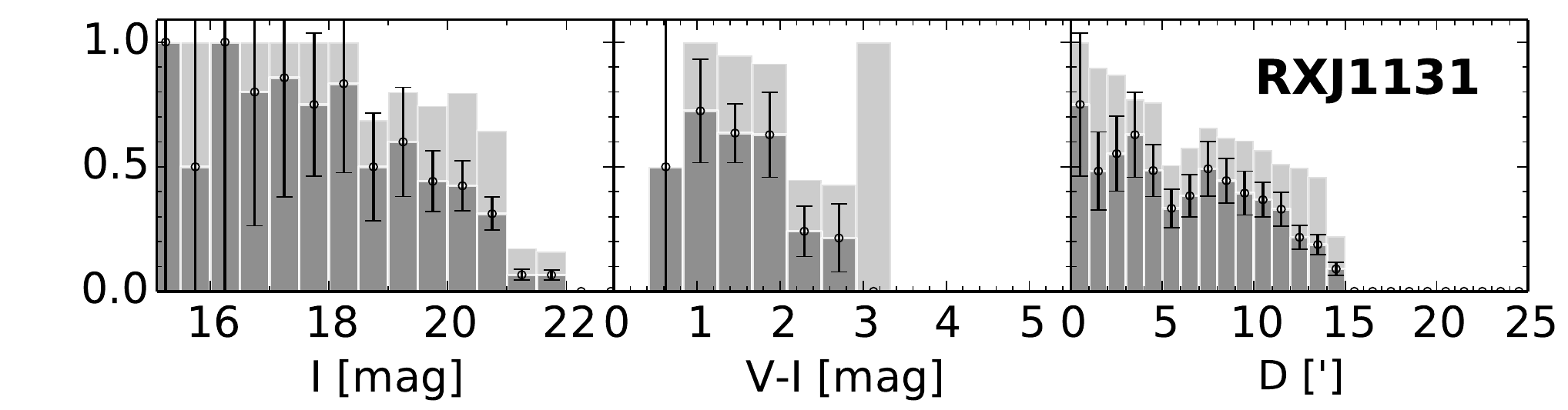} &  \includegraphics[width=0.52\textwidth]{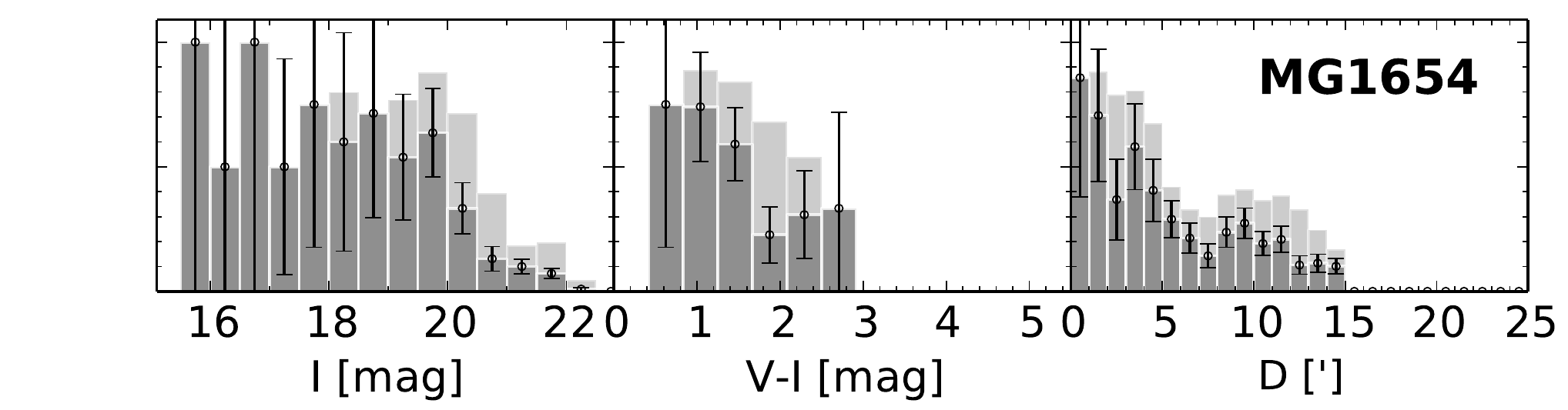} \\[-22pt]
\end{tabular}
\vspace{7mm}
\caption{\footnotesize Same as Figure \ref{complete1} for the fields with $V-I$ colors. \label{complete2}}
\end{figure*}

In this section we consider the completeness of the observations and their success rate. For this purpose, we compare the contents of three different catalogs: the master galaxy photometric catalog, the catalog of targeted objects and the list of objects with successful spectroscopic redshifts. Some targeted galaxies with measured redshifts are excluded from the completeness analysis in Figures \ref{gals_figure}, \ref{complete1} and \ref{complete2} because of poor photometry, usually due to defects in the detector or bleed trails from nearby stars ($\sim265$ objects or $\sim2.6\%$ of all redshifts).  In addition, objects classified as ``unresolved" in the final photometric catalogs (such as stars, compact galaxies, QSOs) are also excluded from the completeness analysis ($\sim250$ objects or $2.4\%$ of all redshifts). Figures  \ref{gals_figure}, \ref{complete1} and \ref{complete2} also do not include the 964 objects from NED.

Figure  \ref{gals_figure} demonstrates the distributions of objects in the spectroscopic, targeted and master photometric catalogs as a function of magnitude (first and third columns) and color ($I<22.0$, second and fourth columns), with only $R-I$ colors (left two columns) and only $V-I$ colors (right two columns). The top row of panels show the direct object distributions. The magnitude distribution of the spectroscopic catalog qualitatively mirrors the distribution of the photometric catalog down to $I=20.0$. The color distributions of the galaxies in the three catalogs are also qualitatively similar, although discrepancies appear at the red and blue ends. Very red and very blue galaxies were targeted at lower priority because they were likely to be at too high or too low redshift relative to the lens plane. 

The bottom row of Figure  \ref{gals_figure} shows the targeted completeness (dark grey) and the spectroscopic success rate (light grey). We define the targeted completeness as the fraction of galaxies from the full photometric catalog on which we placed slits/fibers and the spectroscopic success rate as the fraction of galaxies from the targeted catalog for which we have measured redshifts. The targeted completeness as a function of magnitude is uniformly $\sim20\%$ down to $I=20$ for the $R-I$ sample and then slowly decreases. The targeted completeness as a function of magnitude for the $V-I$ sample is higher: $70\%$ at $I<16$ and between 30 and $40\%$ down to $I=20$. This higher completeness is due to the fact that the six $V-I$ fields were typically observed during multiple observing runs (as many as six runs for MG1654).  Fainter than $I=20$, the targeted completeness decreases as we approach the observed limiting magnitude. The targeted completeness as function of color (for $I<22.0$) is lower at the blue and red ends of the color distributions for both samples. This is a reflection of the targeting priorities, which gave higher weight to targets with colors similar to suspected red sequences, therefore avoiding galaxies with very blue and very red colors which are at very low or high redshift, respectively. 

The success rates for obtaining spectroscopic redshifts (Figure \ref{gals_figure}, bottom row, light gray histograms, first and third panel) as a function of magnitude is $80\%$ down to $I=19$,  decreases to $60\%$ at $I=20$ and declines further to $40\%$ at $I=21$, demonstrating the increasing difficulty of acquiring redshifts at fainter magnitudes. The success rate as a function of color (light gray histograms, second and fourth panel, bottom row of Figure \ref{gals_figure}) is $\sim60\%$ at the blue end of the color distribution and progressively decreases towards the red end. Such a difference is likely to arise from our differing ability to obtain redshifts for different types of galaxies. We are more successful at measuring redshifts for objects with strong emission lines (which tend to be blue), than for objects with only continuum spectra (which tend to be red). 

We now look at the spectroscopic catalog on a field-by-field basis. For each of the 26 fields, Figures~\ref{complete1} and \ref{complete2} present the completeness of the spectroscopic (shaded histogram) and targeted (open histogram) catalogs. The targeted completeness is defined as above. The spectroscopic completeness is defined as the fraction of galaxies from the full photometric catalog for which we have measured redshifts. The general level of completeness varies from field to field as a result of our variable sampling. Fields that have been observed several times and for which we have obtained high quality data are better sampled. Here we explore whether any biases have been introduced in our sample as a result of the variable completeness.

For each field the first columns of Figures \ref{complete1} and \ref{complete2} show the targeted and the spectroscopic completeness as a function of I magnitude within 5$\arcmin$\ of the lens. We reach a relatively constant $\gtrsim40$\% completeness down to $I=20.5$ in Q0047, HE0435, MG0751, FBQ0951, BRI0952, PG1115, RXJ1131, B1152, B1422, MG1654, and B2114. Most fields show a clear trend with decreasing completeness towards fainter magnitudes; however, for Q0158, B0712, BRI0952, Q1017, MG1131, H12531 and HE2149 these distributions are rather flat, indicating that completeness is low even at fairly bright magnitudes. The completeness is generally lower for fields that were targeted only during one observing run. If groups are identified in these fields, further follow-up may be beneficial to determine the properties of such structures.

The middle columns of Figures \ref{complete1} and \ref{complete2} show the spectroscopic and targeted completeness as a function of color for objects within 5$\arcmin$\ of the lens galaxy and with $I\le20.5$. While we did apply a color-based priority when designing the multi-object configurations, this selection was very broad and not strict (galaxies outside the prioritized color region were not discarded as targets but just given lower priority). Thus we do not expect to have strong color biases in the targeted completeness and this is manifested in Figures \ref{complete1} and \ref{complete2}. The spectroscopic completeness, however, is skewed towards bluer objects due to our higher success rate of obtaining redshifts for blue galaxies as discussed above. For the individual fields, however, the effect is not as obvious. The colors used in the target selection for the B0712 field were initially incorrect, but we expect that the color selection to be unbiased. The resulting color distribution is peaked at $(R-I)\sim0.7$, likely reflecting the large structure in this field (Figure 11c).

The last columns in Figures \ref{complete1} and \ref{complete2} present the completeness as a function of distance from the lens for objects with $I\le20.5$. Here we see the strongest bias inherent from our target selection; the completeness is highest in the vicinity of the lens galaxy. This effect is less noticeable for fields that have been targeted using fewer masks and/or during a single observing run, such as B0712, MG1131, H1413, SBS1520, MG1549 and B1600 (single run Hectospec observations), Q1017, HE1104, Q1355 (single run IMACS observations). The effect is most pronounced for fields targeted with LDSS-2 and LDSS-3, as the smaller field-of-view of these instruments and our requirement that the lens be within each slit mask results in a denser sampling near the lens. This bias is in fact useful because we expect that the largest shear is caused by structures with a small impact parameter relative to the lens \citep{my}. Therefore the high completeness in the center of the field will allow us to identify sparsely populated structures at small impact parameter, while larger structures can still be discovered at large impact parameters even with sparser coverage. At large radius we notice the effects of the instrument field-of-view: fields observed with IMACS have coverage out to 15\arcmin\ from the lens (FOV: 15\arcmin\ radius), while fields observed with Hectospec have coverage out to the extent of our MOSAIC imaging,  $\sim$25\arcmin\ from the lens (FOV: 30\arcmin\ radius).

\section{Summary}

We have presented a detailed description of the data acquisition and reduction, as well as the redshift catalog, for a spectroscopic survey of the environments and lines of sight of 28 galaxy-mass strong gravitational lenses. The spectroscopic catalog consists of 9768 unique new redshifts acquired over the course of 15 observing runs utilizing more than 40 nights of 6.5-meter telescope time and four different instruments. We test the redshifts both internally and against external redshift measurements to ascertain that no biases are introduced when the different data-sets are combined. The redshift uncertainties are determined based on the scatter in repeat redshift measurements, and they range from $1.3\times10^{-4}$ to $4.3\times10^{-4}$ (i.e., $c\Delta z= 40$ to 130 \kms), depending on the spectral properties of the object. The final spectroscopic catalog includes J2000.0 coordinates, redshifts, and redshift errors. Many structures are visible in the redshift distributions along the lines of sight to these lenses. These structures will be discussed in upcoming work.

\acknowledgements
We thank the staff of the Las Campanas and MMT observatories for their tireless effort on behalf of this project. We are particularly grateful to the instrument specialists, telescope and robot operators and the computer support staff at both facilities. We are also grateful to Nelson Caldwell and Daniel Fabricant for their help with the Hectospec instrument in its early days.  We thank Dan Kelson and Gregory Walth for their assistance with the IMACS data reduction, and Andy Marble for the optimal extraction code. We thank George Rieke, Chris Impey, Romeel Dave, Ben Weiner, Michael Cooper, Yujin Yang and Ken Wong for helpful suggestions and feedback along the way.

I.M., K.A.W and A.I.Z. acknowledge support from NSF grant AST 02-06084. C. K. acknowledges support from NSF through grant AST-0747311. I. M. also acknowledges the support of the Martin F. McCarthy Scholarship in Astrophysics awarded by the Vatican Observatory. K.A.W. also acknowledges the support of NSF Astronomy \& Astrophysics Postdoctoral Fellowship AST-0602288.

{\it Facilities:} Magellan-1 (LDSS-2, LDSS-3), Magellan-2 (IMACS), MMT (Hectospec), Mayall (Mosaic-1), Blanco (Mosaic-II)

\appendix

\section{Notes on Individual Systems}\label{app1}

{\bf Q ER 0047-2808} (herein Q0047) was discovered serendipitously by \citet{warren1} as part of a spectroscopic survey of early type galaxies that included the lens: a massive early type galaxy at $z_{l}=0.485$. Q0047 was the first optical Einstein ring discovered. \citet{warren2} confirmed that the source is  a $z_{s}=3.595$ highly star-forming galaxy and measured the ring radius to be $r=1.35\arcsec$. There is no prior observational work on the environment of the lens galaxy. Based on non-parametric lens models, \citet{wayth} conclude that the SIS+external shear models are ruled out by the data.

{\bf Q J0158-4325} ({\it a.k.a.} CTQ 0414, herein Q0158) is a doubly imaged $z_{s}=1.294$ quasar lens with image separation 1.2\arcsec~ \citep{morgan99, faure09}. The lens redshift is $z_{l}=0.317\pm0.001$ \citep{faure09}.  The lens galaxy spectrum and light distribution are consistent with it being an elliptical galaxy. An over-density of galaxies is found at a photometric redshift $z=0.5\pm0.1$ by \citet{faure04}. 

{\bf HE0435-1223} (herein HE0435) was discovered as part of the Hamburg/ESO (HES) survey of bright quasars \citep{wisotzki1} as a $z=1.689$ QSO. A follow-up high resolution image with the 6.5 m Baade/Magellan I telescope revealed the four lensed images of the quasar \citep{wisotzki2} in a configuration resembling the Einstein Cross lens Q2237+0305 \citep{huchra}. The image separations are 2.3\arcsec~ (B-D) and 2.6\arcsec~ (A-C). The lens has a spectrum and a spatial profile characteristic of an early type galaxy. \citet{morgan} measure the lens redshift $z_l=0.4546\pm0.0002$. \textit{HST} ACS observations \citep{morgan} reveal a spiral rich group of galaxies within 40\arcsec~ of the lens, but their 18 measured redshifts do not reveal a coherent structure. An SIS+shear lens model fits the lens well and requires $\gamma=0.074$ at $\phi_{\gamma}=76.5\deg$. HE0435 is one of the twelve time delay lenses in this sample with $\Delta t_{BA} = 8.4\pm2.1$ days, $\Delta t_{BC} = 7.8\pm0.8$ days and $\Delta t_{BD} = -6.5\pm0.7$ days \citep{he0435-td}. %$\Delta t_{AD}=-14.37^{+0.75}_{-0.75}$ days, $\Delta t_{AB}=-8.00^{+0.73}_{-0.82}$ days and $\Delta t_{AC}=-2.10^{+0.78}_{-0.71}$ days \citep{kochanek1}.

{\bf CLASS B0712+472} (herein B0712) was discovered as part of the JVAS/CLASS \citep{patnaik92b, jackson1995} survey of flat spectrum radio sources. Follow-up imaging revealed a quadruply imaged $z_s\sim1.33$ quasar and a $V=22.2$ early-type lensing galaxy  \citep{jackson1998}. The maximum image separation is 1.27\arcsec. \citet{fassnacht1998} confirm $z_l=0.4060$ and $z_s=1.339$. B0712 is a flux anomaly lens with the major flux density  discrepancy involving the B and D images. \citet{jackson1998, jackson2000} suggest that while the D image discrepancy is probably due to reddening, the B image discrepancy is most likely caused by microlensing. \citet{fassnacht2002} find a foreground group  at $z=0.2909$ \citep[confirmed by ][]{fassnacht08}, spatially coincident with the lens and measure its velocity dispersion $\sigma=306^{+110}_{-58}$ \kms. The shear due to this foreground group is expected to be small: $\gamma=0.03$ to $0.05$ \citep{keeton1998, fassnacht2002}.

{\bf MG0751+2716} (herein MG0571) was discovered as a part of the MIT-Greenbank-VLA survey for strong gravitational lenses. It consists of four images of a $z_s=3.200\pm0.001$ quasar and a partial ring \citep{lehar1993, tonry1999}. The lens galaxy is a $R=21.3$ early type galaxy \citep{lehar1997} at $z_l=0.3502\pm0.0003$ \citep{tonry1999} and is a satellite of a nearby massive elliptical \citep[$R=19.1$, $z=0.3501\pm0.0003$, ][]{ tonry1999}. \citet{tonry1999} also identify a third member of the group: a nearby emission line galaxy at $z=0.3505\pm0.00003$. Lens models \citep{lehar1997} suggest that MG0751 requires more external shear, which may be due to its complex environment. This hypothesis is confirmed by \citet{my}, who identify another 10 members of the group and determine its velocity dispersion $\sigma=320^{+170}_{-110} km\;s^{-1}$. \citet{kurtis} also identify a background red sequence at $z_{RS}=0.48$ consistent with an under-sampled $z=0.5605$ group with 5 members and $\sigma\sim550$ \kms~ found by \citet{my}.

{\bf FBQS J0951+2635} (herein FBQ0951) was the first strong lens discovered by the FBQS \citep[FIRST Bright Quasar Survey, ][]{gregg}. It was originally identified as a $B\sim 16.9$, $z_s=1.24$ quasar, but high-resolution follow-up imaging and spectroscopy \citep{schechter98} revealed two images of the same background quasar, separated by $1.1\arcsec$. \citet{cosmograil} determine the lens redshift $z_l=0.26\pm0.002$. This is coincident with the redshift $z_{RS}=0.27$ of one of the three groups identified by \citet{kurtis} along the line of sight of FBQ0951, the other two red sequences being at $z_{RS}=0.16$ and $z_{RS}=0.43$. \citet{jakobsson} measure the time delay between the two images: $\Delta t = 16\pm2$ days.

{\bf BRI0952-0115} (herein BRI0952) is a doubly imaged $z_s=4.5$ optical quasar discovered by \citet{irwin}. The image separation is $0.9\arcsec$. The quasar was also detected at millimeter wavelengths \citep{omont}. \citet{keeton1998} find that the lens is a flattened early-type galaxy. \citet{my} suggest that the lens galaxy may be associated with a $z_{g}=0.42$ group of 5 galaxies with $\sigma=170^{+150}_{-100}$. This suggestion is rejected by \citet{cosmograil}, who determine the lens redshift to be $z_l=0.632\pm0.002$.

{\bf Q J1017-207} ({\it a.k.a.} CTS J03.13, herein Q1017) was discovered by \citet{claeskens} by careful analysis of the optical images for selected highly luminous quasars, which reveled that the J03.13 quasar consisted of at least two separate components. Further observations \citep{surdej} suggested that the two sources were images of the same $z_s=2.545$ quasar separated by $0.849\pm0.001\arcsec$. ~\citet{lehar2000} detect the galaxy in \textit{HST} NICMOS imaging. \citet{kochanek2000} estimate the
lens galaxy redshift to be $z_l = 0.78^{+0.09}_{-0.05}$ based on Fundamental Plane
fitting. A reliable spectroscopic redshift is not available, but \citet{ofek} estimate $z_{l}=1.088\pm0.001$ based on a Mg II absorption line in the quasar spectra.

{\bf HE1104-1805} (herein HE1104) was serendipitously discovered by \citet{wisotzki1993} as a doubly imaged radio-quiet quasar at $z_{s}=2.319$ with image separation 3\arcsec. The lens redshift $z_l=0.729\pm0.001$ was measured by \citet{lidman}. Based on its spectrum and optical colors, the galaxy is likely an elliptical. HE1104 is unusual because the brighter image is closer to the lens galaxy, and lens models of the system require fairly large shear $\gamma\sim0.125$ to $0.142$. The time delay $\Delta t = 161\pm7$ days was measured by \citet{ofek2}, who also detected residual variation attributed to microlensing.  \citet{faure04} find an overdensity of galaxies in the field of H1104, which, based on the photometric redshifts, might be associated with the background quasar.

{\bf PG 1115+080} (herein PG1115) is a $z_{s}=1.722$ radio-quiet quasar lensed into four images by a $z=0.3098\pm0.0002$ elliptical galaxy \citep{weymann, kundic97, tonry98}. The presence of a small group associated with the lens is suggested by \citet{young} and confirmed by \citet{kundic97} and \citet{tonry98}, who measure the redshifts of 4 galaxies within 20\arcsec~ of the lens and estimate $\sigma=270\pm70$ \kms~ and $\sigma=326$ \kms, respectively. \citet{grant} and \citet{fassnacht08} detect diffuse X-ray emission associated with the group with temperature $kT\sim0.8\pm0.2$ keV. Time delays were measured by \citet{schechter97} and improved by \citet{barkana}: $\Delta t_{BC}=25.0^{+3.3}_{-3.8}$ days and $\Delta t_{AC}/\Delta t_{BA} = 1.13^{+0.18}_{-0.17}$ and more recently confirmed by \citet{artamonov11}, despite contradicting reports \citep{vakulik09,tsvetkova10}. PG1115 has anomalous flux ratios that are probably due to microlensing \citep{chiba05}.

{\bf MG1131+0456} (herein MG1131) was observed with the VLA as part of the MIT -- Green Bank (MG) survey. Even a short exposure revealed its unusual ring-like morphology. Follow-up observations \citep{hewit, chen} identified it as a radio Einstein ring with two compact, embedded sources and revealed an even more complex morphology: a radio jet lensed into an elliptical ring, the radio core doubly-imaged, and the oposite radio jet unlensed. In the optical \citep{kochanek2000a}, there is an incomplete ring image of the AGN host galaxy, while in H-band the ring image of the host galaxy is complete. Ground-based optical and IR  imaging \citep{larkin} showed that both the lens and source were extremely red. Based on broad band colors, \citet{hammer} suggest tentative lens and source redshifts: $z_l\sim0.85$ and $z_s\sim1.13$. \textit{HST} imaging shows that the lens is an early type galaxy, while the source appears to be an extremely red object at $z_s\sim2$ \citep{kochanek2000a}. \citet{tonry1131} measured the lens redshift $z_l=0.8440\pm0.0005$. \citet{larkin} notice a significant excess of objects within 20\arcsec~ of the lens and suggest that they might complicate the lens potential. Based on the colors of the galaxies in the field, \citet{kochanek2000a} suggest that the lens belongs to a group of at least seven galaxies. \citet{tonry1131} measure redshifts for 3 galaxies in the field and find evidence for a foreground group at $z=0.343$ with $\sigma=232$ \kms.

{\bf RXJ1131-1231} (herein RXJ1131) is a four-image lens with an Einstein ring discovered serendipitously by \citet{sluse}. The source and lens redshifts are $z_s=0.658\pm0.001$ and $z_l=0.295\pm0.002$. The lens galaxy has a characteristic early-type spectrum. RXJ1131 has anomalous flux ratios by a factor of two in the optical and three to nine in the X-ray \citep{blackburn}, possibly explained by substructure \citep{morgan2006}. RXJ1131 is also a time delay lens with measured delays of $t_{AB}=0.7\pm1.4$ days, $t_{CB}=-0.4\pm2.0$ days, and $t_{DA}=91.4\pm1.5$ days \citep{morgan2006,tewes13}. \citet{suyu13} measure a velocity dispersion for the lens: $\sigma = 323\pm20$ \kms. There is evidence for significant structure along the line of sight of this lens: a group at $z\sim0.1$ and possibly a group associated with the lens \citep{morgan2006, kurtis}. Both structures are detected as X-ray sources with luminosities $3.1\pm0.5\times10^{43}$ and $2\times10^{43}$ ergs s$^{-1}$ respectively \citep{morgan2006, sluse07, sluse08}.  SIS$+\gamma$ models require a large shear $>0.1$ \citep{morgan2006}. A detailed model of this system is presented in \citet{suyu13}.

{\bf CLASS B1152+200} (herein B1152) is a two-image gravitational lens discovered in the Cosmic Lens All-Sky Survey \citep[CLASS, ][]{browne03, myers03}. A $z_s=1.0189\pm0.0004$ quasar is lensed by a $z_l=0.4386\pm0.0008$ galaxy. The lens galaxy spectrum shows prominent [OII] emission, suggesting that the lens is a late-type galaxy. The image separation is 1.6\arcsec. The lens galaxy is faint and difficult to de-convolve from the lensed images, thus little is known about the lens itself and lens models are very unconstrained. B1152 is a good time delay candidate with an expected time delay $\Delta t=32\pm4\;h^{-1}$ days \citep{munoz}, but it is yet to be measured.

{\bf HST J12531-2914} (herein HST12531) was discovered serendipitously in \textit{HST} WFPC2 observations as part of the Medium Deep Survey \citep[MDS, ][]{ratnatunga}. The four images in an Einstein cross configuration are separated by $\sim1\arcsec$ and are much fainter ($I\sim25-27$) than the early type lens galaxy $I\sim19-22$. Based on fundamental plane fitting, \citet{kochanek2000} estimate $z_l=0.63^{+0.20}_{-0.03}$. Lens models of HST12531 require unusually large shear $\gamma\sim0.2$ \citep{witt}, suggesting a misalignment between the galaxy and its DM halo, or perhaps a complex environment.

{\bf LBQS 1333+0113} ({\it a.k.a.} SDSS J1335$ +$0118, herein LBQ1333) was identified as a quasar in the Large Bright Quasar Survey \citep[LBQS, ][]{hewett}, but not as a lensed source: \citet{hewett1998} looked for lenses in LBQS, but did not discover this one because they were only sensitive to separations larger than 3\arcsec. \citet{oguri04} identified it as a doubly lensed $z_s=1.57\pm0.05$ quasar in the SDSS. The image separation is 1.56\arcsec. The lens redshift is $z_l=0.440\pm0.001$ \citep{eigenbrod}. Based on its spectrum and colors, both \citet{eigenbrod} and \citet{oguri04} conclude that the lens is an early type galaxy.

{\bf Q1355-2257} ({\it a.k.a.} CTQ 0327, herein Q1355) was discovered by \citet{morgan} during an \textit{HST} STIS snapshot campaign to find small separation gravitational lenses. The two images of the $z=1.37$ quasar are separated by 1.22\arcsec, and the lens is an early type galaxy at $z_l\sim0.4-0.6$ based on the Faber-Jackson relationship\citep{morgan}. \citet{eigenbrod07} give a tentative measurement of the lens redshift $z_l=0.702$. The lens redshift is not in NED, but we have added it to our spectroscopic catalog.

{\bf HST J14113+5211} (herein HST14113) is a quadruple lens with maximum separation 2.28\arcsec~ discovered serendipitously by \citet{fischer} in the $z=0.46$ cluster CL 140933+5226 ({\it a.k.a.} 3C 295). The lens is identified as an early type galaxy most probably belonging to the cluster. \citet{lubin} measure the lens and source redshifts: $z_l=0.465$ and $z_s=2.811$. The cluster has velocity dispersion $\sigma=1300$ \kms~ based on 21 members \citep{dressler}.

{\bf [HB89] 1413+117} ({\it a.k.a.} the Cloverleaf, herein H1413) is one of the most widely studied strong gravitational lenses. It was identified as a broad-absorption line quasar at $z_{s}=2.55$ by \citet{hazard} and originally included in the Hewitt-Burbidge QSO catalog \citep{hewitt}. Later \cite{magain} discovered that it is in fact a gravitational lens with 4 images separated by 1.1\arcsec~ and 1.36\arcsec~ along the diagonals, and identify two absorption line systems in the quasar spectra at $z=1.43$ and $1.661$ possibly caused by the lens and/or line-of-sight structures. The lens galaxy was detected by \citet{knieb98a}, who derive a photometric redshift $z_{l}=0.9\pm0.1$ for it and the surrounding group of galaxies. They suggest that the lens lies in a dense environment \citep[also ][]{knieb98b}, which causes the large shear \citep[$\gamma=0.110\pm0.003$, ][]{keeton97} required by lens models. Based on extensive photometric redshifts, \citet{faure04} find two separate over-densities along the line of sight to H1413 at $z=0.8\pm0.3$ and $1.75\pm0.2$. Variations between the image fluxes \citep[e.g., ][]{angonin} have been ascribed to microlensing and/or intrinsic quasar variability. \citet{h1413td} determine time delays $\Delta t_{AB}=-17\pm3$ days, $\Delta t _{AC}=-20\pm4$ days, and $\Delta t_{AD}=23\pm4$ days.

{\bf JVAS B1422+231} (herein B1422) is a four image lens discovered in the Jordell Bank -- VLA Astrometric Survey \citep[JVAS, ][]{patnaik92a, patnaik92b, king99}. The source is a $z_s=3.62$ radio-loud quasar lensed by a $z_{l}=0.338$ luminous elliptical galaxy \citep{impey96, kundic97b}. Using the radio light curves, \citet{patnaik01} measure the time delays between the images to be $1.5\pm1.4$ days (between B and A), $7.6\pm2.5$ days (between A and C), and $8.2\pm2.0$ days (between B and C). The lens belongs to a group of 16 galaxies with $\sigma=470^{+100}_{-90}$ \kms~ \citep{kundic97b, my}. \citet{grant} and \citet{fassnacht08} detect the diffuse X-ray emission of the group. \citet{grant} determine its temperature $kT=1.0$ keV and luminosity $L_X=8\times10^{42}$ erg s$^{-1}$. B1422 is also a lens with anomalous flux ratios \citep{mao, chiba05}.

{\bf SBS 1520+530} (herein SBS1520) is a doubly imaged $z_{s}=1.855$ BAL quasar discovered in the Second Byurakan Survey \citep[SBS, ][]{sbs1520}. The images are separated by 1.005\arcsec. The redshift of the lens galaxy is ambiguous: \citet{burud02} measure $z_l=0.71\pm0.005$, but \citet{auger08} claim $z_{l}=0.761$ is more likely. Because of this ambiguity, neither redshift is added in our catalog. The time delay between the images is $\Delta t=125.8\pm2.1$ days \citep{eulaers11}. Lens models of SBS1520 require unusually large shear $\gamma\sim0.34$, which is attributed to a nearby galaxy and a group at the lens redshift \citep{burud02, faure02, auger08}. \citet{auger08} also find that the lens is best fit with a steeper than isothermal profile, which may arise from a tidal interaction with the nearby galaxy. Short term variations in the image fluxes are probably due to microlensing \citep{gaynullina}.

{\bf MG J1549+3047} (herein MG1549) was recognized as a gravitational lens by \citet{lehar1993a} in a radio source mapped as part of the MIT-Greenbank-VLA survey \citep{lehar1991}. The $z_{l}=0.111$ galaxy lenses one of the radio lobes of a $z_{s}=1.170\pm0.001$ \citep{treu2003} background radio galaxy into an radio Einstein ring. The velocity dispersion of the SB0 lens galaxy is $\sigma_{\ast}=227\pm18$ \kms~ \citep{lehar96}. A third object at $z=0.604\pm0.001$ lies projected between the lens and the radio galaxy producing the jet.

{\bf CLASS B1600+434} (herein B1600) is a two image gravitational lens discovered in the CLASS survey \citep{jackson1995}. A $z_{s}=1.589\pm0.006$ radio source is lensed by a $z_{l}=0.414\pm0.0003$ edge-on spiral galaxy \citep{fassnacht1998, jaunsen, koopmans98}. The images are separated by 1.4\arcsec. \citet{koopmans99} measure the radio time delay of $\Delta t = 47^{+5}_{-6}$ days and also report that short term variability in the A image is probably caused by microlensing. \citet{burud2000} measure the optical time delay $\Delta t = 51\pm4$ days. \citet{kurtis} find a red line-of-sight structure at $z\sim0.5$. \citet{auger} find that the lens belongs to a group of 6 late-type members with $\sigma=100\pm40$ \kms~ as well as identifying two background groups ($z=0.543$, 0.629) or associations that might not be bound. X-ray observations of B1600 \citep{dai} fail to detect extended emission from the intra-group gas and place a limit on its luminosity of $L_X\sim2\times10^{42}$ for a group at the lens redshift.

{\bf CLASS B1608+656} (herein B1608) was the first lens discovered as part of CLASS \citep{browne03, myers03}. A $z_s=1.394$ post-starburst galaxy \citep{fassnacht96} is lensed into four images by a $z_{l}=0.6304$ pair of interacting early-type galaxies. The stellar velocity dispersion of the main lens galaxy is $\sigma_{\ast}=247\pm35$ \kms~, and it has an E+A spectrum \citep{koopmans03}. The maximum separation between the images is 2.1\arcsec. B1608 is a time delay system with radio delays $\Delta t_{BA} =31.5 ^{+2}_{-1}$ days, $\Delta t_{BC} = 36.0^{+1.5}_{-1.5}$ days, and $\Delta t_{BD} = 77.0^{+2.0}_{-1.0}$ days \citep{fassnacht02}. \citet{suyu10} measure the velocity dispersion of the main lens galaxy: $260\pm15$ \kms. \citet{auger08} find a group of eight galaxies at the lens redshift to which \citet{fassnacht08} add two more members. The group velocity dispersion is $\sigma=150\pm30$ \kms~\citep{fassnacht08}. Three other groups have been found along the line of sight \citep[$z=0.265, 0.426$, and $0.52$, ][]{auger08,fassnacht08}. So far none of these groups is detected in X-rays \citep{dai,fassnacht08}.

{\bf MG J1654+1346} (herein MG1654) was recognized as an unusual radio source in the MIT-Greenbank-VLA survey. A $z_l=0.254$ giant elliptical galaxy lenses one of the radio lobes of a $z_{s}=1.74$ radio quasar into a radio Einstein ring \citep{langston88, langston89, kochanek2000}. \citet{langston89} notice an enhancement in the number density of galaxies around the lens, suggesting a complex environment. \citet{my} identify a group of seven galaxies with $\sigma=200^{+120}_{-80}$ \kms~ at the lens redshift.

{\bf PMN J2004-1349} (herein PMN2004), discovered by \citet{winn} in a southern survey for radio lenses, is a two image lens. The radio spectral index of the source is typical for radio-loud quasars, so it is considered a quasar despite the lack of optical confirmation. The lens is a spiral galaxy. Based on photometry, \citet{winn} suggest $0.5<z_l<1.0$, but extinction considerations used to explain the color differences between the images imply lower values of $0.03\lesssim z_l \lesssim 0.36$ \citep{winn03}.

{\bf WFI J2033-4723} (herein WFI2033) is a quadruply imaged $z=1.66$ quasar \citep{morgan2004} discovered as part of a southern hemisphere optical survey for gravitational lenses using the MPG/ESO 2.2 m telescope. The image separation is 2.53\arcsec. \citet{eigenbrod} and \citet{ofek} measure the S0 lens redshift to be $z_{l}=0.661\pm0.001$ and $z_{l}=0.658\pm0.001$, respectively. Lens models of the system require large shear $\gamma=0.225$ due to a nearby galaxy and/or a group at the lens redshift \citep{morgan2004}. Recently \citet{td_wfi2033} measure two independent time delays: $\Delta t_{B-A} = 35.5 \pm 1.4$ days and $\Delta t_{B-C} = 62.6^{+4.1}_{- 2.3}$ days.

{\bf CLASS B2114+022} (herein B2114) was discovered in the Jordell Bank - VLA Astrometric Survey \citep{king99}. The four radio sources are arranged in an atypical lens configuration. The four images can be divided into two distinct groups: A and D are similar to each other, as are B and C, but the pairs are different from one another. Optical observations fail to detect the images, but identify two lens galaxies with early type colors and morphologies at $z_{l1}=0.3157$ and $z_{l2}=0.5883$ \citep{augusto}. The foreground galaxy G1 has a spectrum characteristic of E+A galaxies. Using a two plane lens model, \citet{chae} explain the A and D radio components. There is no lens model to explain the B and C components yet. \citet{my} find a group of five galaxies at $z=0.3141$ associated with the foreground lens galaxy and determine $\sigma=110^{+170}_{-80}$ \kms.

{\bf HE2149-2745} (herein HE2149) is a doubly imaged BAL quasar at $z=2.033$ discovered by \citet{wisotzki96} as part of the Hamburg/ESO survey of bright quasars. The elliptical galaxy lens redshift $z_l=0.495\pm0.01$ was measured by \citet{burud}, who also determine the time delay between the images $\Delta t = 103\pm12$ days. Based on the large number of galaxies in the R band image of the HE2149 field, \citet{lopez} suggest that the lens might be a member of a cluster. \citet{my} and \citet{kurtis}, however, only find several groups along the line of sight (at $z=0.27,\;0.45$, and $0.60$), none of them associated with the lens.

\section{Additional Figures}

\begin{figure*}[ht]
\figurenum{10a}
\begin{tabular}{@{}c@{\hspace{-5.0mm}}@{\hspace{-5.0mm}}c@{\hspace{-5.0mm}}@{\hspace{-5.0mm}}c@{\hspace{-5.0mm}}@{\hspace{-5.0mm}}c@{\hspace{0mm}}}
\includegraphics[width=0.28\textwidth]{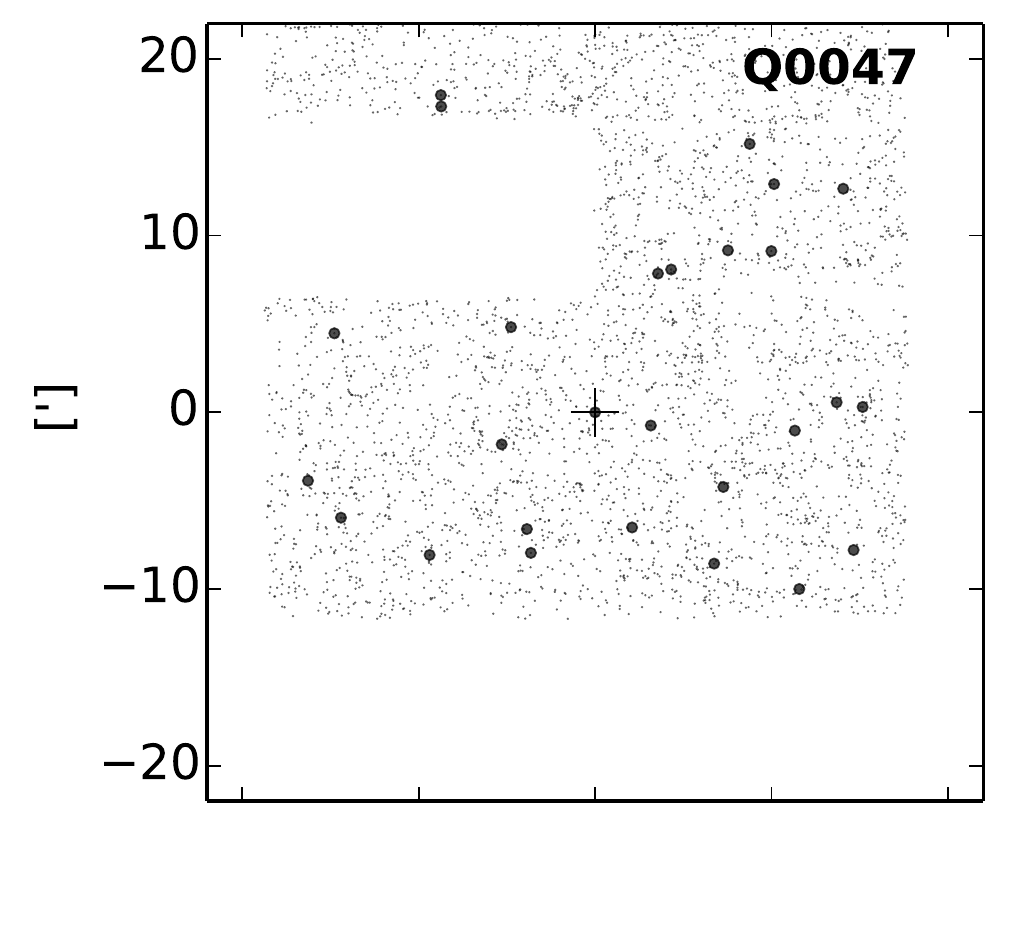} &
\includegraphics[width=0.28\textwidth]{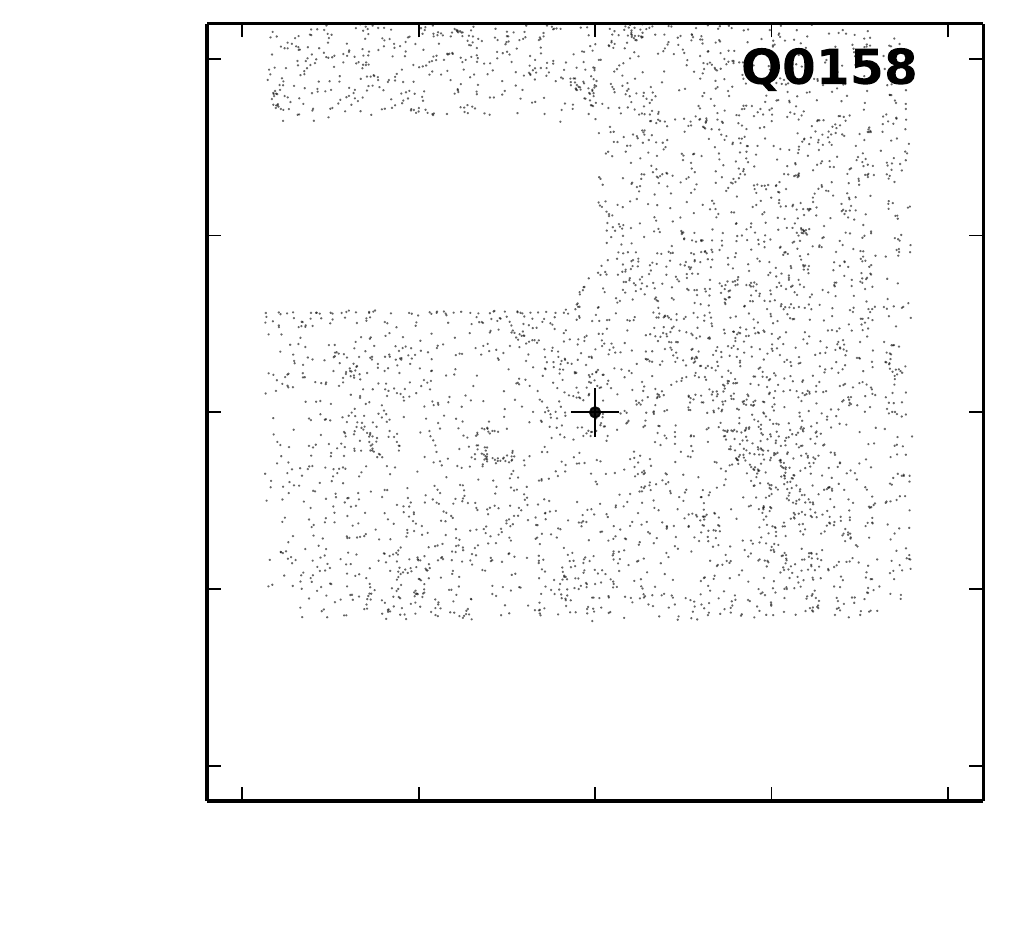} &
\includegraphics[width=0.28\textwidth]{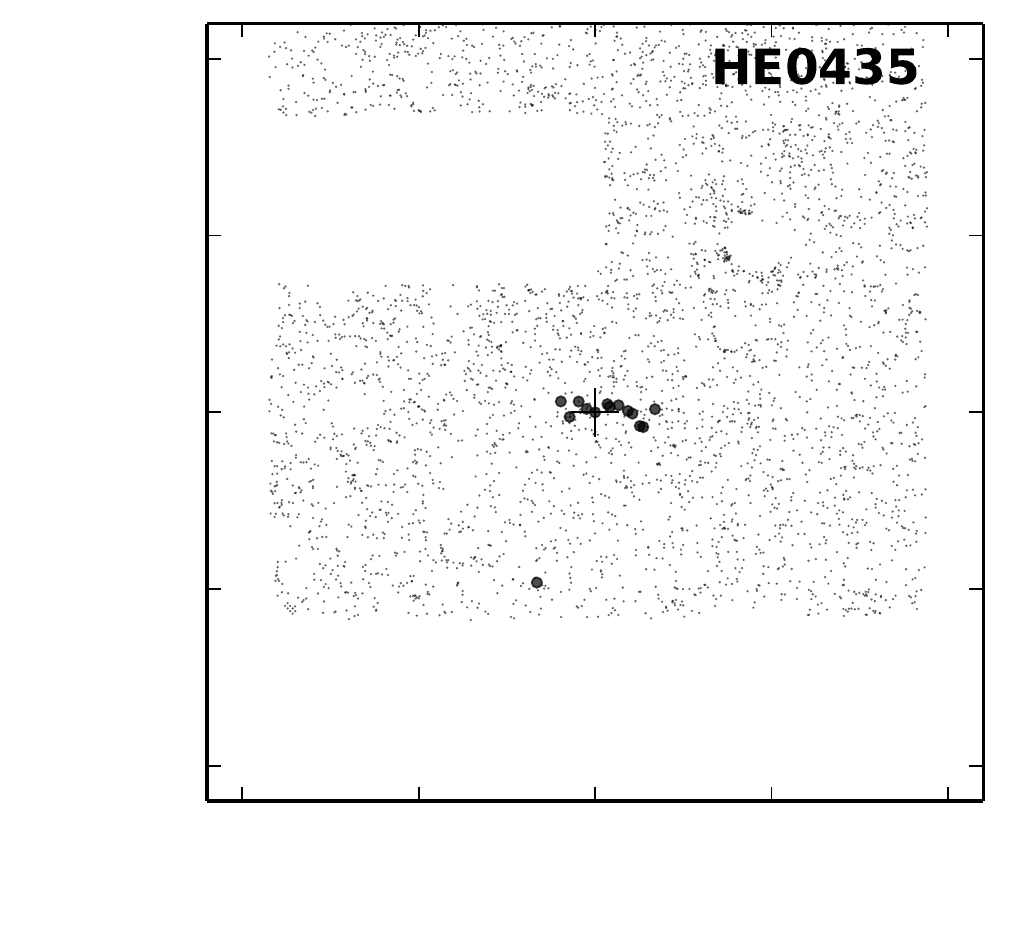} &
\includegraphics[width=0.28\textwidth]{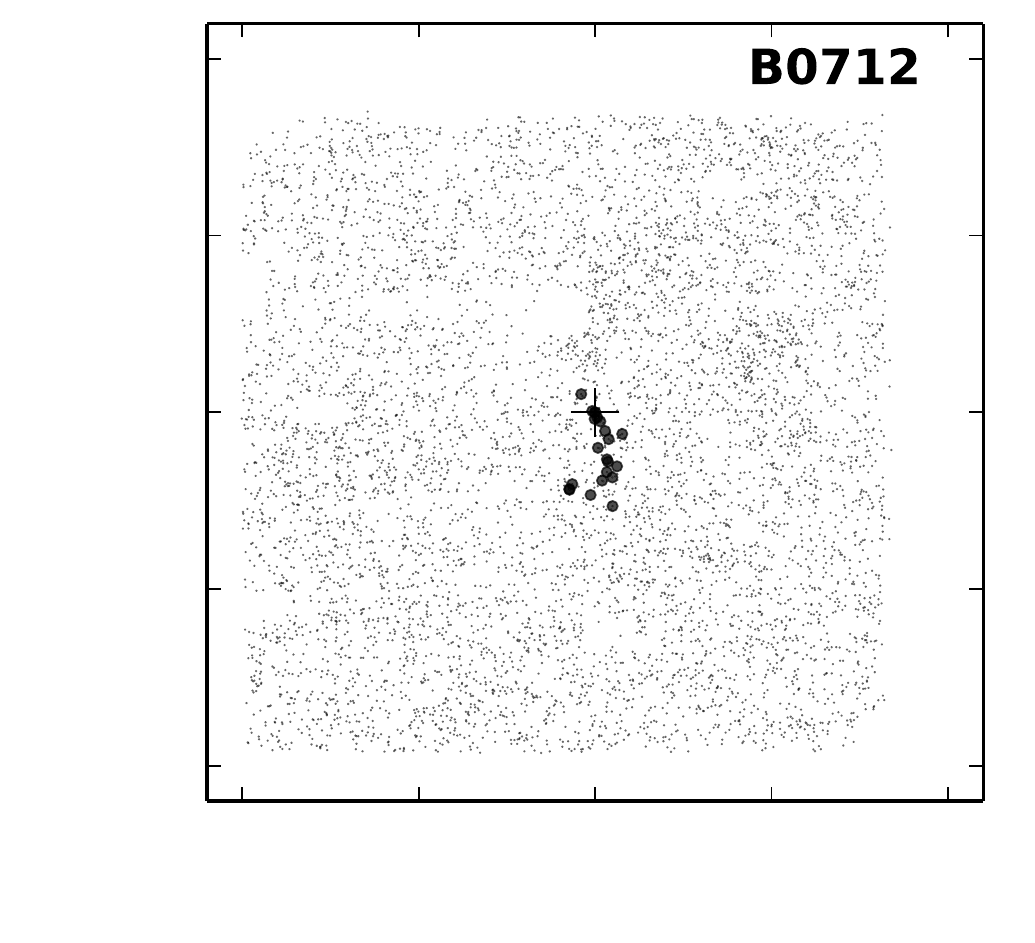} \\ [-22pt]
\includegraphics[width=0.28\textwidth]{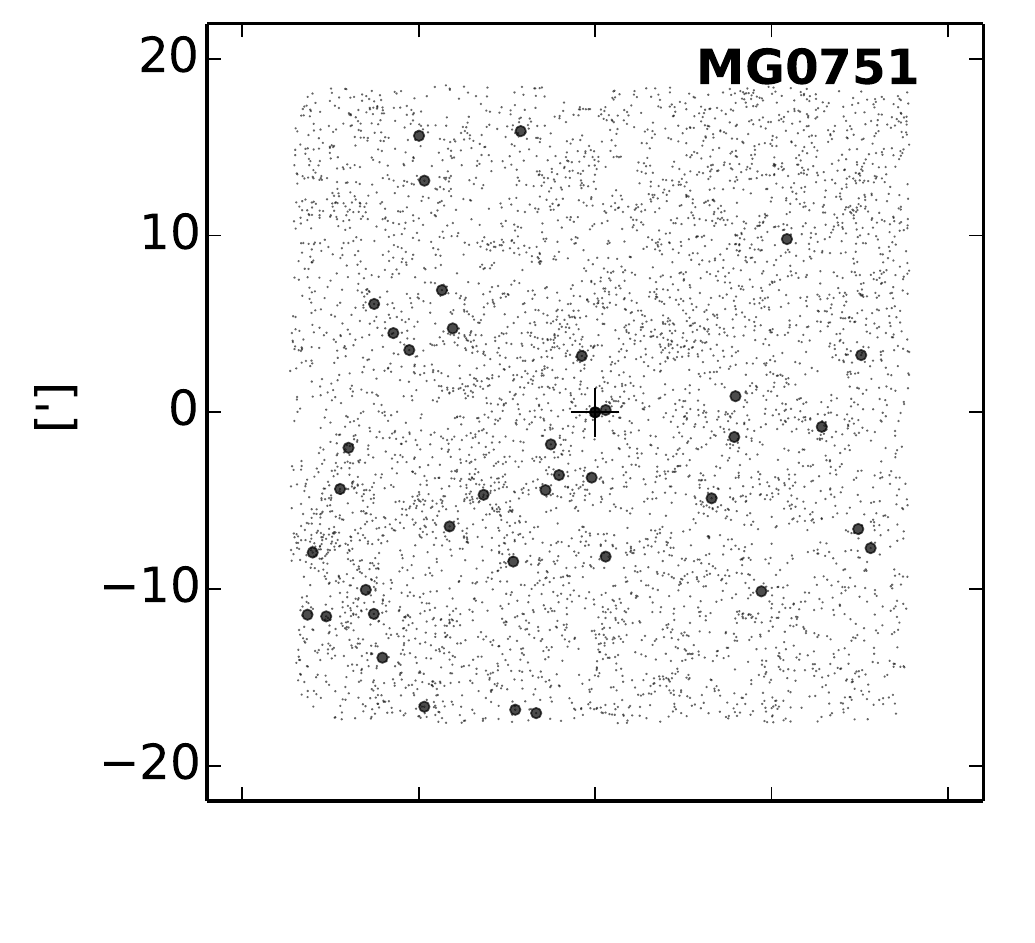} &
\includegraphics[width=0.28\textwidth]{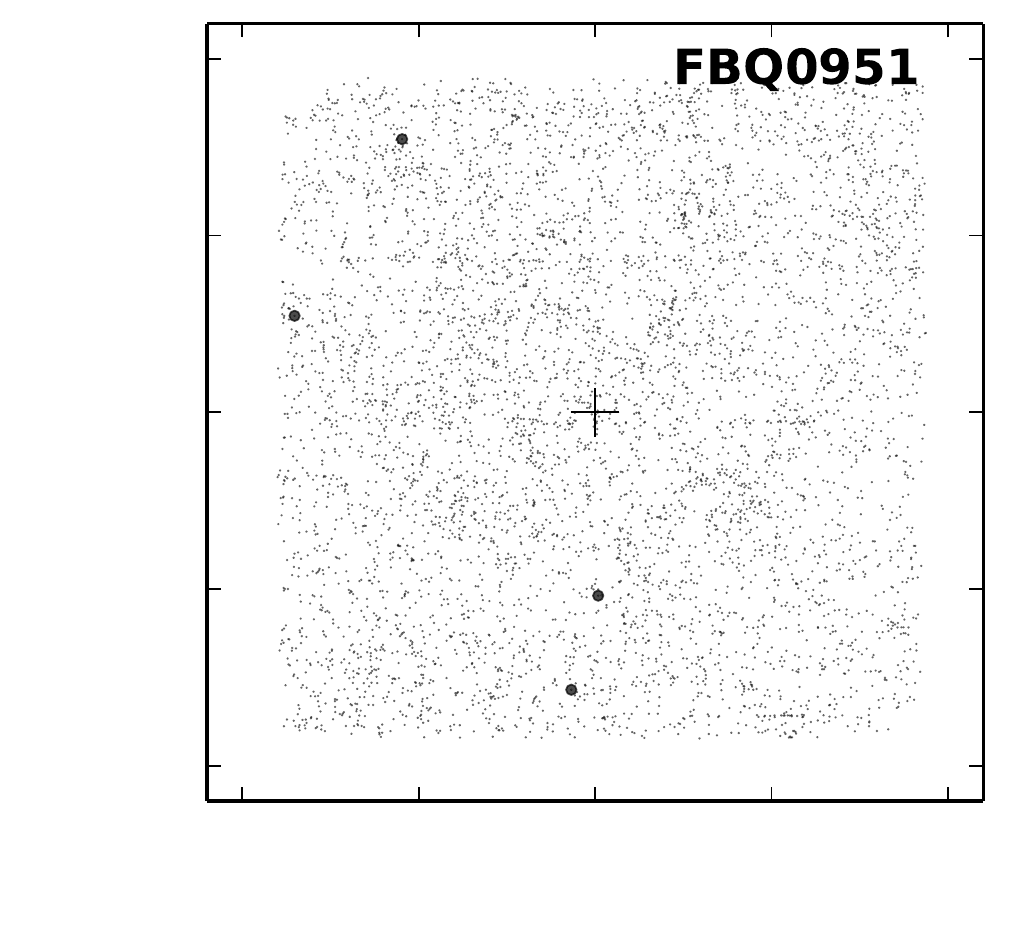} &
\includegraphics[width=0.28\textwidth]{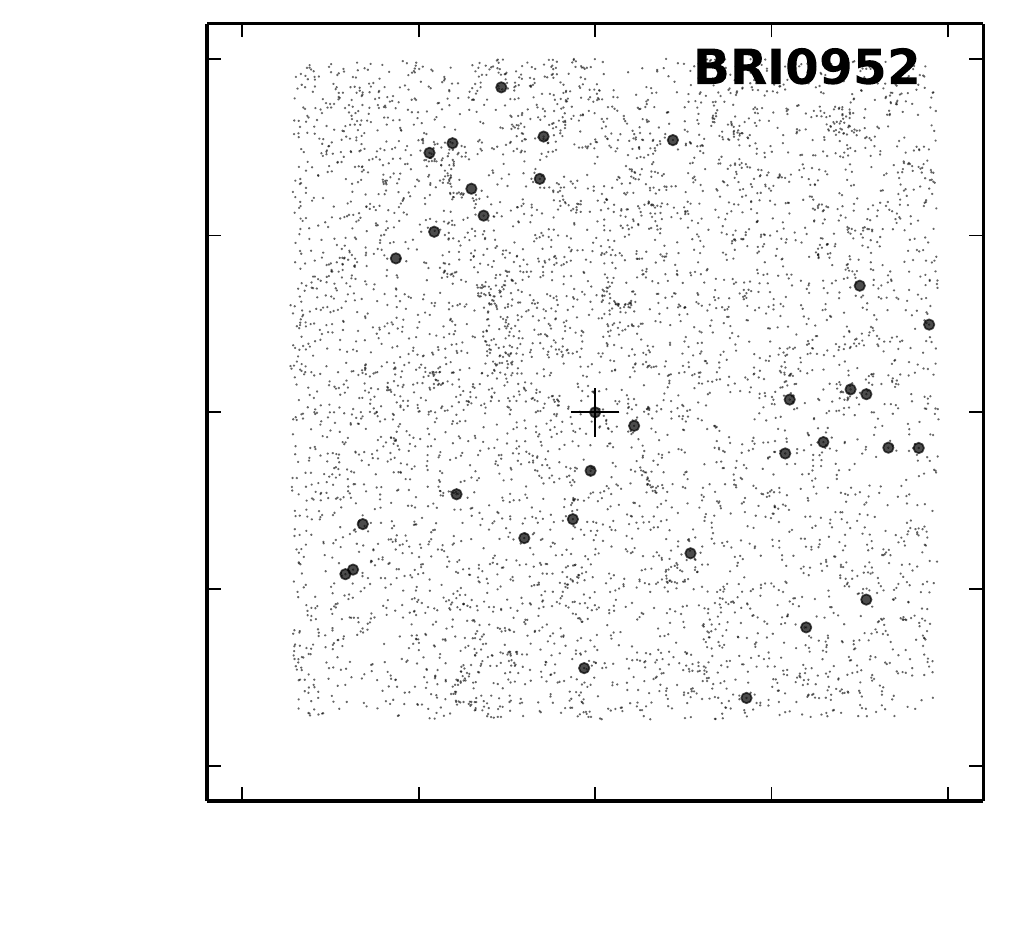} &
\includegraphics[width=0.28\textwidth]{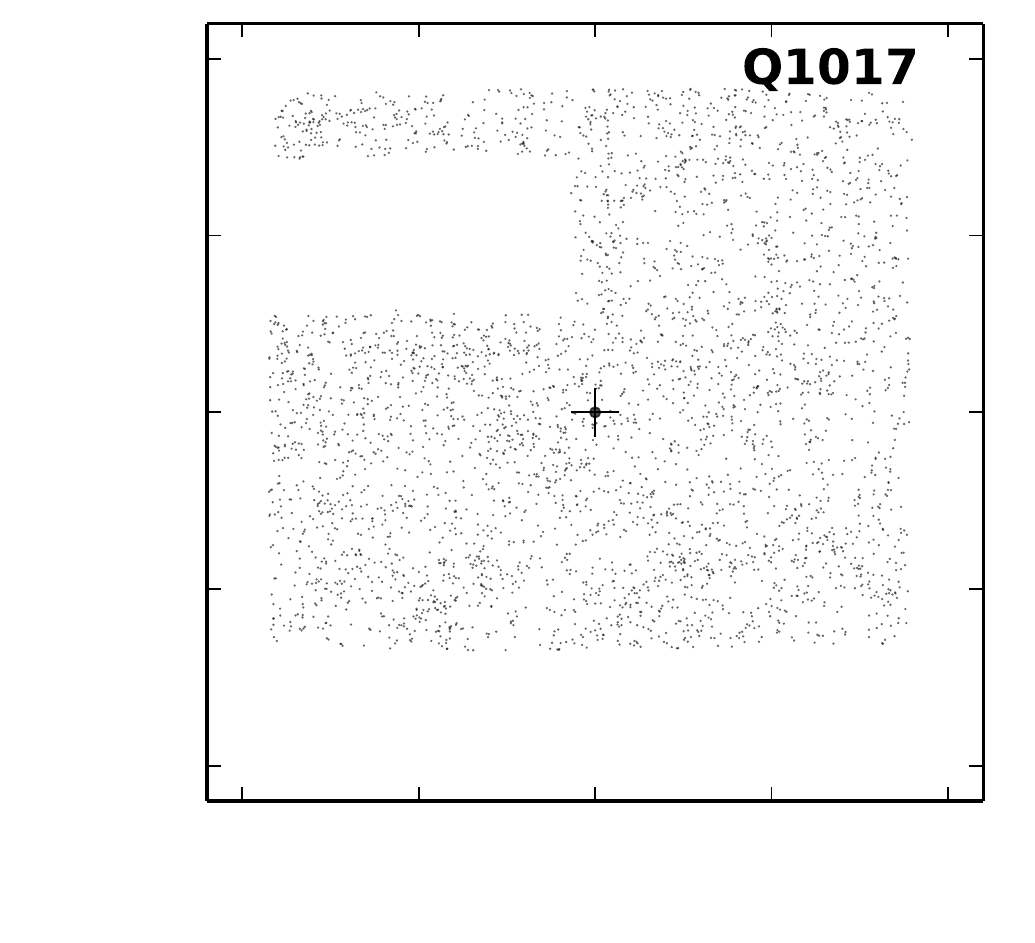} \\ [-22pt]
\includegraphics[width=0.28\textwidth]{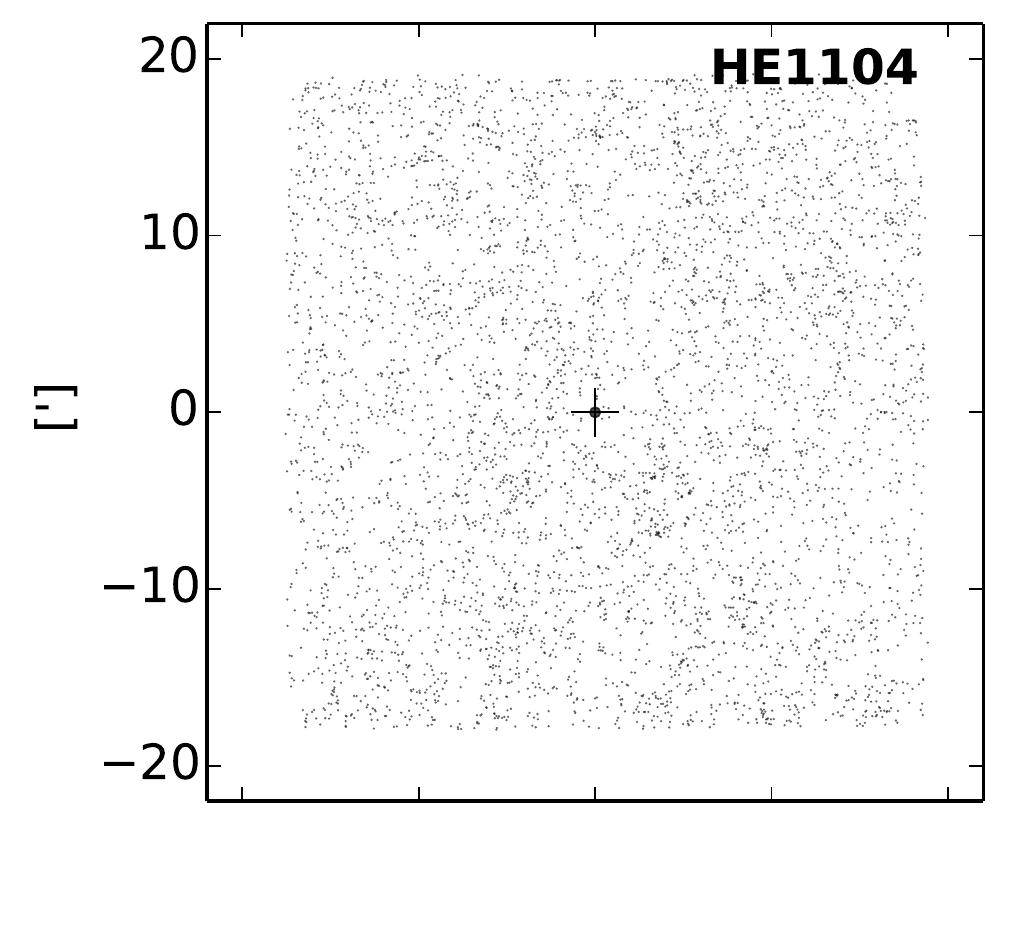} &
\includegraphics[width=0.28\textwidth]{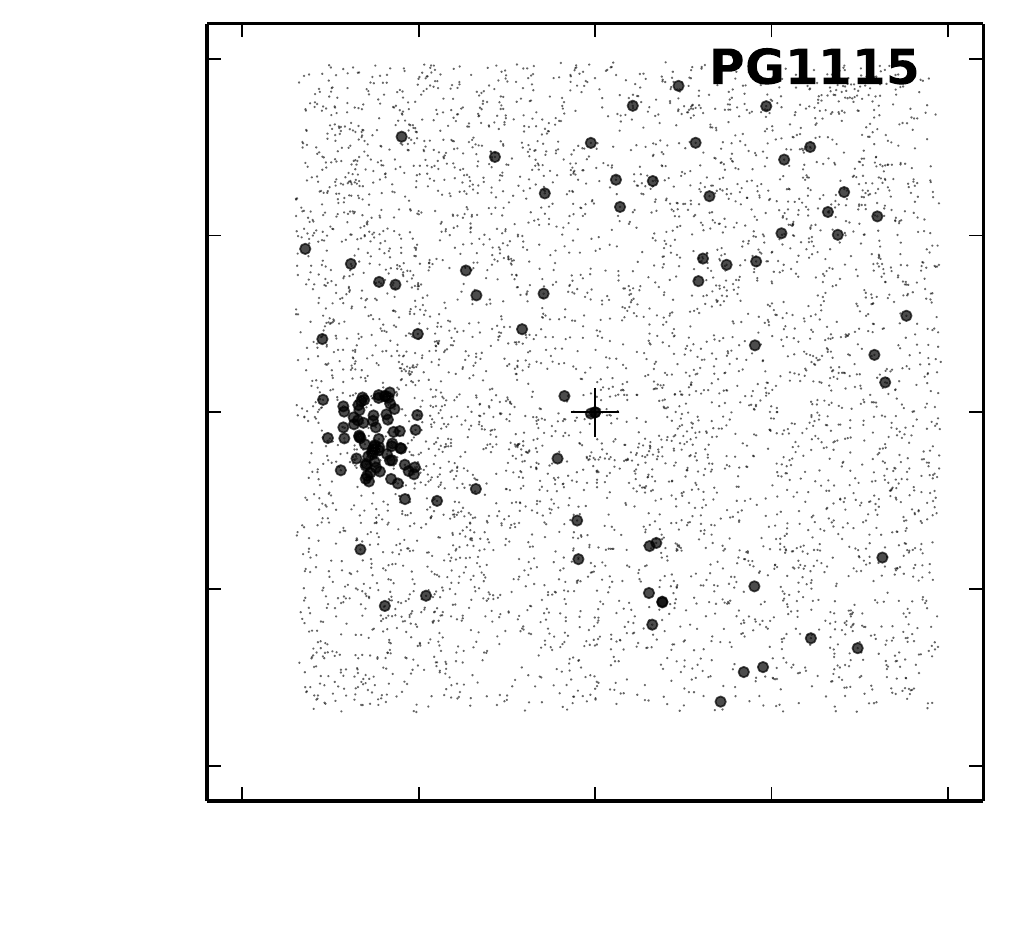} &
\includegraphics[width=0.28\textwidth]{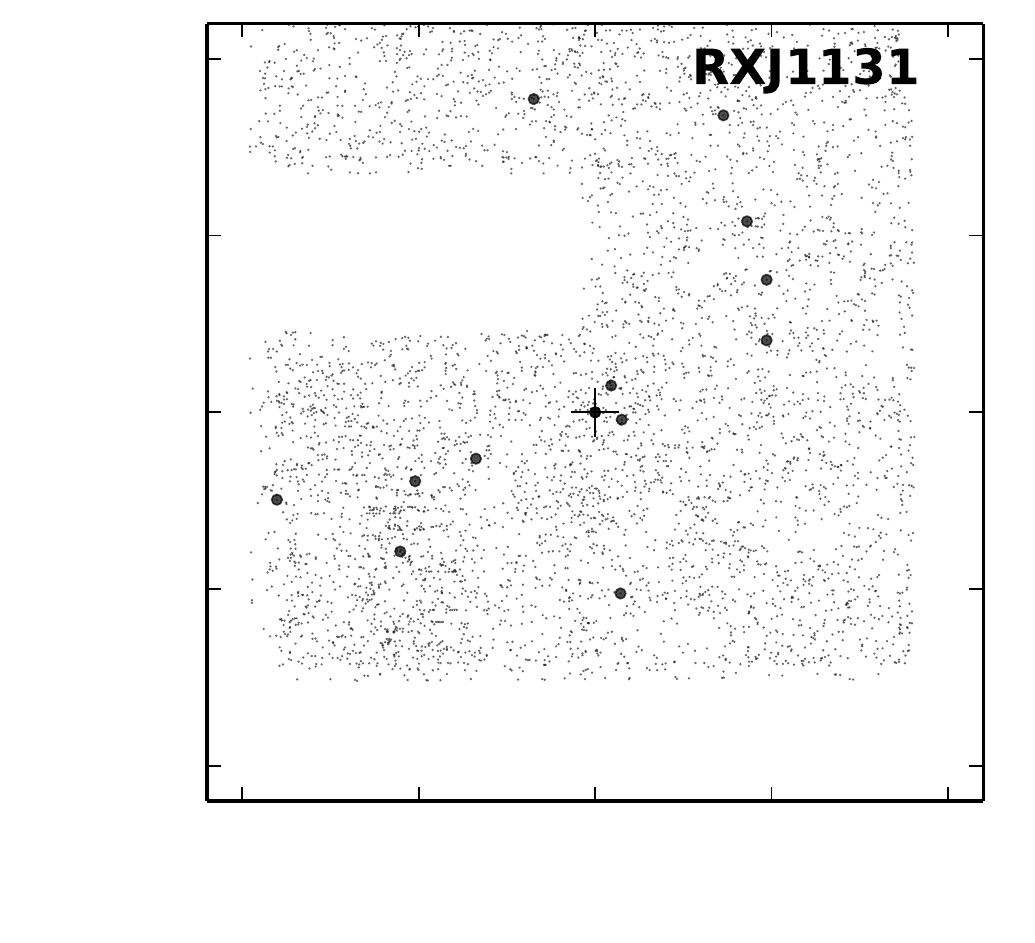} &
\includegraphics[width=0.28\textwidth]{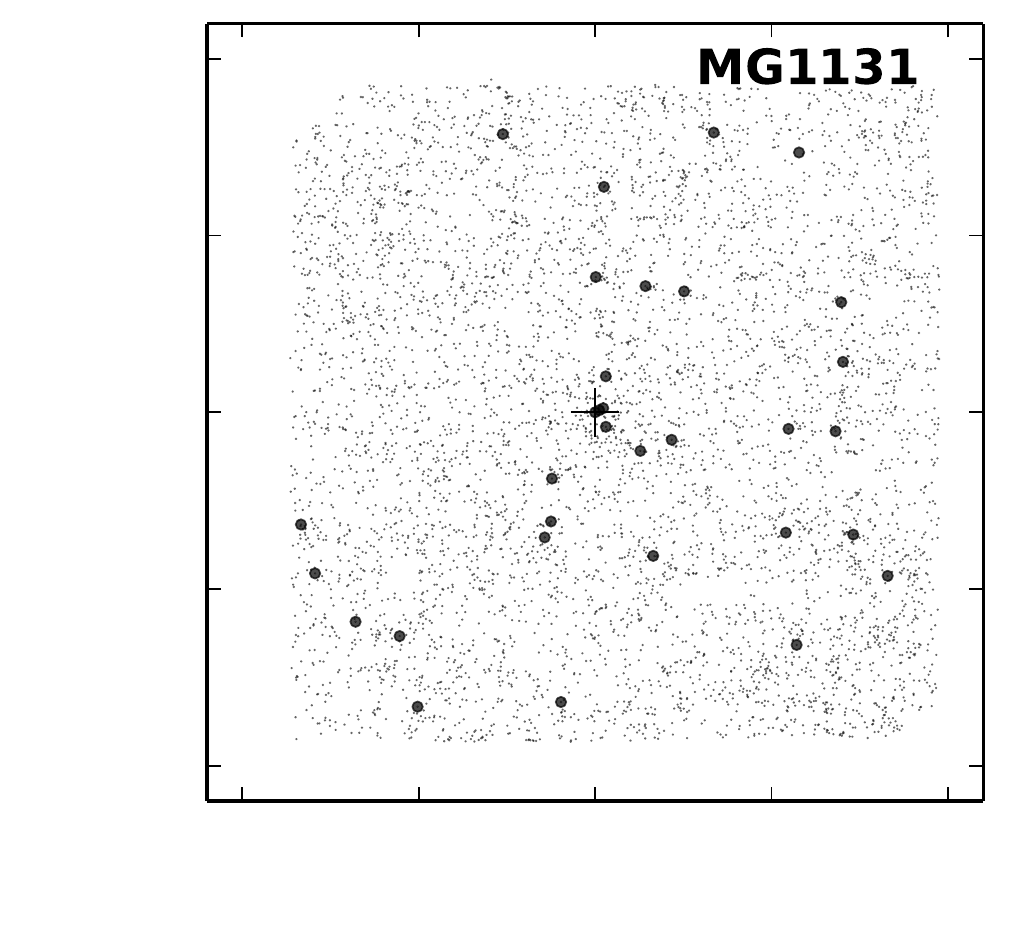} \\ [-22pt]
\includegraphics[width=0.28\textwidth]{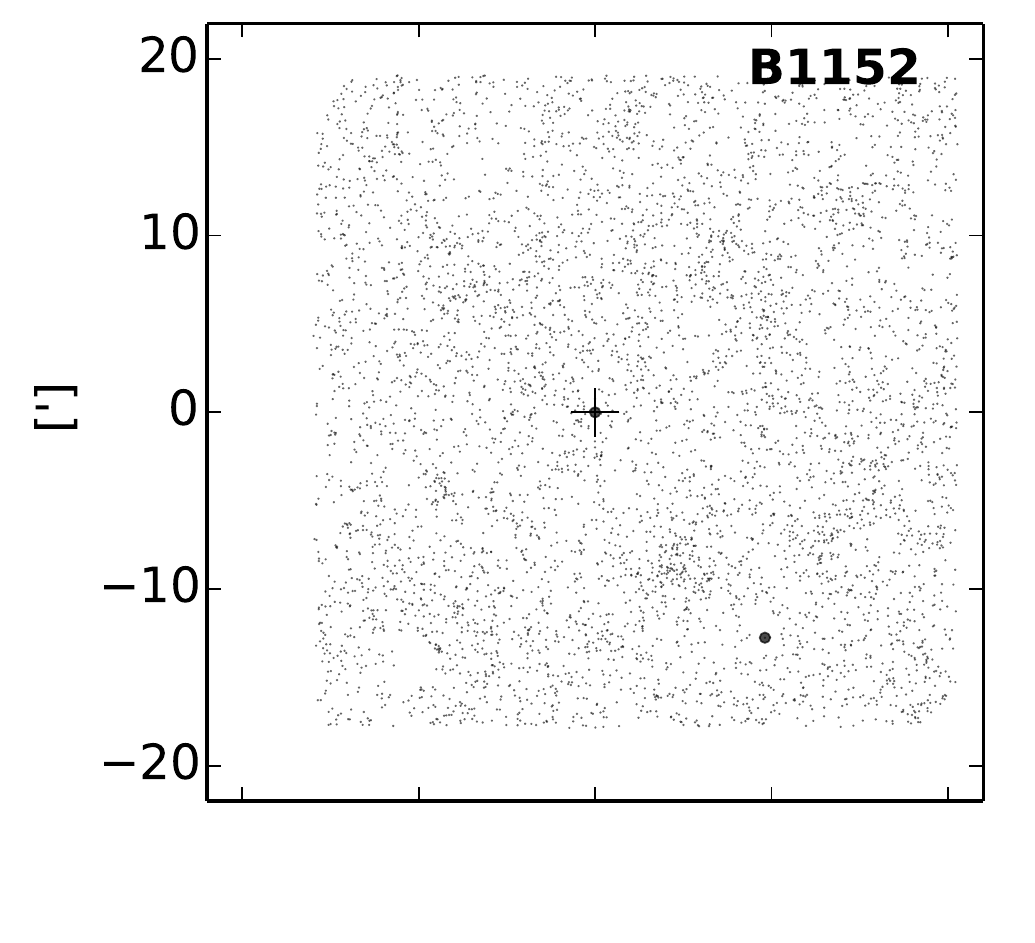} &
\includegraphics[width=0.28\textwidth]{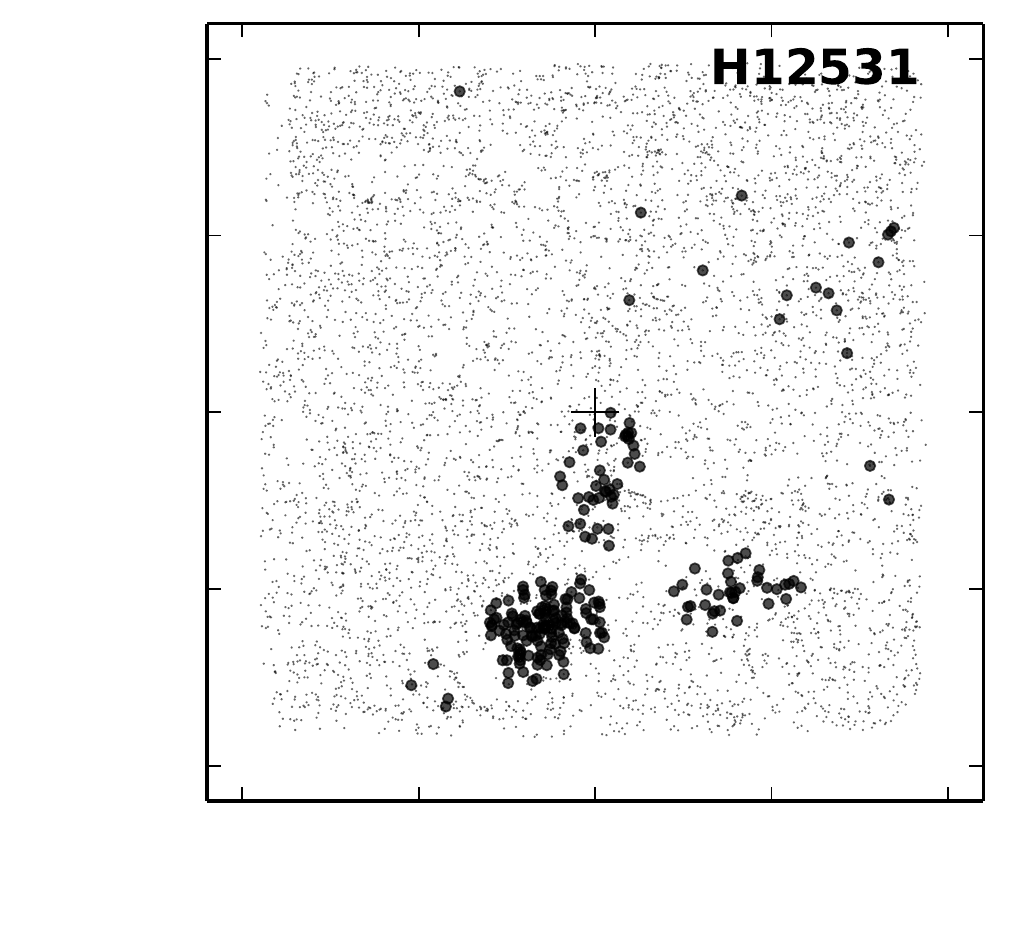} &
\includegraphics[width=0.28\textwidth]{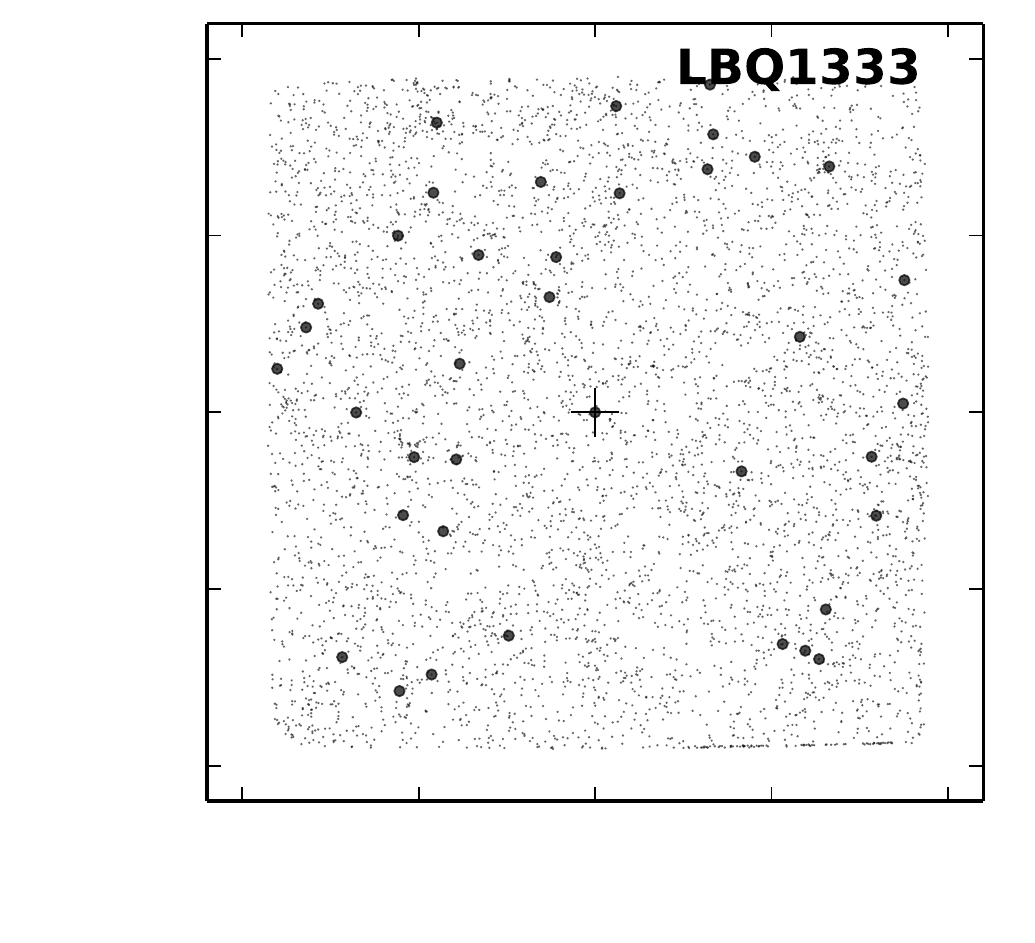} &
\includegraphics[width=0.28\textwidth]{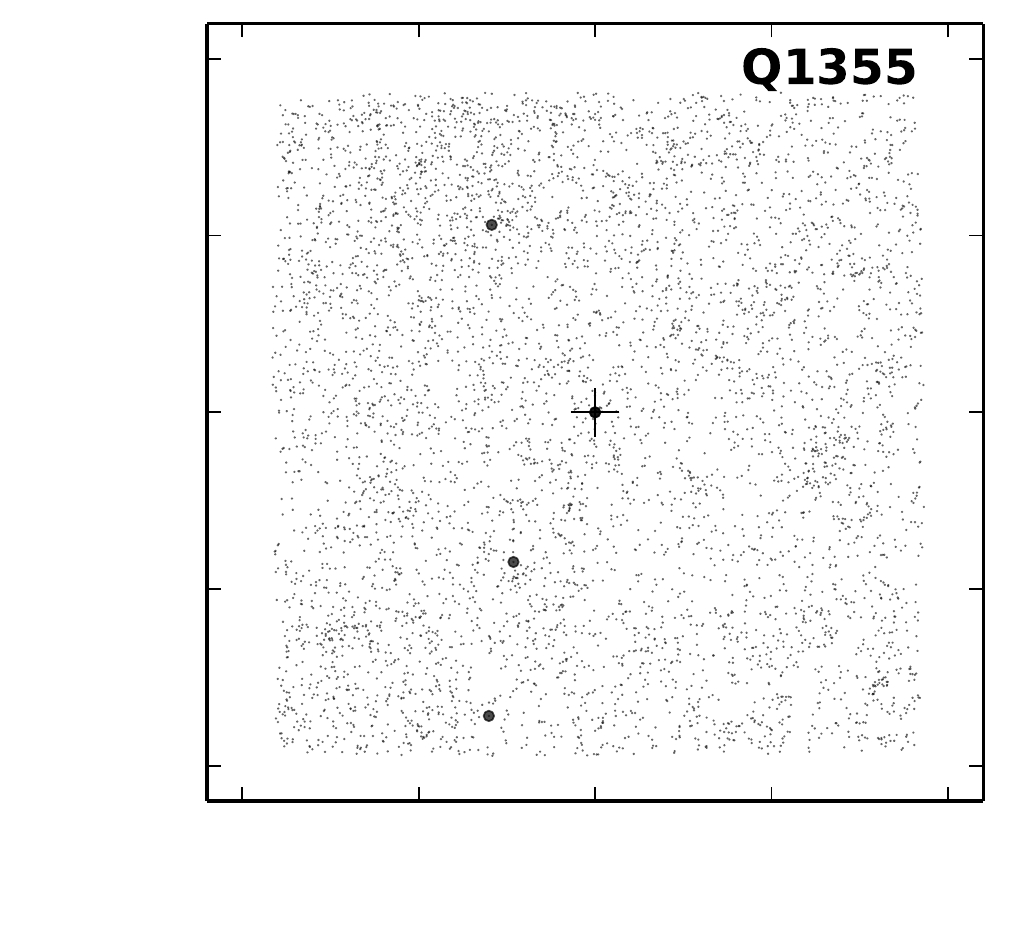} \\ [-22pt]
\includegraphics[width=0.28\textwidth]{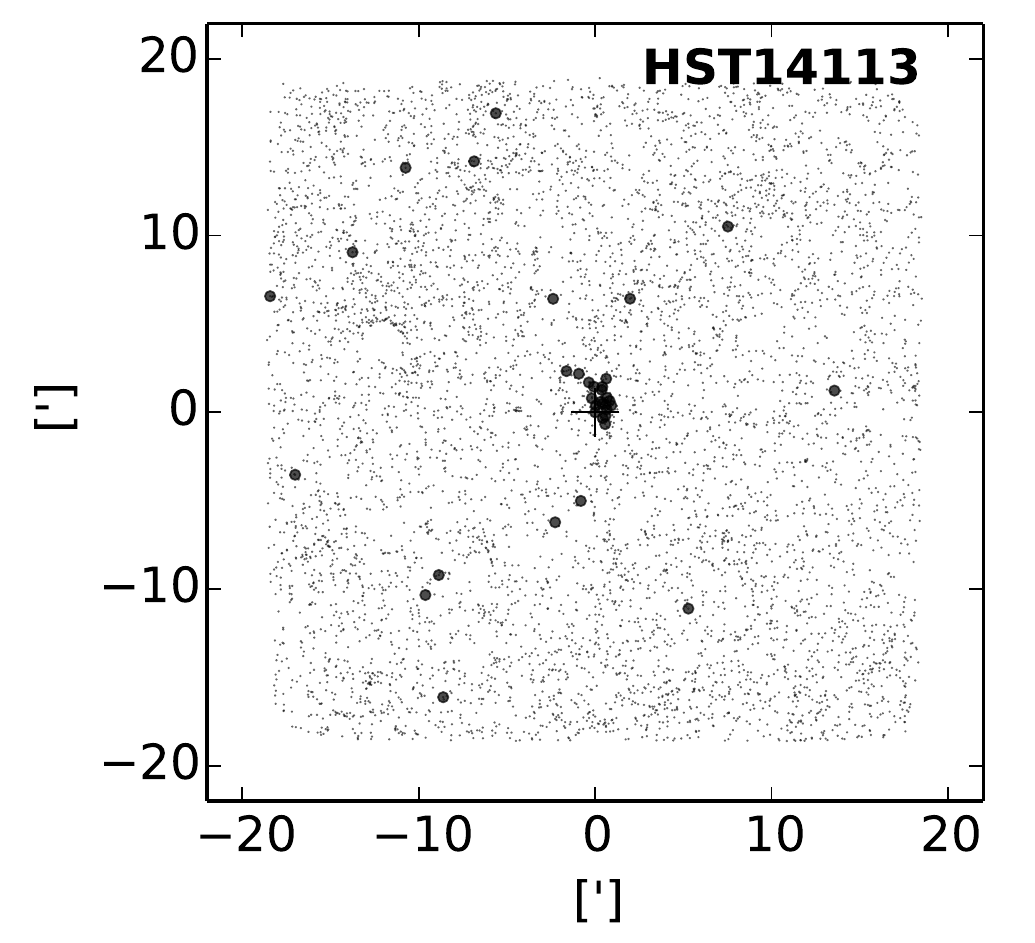} &
\includegraphics[width=0.28\textwidth]{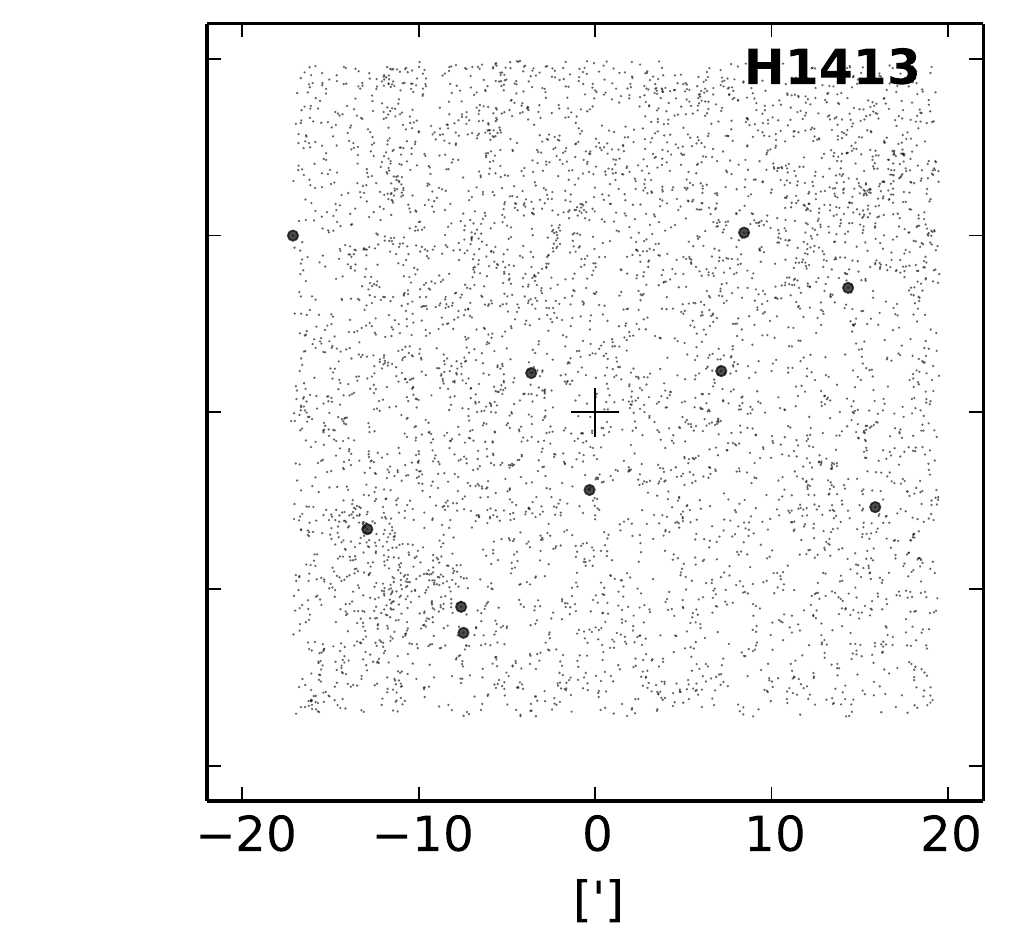} &
\includegraphics[width=0.28\textwidth]{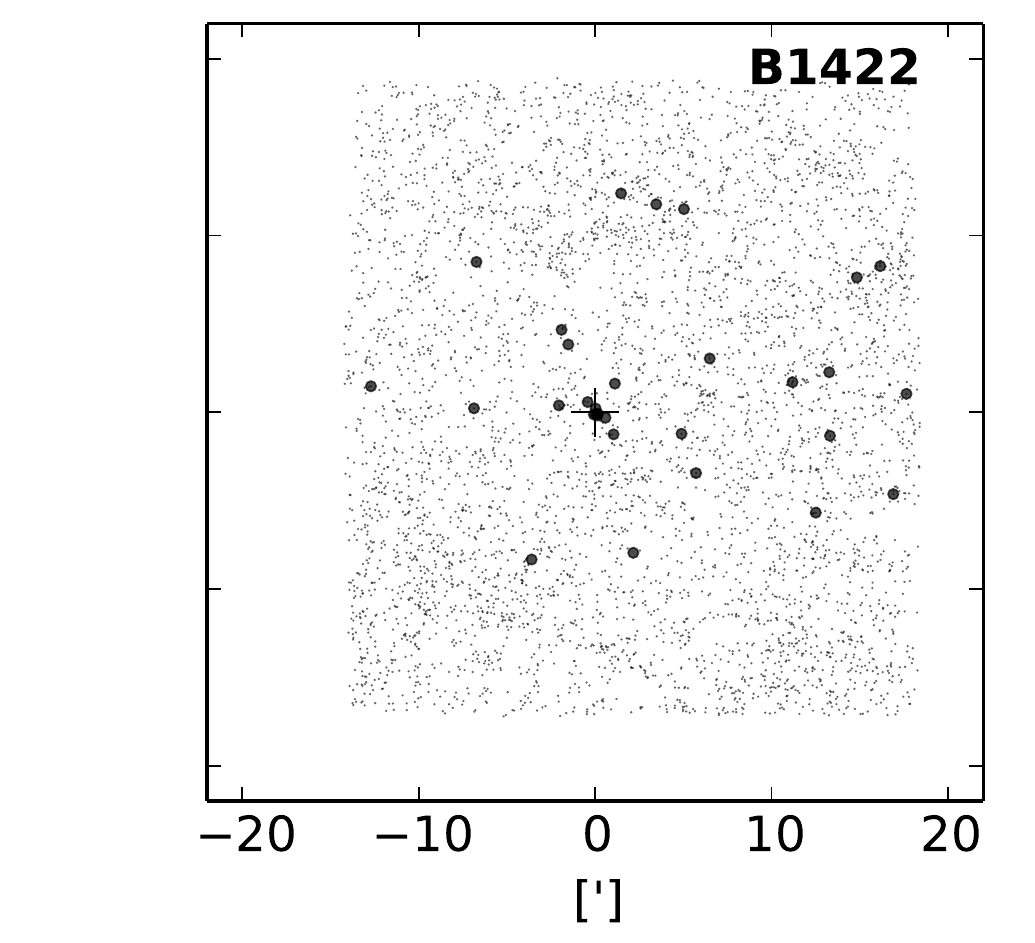} &
\includegraphics[width=0.28\textwidth]{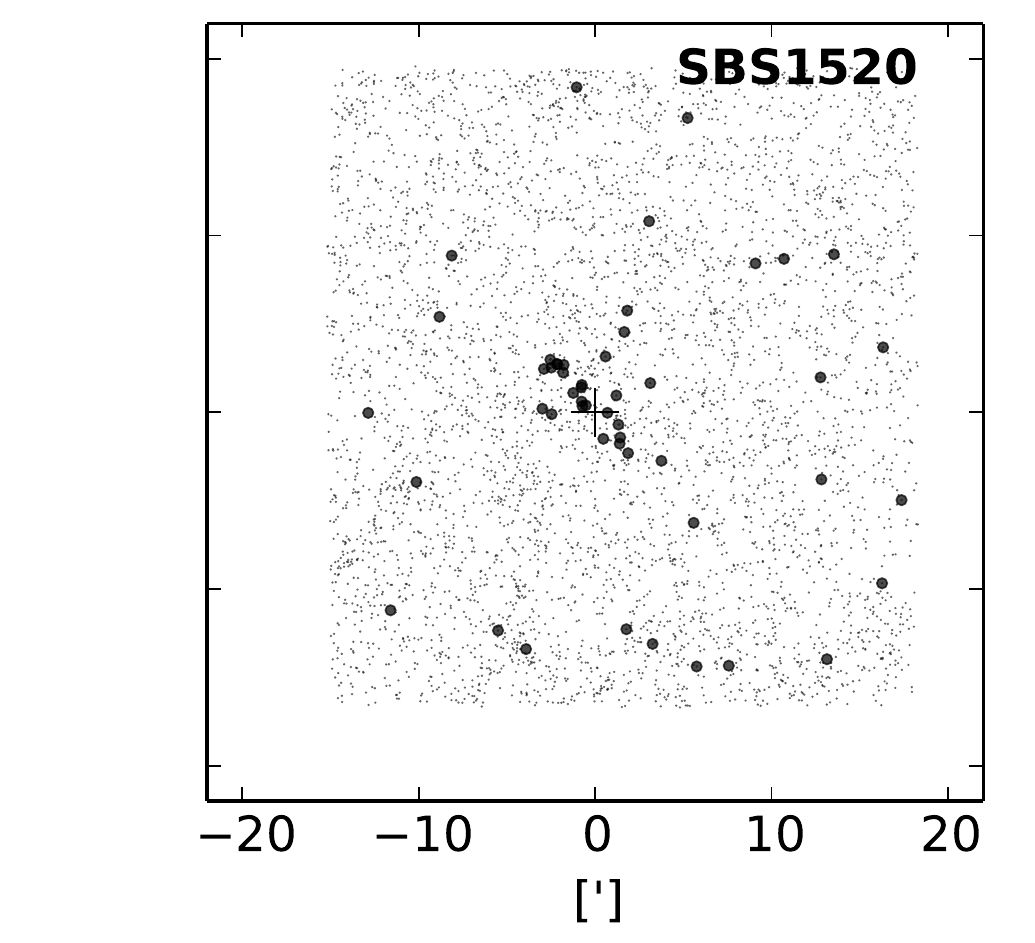} \\ [-22pt]
\end{tabular}
\vspace{7mm}
\caption{\footnotesize Projected distributions of the objects from NED added to the final spectroscopic catalog (black points) relative to the objects in the full photometric catalog (gray points). Each field is centered on the lens galaxy (cross symbol). \label{fig:radec_extra_a}}
\end{figure*}

\begin{figure*}[ht]
\figurenum{10b}
\begin{tabular}{@{}c@{\hspace{-5.0mm}}@{\hspace{-5.0mm}}c@{\hspace{-5.0mm}}@{\hspace{-5.0mm}}c@{\hspace{-5.0mm}}@{\hspace{-5.0mm}}c@{\hspace{0mm}}}
\includegraphics[width=0.28\textwidth]{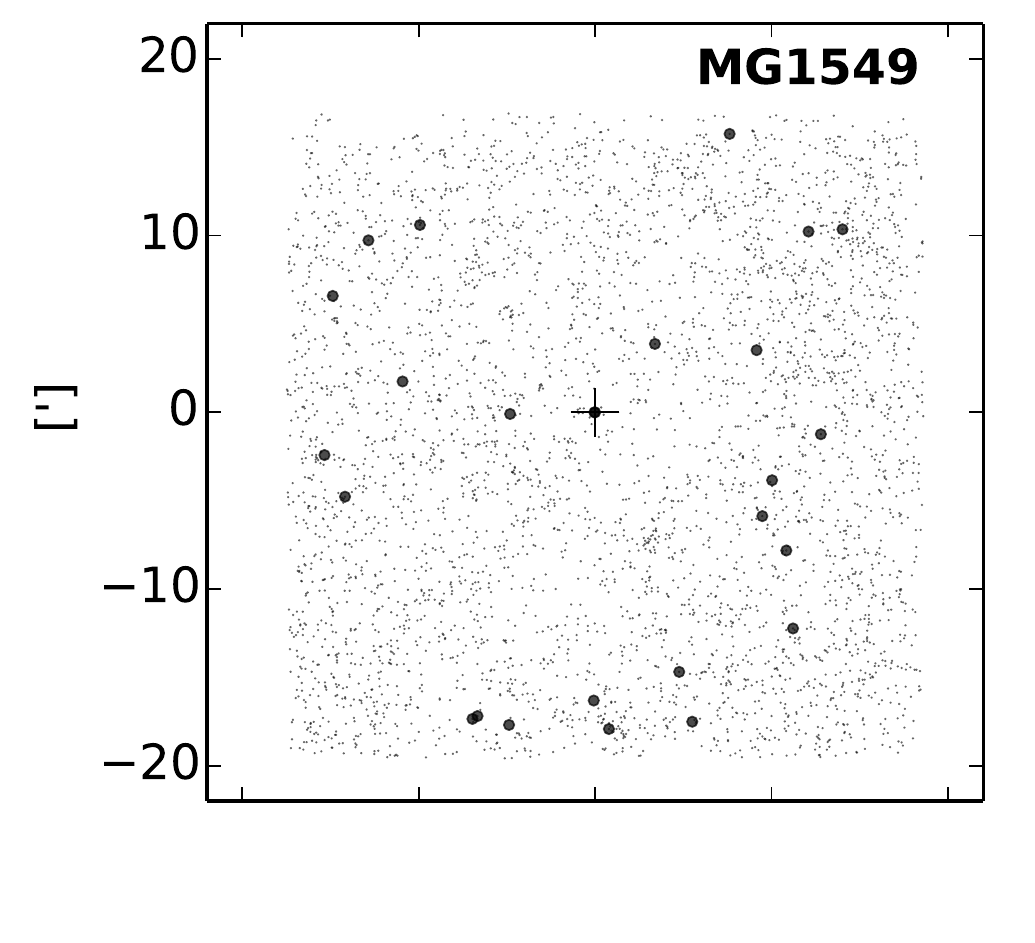} &
\includegraphics[width=0.28\textwidth]{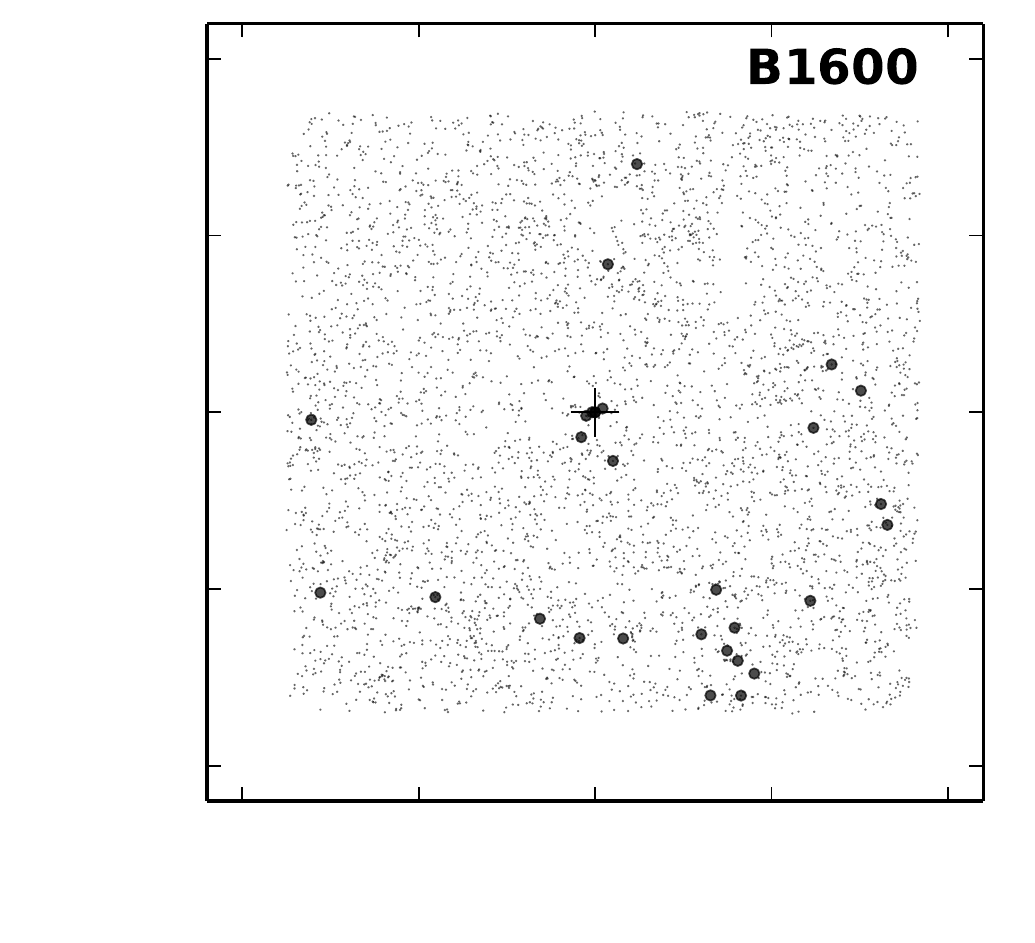} &
\includegraphics[width=0.28\textwidth]{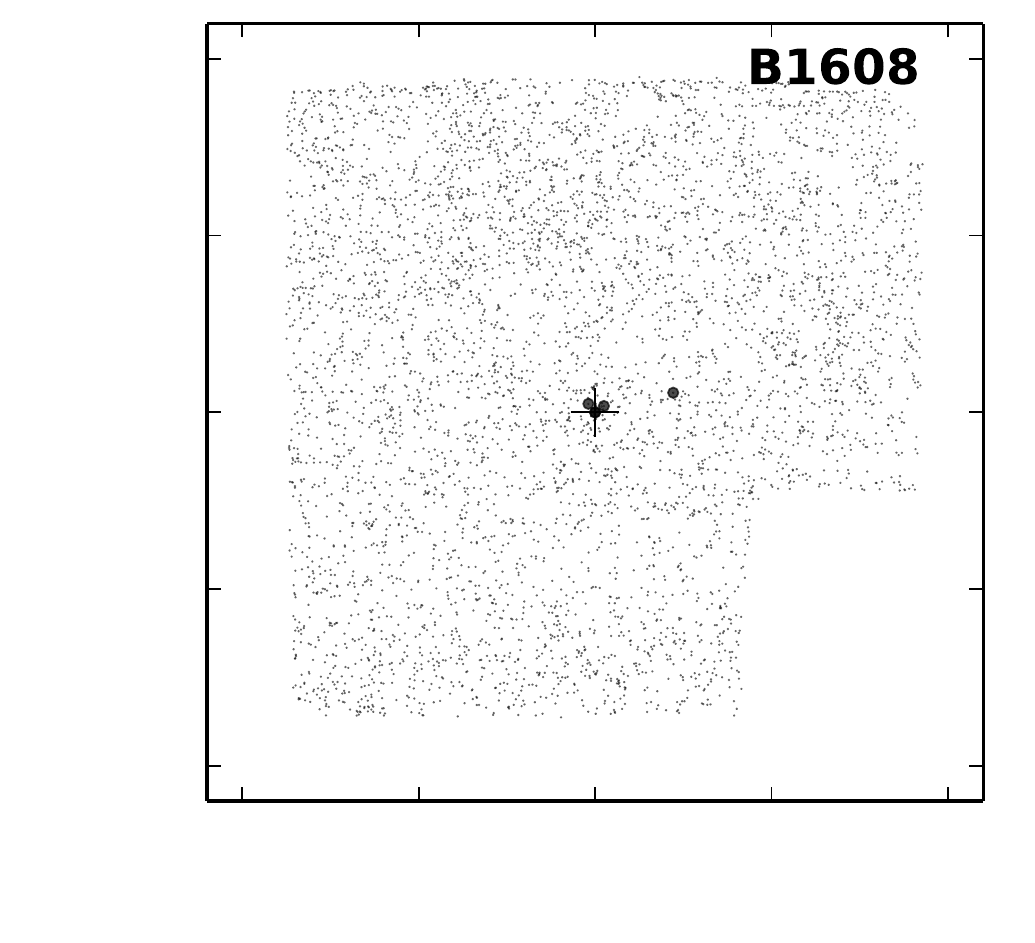} &
\includegraphics[width=0.28\textwidth]{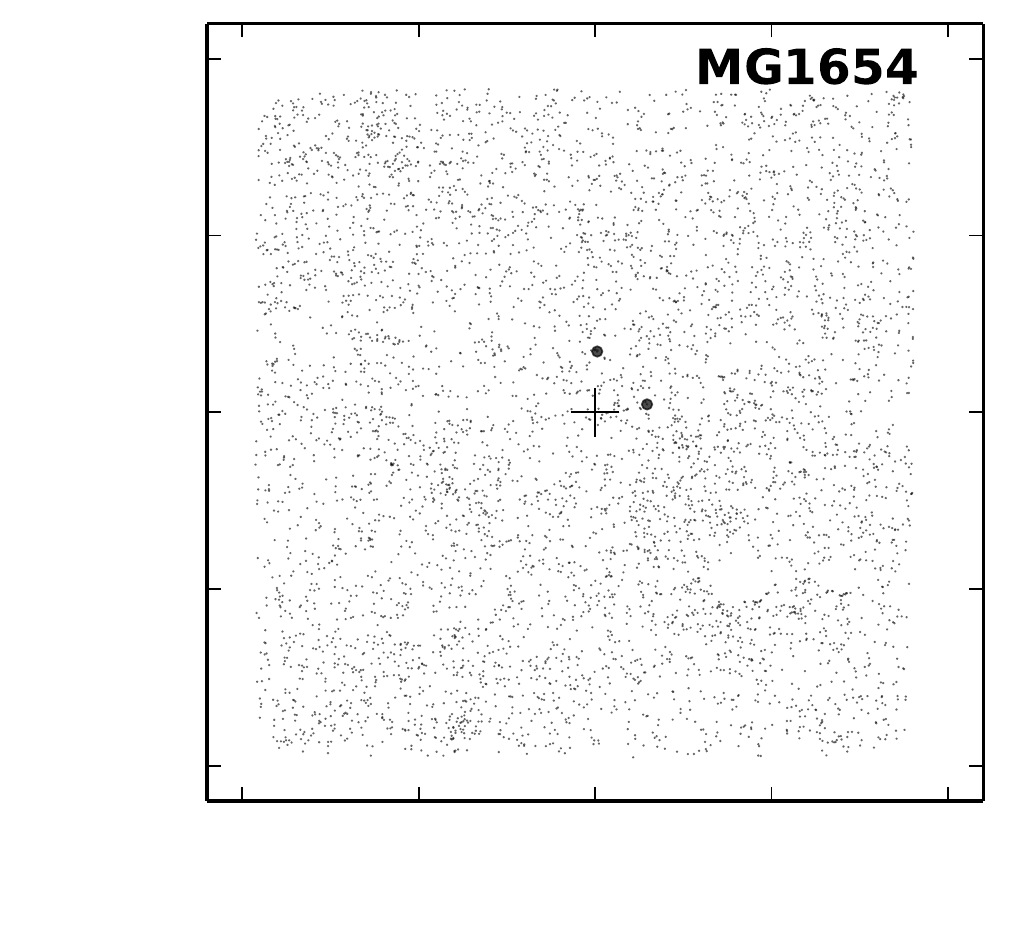} \\ [-22pt]
\includegraphics[width=0.28\textwidth]{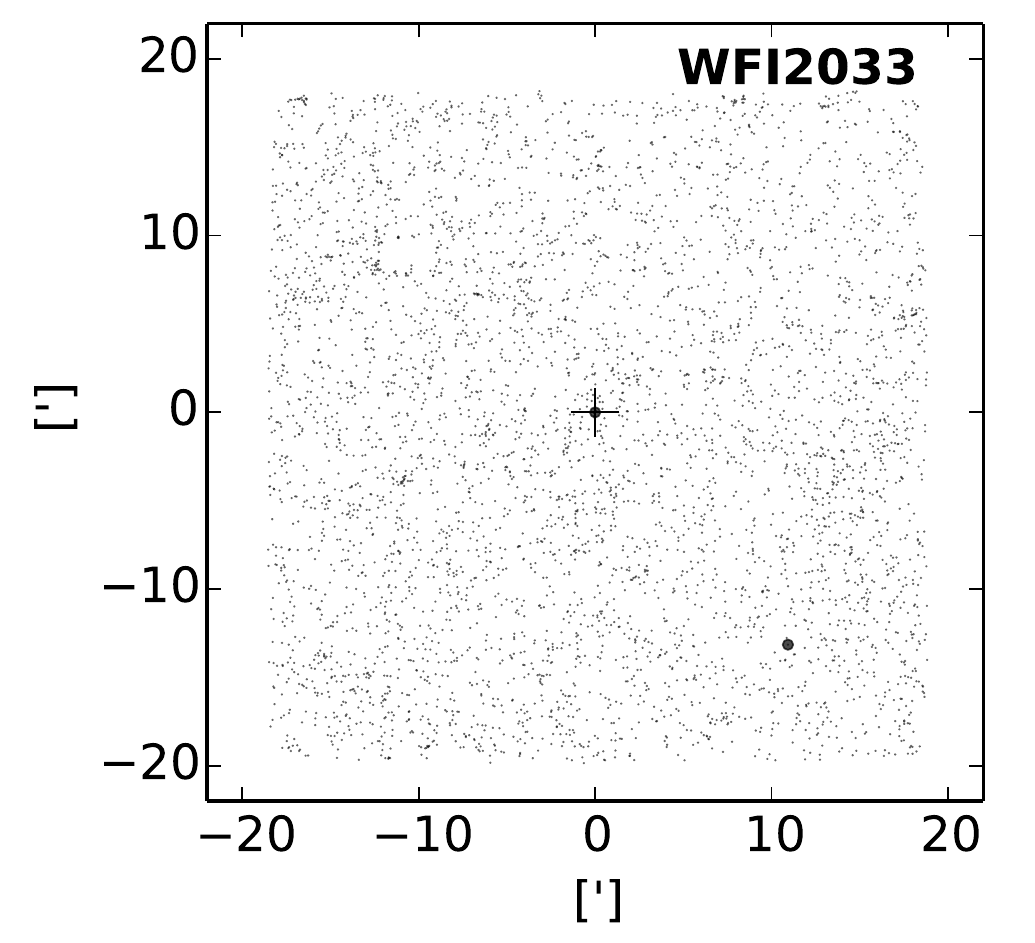} &
\includegraphics[width=0.28\textwidth]{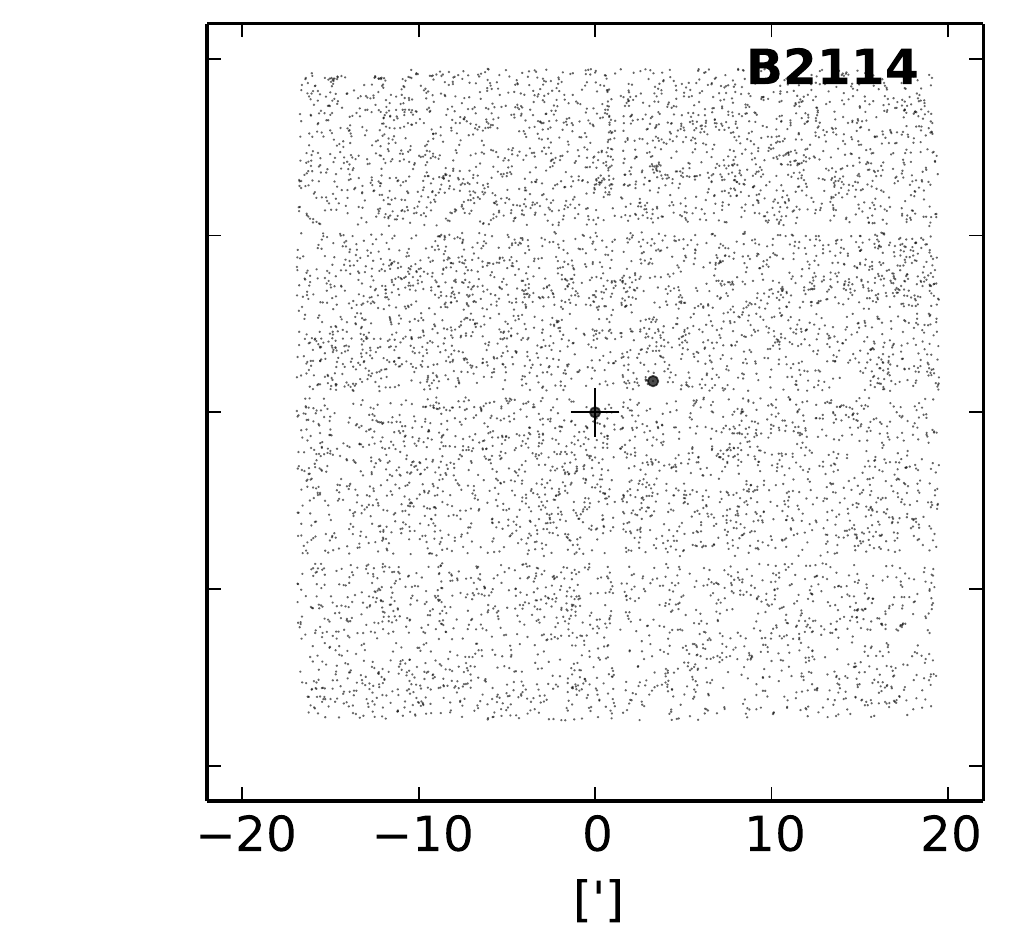} &
\includegraphics[width=0.28\textwidth]{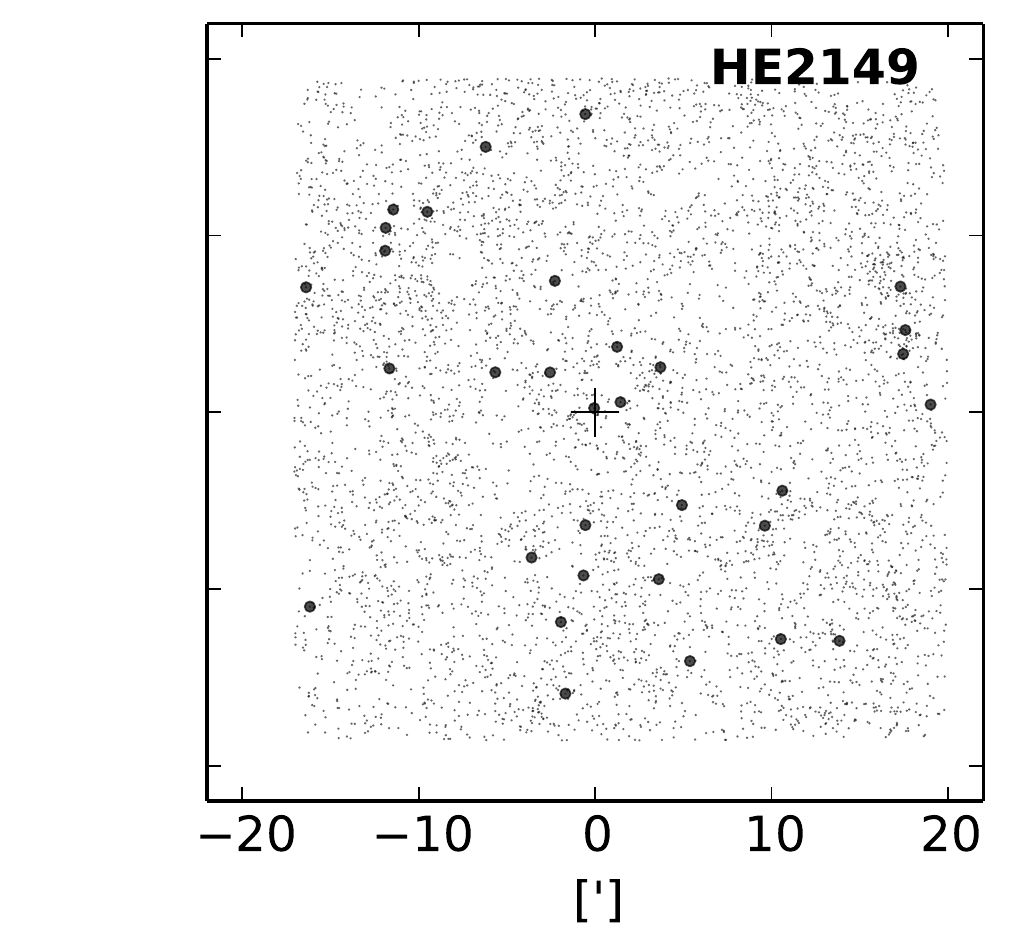} & \\ [-22pt]
\end{tabular}
\vspace{7mm}
\caption{\footnotesize Continued from Figure \ref{fig:radec_extra_a}. No NED redshifts were added in the PMN2004 field therefore this field is not included in the plot. \label{fig:radec_extra_b}}
\end{figure*}

\begin{figure*}
\figurenum{11a}
\epsscale{1.3}
\label{7a}
\plotone{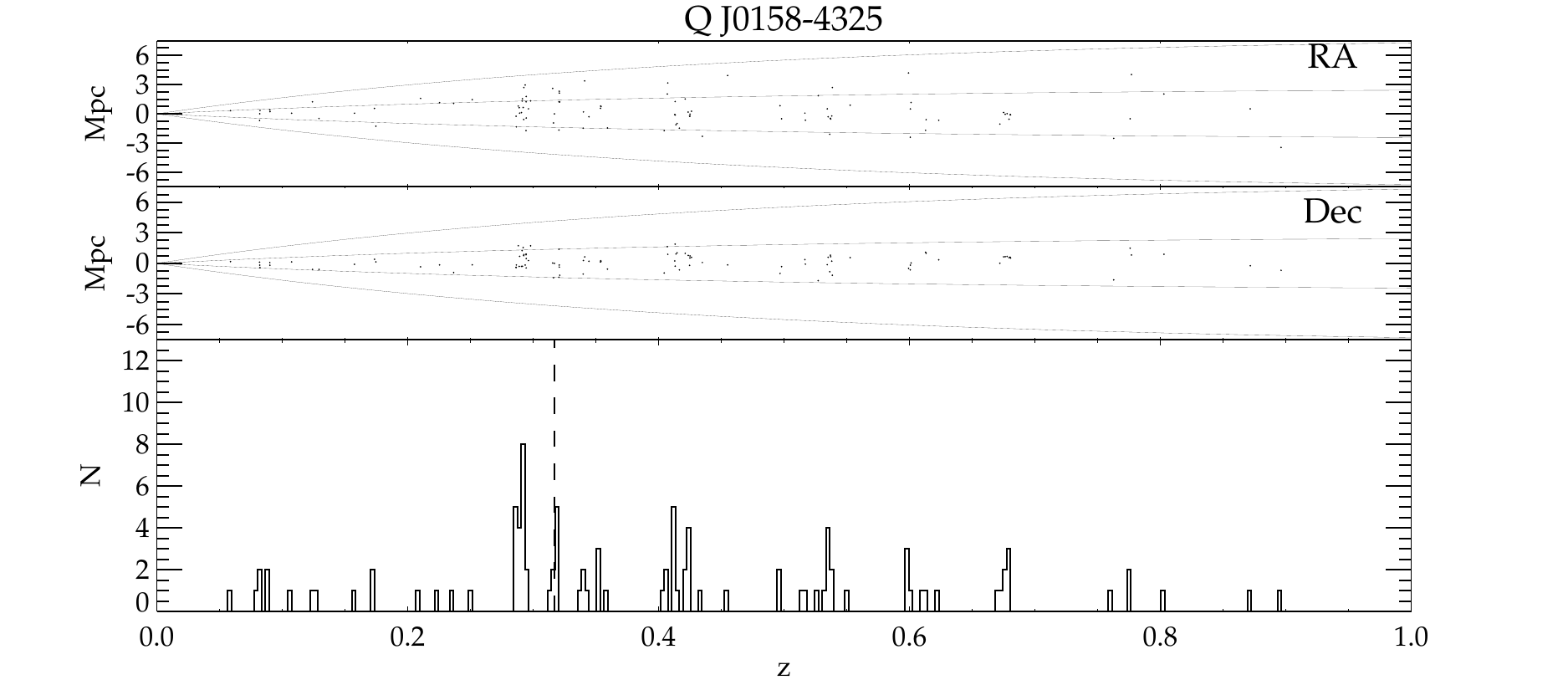}
\caption{\footnotesize Galaxy redshift distribution in the field of Q0158. The top two panels show the redshift distributions of galaxies projected in RA (top) and Dec (middle) in units of proper Mpc from the center of the field. This projection exaggerates the opening angle of the beams ($30\arcmin$), which causes structures to be stretched perpendicular to the redshift direction but makes them easier to see. The bottom panel shows the distribution in the form of a histogram with a binsize of $\delta z = 0.001$. The vertical dashed line indicates the spectroscopic redshift of the lens galaxy. The histogram includes all galaxies with spectroscopic redshifts. }
\end{figure*}

\begin{figure*}
\figurenum{11b}
\epsscale{1.3}
\plotone{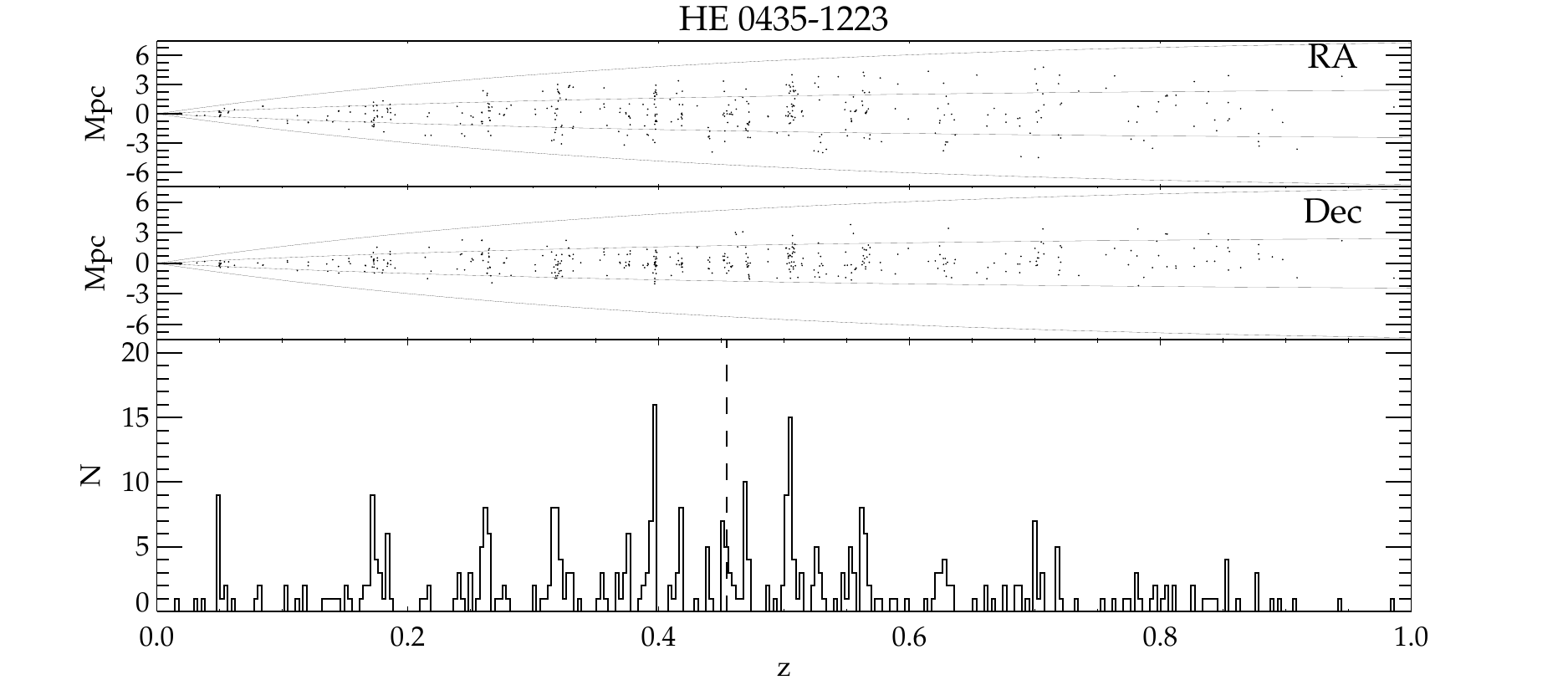}
\caption{\footnotesize Same as Figure  \ref{7a} but for the field of HE0435.}
\end{figure*}

\begin{figure*}
\figurenum{11c}
\epsscale{1.3}
\plotone{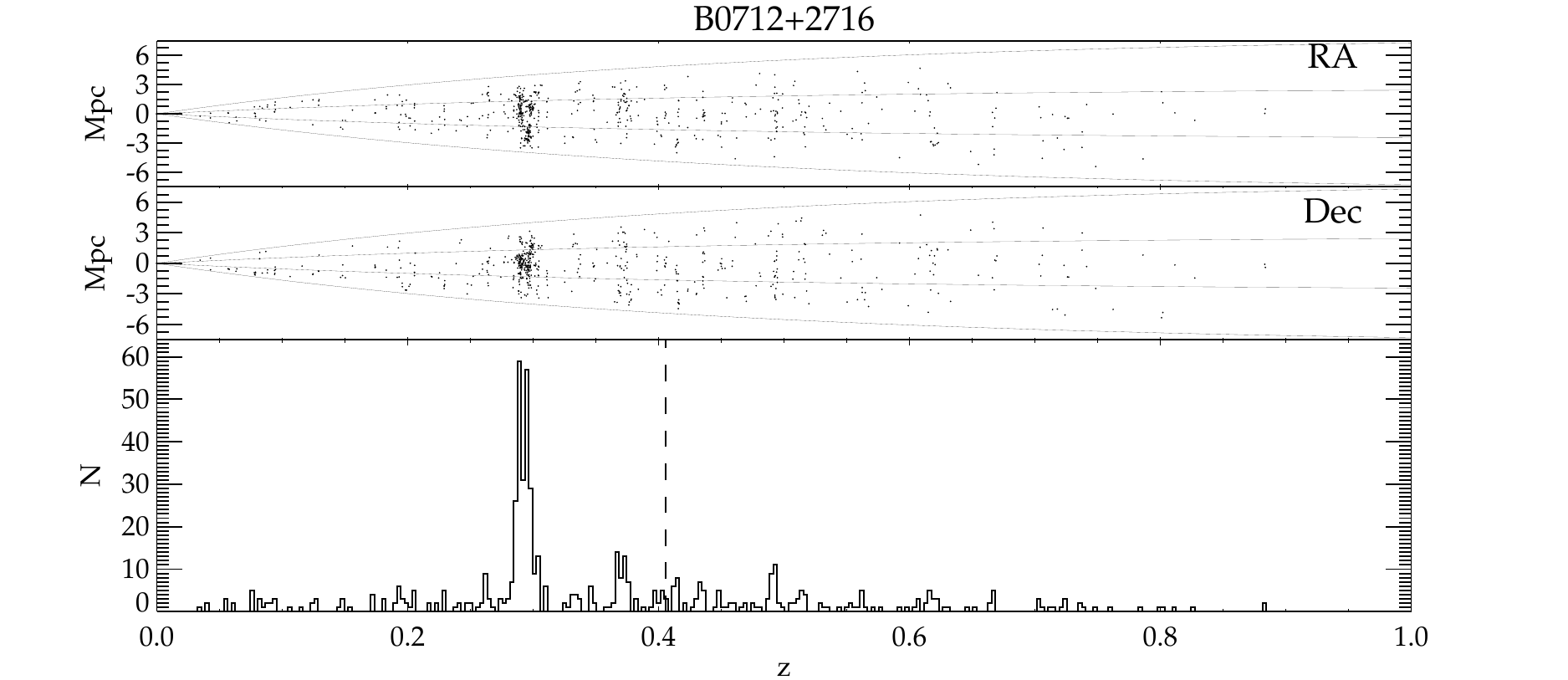}
\caption{\footnotesize Same as Figure  \ref{7a} but for the field of B0721.}
\end{figure*}

\begin{figure*}
\figurenum{11d}
\epsscale{1.3}
\plotone{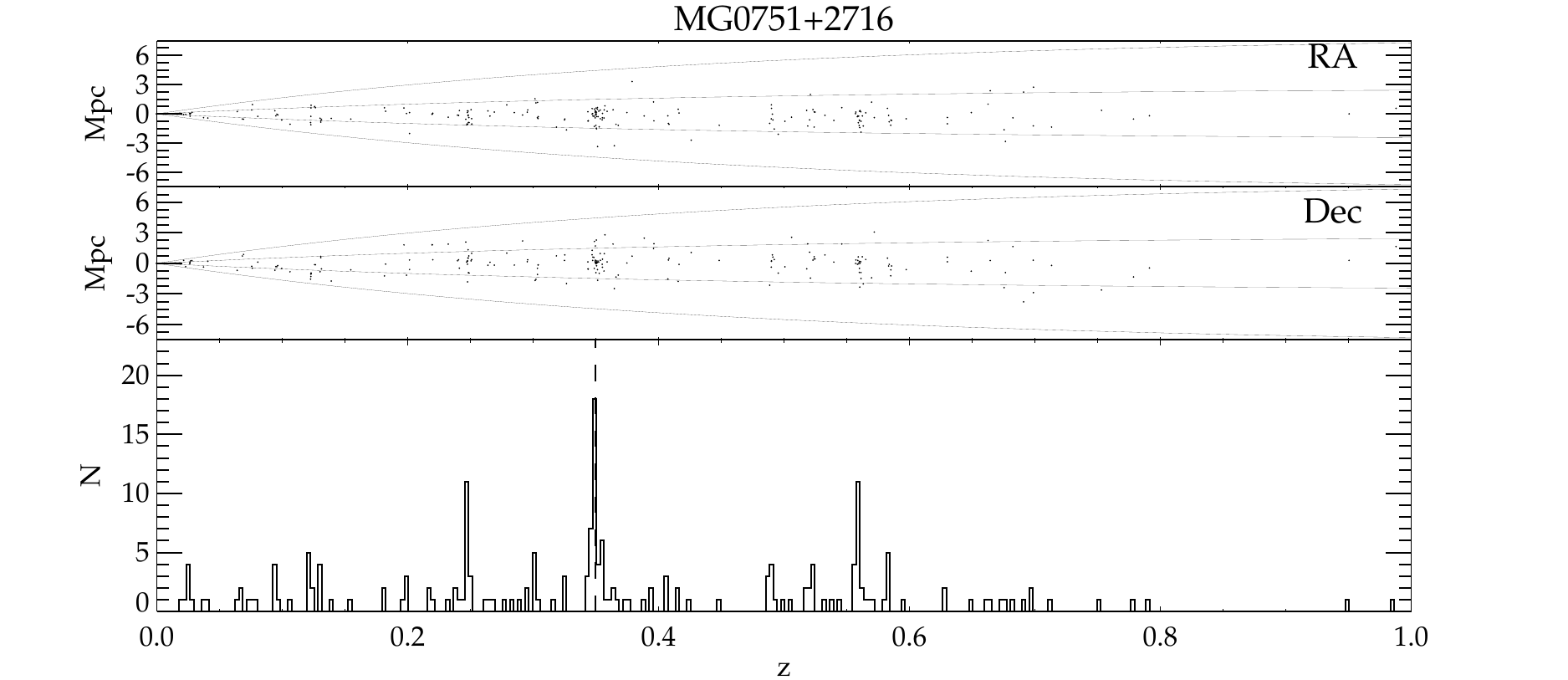}
\caption{\footnotesize Same as Figure  \ref{7a} but for the field of MG0751.}
\end{figure*}

\begin{figure*}
\figurenum{11e}
\epsscale{1.3}
\plotone{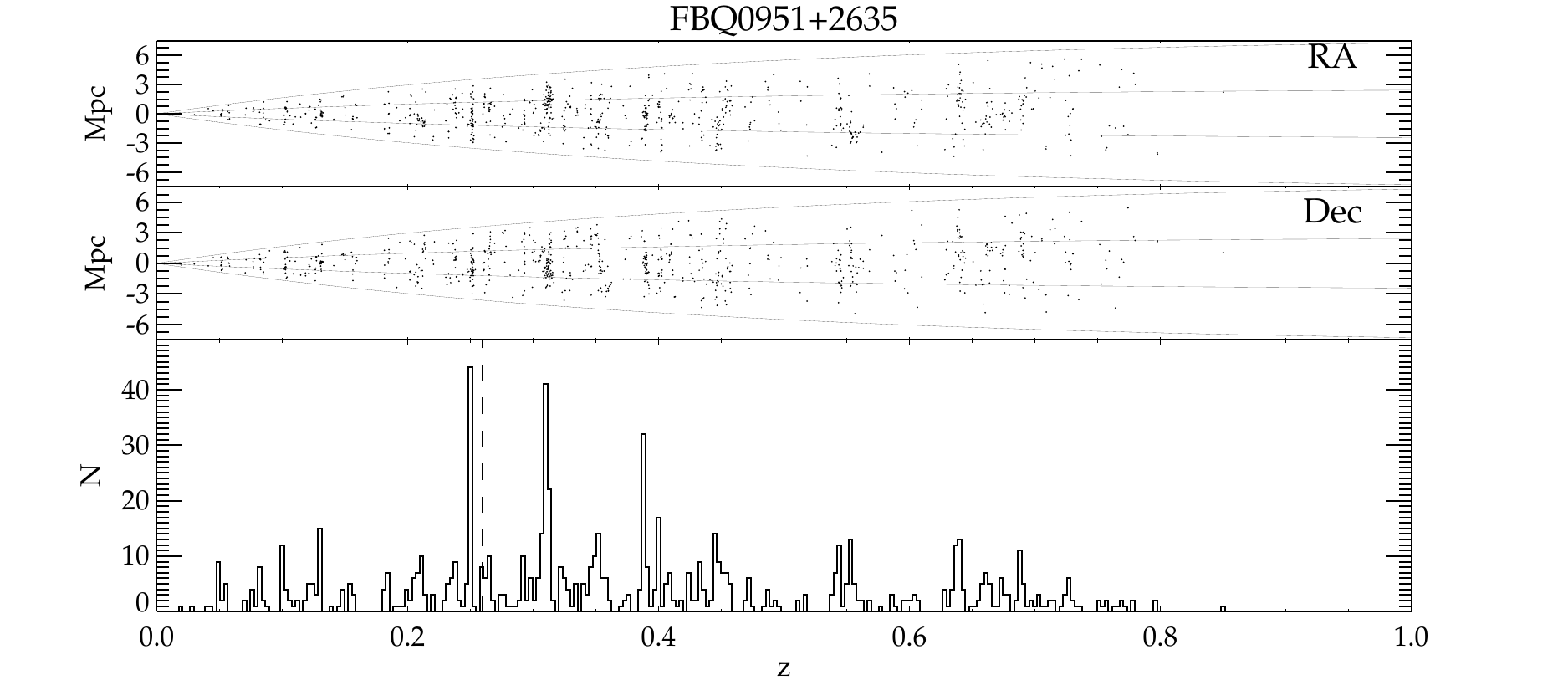}
\caption{\footnotesize Same as Figure  \ref{7a} but for the field of FBQ0951.}
\end{figure*}

\begin{figure*}
\figurenum{11f}
\epsscale{1.3}
\plotone{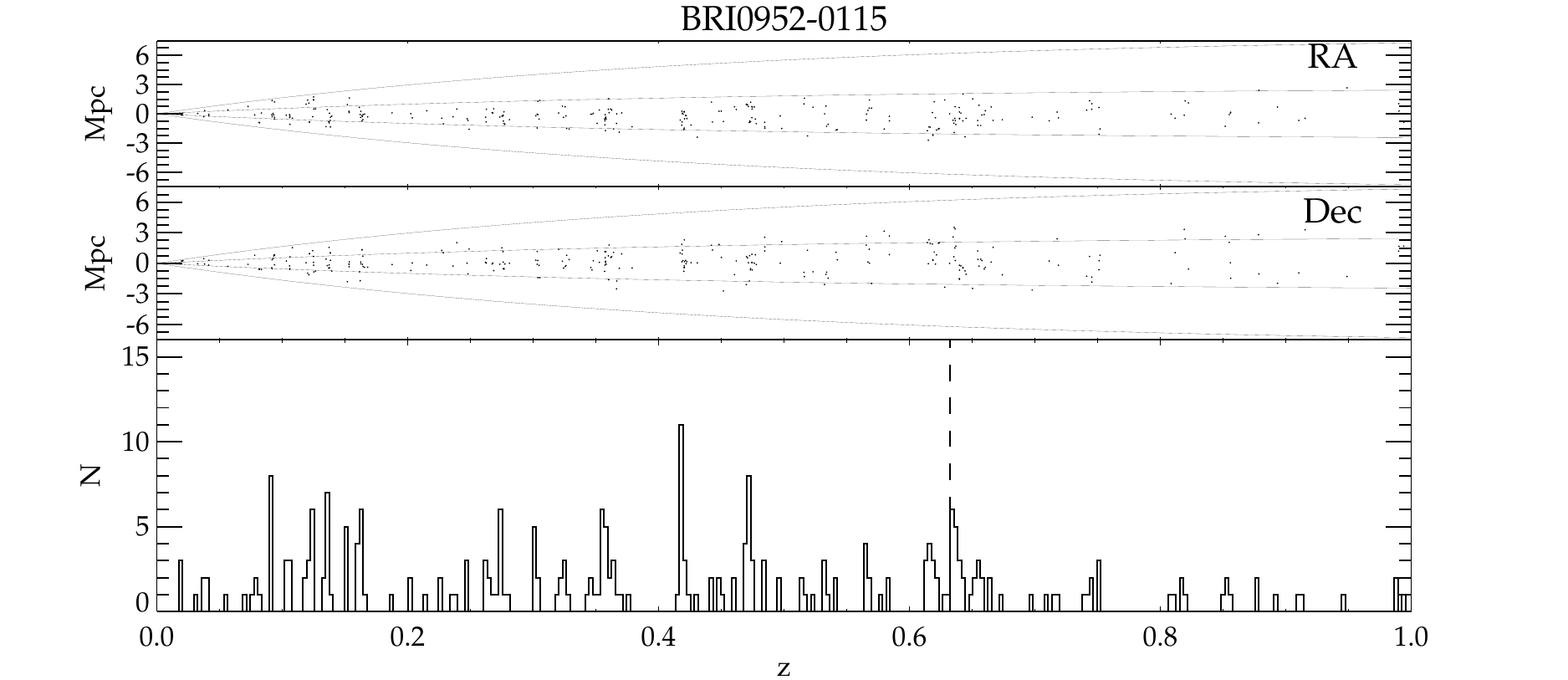}
\caption{\footnotesize Same as Figure  \ref{7a} but for the field of BRI0952.}
\end{figure*}

\begin{figure*}
\figurenum{11g}
\epsscale{1.3}
\plotone{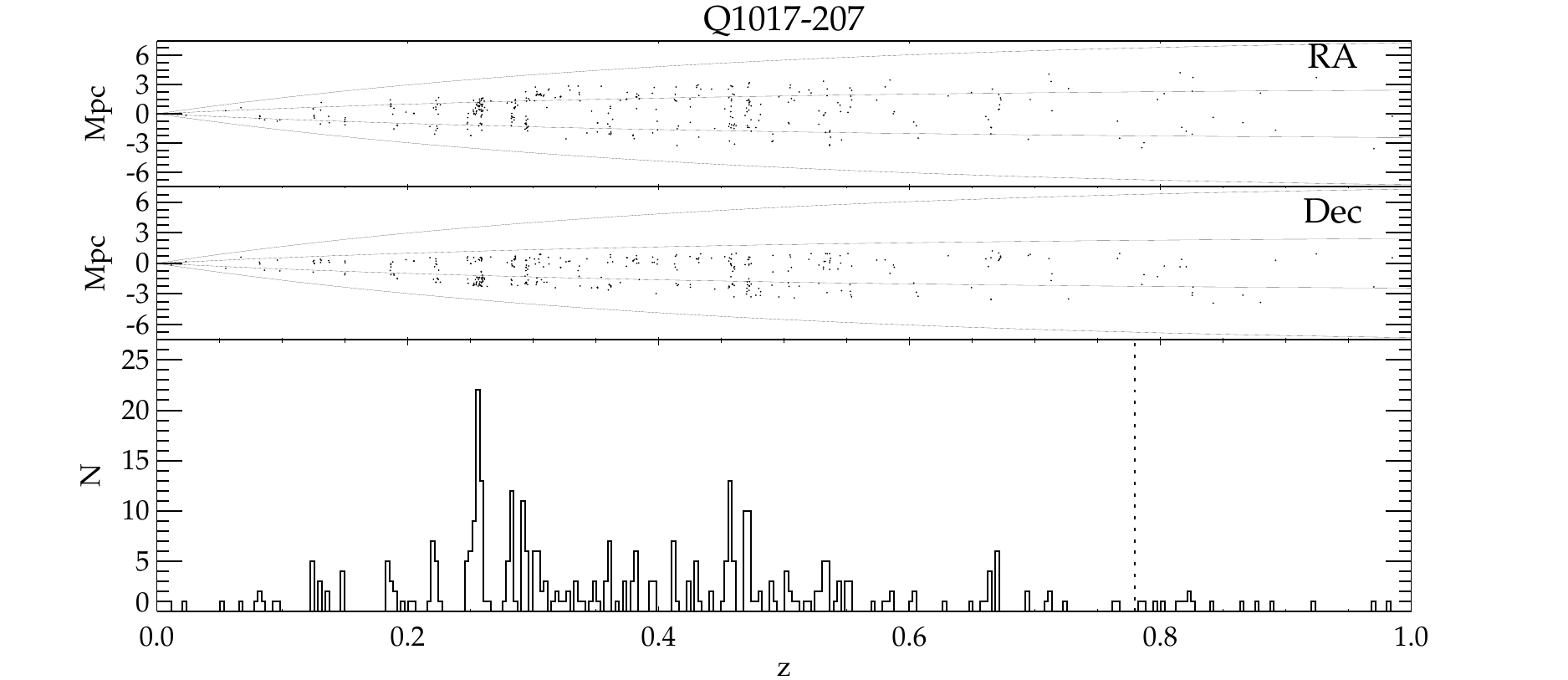}
\caption{\footnotesize Same as Figure  \ref{7a} but for the field of Q1017. The vertical dotted line denotes the photometric redshift of the lens galaxy.}
\end{figure*}

\begin{figure*}
\figurenum{11h}
\epsscale{1.3}
\plotone{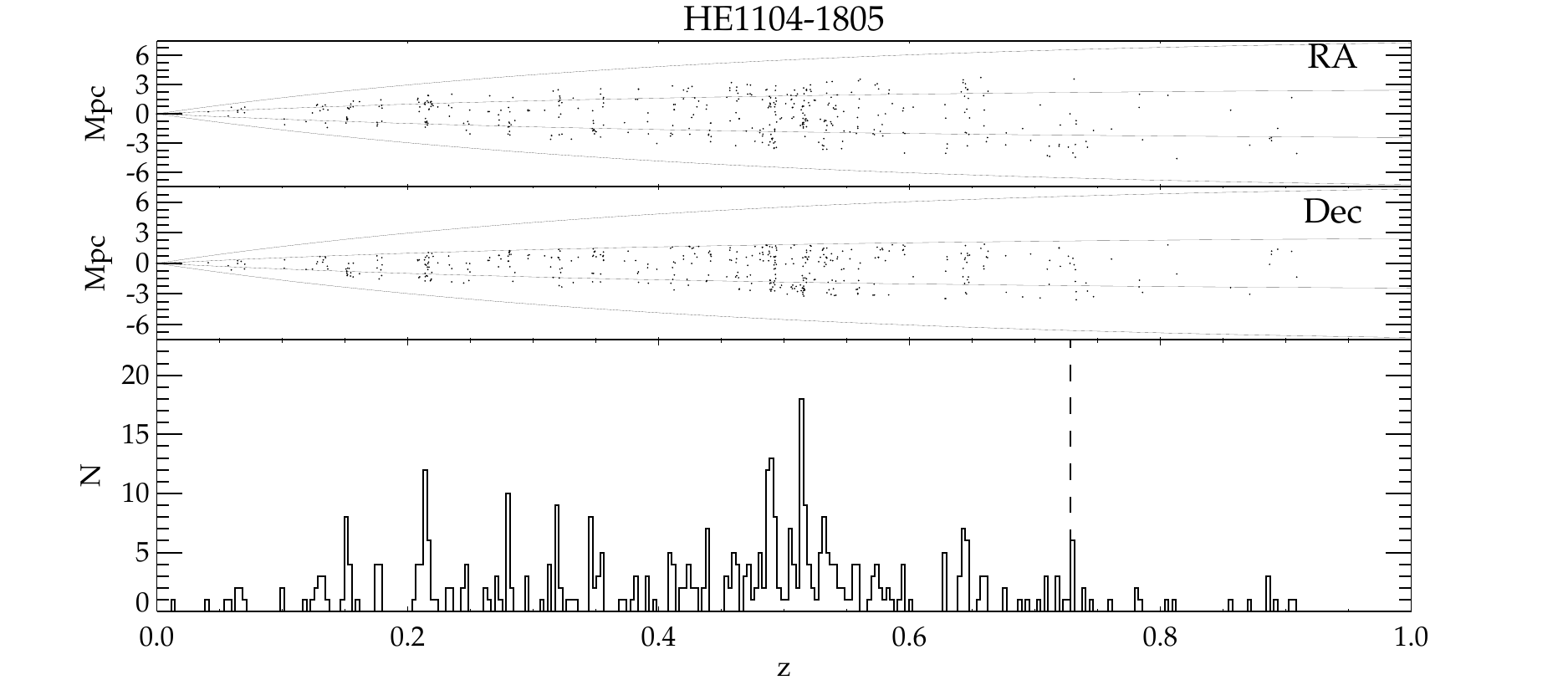}
\caption{\footnotesize Same as Figure  \ref{7a} but for the field of HE1104.}
\end{figure*}

\begin{figure*}
\figurenum{11i}
\epsscale{1.3}
\plotone{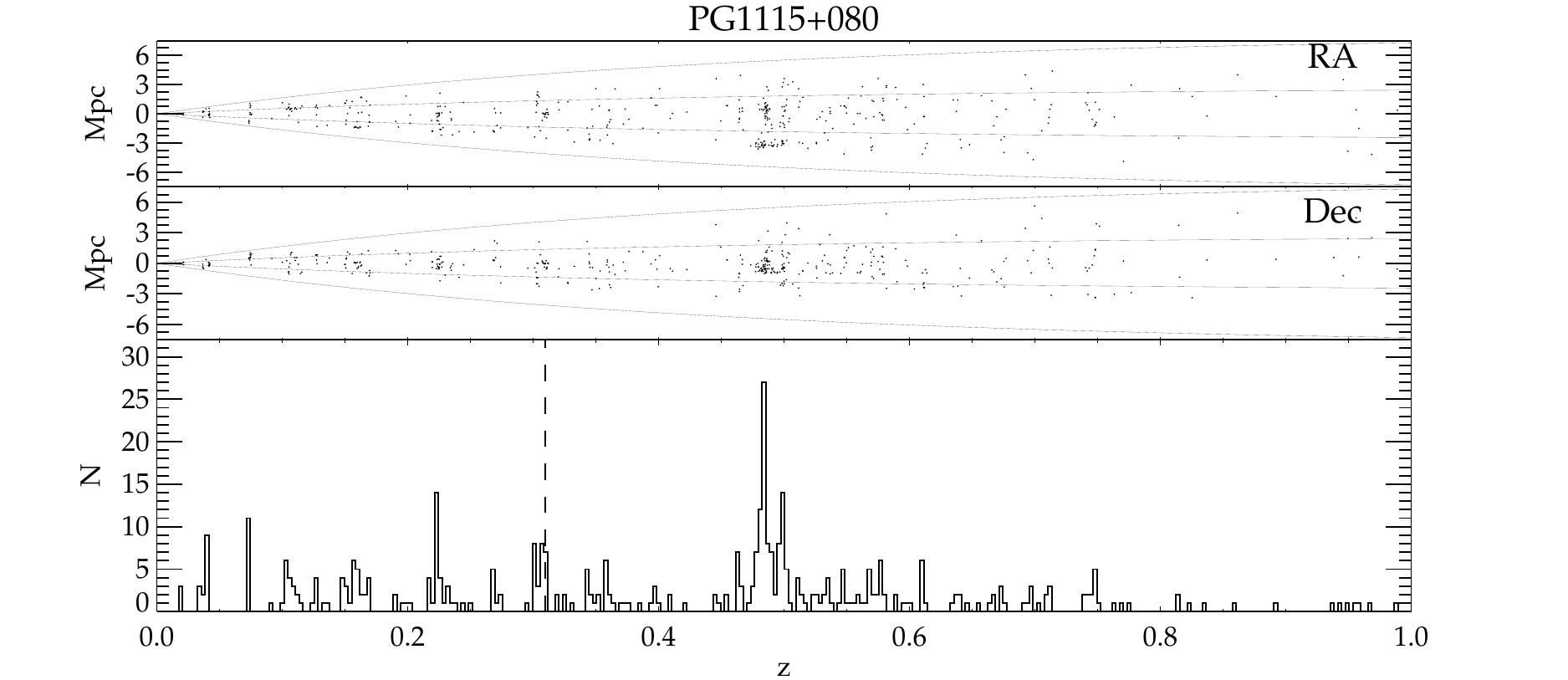}
\caption{\footnotesize Same as Figure  \ref{7a} but for the field of PG1115.}
\end{figure*}

\begin{figure*}
\figurenum{11j}
\epsscale{1.3}
\plotone{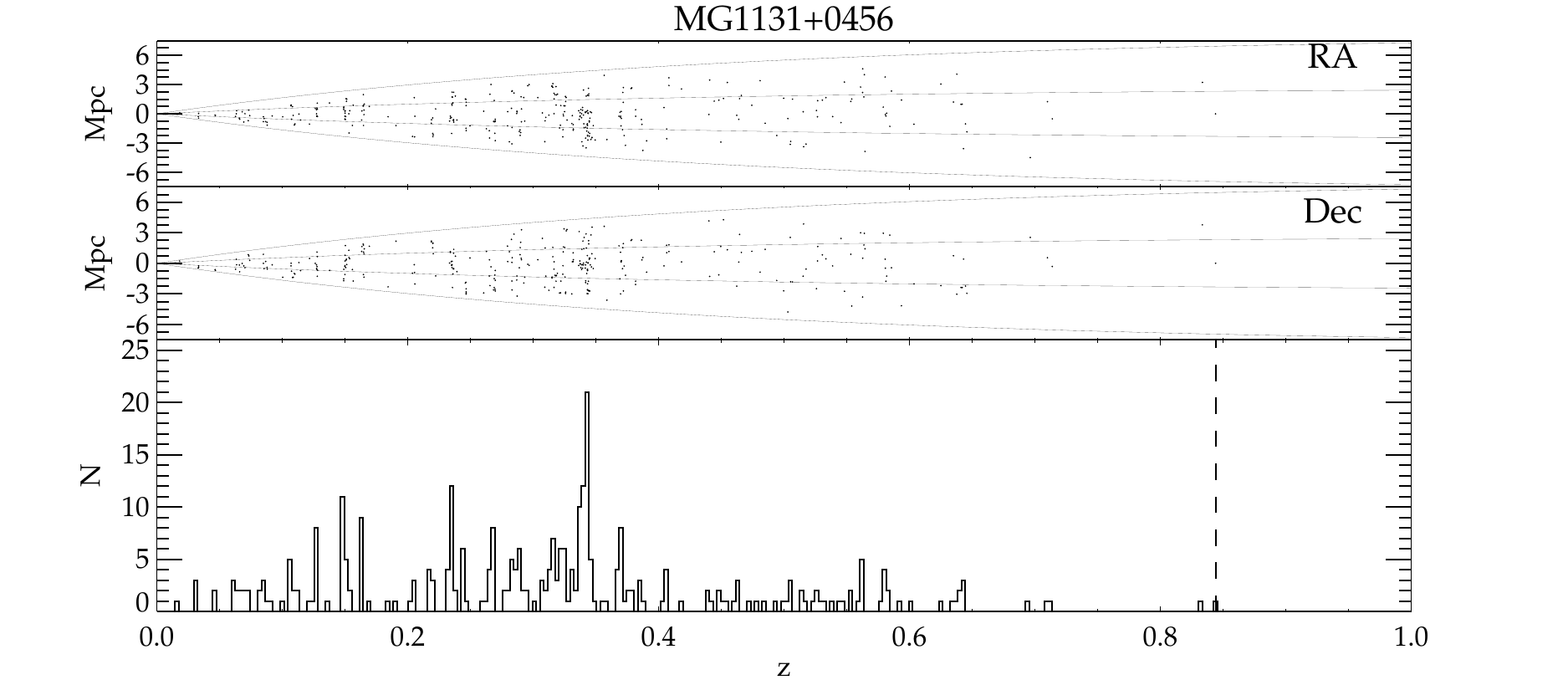}
\caption{\footnotesize Same as Figure  \ref{7a} but for the field of MG1131.}
\end{figure*}

\begin{figure*}
\figurenum{11k}
\epsscale{1.3}
\plotone{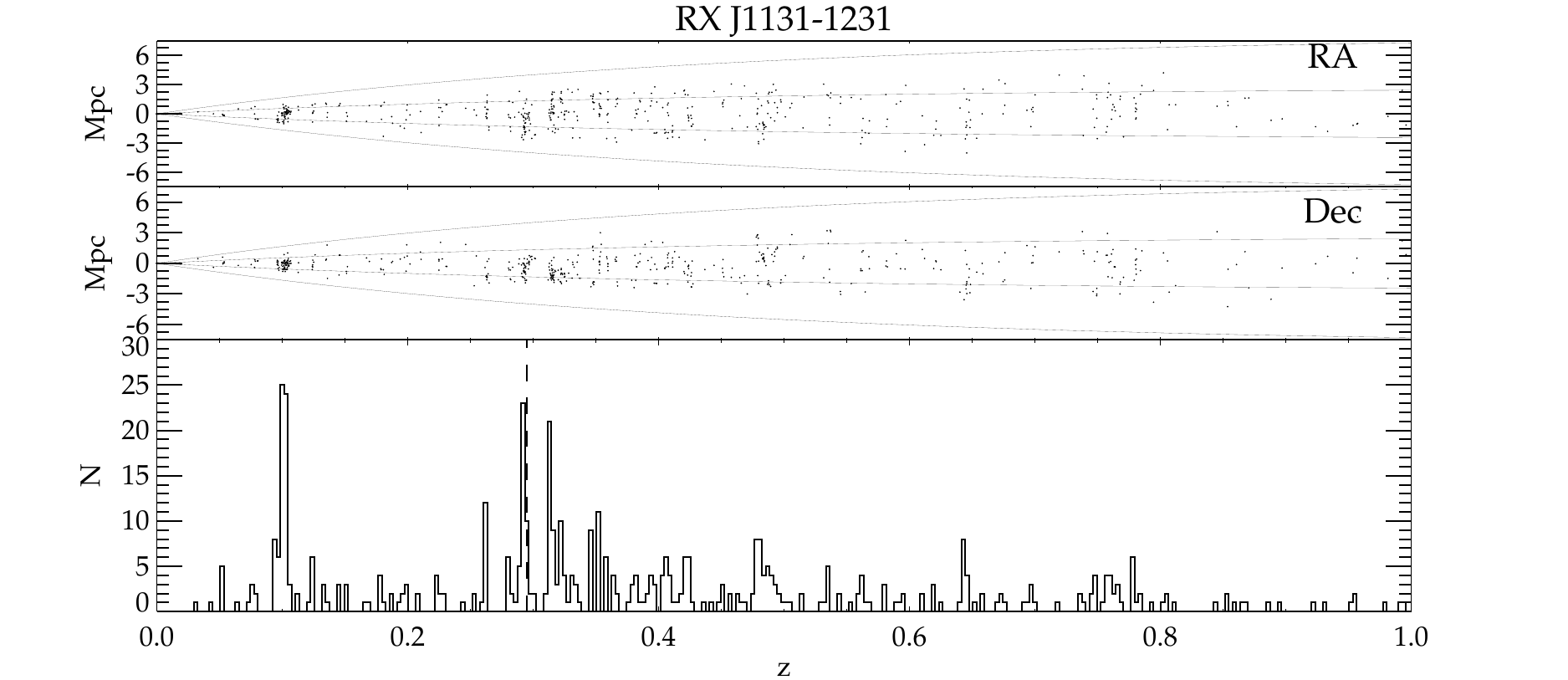}
\caption{\footnotesize Same as Figure  \ref{7a} but for the field of RXJ1131.}
\end{figure*}

\clearpage

\begin{figure*}
\figurenum{11l}
\epsscale{1.3}
\plotone{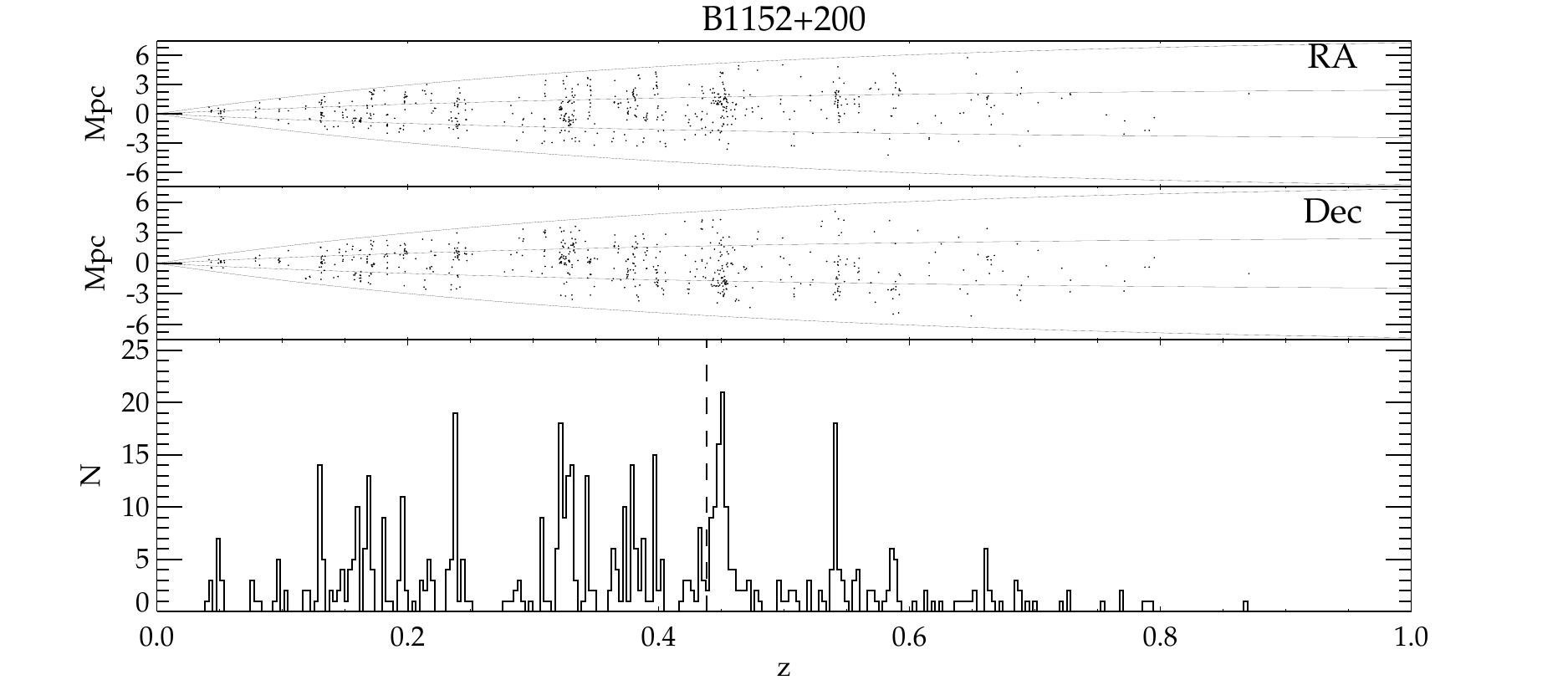}
\caption{\footnotesize Same as Figure  \ref{7a} but for the field of B1152.}
\end{figure*}

\begin{figure*}
\figurenum{11m}
\epsscale{1.3}
\plotone{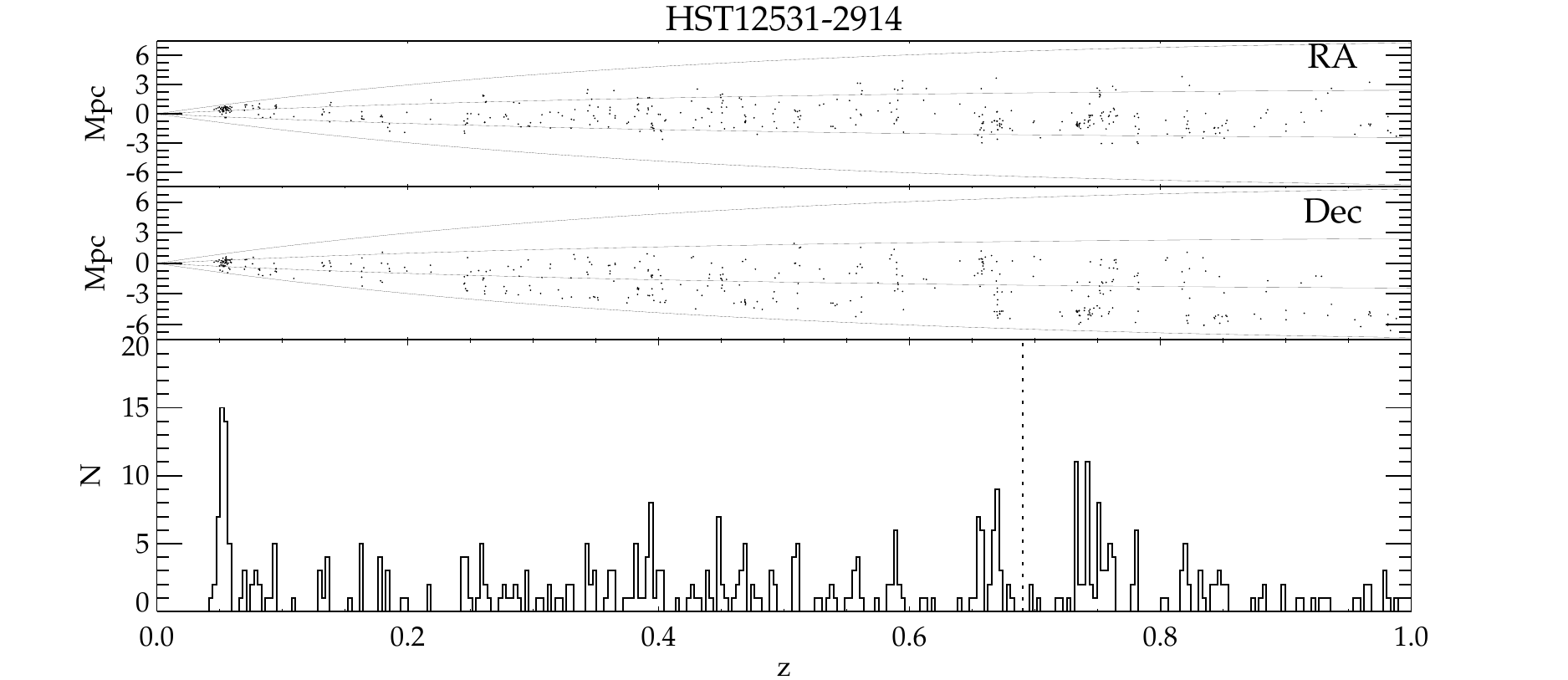}
\caption{\footnotesize Same as Figure  \ref{7a} but for the field of H12531. The vertical dotted line denotes the photometric redshift of the lens galaxy.}
\end{figure*}

\begin{figure*}
\figurenum{11n}
\epsscale{1.3}
\plotone{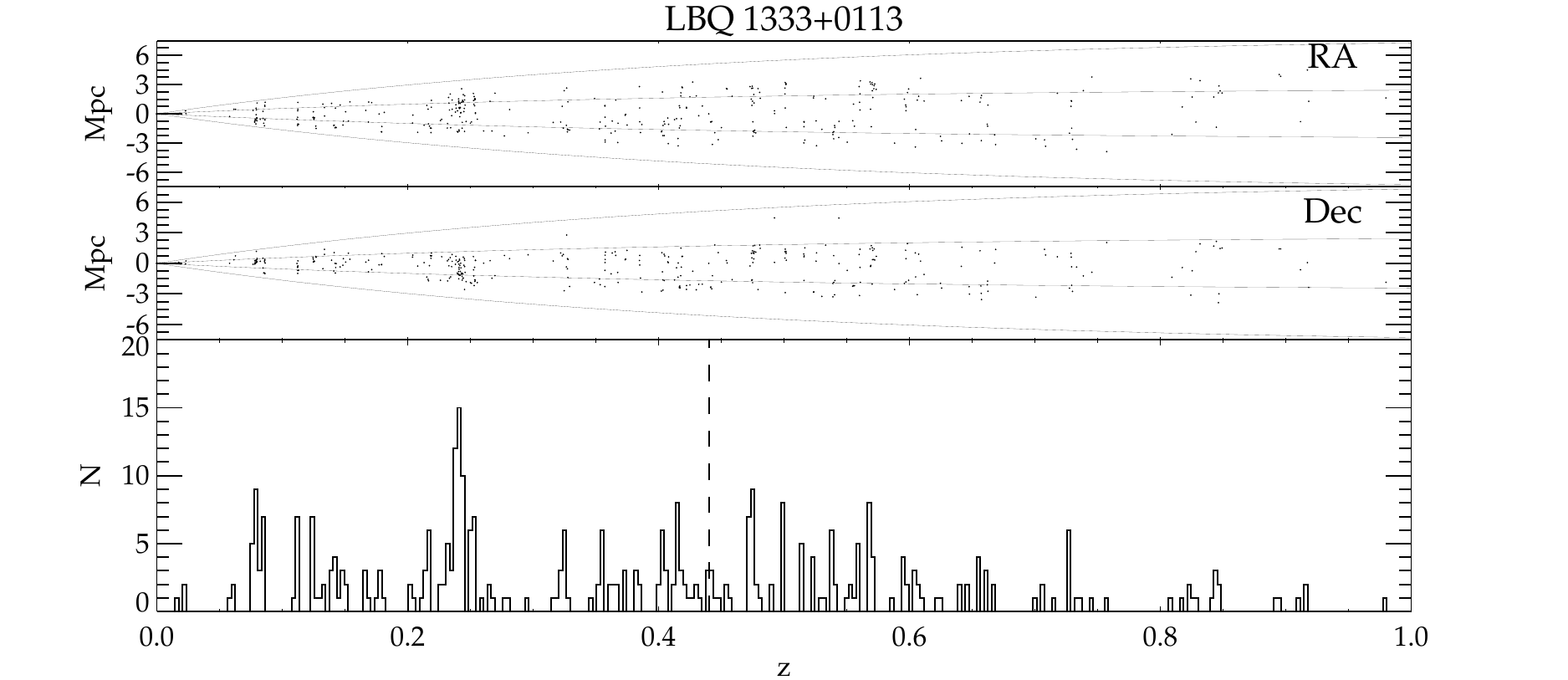}
\caption{\footnotesize Same as Figure  \ref{7a} but for the field of LBQ1333.}
\end{figure*}

\begin{figure*}
\figurenum{11o}
\epsscale{1.3}
\plotone{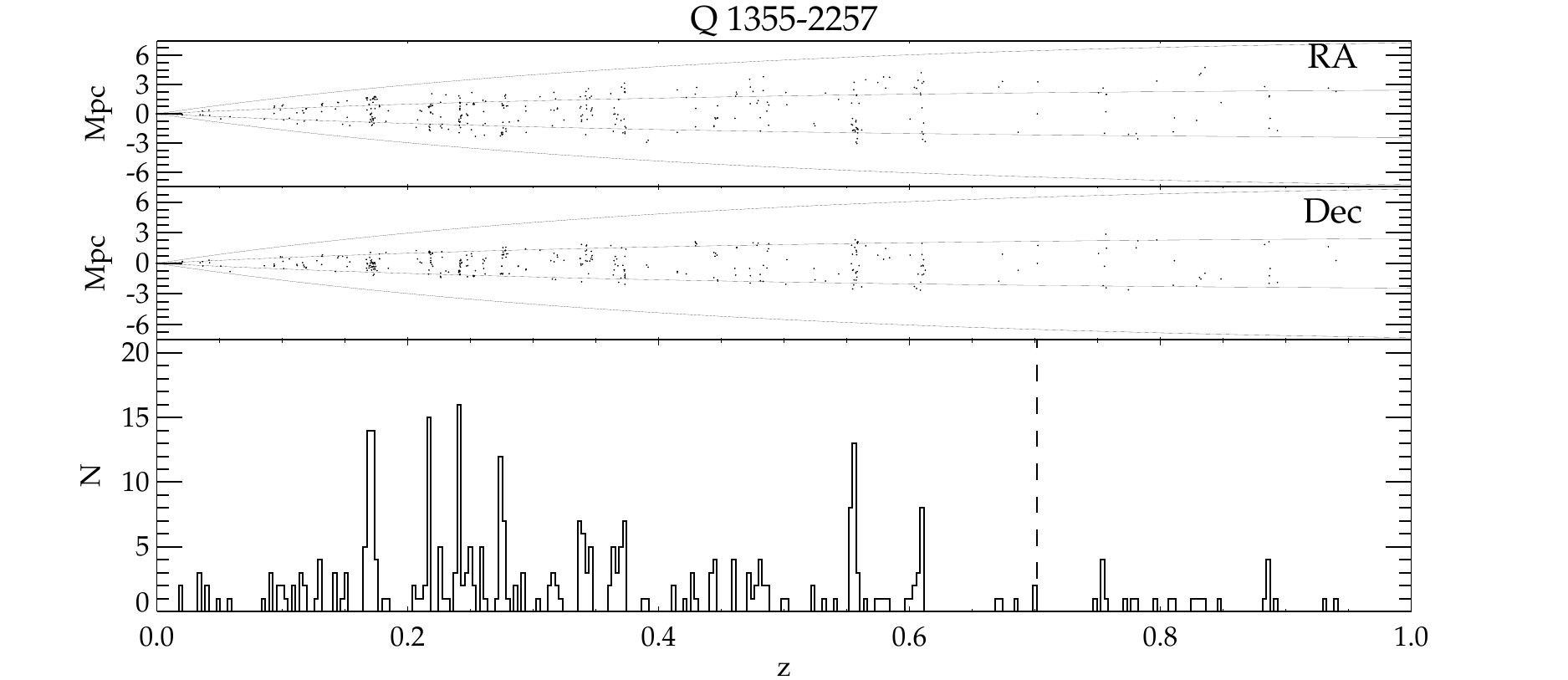}
\caption{\footnotesize Same as Figure  \ref{7a} but for the field of Q1355.}
\end{figure*}

\begin{figure*}
\figurenum{11p}
\epsscale{1.3}
\plotone{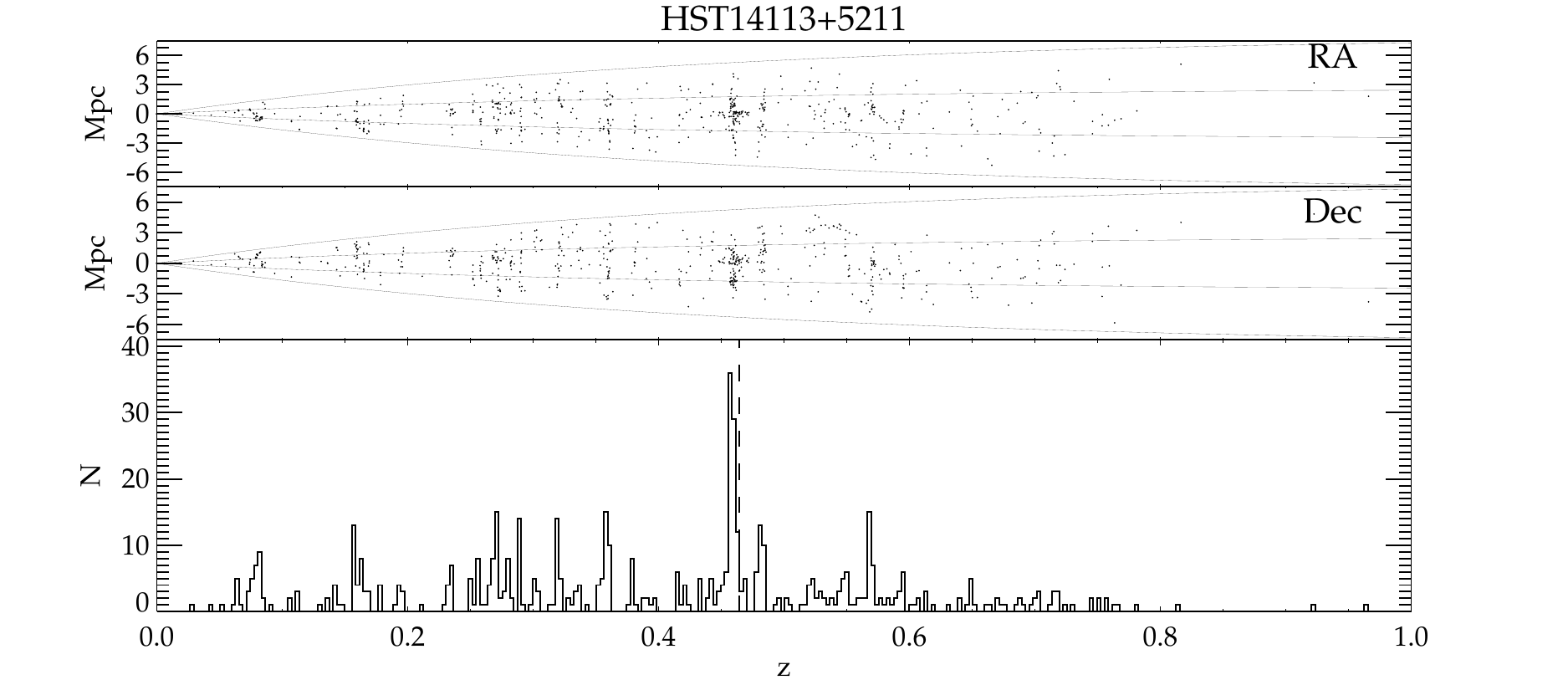}
\caption{\footnotesize Same as Figure  \ref{7a} but for the field of HST14113.}
\end{figure*}

\begin{figure*}
\figurenum{11q}
\epsscale{1.3}
\plotone{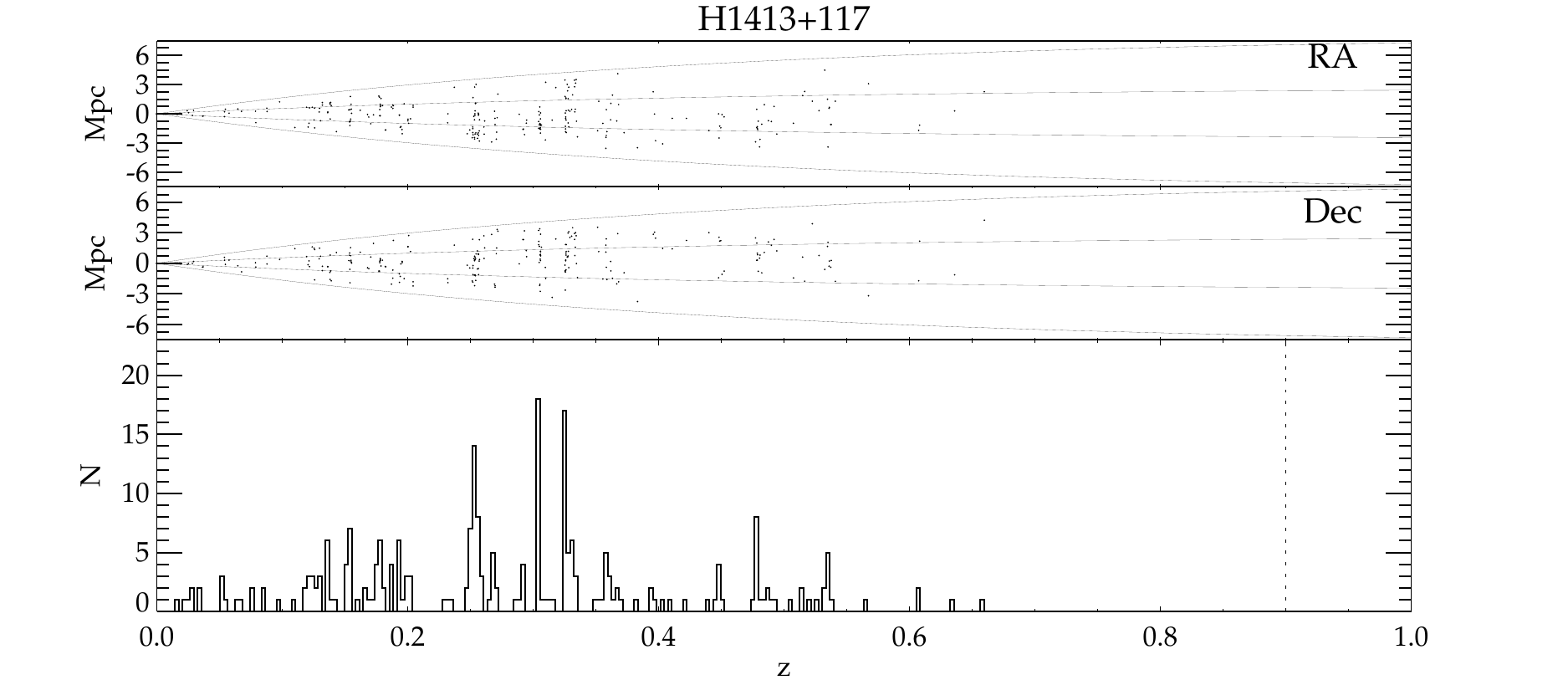}
\caption{\footnotesize Same as Figure  \ref{7a} but for the field of H1413. The vertical dotted line denotes the photometric redshift of the lens galaxy.}
\end{figure*}

\clearpage

\begin{figure*}
\figurenum{11r}
\epsscale{1.3}
\plotone{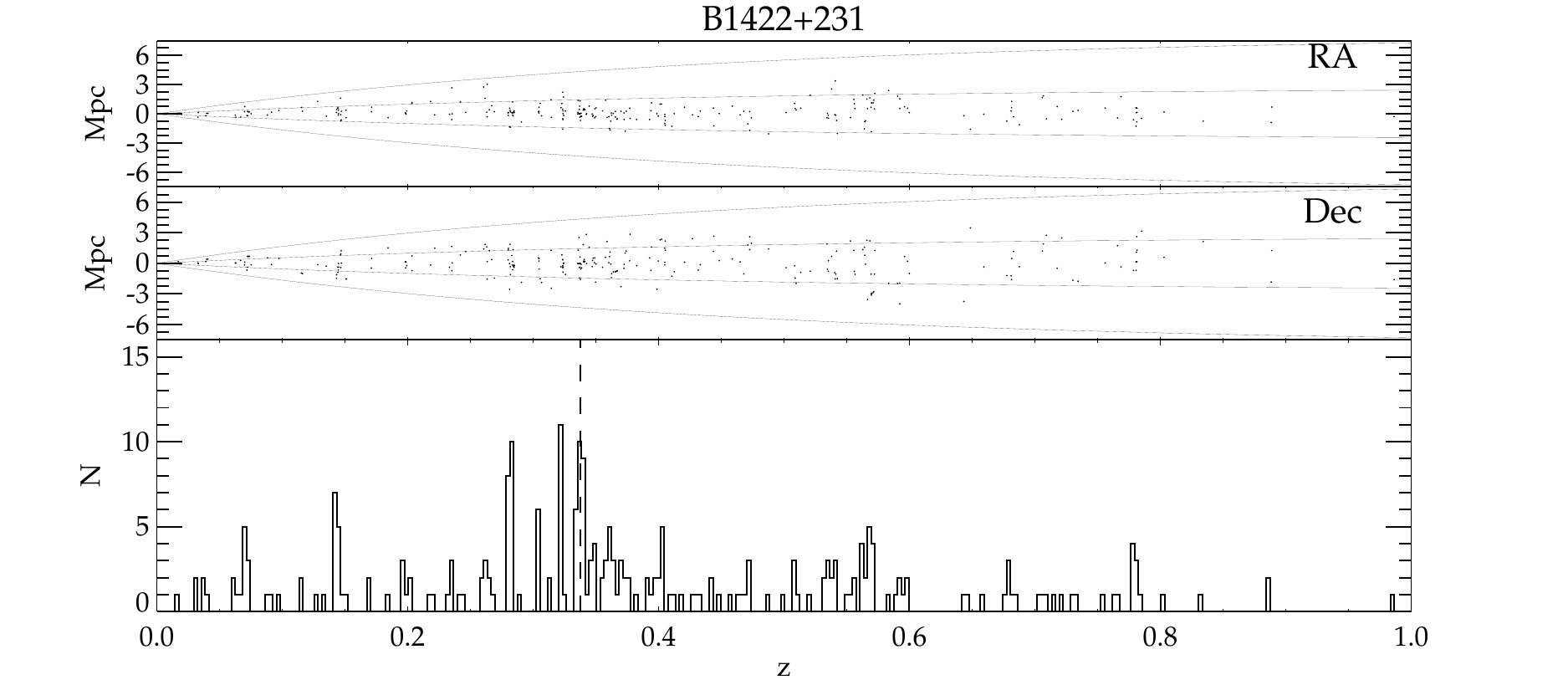}
\caption{\footnotesize Same as Figure  \ref{7a} but for the field of B1422.}
\end{figure*}

\begin{figure*}
\figurenum{11s}
\epsscale{1.3}
\plotone{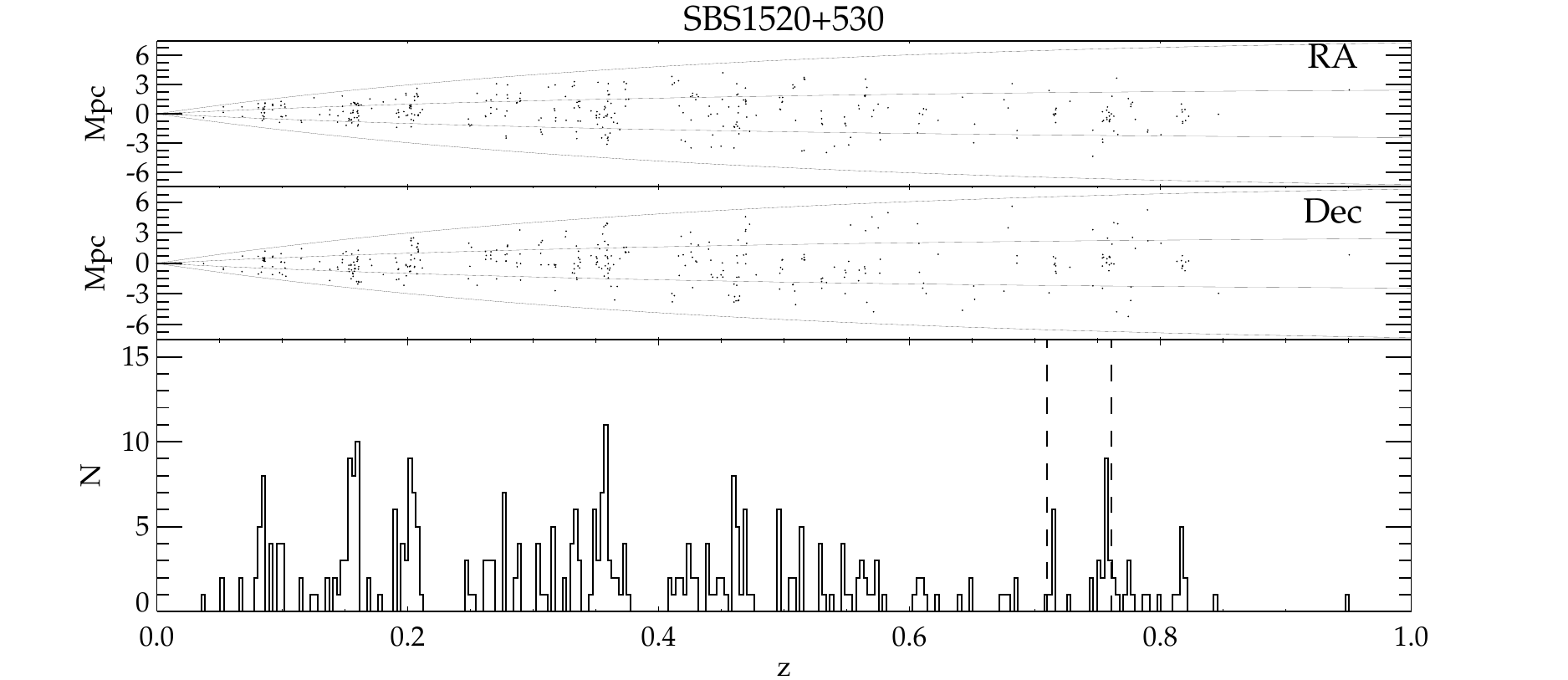}
\caption{\footnotesize Same as Figure  \ref{7a} but for the field of SBS1520. The redshift of the lens galaxy in this system is ambiguous: \citet{burud02} measure $z_l=0.71\pm0.005$, but \citet{auger08} claim $z_{l}=0.761$ is more likely. Both values are shown (dashed vertical lines).}
\end{figure*}

\begin{figure*}
\figurenum{11t}
\epsscale{1.3}
\plotone{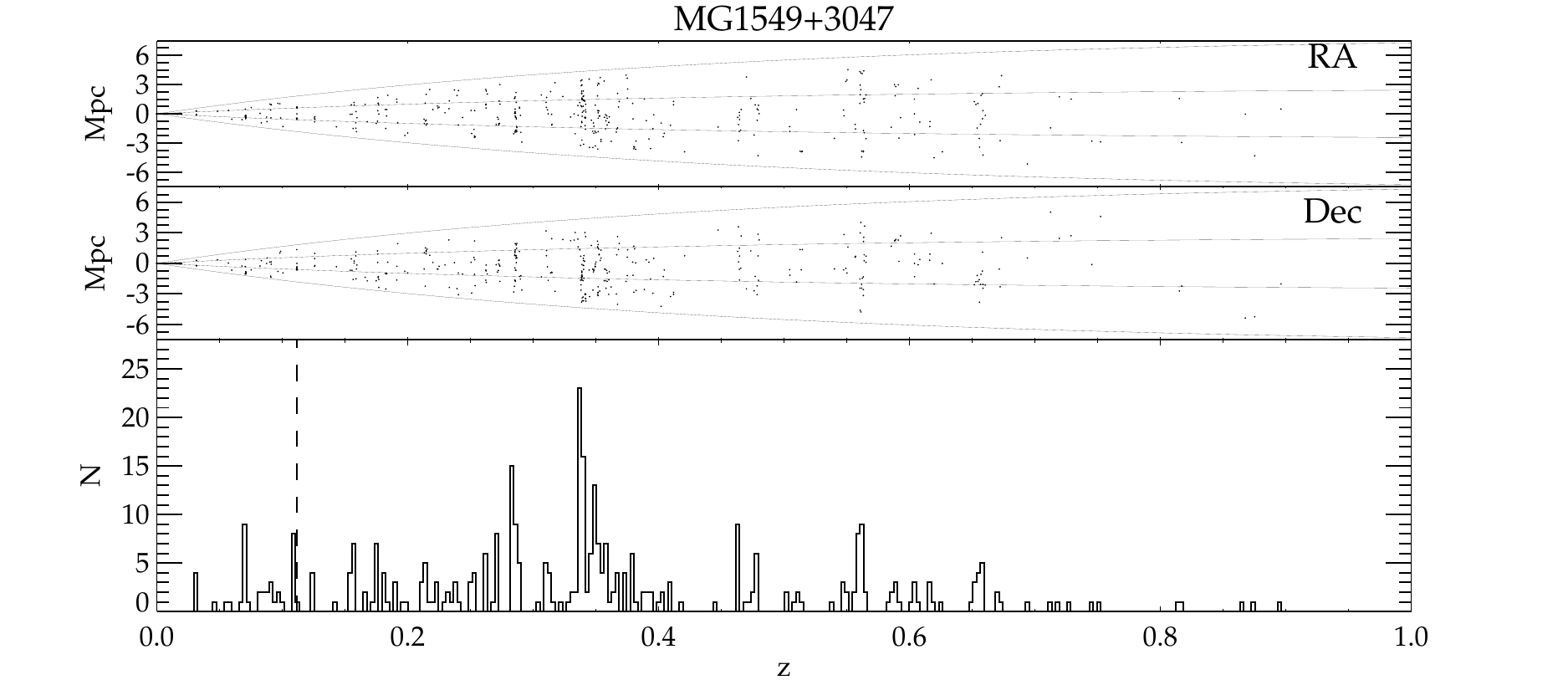}
\caption{\footnotesize Same as Figure  \ref{7a} but for the field of MG1549.}
\end{figure*}

\begin{figure*}
\figurenum{11u}
\epsscale{1.3}
\plotone{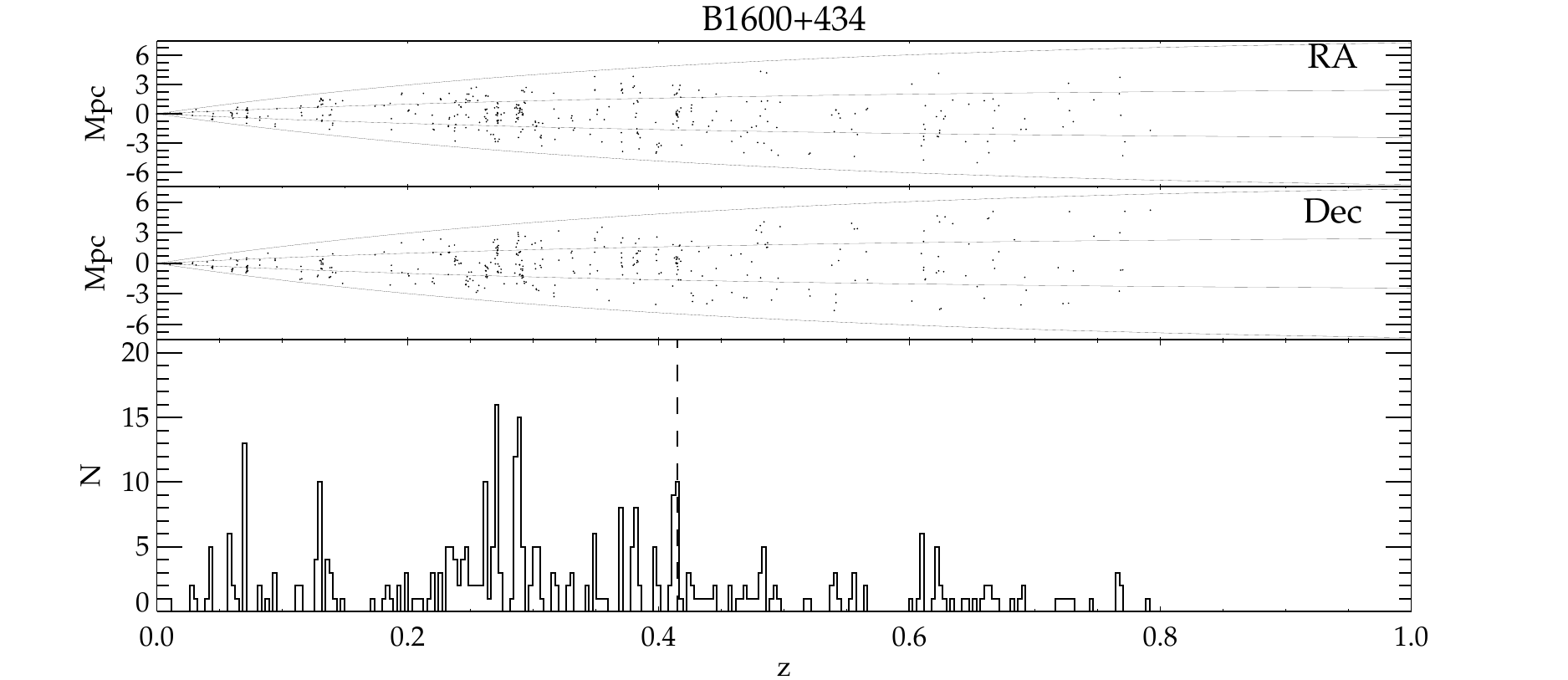}
\caption{\footnotesize Same as Figure  \ref{7a} but for the field of B1600.}
\end{figure*}

\begin{figure*}
\figurenum{11v}
\epsscale{1.3}
\plotone{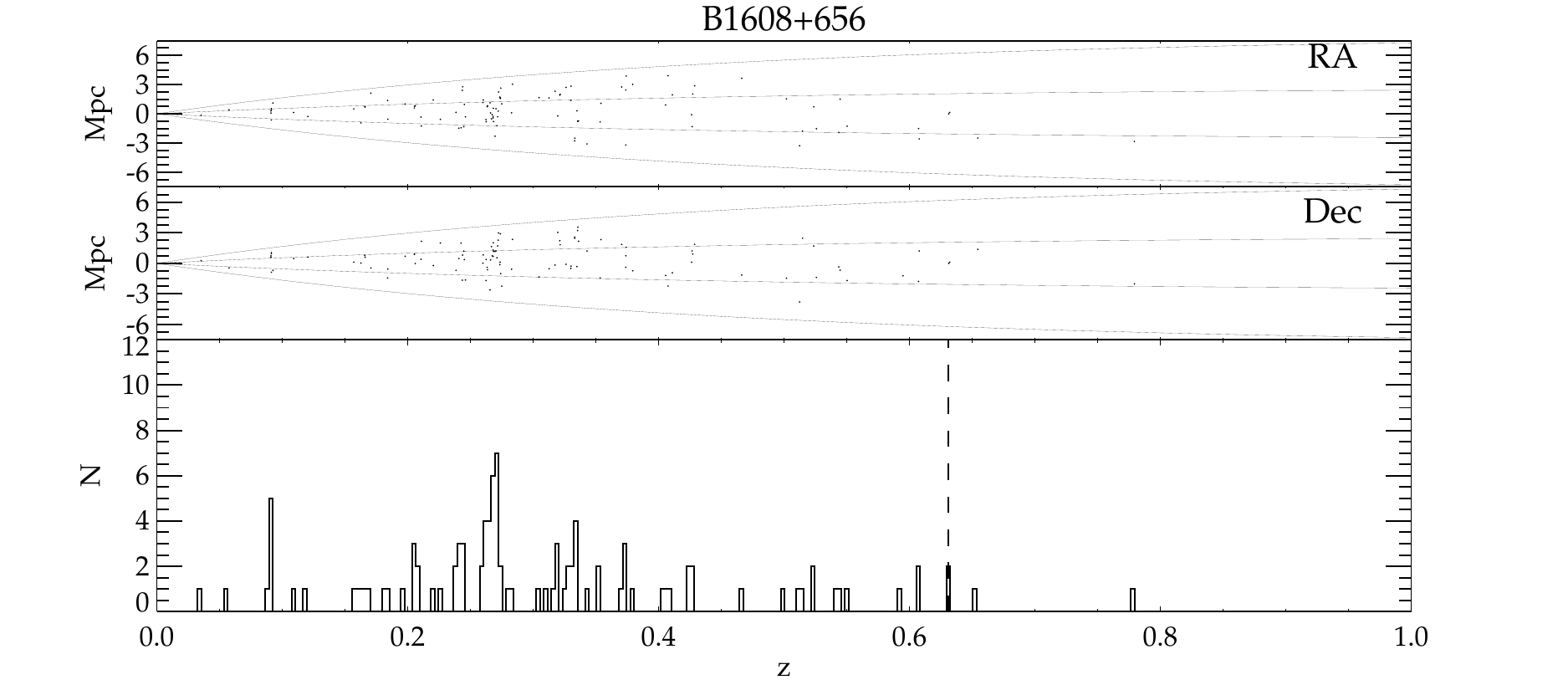}
\caption{\footnotesize Same as Figure  \ref{7a} but for the field of B1608.}
\end{figure*}

\begin{figure*}
\figurenum{11w}
\epsscale{1.3}
\plotone{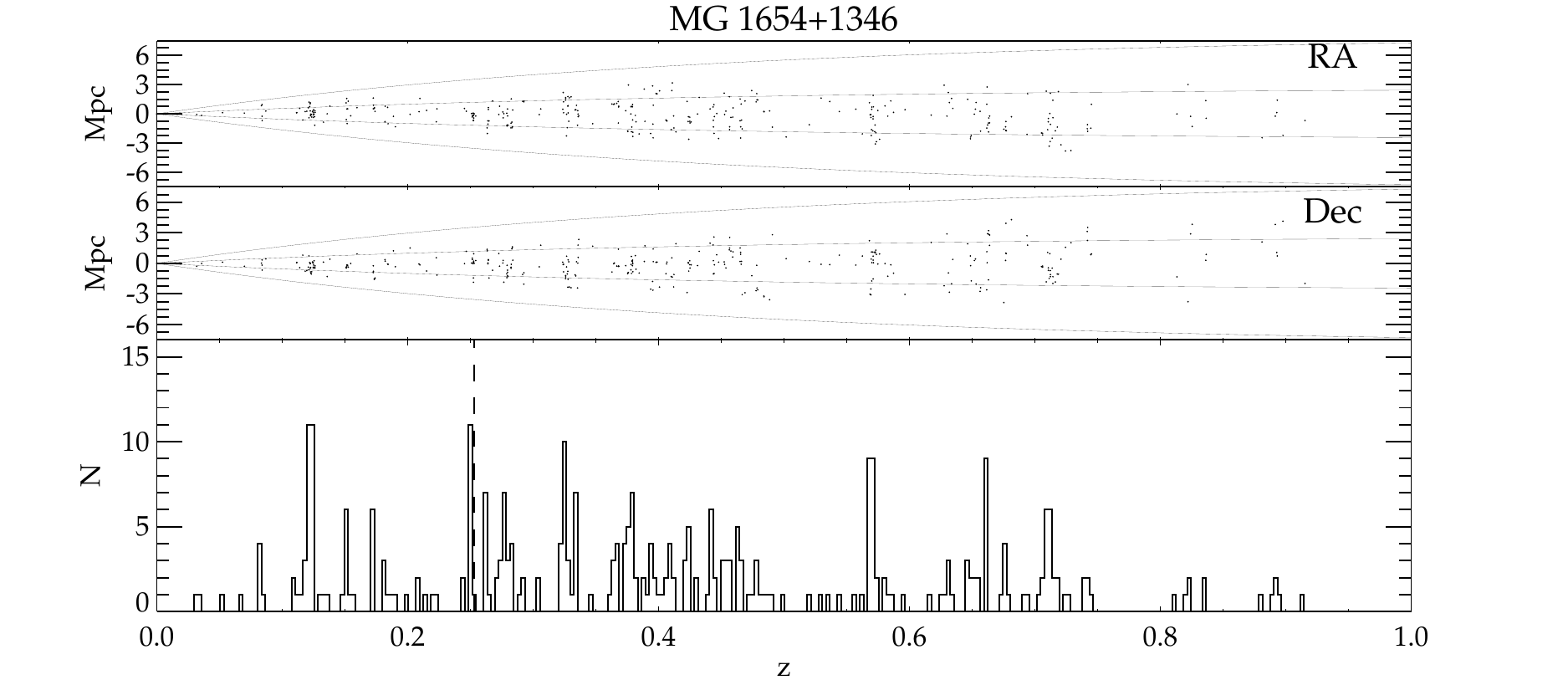}
\caption{\footnotesize Same as Figure  \ref{7a} but for the field of MG1654.}
\end{figure*}

\begin{figure*}
\figurenum{11x}
\epsscale{1.3}
\plotone{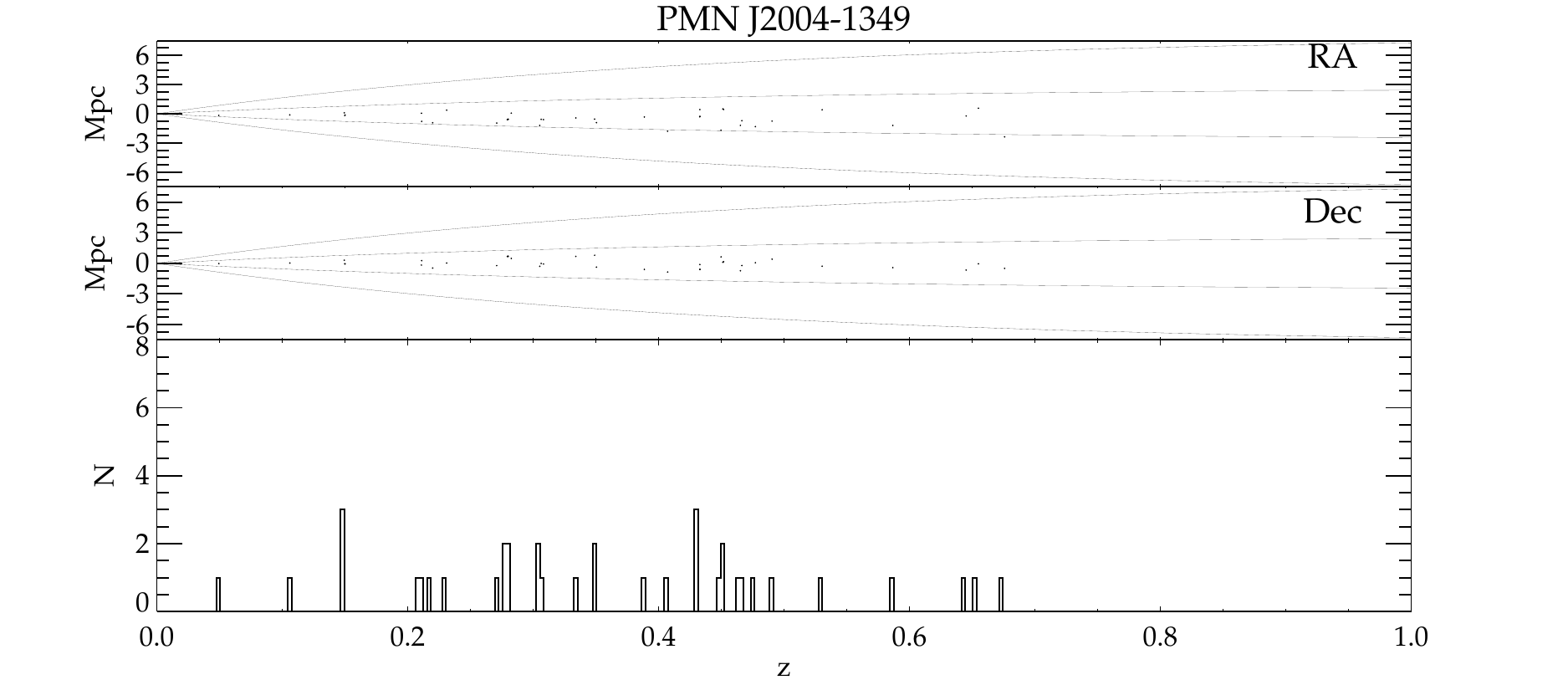}
\caption{\footnotesize Same as  Figure  \ref{7a} but for the field of PMN2004. The redshift of the lens galaxy is unknown.}
\end{figure*}

\begin{figure*}
\figurenum{11y}
\label{}
\epsscale{1.3}
\plotone{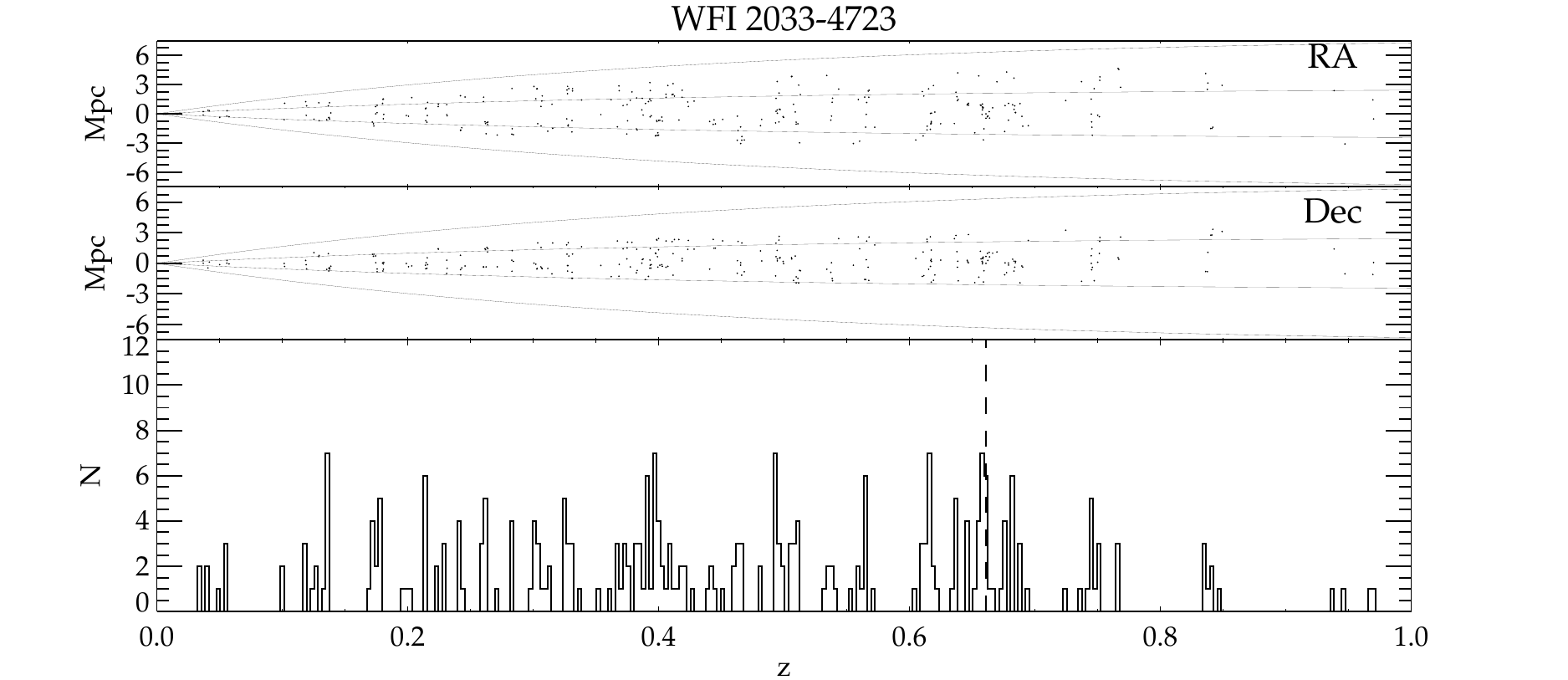}
\caption{\footnotesize Same as Figure  \ref{7a} but for the field of  WFI2033.}
\end{figure*}

\clearpage

\begin{figure*}
\figurenum{11z}
\label{}
\epsscale{1.3}
\plotone{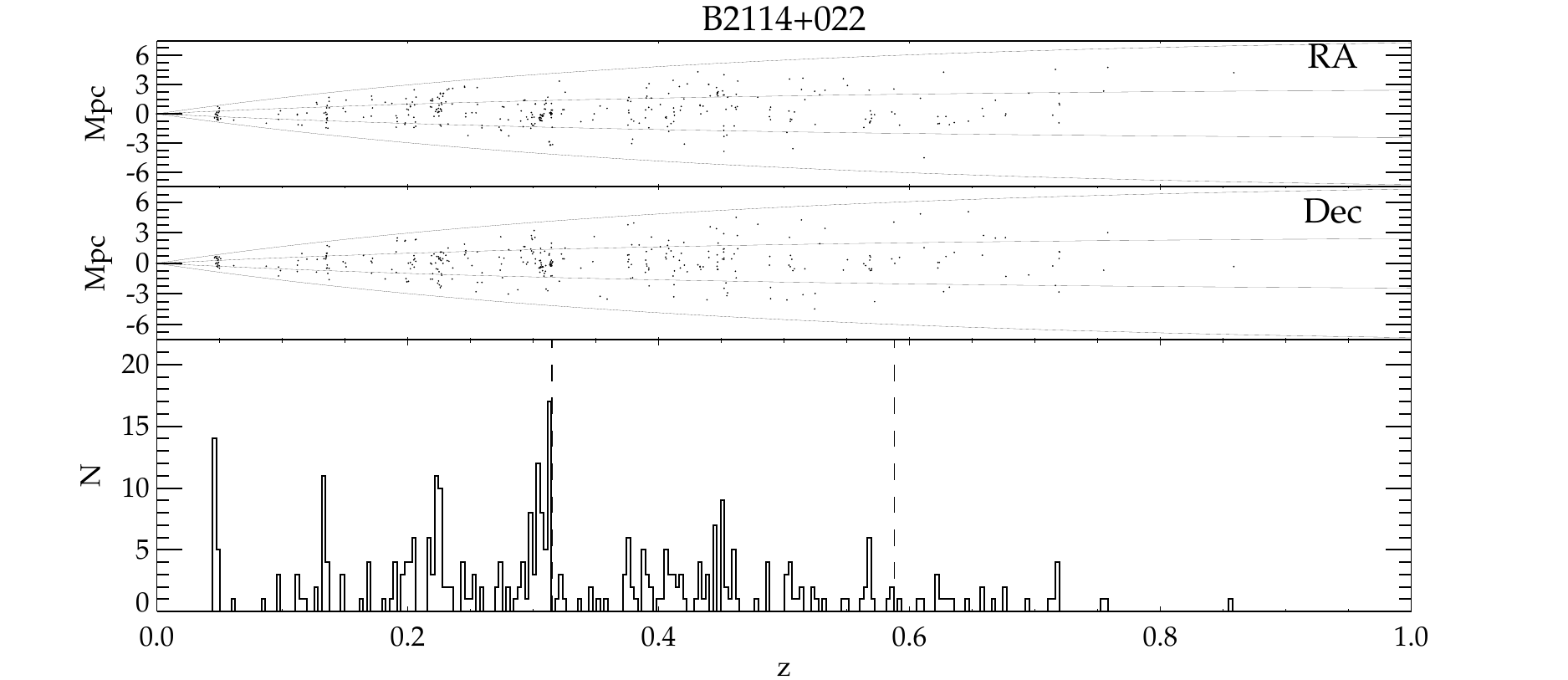}
\caption{\footnotesize Same as Figure  \ref{7a} but for the field of  B2114.}
\end{figure*}

\begin{figure*}
\figurenum{11aa}
\label{field-last}
\epsscale{1.3}
\plotone{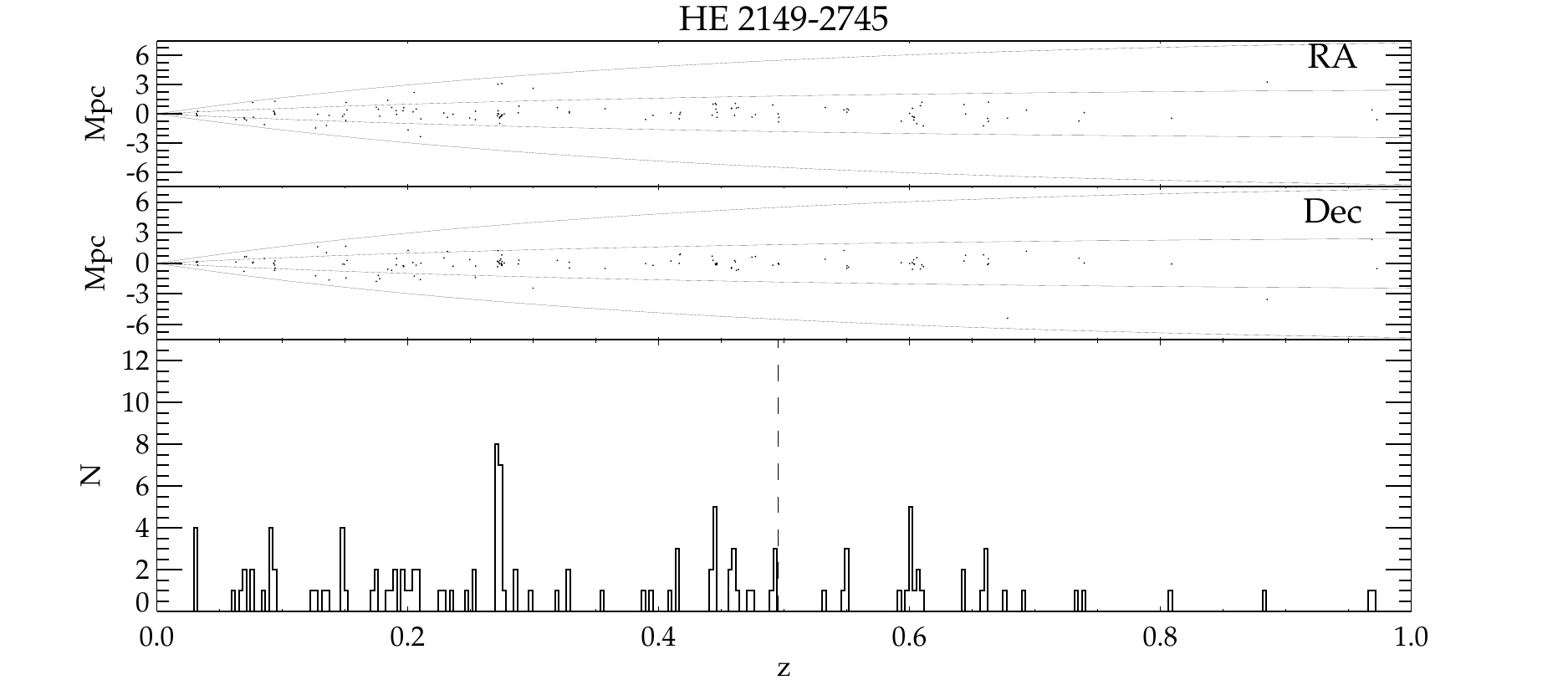}
\caption{\footnotesize Same as Figure  \ref{7a} but for the field of HE2149.}
\end{figure*}

\clearpage

\bibliographystyle{apj}

\begin{onecolumngrid}

%\LongTables
\clearpage
\begin{landscape}
%\LongTables
%\begin{landscape}
\begin{deluxetable*}{l|rrrr|rrrr|rrrr|rrrr|r|rr|r}
\tablecolumns{21}
\tablewidth{0pt}
\tabletypesize{\footnotesize}
\tablecaption{Spectroscopic Data Summary\label{tab:data}}
\tablehead{
\colhead{} & \multicolumn{4}{c}{LDSS2} & \multicolumn{4}{c}{LDSS3} & \multicolumn{4}{c}{IMACS} & \multicolumn{4}{c}{Hectospec} &  \colhead{$N_z$\tablenotemark{e}}  & \colhead{$N_z$\tablenotemark{f}} & \colhead{$N_z$\tablenotemark{g}} & \colhead{$N_z$\tablenotemark{h}}\\
\cline{2-17} \\
\colhead{Lens ID} & \colhead{$N_m$\tablenotemark{a}} & \colhead{$N_{s}$\tablenotemark{b}} & \colhead{$N_z$\tablenotemark{c}} & \colhead{Date} & \colhead{$N_m$} & \colhead{$N_{s}$} & \colhead{$N_z$} & \colhead{Date}
& \colhead{$N_m$} & \colhead{$N_{s}$} & \colhead{$N_z$} & \colhead{Date} & \colhead{$N_m$} & \colhead{$N_{f}$\tablenotemark{d}} & \colhead{$N_z$} & \colhead{Date} & \colhead{All} & \colhead{Unique}  & \colhead{NED} & \colhead{Catalog}}
\startdata
Q0047 		& \nodata & \nodata   	& \nodata & \nodata     	& 3 & 89 & 62& 09/05 	& 3 & 732&254		& 11/04 	& \nodata & \nodata  		& \nodata & \nodata         	& 338 & 313 & 28 	  & 341 \\ 
                		&   &     	&   &       		& 2 & 20 & 15& 08/06 	&   &     	&   		&       	&   &    		&   &           	&               &     &  & \\
                		&   &     	&   &       		& 1 & 9   & 7 & 09/06 	&   &     	&   		&       	&   &    		&   &           	&    		  &    &  & \\
QJ0158  	& \nodata & \nodata   	& \nodata & \nodata     	& 2 & 38 & 30& 08/06	& 1 & 249& 74		& 11/04 	& \nodata & \nodata  		& \nodata & \nodata        	&  104 & 104 & 2  & 106 \\
HE0435  	& \nodata & \nodata   	& \nodata & \nodata     	& 2 & 29 & 22& 08/06 	& 4 & 991&373		& 11/04 	& \nodata & \nodata  		& \nodata & \nodata         	& 411 & 398 & 15 	  & 413  \\
                		&   &     	&   &      		& 2 & 28 & 16& 09/06 	&   &     	&   		&       	&   &    		&   &           	&    		  &     &   & \\
B0712       		& \nodata & \nodata  	& \nodata & \nodata     	& \nodata & \nodata    & \nodata & \nodata     	   	& \nodata & \nodata   	& \nodata 		&       	& 4 &1059	&618& Fall/04  & 618 & 618 & 23 	  & 641   \\
MG0751    	& 4 & 89  &36 & 03/03 	& 3 & 50 & 41& 02/06 	& 3 & 703&137		& 03/04 	& \nodata & \nodata  		& \nodata &           	& 214 & 211 & 37 	  & 248  \\
FBQ0951   	& \nodata & \nodata   	& \nodata & \nodata     	& \nodata & \nodata    & \nodata &  \nodata        	& \nodata & \nodata   	& \nodata 		& \nodata     	& 3 & 717		&559& Spr/05& 872 & 859 & 9  	  & 868  \\
                		&   &     	&   &       		&   &       &   &           	&   &     	&   		&       	& 2 & 451		&313& Spr/06 &   	  &      &  & \\
BRI 0952 	& 4 & 90  &44 & 03/03 	& 1 & 15 & 12& 02/06 	& 3 & 654 &182	& 03/04 	& \nodata & \nodata  		& \nodata &        \nodata    	& 238 & 238 & 33	   & 271  \\
Q1017    	& \nodata & \nodata   	& \nodata & \nodata     	& \nodata & \nodata    & \nodata & \nodata     	   	& 4 & 720 &442	& 04/05 	& \nodata & \nodata  		& \nodata &       \nodata     	& 442 & 380 & 4  		  & 384  \\
HE1104  	& \nodata & \nodata   	& \nodata & \nodata     	& \nodata & \nodata  	& \nodata & \nodata         	& 3 & 571 &443	& 04/05 	& \nodata & \nodata  		& \nodata &       \nodata     	& 443 & 443 & 2  		  & 445  \\
PG1115     	& 4 & 89  &44 & 03/03 	& 2 & 35 	& 27 & 02/06 	& 3 & 653 &137	& 03/04 	& \nodata & \nodata  		& \nodata &        \nodata    	& 315 & 307 & 132		  & 439   \\
                		&   &     	&   &       		&   &    	&   &           	& 1 & 201 &107	& 03/05 	&   &    		&   &           	&    		  &      &   & \\
MG1131    	& \nodata & \nodata   	& \nodata & \nodata     	& \nodata & \nodata  	& \nodata & \nodata        	& \nodata & \nodata   	& \nodata 		& \nodata     	& 2 & 530		&299& Spr/04 & 299 & 299 &  32 	  & 331 \\
RXJ1131 	& \nodata & \nodata   	& \nodata & \nodata     	& 2 & 35 	& 15& 02/06 	& 3 & 656 &331	& 03/04 	& \nodata & \nodata  		& \nodata &      \nodata      	&  543 & 543 & 15 	  & 558   \\
                		&   &     	&   &       		&   &    	&   &       	    	& 2 & 352 &197	& 03/05 	&   &    		&   &           	&    		  &     &  &  \\
B1152       		& \nodata & \nodata   	& \nodata & \nodata     	& \nodata & \nodata  	& \nodata & \nodata         	& \nodata & \nodata   	& \nodata 		& \nodata     	& 3 & 757		&596& Sum/05 	& 596 & 595 &  2    & 597 \\
H12531 	& \nodata & \nodata   	& \nodata & \nodata     	& \nodata &    	& \nodata & \nodata         	& 3 & 741 &244	& 03/04 	& \nodata & \nodata  		& \nodata &         \nodata   	& 244 & 244 & 240	  & 484  \\
LBQ1333   	& \nodata & \nodata   	& \nodata & \nodata     	& \nodata & \nodata  	& \nodata & \nodata         	& 4 & 751 &478	& 04/05 	& \nodata & \nodata  		& \nodata &        \nodata    	& 478 &378 &  38 	  & 416  \\
Q1355  	& \nodata & \nodata   	& \nodata & \nodata     	& \nodata & \nodata  	& \nodata & \nodata         	& 1 & 217 &104	& 03/05 	& \nodata & \nodata  		& \nodata &        \nodata    	&403 & 403 & 6  		  & 409  \\
                		&   &     	&   &       		&   &    	&   &           	& 2 & 404 &299	& 04/05 	&   &    		&   &           	&    		  &     &  &  \\
HST14113   	& \nodata & \nodata   	& \nodata & \nodata     	& \nodata & \nodata  	& \nodata & \nodata     	& \nodata & \nodata   	& \nodata 		& \nodata     	& 1 & 251		&214& Spr/05 & 586 & 585 & 36 & 621  \\
                		&  &    	& &     	&  &  	&     &         	&     &       	&     		&         	& 2 & 474		&372& Spr/06 &    	  &     &   & \\
H1413      	& \nodata & \nodata   	& \nodata & \nodata     	& \nodata & \nodata  	& \nodata & \nodata     	& \nodata & \nodata   	& \nodata 		& \nodata     	& 2 & 451		&263& Spr/05 &  263 & 263 & 16	  & 279  \\
B1422       		& 4 & 93  &50 & 03/03 	& 1 & 20 	& 14& 02/06 	& 3 & 690&181		& 03/04 	& \nodata & \nodata  		& \nodata &        \nodata    	& 245 & 243 &  32 	  & 275 \\
SBS1520    	& \nodata & \nodata   	& \nodata & \nodata     	& \nodata & \nodata  	& \nodata & \nodata     	& \nodata & \nodata   	& \nodata 		& \nodata     	& 2 & 521		&333& Spr/04 & 333 & 333 & 53 	  & 386  \\
MG1549    	& \nodata & \nodata   	& \nodata & \nodata     	& \nodata & \nodata  	& \nodata & \nodata     	& \nodata & \nodata   	& \nodata 		& \nodata     	& 2 & 529		&377& Spr/04 & 377 & 377 & 26 	  & 403 \\
B1600       		& \nodata & \nodata   	& \nodata & \nodata     	& \nodata & \nodata  	& \nodata & \nodata     	& \nodata & \nodata   	& \nodata 		& \nodata     	& 2 & 519		&366& Spr/04 & 366 & 366 & 30 	  & 396  \\
B1608       		& \nodata & \nodata   	& \nodata & \nodata     	& \nodata & \nodata  	& \nodata & \nodata     	& \nodata & \nodata   	& \nodata 		& \nodata     	& 1 & 259		&106& Spr/04 & 106 & 106 & 5  	  & 111  \\
MG1654    	& 4 & 105&38 & 03/03 	& 1 & 13 	& 4 & 02/06 	& 3 & 718 &180	& 03/04 	& \nodata & \nodata  		& \nodata &    \nodata        	& 377 & 365 & 2  		  & 367 \\
                		& 3 & 72  &19 & 08/03 	& 1 & 9  	& 8 & 08/06 	& 1 & 174 &128	& 03/05 	&   &    		&   &           	&    		  &      &  &  \\
PMN2004	& 5 & 119&37 & 08/03 	& \nodata & \nodata  	& \nodata & \nodata     	& \nodata & \nodata   & \nodata 		& \nodata     	& \nodata & \nodata  		&\nodata &        \nodata    	& 37 & 37 & 0  		  & 37   \\
WFI2033 	& \nodata & \nodata   	& \nodata & \nodata     	& 4 & 94 	& 60& 09/05 	& 3 & 535 &260	& 04/05 	& \nodata & \nodata  		& \nodata &        \nodata    	& 320 & 303 & 2  		  & 305   \\
B2114      		& 5 & 85  &36 & 08/03 	& 3 & 50 	& 28& 09/06 	& \nodata & \nodata   	& \nodata 		& \nodata     	& 2 & 517		&327& Spr/04 & 391 & 372 & 4  	  & 376  \\
HE2149  	& 5 &116 &36 & 08/03 	& 2 & 27 	& 20& 08/06 	& \nodata & \nodata   	& \nodata 		& \nodata     	& \nodata & \nodata  		& \nodata &      \nodata      & 85 & 85 & 46 		  & 131  \\
                		&   &     	&   &       		& 3 & 43 	& 29& 09/06 	&   &     	&   		&       	&   &    		&   &           &    		  & & & \\
Total  & 38&   858  &340&       	&35 &605	&410&       	&50 &10713&4551	&       	& 28&7036	&4743 & 	         & 10044 & 9768 & 870	  &10638  \\
\enddata
\tablenotetext{a}{Number of masks or fiber configurations observed on the field.}
\tablenotetext{b}{Number of slits observed on the field.}
\tablenotetext{c}{Number of redshifts obtained. Includes repeat redshift measurements.}
\tablenotetext{d}{Number of fibers targeting objects in the field. Does not include sky fibers.}
\tablenotetext{e}{The sum of the redshifts from different instruments. Includes repeat measurements.}
\tablenotetext{f}{The number of unique redshift measurements.}
\tablenotetext{g}{The number of NED redshift measurements added to our catalog.}
\tablenotetext{h}{The sum of unique redshift measurements and NED redshifts, i.e., all galaxies and stars in our catalog.}
\end{deluxetable*}

\clearpage
\end{landscape}

\end{onecolumngrid}

\LongTables
\begin{deluxetable}{lllllrll}
\tablecolumns{8}
\tablewidth{0pc}
\tabletypesize{\footnotesize}
\tablecaption{Spectroscopic Catalog\label{catalog}}
\tablehead{
\colhead{Lens} & \colhead{ID\tablenotemark{a}} & \colhead{RA} & \colhead{Dec} & \colhead{b} & \colhead{z} & \colhead{$\Delta$z} & \colhead{Flag\tablenotemark{b}}\\
\colhead{} & \colhead{} & \colhead{[J2000]} & \colhead{[J2000]} & \colhead{[\arcmin]} &  \colhead{} & \colhead{} & \colhead{}
}
\startdata     
         Q0047&	10217	&	 12.42451	&	-27.87381	&	 0.00	&	0.48416	&	2.2e-04	&	4	\\ 
&	10217	&	 12.42451	&	-27.87381	&	 0.00	&	3.59500	&	1.0e-01	&	6	\\ 
&	2267	&	 12.15079	&	-27.92057	&	14.78	&	0.23926	&	1.6e-04	&	4	\\ 
&	3127	&	 12.18116	&	-27.88725	&	12.93	&	0.39620	&	1.2e-04	&	4	\\ 
&	3241	&	 12.18753	&	-27.97718	&	14.00	&	0.33659	&	3.0e-04	&	4	\\ 
&	3365	&	 12.19159	&	-27.98285	&	13.97	&	0.36733	&	2.3e-04	&	4	\\ 
&	3422	&	 12.19416	&	-27.96951	&	13.49	&	0.80010	&	1.8e-04	&	4	\\ 
&	3464	&	 12.19584	&	-27.90281	&	12.25	&	0.58873	&	3.0e-04	&	4	\\ 
&	3505	&	 12.19643	&	-27.80256	&	12.84	&	0.42918	&	2.3e-04	&	4	\\ 
&	3534	&	 12.19875	&	-27.88514	&	11.99	&	0.58333	&	1.8e-04	&	4	\\ 
&	3693	&	 12.20341	&	-27.95319	&	12.65	&	0.30984	&	1.8e-04	&	4	\\ 
&	3775	&	 12.20468	&	-27.96061	&	12.76	&	0.11481	&	1.6e-04	&	4	\\ 
&	3748	&	 12.20681	&	-27.83794	&	11.75	&	0.23802	&	1.6e-04	&	4	\\ 
&	3906	&	 12.20864	&	-27.84015	&	11.63	&	0.24151	&	1.6e-04	&	4	\\ 
&	4008	&	 12.21207	&	-27.83905	&	11.46	&	0.23939	&	1.9e-04	&	4	\\ 
&	3931	&	 12.21374	&	-27.85583	&	11.23	&	0.23733	&	2.2e-04	&	4	\\ 
&	4093	&	 12.21538	&	-27.82001	&	11.56	&	0.30709	&	1.8e-04	&	4	\\ 
&	4086	&	 12.21682	&	-27.92271	&	11.39	&	0.63518	&	2.3e-04	&	4	\\ 
&	4116	&	 12.21871	&	-27.99950	&	13.25	&	0.19498	&	1.8e-04	&	4	\\ 
&	4250	&	 12.22098	&	-27.83200	&	11.08	&	0.24111	&	1.8e-04	&	4	\\ 
&	4237	&	 12.22311	&	-27.84106	&	10.86	&	0.71106	&	1.8e-04	&	4	\\ 
&	4359	&	 12.22523	&	-27.83684	&	10.80	&	0.23834	&	2.2e-04	&	4	\\ 
&	4376	&	 12.22898	&	-27.92575	&	10.82	&	0.70888	&	2.3e-04	&	4	\\ 
&	4612	&	 12.23434	&	-27.98648	&	12.13	&	0.43872	&	2.3e-04	&	4	\\ 
&	4598	&	 12.23683	&	-27.98733	&	12.05	&	0.30786	&	3.0e-04	&	4	\\ 
&	4636	&	 12.23825	&	-27.82676	&	10.28	&	0.65420	&	3.0e-04	&	4	\\ 
&	4834	&	 12.24329	&	-28.00076	&	12.25	&	0.19352	&	2.3e-04	&	4	\\ 
&	4842	&	 12.24414	&	-27.82518	&	10.00	&	0.45331	&	1.2e-04	&	4	\\ 
&	4847	&	 12.24470	&	-27.77219	&	11.33	&	0.27697	&	1.8e-04	&	4	\\ 
&	4876	&	 12.24578	&	-27.98167	&	11.47	&	0.30865	&	2.3e-04	&	4	\\ 
&	4912	&	 12.24608	&	-27.82898	&	 9.84	&	0.57034	&	1.2e-04	&	4	\\ 
&	4860	&	 12.24616	&	-27.97461	&	11.22	&	0.81058	&	1.8e-04	&	4	\\ 
&	4933	&	 12.24688	&	-27.90489	&	 9.60	&	0.30805	&	1.9e-04	&	4	\\ 
&	4906	&	 12.24691	&	-28.01426	&	12.63	&	0.65565	&	3.0e-04	&	4	\\ 
&	4952	&	 12.24901	&	-27.95223	&	10.42	&	0.61204	&	1.8e-04	&	4	\\ 
&	5010	&	 12.24985	&	-27.99936	&	11.93	&	0.37581	&	2.2e-04	&	4	\\ 
&	4985	&	 12.25027	&	-27.96143	&	10.62	&	0.71003	&	3.0e-04	&	4	\\ 
&	5024	&	 12.25172	&	-27.87908	&	 9.17	&	0.70729	&	3.0e-04	&	4	\\ 
&	5097	&	 12.25227	&	-27.89299	&	 9.20	&	0.19637	&	1.9e-04	&	4	\\ 
&	5163	&	 12.25613	&	-27.81578	&	 9.59	&	0.29255	&	1.8e-04	&	4	\\ 
&	5254	&	 12.26032	&	-27.99897	&	11.49	&	0.37575	&	2.2e-04	&	4	\\ 
&	5335	&	 12.26330	&	-27.93708	&	 9.35	&	0.25492	&	2.3e-04	&	4	\\ 
&	5479	&	 12.26927	&	-28.00589	&	11.42	&	0.37507	&	1.8e-04	&	4	\\ 
&	5610	&	 12.27396	&	-27.77823	&	 9.84	&	0.43122	&	1.8e-04	&	4	\\ 
&	5663	&	 12.27628	&	-27.83928	&	 8.13	&	0.41219	&	2.3e-04	&	4	\\ 
&	5761	&	 12.27732	&	-27.98761	&	10.36	&	0.55562	&	3.0e-04	&	4	\\ 
&	5837	&	 12.28226	&	-27.99724	&	10.56	&	0.55376	&	3.0e-04	&	4	\\ 
&	5873	&	 12.28276	&	-27.87185	&	 7.52	&	0.53312	&	3.0e-04	&	4	\\ 
&	5909	&	 12.28497	&	-27.91055	&	 7.72	&	0.30879	&	2.2e-04	&	4	\\ 
&	5916	&	 12.28592	&	-27.94979	&	 8.64	&	0.63381	&	3.0e-04	&	4	\\ 
&	5899	&	 12.28606	&	-27.98928	&	10.09	&	0.55833	&	3.0e-04	&	4	\\ 
&	5891	&	 12.28621	&	-27.86650	&	 7.35	&	0.19428	&	2.3e-04	&	4	\\ 
&	5993	&	 12.28957	&	-27.95112	&	 8.52	&	0.62033	&	1.8e-04	&	4	\\ 
&	6030	&	 12.29061	&	-27.80473	&	 8.23	&	0.44744	&	1.8e-04	&	4	\\ 
&	6069	&	 12.29189	&	-27.91300	&	 7.41	&	0.59416	&	3.0e-04	&	4	\\ 
&	6343	&	 12.29425	&	-27.88882	&	 6.97	&	0.19536	&	1.6e-04	&	4	\\ 
&	6226	&	 12.29705	&	-27.83035	&	 7.25	&	0.45505	&	3.0e-04	&	4	\\ 
&	6318	&	 12.29778	&	-27.96538	&	 8.68	&	0.37374	&	2.3e-04	&	4	\\ 
&	6418	&	 12.30197	&	-27.89278	&	 6.60	&	0.45111	&	2.2e-04	&	4	\\ 
&	6470	&	 12.30366	&	-27.89186	&	 6.50	&	0.55688	&	3.0e-04	&	4	\\ 
&	6451	&	 12.30452	&	-27.84052	&	 6.67	&	0.18501	&	1.2e-04	&	4	\\ 
&	6471	&	 12.30507	&	-27.95894	&	 8.13	&	0.70917	&	1.8e-04	&	4	\\ 
&	6568	&	 12.30878	&	-28.01864	&	10.63	&	0.56688	&	3.0e-04	&	4	\\ 
&	6630	&	 12.31023	&	-27.95207	&	 7.66	&	0.40121	&	1.2e-04	&	4	\\ 
&	6604	&	 12.31040	&	-27.91487	&	 6.53	&	0.59446	&	2.3e-04	&	4	\\ 
&	6614	&	 12.31053	&	-27.98168	&	 8.85	&	0.30710	&	2.3e-04	&	4	\\ 
&	6655	&	 12.31113	&	-27.87512	&	 6.01	&	0.88495	&	1.8e-04	&	4	\\ 
&	6786	&	 12.31567	&	-27.82306	&	 6.53	&	0.67993	&	1.8e-04	&	4	\\ 
&	6958	&	 12.31926	&	-27.95023	&	 7.22	&	0.40295	&	1.2e-04	&	4	\\ 
&	6950	&	 12.32000	&	-27.95644	&	 7.43	&	0.71265	&	3.0e-04	&	4	\\ 
&	6899	&	 12.32002	&	-27.94469	&	 6.98	&	0.40293	&	2.3e-04	&	4	\\ 
&	6927	&	 12.32015	&	-27.94728	&	 7.07	&	0.40295	&	1.2e-04	&	4	\\ 
&	7101	&	 12.32228	&	-27.90206	&	 5.68	&	0.88094	&	1.8e-04	&	4	\\ 
&	6976	&	 12.32228	&	-27.84705	&	 5.66	&	0.65539	&	3.0e-04	&	4	\\ 
&	7050	&	 12.32443	&	-27.92019	&	 5.99	&	0.53162	&	1.8e-04	&	4	\\ 
&	7073	&	 12.32565	&	-28.00320	&	 9.36	&	0.56819	&	1.8e-04	&	4	\\ 
&	7209	&	 12.33048	&	-28.04399	&	11.36	&	0.61577	&	1.8e-04	&	4	\\ 
&	7317	&	 12.33291	&	-27.86770	&	 4.87	&	0.58993	&	1.8e-04	&	4	\\ 
&	7338	&	 12.33586	&	-27.89142	&	 4.82	&	0.38290	&	1.9e-04	&	4	\\ 
&	7404	&	 12.33668	&	-27.93955	&	 6.10	&	0.61240	&	3.0e-04	&	4	\\ 
&	7446	&	 12.33993	&	-27.98593	&	 8.08	&	0.62774	&	3.0e-04	&	4	\\ 
&	7492	&	 12.34072	&	-27.96618	&	 7.10	&	0.23758	&	1.9e-04	&	4	\\ 
&	7529	&	 12.34145	&	-27.94023	&	 5.94	&	0.53204	&	3.0e-04	&	4	\\ 
&	7531	&	 12.34191	&	-27.88562	&	 4.44	&	0.44823	&	1.8e-04	&	4	\\ 
&	7590	&	 12.34378	&	-27.88981	&	 4.39	&	0.63251	&	2.3e-04	&	4	\\ 
&	7682	&	 12.34446	&	-27.93345	&	 5.55	&	0.19431	&	2.2e-04	&	4	\\ 
&	7712	&	 12.34690	&	-27.92947	&	 5.30	&	0.30788	&	2.3e-04	&	4	\\ 
&	7728	&	 12.34801	&	-27.89831	&	 4.31	&	0.59559	&	3.0e-04	&	4	\\ 
&	7742	&	 12.34823	&	-27.98398	&	 7.75	&	0.56902	&	3.0e-04	&	4	\\ 
&	7778	&	 12.34947	&	-27.98820	&	 7.93	&	0.32169	&	2.3e-04	&	4	\\ 
&	7809	&	 12.35114	&	-27.89370	&	 4.07	&	0.59267	&	2.3e-04	&	3	\\ 
&	7874	&	 12.35183	&	-27.85640	&	 3.99	&	0.40452	&	1.9e-04	&	3	\\ 
&	7881	&	 12.35280	&	-27.78493	&	 6.55	&	0.74080	&	1.8e-04	&	4	\\ 
&	7986	&	 12.35516	&	-27.89809	&	 3.95	&	0.59733	&	3.0e-04	&	3	\\ 
&	7938	&	 12.35547	&	-28.01996	&	 9.50	&	0.68559	&	1.8e-04	&	4	\\ 
&	8003	&	 12.35752	&	-27.90359	&	 3.98	&	0.65363	&	3.0e-04	&	4	\\ 
&	8052	&	 12.35754	&	-27.95168	&	 5.87	&	0.21733	&	1.8e-04	&	4	\\ 
&	8080	&	 12.35809	&	-27.93124	&	 4.93	&	0.30487	&	2.3e-04	&	4	\\ 
&	8233	&	 12.35984	&	-27.89345	&	 3.63	&	0.34184	&	1.2e-04	&	3	\\ 
&	8118	&	 12.36011	&	-27.96099	&	 6.25	&	0.56186	&	3.0e-04	&	4	\\ 
&	8138	&	 12.36157	&	-27.85349	&	 3.55	&	0.65531	&	2.3e-04	&	3	\\ 
&	8194	&	 12.36254	&	-27.89627	&	 3.55	&	0.59599	&	3.0e-04	&	4	\\ 
&	8234	&	 12.36435	&	-27.90252	&	 3.62	&	0.59714	&	2.3e-04	&	4	\\ 
&	8501	&	 12.36599	&	-27.88721	&	 3.21	&	0.30628	&	1.6e-04	&	4	\\ 
&	8484	&	 12.36717	&	-27.85149	&	 3.32	&	0.17061	&	1.6e-04	&	3	\\ 
&	8314	&	 12.36792	&	-27.98474	&	 7.30	&	0.30894	&	1.8e-04	&	4	\\ 
&	8380	&	 12.37037	&	-27.89271	&	 3.09	&	0.30739	&	1.8e-04	&	4	\\ 
&	8378	&	 12.37040	&	-27.95639	&	 5.72	&	0.37366	&	1.8e-04	&	4	\\ 
&	8443	&	 12.37147	&	-27.85328	&	 3.07	&	0.18334	&	2.3e-04	&	4	\\ 
&	8455	&	 12.37181	&	-27.85694	&	 2.97	&	0.76905	&	2.3e-04	&	4	\\ 
&	8446	&	 12.37296	&	-27.90727	&	 3.39	&	0.31014	&	2.3e-04	&	3	\\ 
&	8587	&	 12.37391	&	-27.93327	&	 4.46	&	0.30975	&	1.6e-04	&	4	\\ 
&	8542	&	 12.37562	&	-27.91275	&	 3.49	&	0.59018	&	2.3e-04	&	3	\\ 
&	8602	&	 12.37689	&	-27.86125	&	 2.64	&	0.23737	&	1.9e-04	&	4	\\ 
&	8650	&	 12.37759	&	-27.87823	&	 2.50	&	0.23722	&	1.9e-04	&	3	\\ 
&	8624	&	 12.37773	&	-27.86422	&	 2.55	&	0.65382	&	2.3e-04	&	3	\\ 
&	8627	&	 12.37876	&	-27.98739	&	 7.23	&	0.61885	&	3.0e-04	&	4	\\ 
&	8743	&	 12.37899	&	-27.89928	&	 2.86	&	0.17071	&	1.1e-04	&	3	\\ 
&	8711	&	 12.38032	&	-27.88378	&	 2.42	&	0.08350	&	3.0e-04	&	3	\\ 
&	8646	&	 12.38077	&	-28.06579	&	11.75	&	0.68411	&	1.8e-04	&	4	\\ 
&	8729	&	 12.38181	&	-27.83870	&	 3.09	&	0.66488	&	3.0e-04	&	3	\\ 
&	8705	&	 12.38207	&	-27.88759	&	 2.40	&	0.53259	&	3.0e-04	&	3	\\ 
&	8828	&	 12.38361	&	-28.05970	&	11.36	&	0.17988	&	2.3e-04	&	4	\\ 
&	8831	&	 12.38576	&	-27.87189	&	 2.06	&	0.23759	&	3.0e-04	&	3	\\ 
&	8876	&	 12.38719	&	-27.85062	&	 2.42	&	0.53741	&	2.3e-04	&	3	\\ 
&	8863	&	 12.38753	&	-28.01361	&	 8.61	&	0.43782	&	1.8e-04	&	4	\\ 
&	8939	&	 12.38868	&	-27.85079	&	 2.35	&	0.53860	&	3.0e-04	&	3	\\ 
&	8956	&	 12.38943	&	-27.88949	&	 2.08	&	0.45495	&	2.3e-04	&	4	\\ 
&	8968	&	 12.39028	&	-27.85369	&	 2.18	&	0.53780	&	3.0e-04	&	3	\\ 
&	9017	&	 12.39148	&	-27.90663	&	 2.64	&	0.30687	&	2.2e-04	&	4	\\ 
&	9152	&	 12.39159	&	-27.92009	&	 3.28	&	0.32670	&	2.3e-04	&	4	\\ 
&	9085	&	 12.39303	&	-27.86202	&	 1.81	&	0.66439	&	2.3e-04	&	3	\\ 
&	9052	&	 12.39309	&	-27.94558	&	 4.62	&	0.66808	&	1.8e-04	&	4	\\ 
&	9210	&	 12.39744	&	-27.88411	&	 1.56	&	0.53664	&	3.0e-04	&	4	\\ 
&	9273	&	 12.39836	&	-27.86292	&	 1.53	&	0.48892	&	2.3e-04	&	3	\\ 
&	9285	&	 12.40132	&	-27.87644	&	 1.24	&	0.82520	&	1.8e-04	&	4	\\ 
&	9325	&	 12.40238	&	-27.94866	&	 4.64	&	0.59790	&	2.3e-04	&	4	\\ 
&	9332	&	 12.40249	&	-27.89165	&	 1.58	&	0.53724	&	2.3e-04	&	4	\\ 
&	9350	&	 12.40379	&	-27.95992	&	 5.28	&	0.44866	&	1.8e-04	&	4	\\ 
&	9395	&	 12.40424	&	-27.86820	&	 1.13	&	0.57359	&	3.0e-04	&	3	\\ 
&	9380	&	 12.40434	&	-27.87875	&	 1.11	&	1.93159	&	3.0e-04	&	3	\\ 
&	9527	&	 12.40748	&	-27.88011	&	 0.98	&	0.55830	&	3.0e-04	&	3	\\ 
&	9540	&	 12.40790	&	-27.89848	&	 1.72	&	0.46727	&	1.8e-04	&	4	\\ 
&	9666	&	 12.40861	&	-27.95743	&	 5.09	&	0.44687	&	1.6e-04	&	3	\\ 
&	9557	&	 12.40885	&	-27.84804	&	 1.76	&	0.82134	&	3.0e-04	&	3	\\ 
&	9563	&	 12.40914	&	-27.96912	&	 5.78	&	0.44987	&	1.8e-04	&	4	\\ 
&	9573	&	 12.40950	&	-27.89170	&	 1.34	&	0.53912	&	2.3e-04	&	4	\\ 
&	9692	&	 12.41009	&	-27.94217	&	 4.17	&	0.18329	&	1.8e-04	&	4	\\ 
&	9673	&	 12.41103	&	-27.89172	&	 1.29	&	0.59534	&	3.0e-04	&	3	\\ 
&	9655	&	 12.41174	&	-27.94756	&	 4.48	&	0.71068	&	3.0e-04	&	3	\\ 
&	9678	&	 12.41229	&	-27.85194	&	 1.46	&	0.60534	&	2.3e-04	&	3	\\ 
&	9691	&	 12.41329	&	-27.85695	&	 1.17	&	0.60573	&	2.3e-04	&	3	\\ 
&	9852	&	 12.41707	&	-27.89282	&	 1.21	&	0.65336	&	2.2e-04	&	3	\\ 
&	9971	&	 12.41783	&	-27.88888	&	 0.97	&	0.48774	&	1.6e-04	&	4	\\ 
&	9914	&	 12.41820	&	-27.89065	&	 1.06	&	0.48386	&	3.0e-04	&	3	\\ 
&	9887	&	 12.41906	&	-27.86015	&	 0.87	&	0.65385	&	2.3e-04	&	3	\\ 
&	10056	&	 12.42208	&	-27.84682	&	 1.62	&	0.48602	&	1.9e-04	&	3	\\ 
&	10088	&	 12.42338	&	-27.80777	&	 3.96	&	0.48611	&	2.2e-04	&	3	\\ 
&	10058	&	 12.42410	&	-27.93142	&	 3.46	&	0.53842	&	2.2e-04	&	3	\\ 
&	10125	&	 12.42443	&	-27.90675	&	 1.98	&	0.40186	&	9.2e-05	&	3	\\ 
&	10063	&	 12.42524	&	-27.89432	&	 1.23	&	0.65211	&	2.3e-04	&	4	\\ 
&	10380	&	 12.42805	&	-27.92918	&	 3.33	&	0.19602	&	1.9e-04	&	4	\\ 
&	10241	&	 12.42823	&	-27.94141	&	 4.06	&	0.59693	&	3.0e-04	&	4	\\ 
&	10268	&	 12.42931	&	-27.93392	&	 3.62	&	0.59732	&	2.2e-04	&	3	\\ 
&	10303	&	 12.43036	&	-27.85339	&	 1.26	&	0.45597	&	3.0e-04	&	3	\\ 
&	10312	&	 12.43106	&	-27.88190	&	 0.60	&	0.48676	&	3.0e-04	&	3	\\ 
&	10367	&	 12.43225	&	-27.88928	&	 1.01	&	0.66372	&	3.0e-04	&	4	\\ 
&	10407	&	 12.43264	&	-27.84271	&	 1.92	&	0.53704	&	3.0e-04	&	3	\\ 
&	10381	&	 12.43338	&	-27.84458	&	 1.82	&	0.46583	&	1.8e-04	&	4	\\ 
&	10420	&	 12.43341	&	-27.85411	&	 1.27	&	0.65439	&	1.8e-04	&	4	\\ 
&	10483	&	 12.43409	&	-27.79058	&	 5.02	&	0.49328	&	1.8e-04	&	4	\\ 
&	10585	&	 12.43577	&	-27.82096	&	 3.23	&	0.44174	&	2.2e-04	&	3	\\ 
&	10535	&	 12.43613	&	-27.65582	&	13.09	&	0.81882	&	1.8e-04	&	4	\\ 
&	10552	&	 12.43699	&	-27.88738	&	 1.05	&	0.36582	&	2.3e-04	&	3	\\ 
&	10691	&	 12.43833	&	-27.78212	&	 5.55	&	0.56688	&	3.0e-04	&	3	\\ 
&	10731	&	 12.43875	&	-27.79214	&	 4.96	&	0.54518	&	3.0e-04	&	3	\\ 
&	10974	&	 12.43917	&	-27.97851	&	 6.33	&	0.44816	&	2.2e-04	&	4	\\ 
&	10711	&	 12.43922	&	-27.90101	&	 1.81	&	0.85213	&	3.0e-04	&	3	\\ 
&	10789	&	 12.44109	&	-27.83917	&	 2.26	&	0.53666	&	2.3e-04	&	4	\\ 
&	10793	&	 12.44137	&	-27.92563	&	 3.23	&	0.53260	&	2.2e-04	&	3	\\ 
&	10896	&	 12.44227	&	-27.99160	&	 7.13	&	0.39324	&	3.0e-04	&	4	\\ 
&	10821	&	 12.44273	&	-27.83615	&	 2.46	&	0.49294	&	2.3e-04	&	3	\\ 
&	11060	&	 12.44452	&	-27.92032	&	 2.99	&	0.37518	&	1.6e-04	&	3	\\ 
&	11028	&	 12.44518	&	-27.87496	&	 1.10	&	0.19473	&	2.2e-04	&	3	\\ 
&	11122	&	 12.44675	&	-27.64084	&	14.03	&	0.46780	&	1.8e-04	&	4	\\ 
&	11160	&	 12.44702	&	-27.64532	&	13.76	&	0.21024	&	1.8e-04	&	4	\\ 
&	11051	&	 12.44760	&	-27.85812	&	 1.54	&	0.48633	&	2.3e-04	&	3	\\ 
&	11085	&	 12.44775	&	-27.94622	&	 4.52	&	0.71238	&	2.3e-04	&	3	\\ 
&	11179	&	 12.45037	&	-27.86672	&	 1.44	&	0.73776	&	2.3e-04	&	3	\\ 
&	11211	&	 12.45091	&	-27.86211	&	 1.57	&	0.48572	&	1.8e-04	&	4	\\ 
&	11251	&	 12.45172	&	-27.84330	&	 2.33	&	0.61135	&	1.8e-04	&	4	\\ 
&	11320	&	 12.45334	&	-27.92658	&	 3.52	&	0.57479	&	3.0e-04	&	4	\\ 
&	11398	&	 12.45447	&	-27.83591	&	 2.77	&	0.65709	&	3.0e-04	&	3	\\ 
&	11444	&	 12.45531	&	-27.94430	&	 4.53	&	0.70912	&	3.0e-04	&	4	\\ 
&	11580	&	 12.45628	&	-27.80953	&	 4.21	&	0.54591	&	3.0e-04	&	3	\\ 
&	11525	&	 12.45711	&	-27.79146	&	 5.24	&	0.19800	&	1.2e-04	&	3	\\ 
&	11551	&	 12.45716	&	-27.87392	&	 1.73	&	0.35211	&	1.8e-04	&	4	\\ 
&	11517	&	 12.45811	&	-27.88943	&	 2.01	&	0.19475	&	2.2e-04	&	3	\\ 
&	11522	&	 12.45846	&	-27.87548	&	 1.80	&	0.54686	&	2.3e-04	&	3	\\ 
&	11576	&	 12.45989	&	-27.83361	&	 3.06	&	0.53655	&	3.0e-04	&	4	\\ 
&	11631	&	 12.46025	&	-27.92280	&	 3.50	&	0.59680	&	3.0e-04	&	3	\\ 
&	11646	&	 12.46034	&	-27.84852	&	 2.43	&	0.19375	&	1.6e-04	&	4	\\ 
&	11591	&	 12.46057	&	-27.84141	&	 2.73	&	0.61138	&	2.3e-04	&	3	\\ 
&	11682	&	 12.46086	&	-27.82635	&	 3.44	&	0.19384	&	1.6e-04	&	4	\\ 
&	11677	&	 12.46223	&	-27.83313	&	 3.16	&	0.66227	&	2.3e-04	&	3	\\ 
&	11739	&	 12.46241	&	-27.64000	&	14.17	&	0.62088	&	3.0e-04	&	4	\\ 
&	11912	&	 12.46325	&	-27.80845	&	 4.43	&	0.23903	&	1.9e-04	&	4	\\ 
&	11732	&	 12.46362	&	-27.88112	&	 2.12	&	0.65260	&	3.0e-04	&	3	\\ 
&	11833	&	 12.46372	&	-27.84188	&	 2.83	&	0.25447	&	1.9e-04	&	3	\\ 
&	11784	&	 12.46443	&	-27.80382	&	 4.70	&	0.42874	&	2.3e-04	&	4	\\ 
&	11915	&	 12.46657	&	-27.88479	&	 2.33	&	0.19494	&	1.2e-04	&	3	\\ 
&	11993	&	 12.46883	&	-27.81758	&	 4.11	&	0.74534	&	1.8e-04	&	4	\\ 
&	12126	&	 12.47158	&	-27.91178	&	 3.38	&	0.61683	&	2.3e-04	&	3	\\ 
&	12132	&	 12.47192	&	-27.80694	&	 4.74	&	0.70539	&	1.8e-04	&	4	\\ 
&	12318	&	 12.47477	&	-27.79864	&	 5.24	&	0.19626	&	1.9e-04	&	4	\\ 
&	12307	&	 12.47630	&	-27.66514	&	12.82	&	0.76400	&	2.3e-04	&	4	\\ 
&	12391	&	 12.47708	&	-27.71222	&	10.09	&	0.18314	&	1.8e-04	&	4	\\ 
&	12604	&	 12.48292	&	-27.91953	&	 4.14	&	0.14467	&	9.2e-05	&	3	\\ 
&	12660	&	 12.48456	&	-27.75537	&	 7.79	&	0.65535	&	1.8e-04	&	4	\\ 
&	12914	&	 12.48849	&	-27.88781	&	 3.49	&	0.28257	&	2.2e-04	&	3	\\ 
&	12924	&	 12.48908	&	-27.87178	&	 3.43	&	0.29252	&	1.8e-04	&	3	\\ 
&	12962	&	 12.49099	&	-27.82055	&	 4.76	&	0.56396	&	1.8e-04	&	4	\\ 
&	12985	&	 12.49178	&	-27.86716	&	 3.59	&	0.49384	&	2.3e-04	&	3	\\ 
&	13084	&	 12.49326	&	-27.92155	&	 4.63	&	0.42810	&	1.8e-04	&	3	\\ 
&	13223	&	 12.49485	&	-27.94785	&	 5.80	&	0.25512	&	2.3e-04	&	4	\\ 
&	13188	&	 12.49535	&	-27.71084	&	10.48	&	0.65741	&	2.3e-04	&	4	\\ 
&	13171	&	 12.49637	&	-27.84837	&	 4.11	&	0.49169	&	3.0e-04	&	3	\\ 
&	13329	&	 12.49740	&	-27.78777	&	 6.45	&	0.32700	&	2.3e-04	&	4	\\ 
&	13206	&	 12.49771	&	-27.87480	&	 3.88	&	0.82690	&	1.8e-04	&	4	\\ 
&	13350	&	 12.49912	&	-27.78604	&	 6.59	&	0.32632	&	1.8e-04	&	4	\\ 
&	13591	&	 12.50013	&	-27.82742	&	 4.88	&	0.19155	&	1.9e-04	&	4	\\ 
&	13445	&	 12.50057	&	-27.89994	&	 4.33	&	0.71416	&	3.0e-04	&	3	\\ 
&	13664	&	 12.50341	&	-27.74803	&	 8.63	&	0.43334	&	2.2e-04	&	4	\\ 
&	13546	&	 12.50394	&	-27.82010	&	 5.31	&	0.19109	&	1.9e-04	&	4	\\ 
&	13723	&	 12.50802	&	-27.87382	&	 4.43	&	0.42882	&	1.8e-04	&	4	\\ 
&	13842	&	 12.51050	&	-27.83001	&	 5.26	&	0.19671	&	2.3e-04	&	4	\\ 
&	13877	&	 12.51130	&	-27.78889	&	 6.87	&	0.74625	&	1.8e-04	&	4	\\ 
&	13884	&	 12.51278	&	-27.95857	&	 6.91	&	0.83247	&	1.8e-04	&	4	\\ 
&	13961	&	 12.51284	&	-27.82227	&	 5.61	&	0.36422	&	1.2e-04	&	4	\\ 
&	13980	&	 12.51350	&	-27.86092	&	 4.78	&	0.49068	&	2.3e-04	&	4	\\ 
&	14245	&	 12.51422	&	-27.86873	&	 4.77	&	0.19397	&	1.6e-04	&	4	\\ 
&	14000	&	 12.51481	&	-27.83677	&	 5.28	&	0.18386	&	2.3e-04	&	4	\\ 
&	14057	&	 12.51522	&	-27.88490	&	 4.86	&	0.19645	&	2.2e-04	&	4	\\ 
&	14060	&	 12.51659	&	-27.88813	&	 4.96	&	0.19637	&	2.2e-04	&	3	\\ 
&	14128	&	 12.51796	&	-27.80841	&	 6.32	&	0.65734	&	1.8e-04	&	4	\\ 
&	14186	&	 12.51871	&	-27.74855	&	 9.03	&	0.18280	&	1.8e-04	&	4	\\ 
&	14698	&	 12.52732	&	-27.89445	&	 5.59	&	0.18338	&	1.1e-04	&	4	\\ 
&	14682	&	 12.52808	&	-27.68174	&	12.77	&	0.11830	&	1.2e-04	&	4	\\ 
&	14882	&	 12.52925	&	-27.84369	&	 5.84	&	0.19729	&	1.6e-04	&	4	\\ 
&	14712	&	 12.53031	&	-27.85688	&	 5.70	&	0.19783	&	2.2e-04	&	4	\\ 
&	14746	&	 12.53121	&	-27.65087	&	14.53	&	0.67948	&	2.3e-04	&	4	\\ 
&	14869	&	 12.53175	&	-27.81449	&	 6.71	&	0.54584	&	2.3e-04	&	4	\\ 
&	14801	&	 12.53275	&	-27.78620	&	 7.79	&	0.66229	&	3.0e-04	&	4	\\ 
&	14880	&	 12.53398	&	-27.79244	&	 7.59	&	0.18253	&	1.8e-04	&	4	\\ 
&	15127	&	 12.53856	&	-27.72920	&	10.58	&	0.65074	&	3.0e-04	&	4	\\ 
&	15114	&	 12.53888	&	-27.83538	&	 6.49	&	0.17119	&	1.8e-04	&	4	\\ 
&	15135	&	 12.53888	&	-27.82490	&	 6.74	&	0.28005	&	1.9e-04	&	4	\\ 
&	15112	&	 12.53928	&	-27.83892	&	 6.44	&	0.22095	&	2.3e-04	&	4	\\ 
&	15203	&	 12.53968	&	-27.67132	&	13.60	&	0.47986	&	3.0e-04	&	4	\\ 
&	15345	&	 12.54202	&	-27.67062	&	13.70	&	0.53154	&	2.3e-04	&	4	\\ 
&	15422	&	 12.54418	&	-27.67051	&	13.76	&	0.55304	&	1.8e-04	&	4	\\ 
&	15492	&	 12.54653	&	-27.88252	&	 6.49	&	0.12240	&	1.8e-04	&	4	\\ 
&	15498	&	 12.54821	&	-27.86563	&	 6.58	&	0.92221	&	3.0e-04	&	4	\\ 
&	15585	&	 12.54924	&	-27.85636	&	 6.70	&	0.32387	&	2.3e-04	&	4	\\ 
&	15628	&	 12.55019	&	-27.80403	&	 7.87	&	0.45598	&	2.3e-04	&	4	\\ 
&	15706	&	 12.55233	&	-27.79783	&	 8.17	&	0.48937	&	2.3e-04	&	4	\\ 
&	15771	&	 12.55238	&	-27.93911	&	 7.83	&	0.19685	&	2.2e-04	&	4	\\ 
&	15866	&	 12.55422	&	-27.79519	&	 8.34	&	0.49079	&	3.0e-04	&	4	\\ 
&	15941	&	 12.55676	&	-27.79328	&	 8.52	&	0.49080	&	2.3e-04	&	4	\\ 
&	15909	&	 12.55684	&	-27.75779	&	 9.89	&	0.59267	&	1.8e-04	&	4	\\ 
&	15883	&	 12.55699	&	-27.81785	&	 7.79	&	0.19114	&	1.8e-04	&	4	\\ 
&	15940	&	 12.55809	&	-27.83231	&	 7.51	&	0.32598	&	1.2e-04	&	4	\\ 
&	16118	&	 12.56216	&	-27.74564	&	10.61	&	0.65609	&	1.8e-04	&	4	\\ 
&	16152	&	 12.56221	&	-27.79243	&	 8.79	&	0.32591	&	1.6e-04	&	4	\\ 
&	16321	&	 12.56317	&	-28.03472	&	12.13	&	0.19816	&	2.2e-04	&	4	\\ 
&	16156	&	 12.56358	&	-27.82120	&	 8.03	&	0.54489	&	1.8e-04	&	4	\\ 
&	16269	&	 12.56586	&	-27.86822	&	 7.50	&	0.49142	&	2.3e-04	&	4	\\ 
&	16544	&	 12.57113	&	-27.92638	&	 8.39	&	0.37669	&	1.6e-04	&	4	\\ 
&	16649	&	 12.57464	&	-27.84398	&	 8.16	&	0.49702	&	1.8e-04	&	4	\\ 
&	16787	&	 12.57724	&	-27.69586	&	13.41	&	0.43644	&	1.8e-04	&	4	\\ 
&	16901	&	 12.57929	&	-27.96442	&	 9.84	&	0.48881	&	2.2e-04	&	4	\\ 
&	17009	&	 12.58134	&	-27.91252	&	 8.63	&	0.19347	&	1.9e-04	&	4	\\ 
&	17772	&	 12.58405	&	-27.79331	&	 9.75	&	0.04018	&	2.3e-04	&	4	\\ 
&	17433	&	 12.59081	&	-27.88066	&	 8.83	&	0.42129	&	1.8e-04	&	4	\\ 
&	17481	&	 12.59309	&	-27.87708	&	 8.94	&	0.08346	&	2.3e-04	&	4	\\ 
&	17552	&	 12.59549	&	-27.87350	&	 9.07	&	0.89252	&	2.3e-04	&	4	\\ 
&	17748	&	 12.59845	&	-27.82591	&	 9.66	&	0.36419	&	2.3e-04	&	4	\\ 
&	18069	&	 12.60364	&	-27.93072	&	10.09	&	0.37643	&	2.2e-04	&	4	\\ 
&	18219	&	 12.60789	&	-27.97964	&	11.61	&	0.44029	&	2.2e-04	&	4	\\ 
&	18244	&	 12.60946	&	-27.84274	&	 9.99	&	0.59480	&	2.3e-04	&	4	\\ 
&	18524	&	 12.61687	&	-27.86658	&	10.21	&	0.89686	&	3.0e-04	&	4	\\ 
&	18571	&	 12.61700	&	-27.92806	&	10.71	&	0.66021	&	3.0e-04	&	4	\\ 
&	18837	&	 12.61952	&	-27.98949	&	12.44	&	0.23964	&	1.6e-04	&	4	\\ 
&	18812	&	 12.62260	&	-27.90800	&	10.70	&	0.42821	&	2.2e-04	&	4	\\ 
&	18898	&	 12.62504	&	-27.94313	&	11.41	&	0.41230	&	1.2e-04	&	4	\\ 
&	19073	&	 12.62781	&	-27.72800	&	13.90	&	0.47770	&	3.0e-04	&	4	\\ 
&	19256	&	 12.63024	&	-27.97806	&	12.57	&	0.36128	&	1.2e-04	&	4	\\ 
&	19261	&	 12.63248	&	-27.89197	&	11.08	&	0.42122	&	2.3e-04	&	4	\\ 
&	19350	&	 12.63482	&	-27.99644	&	13.35	&	0.56412	&	3.0e-04	&	4	\\ 
&	19517	&	 12.63881	&	-27.82623	&	11.72	&	0.59130	&	2.3e-04	&	4	\\ 
&	19837	&	 12.64113	&	-27.97492	&	12.98	&	0.23993	&	1.1e-04	&	4	\\ 
&	19741	&	 12.64387	&	-27.85464	&	11.69	&	0.30514	&	1.8e-04	&	4	\\ 
&	19724	&	 12.64393	&	-28.01362	&	14.33	&	0.61807	&	2.3e-04	&	4	\\ 
&	19919	&	 12.64403	&	-27.78508	&	12.81	&	0.11924	&	1.9e-04	&	4	\\ 
&	19713	&	 12.64456	&	-27.95411	&	12.62	&	0.42901	&	1.8e-04	&	4	\\ 
&	19847	&	 12.64698	&	-27.82707	&	12.13	&	0.49058	&	3.0e-04	&	4	\\ 
&	20031	&	 12.64930	&	-27.97100	&	13.26	&	0.24084	&	2.3e-04	&	4	\\ 
&	20059	&	 12.65092	&	-27.79436	&	12.93	&	0.11640	&	2.3e-04	&	4	\\ 
&	20327	&	 12.65118	&	-27.77308	&	13.46	&	0.34575	&	2.2e-04	&	4	\\ 
&	20332	&	 12.65596	&	-27.81540	&	12.77	&	0.05935	&	2.3e-04	&	4	\\ 
&	21093	&	 12.67754	&	-27.92503	&	13.76	&	0.23381	&	1.8e-04	&	4	\\ 
\cutinhead{Stars}
&	10878	&	 12.44110	&	-27.93995	&	 4.06	&	...	&	...	&	4	 \\ 
&	10966	&	 12.44593	&	-27.85008	&	 1.82	&	...	&	...	&	3	 \\ 
&	21098	&	 12.67174	&	-27.76816	&	14.57	&	...	&	...	&	4	 \\ 
\cutinhead{NED}
&	1365	&	 12.11765	&	-27.93831	&	16.72	&	1.68900	&	5.0e-03	&	6	\\ 
&	2160	&	 12.14584	&	-27.79925	&	15.45	&	0.19630	&	2.3e-04	&	6	\\ 
&	2299	&	 12.15291	&	-27.97317	&	15.58	&	0.03100	&	2.3e-04	&	6	\\ 
&	5022	&	 12.24762	&	-28.00831	&	12.36	&	0.16496	&	3.0e-04	&	6	\\ 
&	5258	&	 12.25966	&	-27.57464	&	19.98	&	0.25230	&	5.0e-03	&	6	\\ 
&	5235	&	 12.26003	&	-27.58533	&	19.39	&	0.17166	&	3.0e-04	&	6	\\ 
&	6224	&	 12.29721	&	-27.87505	&	 6.75	&	0.75150	&	5.0e-03	&	6	\\ 
&	7177	&	 12.32466	&	-27.90401	&	 5.60	&	0.04040	&	3.0e-04	&	6	\\ 
&	7542	&	 12.33464	&	-27.79344	&	 6.78	&	0.13812	&	3.0e-04	&	6	\\ 
&	7850	&	 12.35160	&	-27.98389	&	 7.65	&	3.59500	&	2.0e-04	&	6	\\ 
&	8124	&	 12.35585	&	-28.00636	&	 8.74	&	0.17280	&	2.3e-04	&	6	\\ 
&	11761	&	 12.46401	&	-27.98238	&	 6.84	&	1.43300	&	5.0e-03	&	6	\\ 
&	12818	&	 12.48404	&	-27.88624	&	 3.24	&	0.16470	&	3.0e-04	&	6	\\ 
&	13651	&	 12.49168	&	-27.74291	&	 8.63	&	0.19521	&	3.0e-04	&	6	\\ 
&	13783	&	 12.50577	&	-27.73910	&	 9.16	&	0.14341	&	3.0e-04	&	6	\\ 
&	15970	&	 12.55173	&	-28.01646	&	10.89	&	0.19398	&	3.0e-04	&	6	\\ 
&	16036	&	 12.56152	&	-27.94420	&	 8.40	&	1.74600	&	5.0e-03	&	6	\\ 
&	16422	&	 12.56647	&	-27.72104	&	11.87	&	2.13000	&	5.0e-03	&	6	\\ 
&	17788	&	 12.58970	&	-27.62065	&	17.54	&	0.04076	&	3.0e-04	&	6	\\ 
&	18562	&	 12.61284	&	-27.72174	&	13.54	&	0.11816	&	3.0e-04	&	6	\\ 
&	18618	&	 12.61576	&	-27.65853	&	16.43	&	0.07360	&	5.0e-03	&	6	\\ 
&	19451	&	 12.63797	&	-27.89104	&	11.36	&	1.75540	&	5.0e-03	&	6	\\ 
&	19665	&	 12.64270	&	-28.04045	&	15.28	&	0.58670	&	5.0e-03	&	6	\\ 
&	21590	&	 12.68266	&	-27.86440	&	13.70	&	0.11783	&	3.0e-04	&	6	\\ 
&	21638	&	 12.68966	&	-27.66291	&	18.94	&	2.14600	&	1.0e-03	&	6	\\ 
&	21958	&	 12.70075	&	-28.00368	&	16.58	&	1.59300	&	4.8e-03	&	6	\\ 
&	23954	&	 12.71043	&	-27.86877	&	15.17	&	0.03190	&	3.0e-04	&	6	\\ 
Total: & 341 & & & & & & \\ 
\enddata\tablenotetext{a}{IDs are unique within the field of a given lens but not over the entire sample.}\tablenotetext{b}{Spectroscopic flags: Flag=1 for objects with redshifts that are not in our final photometric catalog; Flag=2 for data obtained with LDSS-2; Flag=3 for data obtained with LDSS-3; Flag=4 for data obtained with IMACS; Flag=5 for data obtained with Hectospec; Flag=6 for NED objects.}\end{deluxetable}

\begin{deluxetable}{llll}
\tablecolumns{4}
\tablewidth{0pc}
\tabletypesize{\footnotesize}
\tablecaption{Redshift Errors\label{damn_errors}}
\tablehead{
\colhead{} & \multicolumn{3}{c}{$\sigma_{\Delta z}/\sqrt{2}$}\\
\multicolumn{4}{c}{}\\
\colhead{Sample} & \colhead{S/N$<10$} & \colhead{10$<$S/N$<$20} & \colhead{S/N$>20$}
}
\startdata
Full Sample	&	2.55e-04	&	2.05e-04	&	1.33e-04	\\ 
Non-emission	&	3.00e-04	&	2.20e-04	&	1.57e-04	\\ 
1 line, $>$10A	&	2.34e-04	&	1.94e-04	&	1.08e-04	\\ 
2 lines, $>$10A	&	1.80e-04	&	1.22e-04	&	9.16e-05	\\ 
\enddata
\end{deluxetable}

%%%\begin{deluxetable}{llll}
%\tablecolumns{4}
%\tablewidth{0pc}
%\tabletypesize{\footnotesize}
%\tablecaption{Redshift Errors\label{damn_errors}}
%\tablehead{
%\colhead{} & \multicolumn{3}{c}{$\sigma_{\Delta z}$}\\
%\multicolumn{4}{c}{}\\
%\colhead{Sample} & \colhead{S/N$<10$} & \colhead{10$<$S/N$<$20} & \colhead{S/N$>20$}
%}
%\startdata
%Full Sample     &   2.588E-4 &  2.086E-4 &   1.64E-4 \\
%Non-emission    &   3.06E-4 &   1.87E-4 &   1.64E-4 \\
%1 line, $>$10A    &   2.10E-4 &   2.64E-4 &   1.60E-4 \\
%2 lines, $>$10A   &   1.85E-4 &   1.31E-4 &   1.03E-4 \\
%\enddata
%\end{deluxetable}

\end{document}